\documentclass[11pt]{amsart}
\usepackage{amsmath,amssymb}
\usepackage{graphicx}
\usepackage{placeins}
\usepackage{xcolor}
\graphicspath{{assets/}}
\definecolor{regionYellowGreen}{RGB}{154,205,50}
\definecolor{geomCobalt}{HTML}{0047AB}
\definecolor{detectorPlum}{HTML}{6F2C91}
\definecolor{coneCyan}{HTML}{008C9E}
\definecolor{regionPine}{HTML}{176A3A}
\usepackage{bm}
\newtheorem*{ithm}{Theorem}

\theoremstyle{definition}

\theoremstyle{plain}
\newtheorem{thm}{Theorem}[section]
\newtheorem{lem}[thm]{Lemma}
\newtheorem*{lemnonum}{Lemma}

\newtheorem{prop}[thm]{Proposition}
\theoremstyle{definition}

\theoremstyle{plain}
\newtheorem{cor}[thm]{Corollary}

\theoremstyle{definition}

\newtheorem{notation}[thm]{Notation}
\newtheorem{setting}[thm]{Setting}
\newtheorem{defn}[thm]{Definition}

\theoremstyle{remark}
\newtheorem{rem}[thm]{Remark}
\theoremstyle{definition}

\usepackage{tikz}
\usetikzlibrary{decorations.pathmorphing}
\usepackage[hidelinks]{hyperref}
\numberwithin{equation}{section}
\numberwithin{figure}{section}

\theoremstyle{plain}
\newtheorem{lemma}[thm]{Lemma}
\newtheorem{theorem}[thm]{Theorem}

\newtheorem{proposition}[thm]{Proposition}

\theoremstyle{definition}

\theoremstyle{plain}
\newcommand{\td}{\tilde}
\newcommand{\A}{\mathcal A}
\newcommand{\LL}{\mathbb L}

\newcommand{\caA}{{\mathcal A}}
\newcommand{\caB}{{\mathcal B}}
\newcommand{\caC}{{\mathcal C}}
\newcommand{\caD}{{\mathcal D}}

\newcommand{\caF}{{\mathcal F}}

\newcommand{\caH}{{\mathcal H}}

\newcommand{\caM}{{\mathcal M}}
\newcommand{\cN}{{\mathcal N}}
\newcommand{\caO}{{\mathcal O}}

\newcommand{\caR}{{\mathcal R}}
\newcommand{\caS}{{\mathcal S}}

\newcommand{\bbA}{{\mathbb A}}

\newcommand{\bbC}{{\mathbb C}}

\newcommand{\bbL}{{\mathbb L}}

\newcommand{\bbN}{{\mathbb N}}

\newcommand{\bbR}{{\mathbb R}}

\newcommand{\lV}{\left \Vert}
\newcommand{\rV}{\right \Vert}

\newcommand{\lmk}{\left (}
\newcommand{\rmk}{\right )}

\newcommand{\id}{\mathop{\mathrm{id}}\nolimits}
\newcommand{\Tr}{\mathop{\mathrm{Tr}}\nolimits}

\newcommand{\supp}{\mathop{\mathrm{supp}}\nolimits}
\newcommand{\rank}{\mathop{\mathrm{rank}}\nolimits}

\newcommand{\rr}{\mathbb{R}}

\newcommand{\braket}[2]{\left\langle#1,#2\right\rangle}

\theoremstyle{definition}
\newtheorem{nota}[thm]{Notation}
\theoremstyle{plain}
\def\proofof[#1]{\proof[Proof of #1]}

\newcommand{\bone}{{B_1}}
\newcommand{\btwo}{{B_2}}
\newcommand{\ceb}{{\mathcal C^{\mathrm{EB}}}}

\newcommand{\unit}{\mathbf{1}}

\newcommand{\ld}{\Lambda}

\newtheorem{haaglemma}[thm]{Lemma}
\theoremstyle{definition}
\newtheorem{haagdefinition}[thm]{Definition}

\theoremstyle{plain}

\usepackage{enumerate}
\theoremstyle{definition}
\newtheorem{defi}[thm]{Definition}
\newtheorem{definition}[thm]{Definition}
\theoremstyle{remark}
\newtheorem{remark}[thm]{Remark}
\theoremstyle{plain}

\title{From Entanglement Bootstrap to DHR Theory}

\author{Kohtaro Kato}
\address{Department of Mathematical Informatics,
Graduate School of Informatics,
Nagoya University,
Furo-cho, Chikusa-ku, Nagoya 464-8601, Japan}

\author{Yoshiko Ogata}
\address{Research Institute for Mathematical Sciences,
Kyoto University,
Kyoto 606-8502, Japan}
\providecommand{\tr}{{\rm Tr}}

\begin{document}
\begin{abstract}
We establish a direct connection between the entanglement
bootstrap and the Doplicher--Haag--Roberts theory of
superselection sectors for two-dimensional quantum spin systems.
Under the entanglement-bootstrap assumptions, we prove Haag duality
and establish a one-to-one correspondence between simple
entanglement-bootstrap anyon types and irreducible
superselection sectors.
For each fusion channel, we further construct an affine
bijection from density matrices on the entanglement-bootstrap
fusion space to normalized positive intertwiners in the
corresponding operator-algebraic fusion channel.
These results connect the entropic and operator-algebraic
descriptions of anyons directly, without passing through
exactly solvable lattice models.
\end{abstract}
\maketitle
\section*{Introduction}

Anyons and the chiral central charge are fundamental data
characterizing two-dimensional gapped topological phases,
as advocated in Kitaev's work~\cite{Kitaev2006}.
In exactly solvable models, such as Kitaev's toric code and
quantum double models~\cite{Kitaev2003}, Levin--Wen string-net
models~\cite{LevinWen2005}, and related models based on weak
$C^*$-Hopf algebras~\cite{MolnarEtAl2026,OPR}, anyons can be
described explicitly as localized excitations, and their
fusion and braiding can be computed.
For a general microscopic model, however, identifying the
mathematical objects that describe anyons remains a basic
problem.
The problem is to identify these objects in the microscopic system
and establish their invariance within a phase.

Two approaches to this problem are the entanglement bootstrap
and operator-algebraic superselection theory.
The entanglement bootstrap develops a description of
topological charges from local properties of a reference state
\cite{ShiLu2019,Shi2019,SKK}.
Its starting point consists of entropy equalities on small
contractible regions, which follow from a suitable strict
form of the area law.
From these local conditions, information convex sets encode
charge labels on annuli and fusion spaces on regions with
multiple holes.
In particular, Shi, Kato, and Kim derived the fusion rules
and their consistency relations from these entropic
assumptions~\cite{SKK}.
This approach belongs to the broader program of extracting
topological information from a single ground state, also
pursued in Haah's work~\cite{Haah2016}.

The operator-algebraic approach originates in the
Doplicher--Haag--Roberts (DHR) theory of superselection sectors
in algebraic quantum field theory~\cite{DHR1971}, together
with its extension to charges localized in unbounded
regions~\cite{BF1982}.
The relevance of superselection theory to anyons was already
discussed by Kitaev~\cite{Kitaev2006}.
A rigorous implementation for lattice systems was given
by Naaijkens in his analysis of the toric code~\cite{Naaijkens2011}.
In this framework, charges are described by representations
of the quasi-local observable algebra satisfying a
localization and transportability criterion.
Haag duality, or an appropriate approximate version of it~\cite{Ogata},
allows these representations to be organized into a braided
$C^*$-tensor category.
An important feature of this approach is its stability:
under suitable locality and gap assumptions, the category
is preserved, up to braided equivalence, by the quasi-local
automorphisms relating ground states within a gapped
phase~\cite[Theorem~6.1]{Ogata}.
This provides a precise sense in which anyon data are
invariants of the phase.

Information convex sets and superselection sectors share
the same underlying philosophy: objects that look the same
on suitable parts of a system may nevertheless be globally
distinct.
States in an information convex set agree with the reference state on prescribed local
patches, while different charge labels can be distinguished
on an annulus.
In DHR theory, representations are unitarily
equivalent to the reference representation outside any
chosen localization cone but may
be inequivalent on the full observable algebra.
These are two mathematical implementations of a common
idea: global charge can distinguish objects whose
restrictions satisfy the same local conditions.
This suggests a direct relationship between the charge
structures obtained by the two approaches.

Recent results have clarified several aspects of this
relationship.
Kim and Ranard showed that states satisfying the
entanglement bootstrap axioms with suitable gappable
boundary conditions can be mapped to Levin--Wen string-net
states by local circuits of constant depth~\cite{KimRanard2024}.
Ogata, P\'erez-Garc\'ia, and Ruiz-de-Alarc\'on established
Haag duality for a class of models based on weak
$C^*$-Hopf algebras that includes Kitaev quantum double
and Levin--Wen string-net models~\cite{OPR}.
The DHR-theory of Levin--Wen string-net models was analyzed in \cite{BK1} and \cite{BK2}.
More recently, Liu and van Luijk obtained an entropic
characterization of Haag duality in terms of the asymptotic
vanishing of conditional mutual information.
Their result yields cone Haag duality under suitable
area-law assumptions, without passing through a particular
exactly solvable model~\cite{LiuVanLuijk2026}.

The purpose of the present paper is to connect the
entanglement bootstrap directly with DHR theory, without
using Levin--Wen models as an intermediary.
Working in the infinite-volume setting, we construct
superselection sectors from entanglement-bootstrap charge
labels and relate merging of states to the categorical
tensor product.
We establish the existence of conjugates for the resulting
sectors and compare the entanglement-bootstrap fusion
spaces with the corresponding operator-algebraic
multiplicity spaces.
The central issue is thus to make the common philosophy
of local agreement and global distinction into a
mathematical correspondence compatible with fusion
and conjugation.

Our main results can be summarized informally as follows.

\begin{ithm}
Under the entanglement-bootstrap assumptions, the simple anyon
types of the entanglement bootstrap are in one-to-one
correspondence with the simple DHR sectors.

For each choice of two incoming anyon types and one
outgoing anyon type, we construct an affine bijection
from the density matrices on the entanglement-bootstrap
fusion space to normalized positive operators on the
corresponding DHR fusion space.
\end{ithm}

The correspondence between simple anyon types is
Theorem \ref{anyontheorem}.
The precise formulation of the fusion correspondence
is given in Theorem~\ref{thm:fusion-endomorphisms}.

We also derive Haag duality from the entanglement-bootstrap assumptions. Independently and concurrently with the present work,
Liu and van Luijk~\cite{LiuVanLuijk2026} obtained a stronger
result on Haag duality.
Their proof is substantially different from ours.

The paper is organized as follows.
Section~{{\ref{sec:entanglement-bootstrap}}} introduces
the entanglement-bootstrap setting,
information convex sets, and the commuting parent
Hamiltonian, together with the consistency and local
uniqueness results used later.
Section~{{\ref{sec:haag-duality}}} proves Haag duality
for lattice cones satisfying the stated geometric conditions.
Section~{{\ref{sec:superselection-framework}}} recalls
the operator-algebraic superselection framework
for two-dimensional quantum spin systems.
Section~{{\ref{sec:infinite-volume}}} constructs
infinite-volume charged states, proves that their
representations satisfy the superselection criterion, and establishes the correspondence in Theorem \ref{anyontheorem}.
Section~{{\ref{sec:merging}}} develops merging
for configurations with two and three holes
and relates it to the DHR tensor product.
Section~{{\ref{sec:cone-approximation}}} proves
the existence of DHR conjugates.
Section~{{\ref{sec:cyclicity}}} establishes cyclicity
of the reference vector for tensor products
of suitably separated charged representations,
a property used in the proof of Theorem~\ref{thm:fusion-endomorphisms}.
Finally, Section~{{\ref{sec:fusion}}} compares
the fusion spaces obtained from information convex sets
with the corresponding endomorphism algebras
in superselection theory, showing Theorem~\ref{thm:fusion-endomorphisms}.

While completing this manuscript, we became aware of independent
and concurrent work on closely related topics~\cite{BKprep}.

\section{Entanglement bootstrap}\label{sec:entanglement-bootstrap}

\begingroup
\newcommand{\ot}{\otimes}
\providecommand{\Qsi}{\Psi} %
\def\supp{{\rm supp}}
\def\mL{{\mathbb L}}
\def\cA{{\mathcal A}}
\def\cB{{\mathcal B}}
\def\cC{{\mathcal C}}
\def\cD{{\mathcal D}}
\def\cH{{\mathcal H}}
\def\cN{{\mathcal N}}
\def\cL{{\mathcal L}}
\def\cS{{\mathcal S}}
\def\cV{{\mathcal V}}
\def\tr{{\rm Tr}}
\def\rank{{\rm rank}}
\def\dim{{\rm dim}}

\subsection{Setting and notation}\label{overallsetting}
We introduce the entanglement-bootstrap setting and notation used throughout the paper.

\subsubsection{Lattice regions}
Let $\bbL$ be the set of closed hexagonal faces in a fixed hexagonal tiling of $\mathbb R^2$, with the origin at a face center. 
Let $d_{\mathrm{hex}}$ denote the Euclidean diameter of a face
of the fixed regular hexagonal tiling of $\mathbb R^2$.
Let $f_{{\mathrm{org}}}$ be the face centered at the origin. 

Equip $\bbL$ with the graph distance $d$, where two faces are adjacent if they share an edge. A \emph{lattice region} is a subset of $\bbL$. We write $AB:=A\cup B$ for disjoint lattice regions $A$ and $B$, $X\Subset\bbL$ when $X$ is finite, and $X^c:=\bbL\setminus X$.

For non-empty face lattice regions $X,Y$ and $f\in\bbL$, write
\[
 d(X,Y)=\inf\{d(f,g):f\in X,\ g\in Y\},
 \qquad
 d(f,X)=\inf_{g\in X}d(f,g).
 \]
 For $r\in{\mathbb Z}_{\geq0}$, define
\begin{equation}\label{def:lattice regionplusr}
\begin{split}
&X^{+r}:=\{f\in\bbL:d(f,X)\le r\},
\qquad
\cN_r(X):=X^{+r}\setminus X,\\
&  X^{-r}=\{f\in X:d(f,X^c)>r\}.
\end{split}
\end{equation}
We occasionally abbreviate $X^{+1}$ and $\cN_1(X)$ as $X^+$ and $\cN(X)$, respectively. For a face $f$, we write $\cN(f):=\cN(\{f\})$ and call $D_f:=\{f\}\cup \cN(f)$ the \emph{elementary disk} centered at $f$.

The geometric realization $|X|\subset\mathbb{R}^2$ of a lattice region $X\subset \mL$ is the union of its
closed faces. A lattice region $X\subset \mL$ is \emph{connected} if $|X|$ is connected.  Two lattice regions have the same topology if their geometric realizations are homeomorphic. We use the following basic types of lattice regions. 
\begin{itemize}
    \item A \emph{disk} is a lattice region whose geometric realization is homeomorphic to the closed unit disk.

    \item  A \emph{$k$-hole disk} ($k\geq1$) is a lattice region of the form $D\setminus(D_1\sqcup\cdots\sqcup D_k)$ where $D, D_1,\ldots,D_k$ are disks with $D_i\subset D$ and $d(D^c, D_i) > 1$ for all $i$, and $d(D_i, D_j) > 1$ for $i\neq j$. The disks $D_1,\ldots,D_k$ are called its holes. A 1-hole disk is called an \emph{annulus}.
    \item Write $K_f(s)=\{f\}^{+s}$ and
$A_f(s,t)=K_f(t)\setminus K_f(s)$ for a face $f$ and integers $s<t$.
The set $A_f(s,t)$ is called a \emph{graph annulus} if $19\le s$ and $t-s\ge25$.

\item 
Start with a region
\[
K_o(t)\setminus\bigcup_i K_{f_i}(s_i),
\]
where $K_f(r)$ denotes the graph disk of radius $r$
centered at the face $f$.
There are a positive finite number of holes, each contained
in the outer disk and having integer radius at least $19$.
Distinct holes, and each hole and the outer complement,
have graph distance at least $25$.

A \emph{shell domain} is a region as above, possibly modified
by expanding the outer disk or shrinking one hole by one layer.
This modification adds faces one at a time, starting at a fixed
corner and following a fixed cyclic order, and may stop
partway through the layer. The radius and separation bounds apply to the initial region.
\end{itemize}
\begin{remark}
   The numbers 19 and 25 appearing in the definitions are chosen to satisfy the geometric conditions that will be frequently used later. 
\end{remark}

For nonempty finite $X$, its thickness $\operatorname{th}(X)$ is the largest integer
$r\geq 0$ such that the inclusion
\begin{equation}
    |X^{-j}|\hookrightarrow |X|
\end{equation}
is a homotopy equivalence for every $j=0,\ldots,r$.

We call a finite lattice region $X\Subset \mL$ \emph{regular} if the inclusion
\[
 |X|\hookrightarrow|X^+|
\]
is a homotopy equivalence. In particular, one-step thickening neither merges connected components nor creates or fills holes.
We call an inclusion $X\subset Y$ of regular annuli
\emph{essential} if the inclusion of their geometric realizations
$|X|\hookrightarrow|Y|$ is a homotopy equivalence, i.e, $X$ surrounds the hole of $Y$.

For $N\in\mathbb N$, set
\[
 \Gamma_N:= \{f_{{\mathrm{org}}}\}^{+N}.
\]
These concentric graph-distance disks exhaust $\mL$ and satisfy
$\Gamma_N^{+j}=\Gamma_{N+j}$ for every integer $j\ge0$.

For each subset $R\subset \bbL$,
we set
\[
R_N:=R\cap \Gamma_N.
\]

For a set $S\subset\mathbb R^2$, define the face lattice regions
\begin{align}
\begin{split}
& \mathfrak F(S):=
 \{f\in\mathbb L:|f|\cap S\ne\varnothing\},\\
&\mathfrak G(S):=
 \{f\in\mathbb L:|f|\subset S\ne\varnothing\}.
\end{split}
\end{align}
${\mathfrak G}(S)\subset{\mathfrak F}(S)$ by definition.

\subsubsection{Local and quasi-local algebras}

Each face $f\in\mL$ carries a Hilbert space
$\cH_f\cong\mathbb{C}^{q_f}$, where
\[
 2\leq q_f\leq q_{\max}<\infty\,,
\]
with some finite $q_{\max}$.
For $X\Subset\mL$, set
\[
 \cH_X:=\bigotimes_{f\in X}\cH_f \quad\rm{and}\quad \cA_X:=\cB(\cH_X).
\]
If $X\subset Y\Subset\mL$, we identify $\cA_X$ {with} the unital subalgebra
\[
 \cA_X\otimes I_{Y\setminus X}\subset\cA_Y.
\]
For an arbitrary region $R\subset\mL$, we consider
the inductive limit with respect to these inclusion maps:
\begin{equation}
 \cA_R:=
 \overline{\bigcup_{X\Subset R}\cA_X}^{\|\cdot\|}.   
\end{equation}
The quasi-local observable algebra is 
\begin{equation}
 \cA:= \cA_\mL.
\end{equation}

Let $\mathcal D_X$ denote the set of
density matrices on $\mathcal H_X$ for $X\Subset\bbL$.
If $\rho$ is a state on $\mathcal A_R$ and $X\Subset R$, we write
$\rho_X\in\mathcal D_X$ for the density matrix representing
$\rho|_{\mathcal A_X}$.

For $X,Y\Subset\bbL$, let
$\mathcal R:\mathcal A_X\to\mathcal A_Y$ be a completely positive
trace-preserving (CPTP) map.
We denote its trace adjoint by
$\mathcal R^T:\mathcal A_Y\to\mathcal A_X$.
We write $\mathrm{id}_X$ for the identity map on $\mathcal A_X$
and omit identity maps on remaining tensor factors
when their domains are clear.

\subsubsection{Entropic quantities}
For a density matrix $\rho$ on a finite-dimensional Hilbert space,
we denote by 
$S(\rho):=-\tr(\rho\log\rho)$ its
von Neumann entropy.
For a finite lattice region $X$ and a state $\rho$ on $\caA_R$ with
 $X\subset R\subset \bbL$, set
$S(X)_\rho:=S(\rho_X)$. 
For pairwise disjoint finite lattice regions, define the conditional entropy,
mutual information, and conditional mutual information by
\begin{align}
  S(X|Y)_\rho
    &:= S(XY)_\rho-S(Y)_\rho,\\
  I(X:Y)_\rho
    &:= S(X)_\rho+S(Y)_\rho-S(XY)_\rho,\\
  I(X:Z|Y)_\rho
    &:= S(XY)_\rho+S(YZ)_\rho-S(Y)_\rho-S(XYZ)_\rho,
\end{align}
respectively.

For a state $\sigma$ on a finite system $ABD$,
we introduce
\begin{align}
{\bbA}_1^\sigma(B, A, D):=S(A|B)_\sigma+S(A|D)_\sigma.
\end{align}

\subsubsection{Quantum Markov states}
A tripartite finite-dimensional quantum state $\rho_{ABC}$ is called a quantum Markov state if 
\begin{equation}
    I(A:C|B)_\rho=0\,.
\end{equation}
For any such state, \cite[Theorem 6]{HJPW} provides a decomposition $\supp(\rho_B)=\bigoplus_i\cH_{B_i^L}\ot\cH_{B_i^R}$ such that
\begin{equation}\label{eq:Markov decomposition}
    \rho_{ABC}=\bigoplus_ip_i\rho_{AB_i^L}\ot\rho_{B_i^RC}\,,
\end{equation}
where $\{p_i\}$ is a probability distribution. 

For each finite-dimensional Hilbert space $K$ and its subspace
$V$, let $\operatorname{proj}_V$ denote the orthogonal projection onto $V$.
The following factorization follows from \eqref{eq:Markov decomposition}.
\begin{lem}[Factorization of support projections]\label{lem:supportprojection}
 For a quantum Markov state $\rho_{ABC}$, set $p_{X}:=\operatorname{proj}_{\supp(\rho_{X})}$. Then
    \begin{align}
p_{AB}p_{BC}=p_{BC}p_{AB}=p_{ABC}\,.
    \end{align}
\end{lem}
\begin{lem}[Modular cocycle identity]\cite{Petz2008}
\label{lem:petz-markov-modular}
Let $\rho_{ABC}$ be a quantum Markov state. Then, for every $t\in\rr$,
\begin{equation}
\rho_{ABC}^{it}\rho_{BC}^{-it}
=
\rho_{AB}^{it}\rho_{B}^{-it}
\end{equation}
on $\mathcal H_A\otimes\operatorname{supp}(\rho_{BC})$. All powers are defined on their own supports and
extended by zero.
\end{lem}
This is the finite-dimensional Markov-state specialization of
Petz's characterization of sufficiency by the modular cocycle. 
The nonfaithful case follows blockwise from Eq.~\eqref{eq:Markov decomposition}.

Any quantum Markov state is locally recoverable by the {\it Petz recovery map} $R_{B\to BC}^{\rho_{BC}}$~\cite[Sec. 9.6]{Petz2008}, given by
\begin{equation}\label{eq:Petzrecoverymap}
    R_{B\to BC}^{\rho_{BC}}(X_B)=\rho_{BC}^{\frac{1}{2}}\rho_B^{-\frac{1}{2}}X_B\rho_B^{-\frac{1}{2}}\rho_{BC}^{\frac{1}{2}}
\end{equation}
for $X_B\in \cB(\supp(\rho_B))$, with the inverse taken on $\supp(\rho_B)$. Throughout, we fix a CPTP extension to the full input algebra for each Petz map. Then
for a quantum Markov state $\rho_{ABC}$,
\begin{equation}
    \rho_{ABC}=R_{B\to BC}^{\rho_{BC}}(\rho_{AB})
    =R_{B\to AB}^{\rho_{AB}}(\rho_{BC})\,.
\end{equation}

The following lemma allows two overlapping quantum Markov
states with a common marginal to be merged~\cite{Kato2016PRA,SKK}.
\begin{lem}[Merging lemma]\label{lem:merginglemma}
Let $A,B,C,D$ be pairwise disjoint finite regions, and let
$\rho_{ABC}\in\mathcal D_{ABC}$ and
$\eta_{BCD}\in\mathcal D_{BCD}$ satisfy
\begin{equation}
  \rho_{BC}=\eta_{BC},
  \qquad
  I(A:C|B)_\rho=I(B:D|C)_\eta=0.
\end{equation}
Then there exists a state $\rho\bowtie\eta\in\mathcal D_{ABCD}$
with marginals $\rho_{ABC}$ and $\eta_{BCD}$.
It is given by the Petz maps:
\begin{equation}
  \rho\bowtie\eta
  =R_{B\to AB}^{\rho_{AB}}(\eta_{BCD})
  =R_{C\to CD}^{\eta_{CD}}(\rho_{ABC}),
\end{equation}
and satisfies
\begin{equation}\label{eq:mergedstatepro}
  I(A:CD|B)_{\rho\bowtie\eta}
  =I(AB:D|C)_{\rho\bowtie\eta}=0.
\end{equation}
These marginals and Markov conditions uniquely determine
$\rho\bowtie\eta$.

For fixed $\eta_{BCD}$, any two states $\rho_{ABC}$ and
$\rho'_{ABC}$ satisfying the above assumptions obey
\begin{equation} \label{eq:merging-entropy-difference}
  S(\rho\bowtie\eta)-S(\rho'\bowtie\eta)
  =S(\rho)-S(\rho').
\end{equation}
\end{lem}

\subsubsection{Axioms of entanglement bootstrap}
For a pure state $\omega$ on $\cA$, we consider 
 the following local entanglement-bootstrap axioms (see~\cite{SKK}):
\begin{itemize}
    \item Axiom A0: for every face $A\in\mL$, with $B=\cN(A)$,  
    \begin{equation}\label{eq:A0}
        S(A)_\omega+S(A|B)_\omega=0\,.
    \end{equation}
    \item Axiom A1: for every face $A\in\mL$ and every partition $\cN(A)=B\sqcup C$ into two nonempty connected regions,
        \begin{equation}\label{eq:A1}
        S(A|B)_\omega+S(A|C)_\omega=0\,.
    \end{equation}
 \begin{figure}[ht]
 \centering
 \begin{minipage}[t]{0.2\linewidth}
  \centering
  \includegraphics[width=\linewidth]{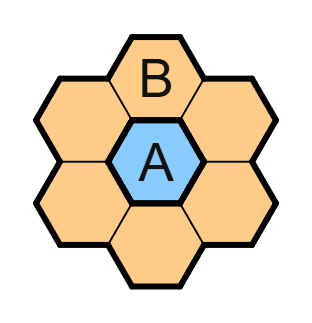}\\[-1mm]
  \textbf{(a)}
 \end{minipage}
 \hspace{0.02\linewidth}
 \begin{minipage}[t]{0.2\linewidth}
  \centering
  \includegraphics[width=\linewidth]{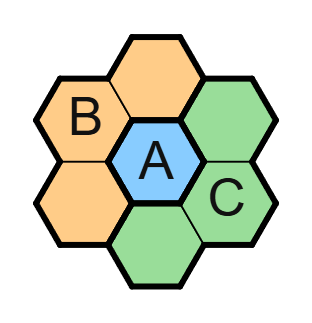}\\[-1mm]
  \textbf{(b)}
 \end{minipage}
 \caption{Local configurations for Axioms A0 and A1, with $A$ a single face. (a) A0: $B = \cN(A)$. (b) A1: a partition $N(A)=B\sqcup C$ into two nonempty connected regions.}
 \label{fig:two-panels}
\end{figure}
\end{itemize}

\begin{nota}
Throughout this paper, we fix a pure reference state $\omega$ on $\cA$ satisfying
A0 and A1.
Let $(\mathcal H,\pi,\Omega)$ be the GNS triple of $\omega$,
where $\pi:\mathcal A\to\mathcal B(\mathcal H)$ and
$\Omega\in\mathcal H$ is a cyclic unit vector.

\end{nota}

Axiom A0 implies 
\begin{equation}
    I(A:E)_\sigma=0
\end{equation}
for every finite-dimensional extension $\sigma_{ABE}$ of $\omega_{AB}$. Similarly, Axiom A1 implies
\begin{equation}
    I(A:E|B)_\sigma=I(A:E|C)_\sigma=0
\end{equation}
for every finite-dimensional extension $\sigma_{ABCE}$ of $\omega_{ABC}$.

These axioms can be extended to larger regions. An {\it A0-type configuration} consists of disjoint finite regions $A$ and $B$ such that $A$ and $AB$ are disks and $B$ is an annulus. An {\it A1-type configuration} consists of pairwise disjoint finite regions $A,B,C$ such that $A,B,C$ and $ABC$ are non-empty disks and $BC$ is an annulus. 
\begin{lem}[Extension of A0 and A1]~\cite[Proposition 3.3]{SKK}\label{lem:extended-a0-a1}
Eq.~\eqref{eq:A0} holds for every A0-type configuration and Eq.~\eqref{eq:A1} holds for every A1-type configuration.
\end{lem}
This lemma has the following consequences for the reference state.
\begin{itemize}
    \item
    Let $(A,C)$ be an A0-type configuration. For every finite region $B\subset(AC)^c$,
    \begin{equation}\label{eq:zeromutualinfo}
        I(A:B)_\omega=0\quad \Leftrightarrow\quad \omega_{AB}=\omega_A\ot\omega_B.
    \end{equation}
    This follows from extended A0 and the data processing inequality.
\item
    Let $A,B,C$ be pairwise disjoint finite regions with $d(A,C)>1$. If $X$, $B$, and $XB$ are disks for some $X\in\{A,C\}$, then
\begin{equation}
  I(A:C|B)_\omega=0.
\end{equation}
    This follows from the extended Axiom A1, applied with the disk $X$ in the hypothesis, and the chain rule and nonnegativity of conditional mutual information.
\end{itemize}

\subsection{Information convex set}

\subsubsection{Definition and elementary properties}

\begin{definition}[Information convex set]
Let $\Omega\Subset\mL$ be a finite regular region. Define the locally consistent state space on $\Omega^+$ by
\begin{equation}
 \widetilde\Sigma_\omega(\Omega)
 :=\left\{
\rho_{\Omega^+}\in\cD_{\Omega^+}:
\rho_{D_f}=\omega_{D_f}
 \text{ for every }f\in \Omega
 \right\}.
 \label{eq:locally-consistent-space}
\end{equation}
The information convex set of $\Omega$ relative to a reference state $\omega$ is
\begin{align}
 \Sigma_\omega(\Omega)
 &:=\left\{
 \rho_\Omega:
 \rho_{\Omega^+}
 \in\widetilde\Sigma_\omega(\Omega)
 \right\}\nonumber\\
 &=\tr_{\cN(\Omega)}
 \bigl(\widetilde\Sigma_\omega(\Omega)\bigr)
 \subset\cD_\Omega.
 \label{eq:information-convex}
\end{align}
We occasionally write $\widetilde\Sigma(\Omega)$ and
$\Sigma(\Omega)$.
\end{definition}

\begin{figure}[ht]
 \centering
 \IfFileExists{hexannulus.png}
  {\includegraphics[width=0.25\linewidth]{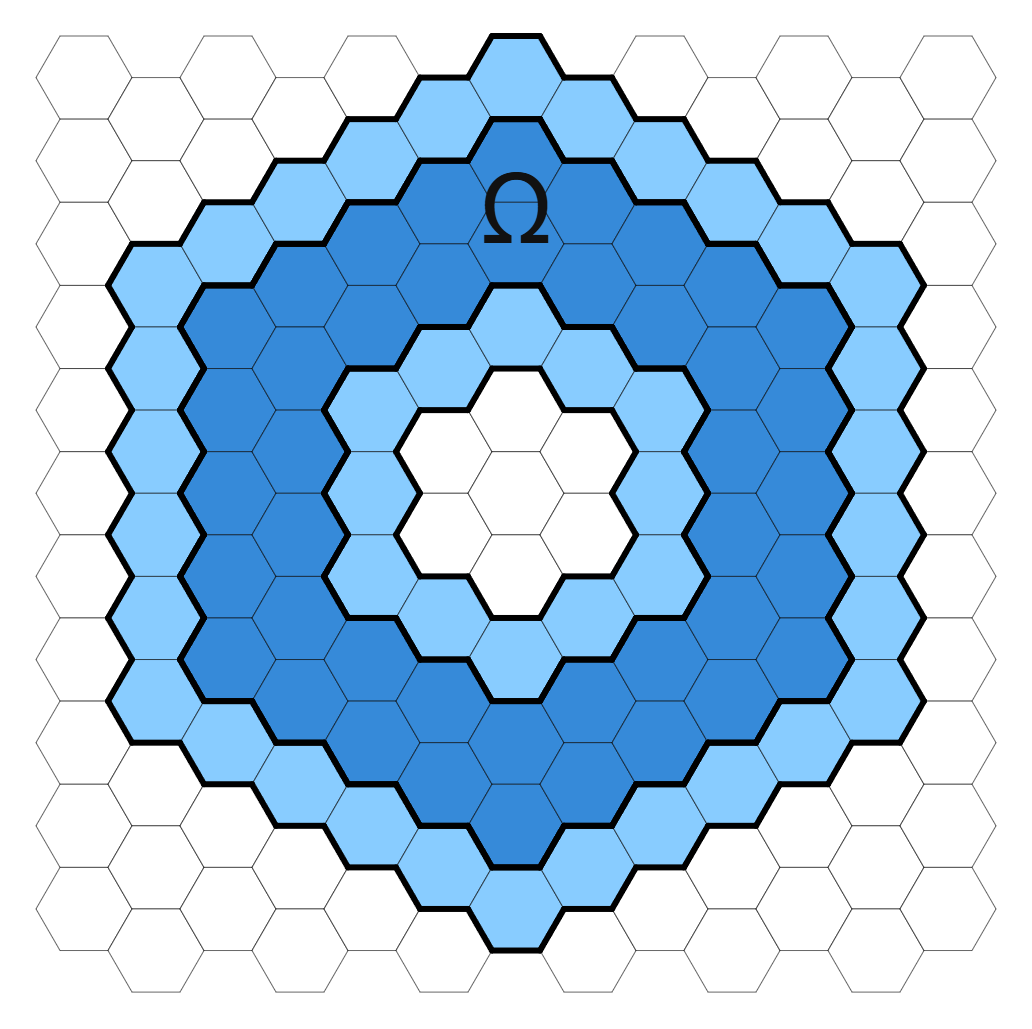}}
  {\fbox{\parbox[c][2.5cm][c]{0.32\linewidth}{\centering
     \texttt{hexannulus.png}}}}
 \caption{A region $\Omega$ and its one-step thickening $\Omega^+$. Dark blue indicates $\Omega$, and light blue indicates $\cN(\Omega)=\Omega^+\setminus\Omega$.}
 \label{fig:region-thickening}
\end{figure}

\begin{remark}
The thickening $\Omega^+$ allows the constraints in Eq.~\eqref{eq:locally-consistent-space} to be imposed on the full elementary disks $D_f$ even when $f$ lies at the boundary of $\Omega$.
\end{remark}

\begin{proposition}[Elementary properties]
Let $\Omega\Subset \bbL$ be regular.
\begin{enumerate}
\item[(i)]
$\Sigma(\Omega)$ is compact and convex and contains $\omega_\Omega$.

\item[(ii)]
Every $\rho_\Omega\in\Sigma(\Omega)$ satisfies
\begin{equation}
  \rho_{D_f\cap\Omega}=\omega_{D_f\cap\Omega}
  \qquad (f\in\Omega).
\end{equation}

\item[(iii)]
If $X\subset Y$ are regular finite regions, the partial trace
induces a map
\begin{equation}  \label{eq:restriction-map}
  \tr_{Y\to X}:\Sigma(Y)\longrightarrow\Sigma(X),
  \qquad
  \tr_{Y\to X}(\rho_Y):=\rho_X.
\end{equation}
\end{enumerate}
\end{proposition}

\begin{proof}
Parts (i) and (ii) follow from the defining marginal constraints
and the continuity and affinity of the partial trace.
For (iii), a state on $Y^+$  satisfying the defining marginal constraints
restricts to a state on $X^+$ satisfying the corresponding constraints,
since $X^+\subset Y^+$.
\end{proof}

\begin{defn}\label{face-order}
Let $B,C$ be disjoint finite lattice regions. We say that $(B,C)$ satisfies
the \emph{face-addition condition} if there is an enumeration
$C=\{f_1,\ldots,f_M\}$ such that, setting
\[
 R_i:=B\cup\{f_1,\ldots,f_i\},\qquad
 D_i:=\mathcal N(f_{i+1})\cap R_i,\qquad
 F_i:=\mathcal N(f_{i+1})\setminus D_i,
\]
for $0\le i<M$, both $D_i$ and $F_i$ are nonempty consecutive arcs
of $\mathcal N(f_{i+1})$.
\end{defn}

\begin{lem}\label{lem:merging-general}
 Let $A,B,C$ be pairwise disjoint
finite lattice regions with $d(A,C)\ge2$. Suppose that $(B,C)$ satisfies the face-addition condition
of Definition~\ref{face-order}. If a state $\varphi$ on $(ABC)^+$ satisfies
\[
 \varphi_{D_f}=\omega_{D_f}\qquad(f\in C),
\]
then $I(A:C| B)_\varphi=0$. In particular, if $ABC$ is regular,
this conclusion holds for every $\tau\in\Sigma(ABC)$.
\end{lem}

\begin{proof}
    The proof is {the} same as in \cite{SKK}.
Choose an ordering $C=\{f_1,\ldots,f_M\}$ satisfying the
face-addition condition, where $M$ is the number of faces in $C$.
Set
\[
 C_0:=\varnothing,\qquad
 C_i:=\{f_1,\ldots,f_i\}\quad(1\le i\le M),
 \qquad \delta C_{i+1}:=\{f_{i+1}\}\quad(0\le i<M).
\]
For each $0\le i<M$, let
\[
 D_i:=\mathcal N(f_{i+1})\cap(B\cup C_i),
 \qquad
 X_i:=\mathcal N(f_{i+1})\setminus D_i.
\]
By the face-addition condition, $D_i$ and $X_i$
are nonempty consecutive arcs of $\mathcal N(f_{i+1})$.
Since $d(A,C)\ge2$, no face of $A$ is adjacent to $f_{i+1}$.
Hence
\[
 D_i\subset B\cup C_i,
 \qquad X_i\cap(A\cup B\cup C_i)=\varnothing.
\]
All these regions are contained in $(ABC)^+$.

The elementary disk centered at $f_{i+1}$ is
\[
 D_{f_{i+1}}=X_i\cup\delta C_{i+1}\cup D_i.
\]
By local consistency, $\varphi_{D_{f_{i+1}}}=\omega_{D_{f_{i+1}}}$.
Therefore, by A1 for the reference state $\omega$,
\begin{align}
 \bbA_1^\varphi(X_i,\delta C_{i+1},D_i)
 =\bbA_1^\omega(X_i,\delta C_{i+1},D_i)=0.
\end{align}

Since $C_0=\varnothing$ and $C_M=C$, a telescoping sum gives
\begin{align}
 I(A:C| B)_\varphi
 &=S(\varphi_{AB})-S(\varphi_B)
   -S(\varphi_{ABC})+S(\varphi_{BC}) \\
 &=\sum_{i=0}^{M-1}
   \bigl[
     S(\varphi_{ABC_i})-S(\varphi_{BC_i})
     -S(\varphi_{ABC_{i+1}})+S(\varphi_{BC_{i+1}})
   \bigr] \\
 &=\sum_{i=0}^{M-1}
   I(A:\delta C_{i+1}| BC_i)_\varphi.
\end{align}
By the definition of $\bbA_1^\varphi$,
\begin{align}
 I(A:\delta C_{i+1}| BC_i)_\varphi
 &=
 \bbA_1^\varphi(X_i,\delta C_{i+1},BC_i)
 -
 \bbA_1^\varphi(X_i,\delta C_{i+1},ABC_i).
\end{align}
By weak monotonicity of the von Neumann entropy, both terms on the right are nonnegative. By strong subadditivity and $D_i\subset BC_i\subset ABC_i$, 
\begin{align}
 0
 &\le \bbA_1^\varphi(X_i,\delta C_{i+1},BC_i)
 \le \bbA_1^\varphi(X_i,\delta C_{i+1},D_i)
 =0,\\
 0
 &\le \bbA_1^\varphi(X_i,\delta C_{i+1},ABC_i)
 \le \bbA_1^\varphi(X_i,\delta C_{i+1},D_i)
 =0.
\end{align}
Consequently,
\[
 I(A:\delta C_{i+1}| BC_i)_\varphi=0
 \qquad(0\le i<M).
\]
Summing these equalities yields
\[
 I(A:C| B)_\varphi=0.
\]

Finally, if $ABC$ is regular and $\tau\in\Sigma(ABC)$,
the definition of the information convex set provides a state
$\varphi$ on $\mathcal A_{(ABC)^+}$ with marginal
$\varphi_{ABC}=\tau$ and
$\varphi_{D_f}=\omega_{D_f}$ for every $f\in ABC$.
Applying the preceding argument to this state gives
$I(A:C| B)_\tau=0$.
\end{proof}
We will repeatedly combine this lemma with Lemma~\ref{lem:merginglemma}.

Note that the face-addition condition is satisfied for the adjacent strip-shaped regions
and the successive annular collars around a common disk used below,
provided that the strips and collars are sufficiently wide. In the annular
case, faces may be added either outwards or inwards. It is also satisfied
for the corresponding local configurations in the two-hole and three-hole
merging constructions, provided that the holes are sufficiently far apart and
sufficiently far from the outer boundary, and that the intervening strips and
collars are sufficiently wide.

\begin{lemma}[Local uniqueness on a finite disk]\label{lem:finite-disk-local-uniqueness}
Let $K\subset\mathbb L$ be a finite disk. If a state $\rho$ on
$\mathcal A_{K^+}$ satisfies
\[
 \rho_{D_f}=\omega_{D_f}\qquad(f\in K),
\]
then
\[
 \rho_K=\omega_K.
\]
Here subscripts denote the density matrices of restrictions to the
indicated finite regions.
\end{lemma}\begin{proof}
Along the ordering of Lemma~\ref{lem:disk-face-ordering}, Lemma~\ref{lem:merging-general} and the Petz maps fixed by the elementary-disk marginals determine each extension uniquely, proving the claim by induction.
\end{proof}

In particular, for disks, the information convex {set contains only} the reference state:
\begin{lem}\cite[Proposition 3.5]{SKK}\label{lem:localuniqueness}
     For any regular disk $D$, $\Sigma(D)=\{\omega_D\}$.
\end{lem}

\subsubsection{Deformation invariance}

\begin{defi}[Elementary enlargement]
Let $X\subsetneq X'$ be finite regular regions. We say that $X'$ is an
\emph{elementary enlargement} of $X$ if there exist pairwise disjoint
nonempty regions $A,B,C,D$ such that
\begin{equation}\label{eq:elementaryenlargement}
 X=A\sqcup B\sqcup C,
 \qquad
 X'=A\sqcup B\sqcup C\sqcup D,
\end{equation}
and the following conditions hold:
\begin{enumerate}[(i)]
 \item $X$ and $X'$ have the same topology;
 \item $B,BC$ and $C_1:=C\cap\cN(D)$ are disks;
 \item $C_1D$ is contained in an elementary disk $D_f$;
 \item $d(A, B)=1$ and $d(A,C)>1$;
 \item $d(AB, D)>3$.
\end{enumerate}
\end{defi}
In this case, $X$ is called an \emph{elementary reduction} of $X'$. An elementary
deformation is an elementary enlargement or reduction. 
A \emph{deformation path} from $X$ to $Y$ is a finite sequence
\begin{equation}
 \gamma=(X=X_0,X_1,\ldots,X_n=Y)
 \label{eq:deformation-path}
\end{equation}
such that each consecutive pair is related by an elementary deformation. 

We frequently consider the following type of
deformation path.
Starting at a fixed corner, we add the faces of a boundary
layer one at a time in a fixed cyclic order.
Each such addition is called a shell addition, and its reverse
is called a shell reduction.
Only one boundary may have a partially added layer at a time.
We consider finite sequences of these operations for which
every intermediate region remains a shell domain.
\begin{thm}[Isomorphism along shell paths]
\label{thm:information-convex-isomorphism}

Let
\begin{equation}
\gamma:\quad X=X_0\leftrightarrow X_1\leftrightarrow
\cdots\leftrightarrow X_n=Y
\end{equation}
be a finite path of shell domains, each step being a shell addition or its reverse.
Partial traces and the Petz maps along
$\gamma$ induce a bijection
\begin{equation}
\Phi_\gamma:\Sigma(X)\xrightarrow{\sim}\Sigma(Y),
\qquad
\Phi_{\bar\gamma}=\Phi_\gamma^{-1},
\end{equation}
where $\bar\gamma$ is the reversed path.
Both maps are restrictions of CPTP maps.
For all $\rho,\sigma\in\Sigma(X)$,
\begin{equation}
\begin{aligned}
\|\Phi_\gamma(\rho)-\Phi_\gamma(\sigma)\|_1
&=\|\rho-\sigma\|_1,\\
S(\Phi_\gamma(\rho))-S(\Phi_\gamma(\sigma))
&=S(\rho)-S(\sigma).
\end{aligned}
\end{equation}
\end{thm}
\begin{proof}
See Appendix~\ref{isomsec} for the proof, which adapts the strategy of
\cite[Theorem~3.10]{SKK} to shell paths on the lattice.
\end{proof}
In particular, we have
\begin{cor}[Nested isomorphism]\label{nesti}
If $X$ is obtained from $Y$ by a sequence of the shell reductions described
above, the partial trace $\tr_{Y\to X}:\Sigma(Y)\to\Sigma(X)$ is a bijection. Its inverse is the {reversed} composition of Petz maps.
\end{cor}

\begin{lem}[Comparison of graph annuli]
\label{lem:standard-annulus-comparison}
Fix a reference graph annulus $A_0$.
There exists a unique family of affine bijections
\[
\Phi_A:\Sigma(A_0)\longrightarrow\Sigma(A),
\]
indexed by all graph annuli $A$, satisfying
\[
\Phi_{A_0}=\operatorname{id}
\]
and
\begin{align}\label{pan}
    \tr_{B\setminus A}\circ\Phi_B=\Phi_A
\end{align}
for every essential inclusion $A\subset B$ of graph annuli.

Each $\Phi_A$ and its inverse are restrictions of CPTP
maps between the corresponding full matrix algebras, and
\[
\Phi_A(\omega_{A_0})=\omega_A.
\]
These bijections preserve trace distance and entropy differences:
for all $\rho,\eta\in\Sigma(A_0)$,
\begin{equation}
S(\Phi_A(\rho))-S(\Phi_A(\eta))
=
S(\rho)-S(\eta),
\end{equation}
where $S$ denotes the von Neumann entropy.
For every essential inclusion $A\subset B$ of graph annuli,
the partial trace is the bijection
\begin{equation}
\tr_{B\setminus A}
=
\Phi_A\circ\Phi_B^{-1}
:\Sigma(B)\longrightarrow\Sigma(A).
\end{equation}
\end{lem}
\begin{proof}
The proof in Appendix~\ref{isomsec} adapts the common exterior annulus
argument of \cite[Lemma~4.3]{SKK} to the present lattice setting. The uniqueness of the normalized, restriction-compatible family follows from Lemma~\ref{lem:annular-restriction-injectivity}.
\end{proof}

\subsubsection{Annuli and charge labels}

The first nontrivial topology is an annulus. The following finite-region
structure theorem is a consequence of A0, A1, and deformation invariance.

\begin{thm}[Annulus simplex]\label{thm:annlusconvex}
Let $A$ be a regular annulus satisfying one of the following
conditions:
\begin{enumerate}
\item
$A=A_f(s,t)$ for some face $f$ and integers $s,t$ with
$s\ge 19$ and $t-s\ge25$.

\item
There exist two annuli $P$ and $B$, each of the form
described in (1), such that
\[
P\subset A\subset B,
\]
where both inclusions $P\subset A$ and $A\subset B$
are essential.
The restriction
\[
\tr_{B\setminus P}:
\Sigma(B)\longrightarrow\Sigma(P)
\]
is a bijection with a CPTP inverse.
\end{enumerate}
Then there is a finite label set
\begin{equation}
 \{1,a,b,c,\ldots\}
\end{equation}
and a family of density matrices
$\{\sigma_A^a\}_{a}\subset\Sigma(A)$ such that
\begin{equation}
 \Sigma(A)
 =\left\{
 \sum_{a}p_a\sigma_A^a:
 p_a\geq0,\ \sum_{a}p_a=1
 \right\}.
 \label{eq:annulus-simplex}
\end{equation}
The states $\sigma_A^a$ are precisely the extreme points of $\Sigma(A)$ and
have mutually orthogonal supports:
\begin{equation}\label{eq:chargeorthogonal}
 \sigma_A^a\sigma_A^b=0
 \qquad(a\neq b).
\end{equation}
The distinguished label $1$ is the vacuum label, characterized by
\begin{equation}
 \sigma_A^1=\omega_A.
\end{equation}
\end{thm}
\begin{proof}
The (1) case follows from the arguments of
\cite[Theorem~4.1, Lemma~D.5, Corollary~D.5.1,
and Proposition~4.2]{SKK}.
The shell isomorphisms give the restriction bijections; A0, A1, and disk uniqueness imply interior decoupling. 
For the (2) case, shell restriction and Lemma~\ref{lem:intrinsic-sandwich}
then transfer the conclusions to every annulus in the stated family.
\end{proof}
\begin{defn}
Fix a graph annulus $A_0$.
By Theorem~\ref{thm:annlusconvex}, the extreme points of
its information convex $\Sigma(A_0)$ can be written as
\[
\{\sigma_{A_0}^a:a\in\mathcal C^{\mathrm{EB}}\},
\]
where $\mathcal C^{\mathrm{EB}}$ is a finite label set
containing a distinguished label $1$, with
$\sigma_{A_0}^1=\omega_{A_0}$, defined for 
the reference-state marginal on $A_0$.

For each graph annulus $A$, let
\[
\Phi_A:\Sigma(A_0)\longrightarrow\Sigma(A)
\]
be the canonical isomorphism of
Lemma~\ref{lem:standard-annulus-comparison}, and define
\[
\sigma_A^a:=\Phi_A(\sigma_{A_0}^a),
\qquad a\in\mathcal C^{\mathrm{EB}}.
\]
Since $\Phi_A$ is affine and both $\Phi_A$ and its
inverse are restrictions of CPTP maps, these states
are precisely the extreme points of $\Sigma(A)$
and satisfy \eqref{eq:annulus-simplex}
and \eqref{eq:chargeorthogonal}.
By (\ref{pan}),
we have
\begin{align}\label{hebi}
    \tr_{B\setminus A}\sigma_B^a=\sigma_A^a,
\end{align}
for every essential inclusion $A\subset B$ of graph annuli.
\end{defn}
\begin{rem}
The states $\sigma_A^a$ also extend to finite regular
annuli $A$ whose holes contain a graph disk of radius
at least $19$. (See  Lemma~\ref{lem:detector-annuli}.)

For an annulus $A$ satisfying condition (2) of
Theorem~\ref{thm:annlusconvex}, choose $P\subset A\subset B$
as in that condition.
By Lemma~\ref{lem:intrinsic-sandwich}, restriction
$\Sigma(B)\to\Sigma(A)$ is an affine bijection.
Define
\[
 \sigma_A^a
 :=\tr_{B\setminus A}\sigma_B^a,
 \qquad a\in\mathcal C^{\mathrm{EB}}.
\]
This definition is independent of the choice of $P,B$
by Lemma~\ref{lem:detector-annuli}, and these states
are precisely the extreme points of $\Sigma(A)$.
\end{rem}

The entropy contribution of a label is~\cite[Definition 4.2]{SKK}
\begin{equation}
 f(a):=\frac{1}{2}
 \left(S(\sigma_A^a)-S(\sigma_A^1)\right).
 \label{eq:sector-entropy-contribution}
\end{equation}
Entropy-difference preservation makes $f(a)$ independent of the annulus under these identifications.
Write $d_a=e^{f(a)}>0$ for the entropy weight of charge $a$. $d_a$ is identified as the quantum dimension of the label $a$ in entanglement bootstrap~\cite[Definition 5.1]{SKK}.

\subsubsection{Charge detecting operator}
For an annulus {$A$ carrying the orthogonal states constructed above,
define its} charge detecting operator {by}

\begin{equation}\label{ame}
    {S_A^a}:={\rm proj}_{\supp(\sigma_A^a)}\,,\quad a\in\ceb\,.
\end{equation}
\begin{lem}\label{hashi}
Let $A\subseteq A'$ be finite regular annuli whose holes
each contain a graph disk of radius at least $19$.
Assume that the inclusion $A\hookrightarrow A'$ is essential.

Then
    \begin{equation}\label{eq:SbSa}
        {S_A^b} {S_{A'}^a}=\delta_{ab}{S_{A'}^a}.
    \end{equation}
\end{lem}

\begin{proof} By (\ref{eq:charge-restriction}) and  positivity, 
    \begin{equation}
        \operatorname{supp}(\sigma_{A'}^a)
        \subseteq
        \operatorname{supp}(\sigma_A^a)
        \ot\cH_{A'\setminus A}.
    \end{equation}
    The claim follows from the orthogonality of the sector supports on $A$.
\end{proof}

\subsubsection{Two-hole disks and fusion spaces}

We describe the fixed-charge sectors and fusion spaces
of the following two-hole shell domains.

\begin{defn}
\label{def:two-hole-shell-family}
Consider a two-hole shell domain
\[
Y=K_o(n)\setminus
\bigl(K_{z_1}(s_1)\cup K_{z_2}(s_2)\bigr),
\qquad
n\ge2000,\quad s_1,s_2\ge50,
\]
where $o,z_1,z_2$ are faces and the radii are integers.
The two holes lie inside $K_o(n)$, their mutual graph
distance is at least $1000$, and each has graph distance
at least $1000$ from $K_o(n)^c$.
The two holes are ordered by their indices.

For integers $25\le w_i\le150$, define the detector annuli
\[
\begin{aligned}
A_1&:=K_{z_1}(s_1+w_1)\setminus K_{z_1}(s_1),\\
A_2&:=K_{z_2}(s_2+w_2)\setminus K_{z_2}(s_2),\\
A_3&:=K_o(n)\setminus K_o(n-w_3).
\end{aligned}
\]
These annuli are pairwise disjoint and contained in $Y$.
The first two surround the respective holes, and the
third surrounds both holes.

Define
\begin{align}
\Sigma_{ab}^c(Y;A_1,A_2,A_3)
:=\{\rho_Y\in\Sigma(Y):\;&
(\rho_Y)_{A_1}=\sigma_{A_1}^a,\quad
(\rho_Y)_{A_2}=\sigma_{A_2}^b,
\nonumber\\
&(\rho_Y)_{A_3}=\sigma_{A_3}^c\},
\label{eq:fixed-charge-information-convex}
\end{align}
where $(\rho_Y)_{A_i}
=\tr_{Y\setminus A_i}\rho_Y$
denotes the marginal on $A_i$.
This sector may be empty.
\end{defn}

\begin{lem}[Independence of collar widths]
\label{lem:shell-sector-independence}
For a fixed $Y$ in
Definition~\ref{def:two-hole-shell-family},
the set $\Sigma_{ab}^c(Y;A_1,A_2,A_3)$ is independent
of the widths $25\le w_i\le150$.
We denote this set by $\Sigma_{ab}^c(Y)$.
\end{lem}

\begin{proof}
Any two allowed collars at the same boundary are
nested concentric graph annuli.
By Lemma~\ref{lem:standard-annulus-comparison},
restriction between their information convex sets
is a bijection carrying each canonical charge state
to the state with the same label.
Thus a marginal on either collar equals its canonical
state with label $a$ if and only if the marginal on
the other collar does.
Applying this to the three boundaries proves the claim.
\end{proof}
\begin{defn}
By the preceding lemma, we may
define 
\[
\Sigma_{ab}^c(Y):=\Sigma_{ab}^c(Y;A_1,A_2,A_3)
\]
for
each $Y$ in
Definition~\ref{def:two-hole-shell-family}.
\end{defn}
The definition of $\Sigma_{ab}^c(Y)$ extends to
more general regions $Y$ (see Lemma~\ref{lem:two-hole-sector-independence}).

\begin{thm}[Fusion-space structure]
\label{thm:fusionspace}

For each triple of charge labels
$a,b,c\in\mathcal C^{\mathrm{EB}}$, there exists a
finite-dimensional Hilbert space $\mathcal V_{ab}^c$
such that the following holds for every two-hole shell
domain $Y$ satisfying
Definition~\ref{def:two-hole-shell-family}.

Here $a,b$ are the ordered inner charges and $c$ is
the outer charge. The corresponding sector
$\Sigma_{ab}^c(Y)$ is empty if and only if
$\mathcal V_{ab}^c=\{0\}$.
For every nonempty sector, there exists an affine bijection
\begin{equation}
 \iota_Y^{abc}:
 \mathcal D(\mathcal V_{ab}^c)
 \longrightarrow \Sigma_{ab}^c(Y),
 \label{eq:fusion-space-isomorphism}
\end{equation}
where $\mathcal D(\mathcal V_{ab}^c)$ denotes the
density matrices on $\mathcal V_{ab}^c$.

Both $\iota_Y^{abc}$ and its inverse on the sector
extend to CPTP maps between
$\mathcal B(\mathcal V_{ab}^c)$ and
$\mathcal B(\mathcal H_Y)$, where $\mathcal H_Y$
is the spin Hilbert space of $Y$.
Moreover, for all
$\tau,\eta\in\mathcal D(\mathcal V_{ab}^c)$,
\[
S\bigl(\iota_Y^{abc}(\tau)\bigr)
-
S\bigl(\iota_Y^{abc}(\eta)\bigr)
=
S(\tau)-S(\eta).
\]

The information convex has the decomposition
\begin{equation}
\Sigma(Y)
=
\left\{
\sum_{a,b,c}p_{ab}^c\rho_{ab}^c:
\rho_{ab}^c\in\Sigma_{ab}^c(Y),\
p_{ab}^c\ge0,\
\sum_{a,b,c}p_{ab}^c=1
\right\},
\label{eq:two-hole-decomposition}
\end{equation}
where the sums range over the nonempty sectors.
States in distinct sectors have mutually orthogonal
supports, so this is an orthogonal direct-sum
decomposition.

\end{thm}

\begin{proof}
The geometric assumptions on $Y$ ensure that the collar
restrictions used in the proof of
\cite[Theorems~4.4--4.5]{SKK}
are covered by Corollary~\ref{nesti}
and have fixed Petz inverses.
With these recovery maps, the conclusion follows from
the arguments of
\cite[Theorems~4.4--4.5, Lemmas~D.2--D.7,
and Appendix~E]{SKK}.
\end{proof}

The Hilbert space $\cV_{ab}^{c}$ is the \emph{information-convex fusion space}, and
\begin{equation}
 N_{ab}^{c}:=\dim\cV_{ab}^{c}
 \label{eq:fusion-multiplicity}
\end{equation}
is the fusion multiplicity. We write the fusion rules formally as
\begin{equation}
 a\times b=\sum_{c\in\ceb}N_{ab}^{c}\,c.
\end{equation}
The arguments of
\cite[Propositions~4.6--4.9]{SKK} give
\[
 N_{1a}^c=N_{a1}^c=\delta_{ac},\qquad
 N_{ab}^c=N_{ba}^c,\qquad
 \Sigma_{11}^1(Y)=\{\omega_Y\}.
\]

\begin{lem}[Sector entropy]
\label{lem:merger-sector-entropy}
Let $Y$ be a two-hole shell domain satisfying the assumptions of Theorem~\ref{thm:fusionspace}.
Every extreme state $x\in\Sigma_{ab}^c(Y)$ satisfies
\[
 S(x)-S(\omega_Y)=f(a)+f(b)+f(c).
\]
\end{lem}
\begin{proof}
Follow the proof of \cite[Lemma~4.8, Appendix~E]{SKK},
using the collar identities established in the proof of
Theorem~\ref{thm:fusionspace} and the entropy-difference
preservation under shell restriction from
Theorem~\ref{thm:information-convex-isomorphism}.
\end{proof}

\subsubsection{Consistency of reduced states}\label{sec:state-consistency}

Fix two ordered graph disks
\[
 D_i:=K_{z_i}(s_i),\qquad i=1,2,
\]
where $z_1,z_2$ are faces and $s_1,s_2\ge50$ are integers,
such that
\[
 d(D_1,D_2)\ge1000.
\]
Choose a face $o$ and an integer $n_0\ge2000$ such that
\[
 D_1\cup D_2\subset K_o(n_0),
 \qquad
 d(D_i,K_o(n_0)^c)\ge1000
 \quad(i=1,2).
\]
For each integer $N\ge0$, set
\[
 Y_N:=K_o(n_0+N)\setminus(D_1\cup D_2).
\]
Then every $Y_N$ satisfies conditions in
Definition~\ref{def:two-hole-shell-family}.

{By Corollary~\ref{nesti} and compatibility of the annular
charge labels, the partial trace from $Y_{N+1}$ to $Y_N$ is a sector-preserving bijection. Let
\[
R_{Y_N\to Y_{N+1}}:
\Sigma(Y_N)\longrightarrow\Sigma(Y_{N+1})
\]
denote its inverse, which is implemented by a CPTP map.

For fixed ordered inner charges $a,b$ and outer charge $c$
with $\Sigma_{ab}^c(Y_0)\ne\varnothing$, we choose an isomorphism
\[
\iota_{Y_0}^{abc}:
\cD(\cV_{ab}^c)\longrightarrow\Sigma_{ab}^c(Y_0)
\]
as in Theorem~\ref{thm:fusionspace}, where
$\cD(\cV_{ab}^c)$ denotes the density matrices on the
fusion space $\cV_{ab}^c$.
For $N=0,1,\ldots$, define recursively
\begin{equation}\label{kumo}
\iota_{Y_{N+1}}^{abc}
:=
R_{Y_N\to Y_{N+1}}\circ\iota_{Y_N}^{abc}.
\end{equation}
Since restriction inverts recovery,
\begin{equation}\label{sora}
\tr_{Y_{N+1}\setminus Y_N}
\circ\iota_{Y_{N+1}}^{abc}
=
\iota_{Y_N}^{abc}.
\end{equation}
The same restriction is onto and preserves entropy differences
on the convex set with fixed inner charges $a,b$.}

\subsubsection{Antilabels}\label{tuchi}
By \cite[Proposition~4.9]{SKK}, each $a\in\ceb$ has
a unique antilabel $\bar a\in\ceb$ satisfying
\[
  N_{ab}^{1}=\delta_{b,\bar a}
  \qquad (b\in\ceb).
\]
The antilabels satisfy
\[
  \bar 1=1,\qquad
  \overline{\bar a}=a,\qquad
  f(\bar a)=f(a).
\]

\subsubsection{Merging two annuli}\label{sougen}

\begin{setting}[Two-annulus merging]
\label{setting:two-annulus-merging}
Let $Y$ be a two-hole shell domain satisfying
Definition~\ref{def:two-hole-shell-family}.
Consider a disjoint partition
\[
 Y=L\sqcup M\sqcup N\sqcup R
\]
into face regions such that $LMN$ and $MNR$ are regular
annuli surrounding the first and second holes, respectively. 
The two inner detector annuli in
Definition~\ref{def:two-hole-shell-family}
are assumed to be essentially contained in $LMN$ and
$MNR$, respectively.

Suppose that the overlap admits another disjoint partition
\[
 MN=M'\sqcup N',
\]
where $M,N,MN,M',N'$ are disks, and that
\begin{equation}\label{eq:annular-merger-geometry}
 \begin{gathered}
 d(L,N),\ d(M,R),\ d(L,N'),\ d(M',R)\ge2,\\
 d(LM',NR)>2.
 \end{gathered}
\end{equation}

Fix labels $a,b$ in the finite charge-label set
$\mathcal C^{\mathrm{EB}}$, and let
\[
 \sigma_{LMN}^a\in\Sigma(LMN),
 \qquad
 \sigma_{MNR}^b\in\Sigma(MNR)
\]
be the corresponding canonical annular states. 

By disk uniqueness and A1 along the relative disk orderings, 
\[
 (\sigma_{LMN}^a)_{MN}
 =
 (\sigma_{MNR}^b)_{MN}
 =
 \omega_{MN},
\]
and
\[
 \begin{aligned}
 I(L:N| M)_{\sigma_{LMN}^a}
 &=I(M:R| N)_{\sigma_{MNR}^b}=0,\\
 I(L:N'| M')_{\sigma_{LMN}^a}
 &=I(M':R| N')_{\sigma_{MNR}^b}=0.
 \end{aligned}
\]
Thus Lemma~\ref{lem:merginglemma} defines the merged state
\[
 \sigma_Y^{a\times b}
 :=
 \sigma_{LMN}^a\bowtie\sigma_{MNR}^b
\]
on $Y$.
\end{setting}

\begin{prop}[Merging two annuli]
\label{prop:two-annulus-merging}
In Setting~\ref{setting:two-annulus-merging},
the merged state $\sigma_Y^{a\times b}$ belongs to
$\Sigma(Y)$ and is the unique entropy maximizer in
\[
 \operatorname{conv}\!
 \left(\bigcup_{c\in\mathcal C^{\mathrm{EB}}}
                  \Sigma_{ab}^c(Y)\right).
\]
It has the orthogonal decomposition
\[
 \sigma_Y^{a\times b}
 =
 \bigoplus_{c:N_{ab}^c>0}p_c\tau_{ab}^c,
\]
where $N_{ab}^c$ is the fusion multiplicity,
$\tau_{ab}^c$ is the unique maximum-entropy state of
$\Sigma_{ab}^c(Y)$, and
\begin{equation}\label{eq:abmergedweight}
 p_c=\frac{N_{ab}^c e^{f(c)}}{e^{f(a)+f(b)}}.
\end{equation}
In particular,
\begin{equation}\label{eq:merger-dimension-identity}
 \sum_cN_{ab}^c e^{f(c)}=e^{f(a)+f(b)}.
\end{equation}

For any entropy-difference-preserving fusion-space
identification
\[
 \iota_Y^{abc}:
 \mathcal D(\mathcal V_{ab}^c)\longrightarrow\Sigma_{ab}^c(Y),
\]
we have
\[
 \tau_{ab}^c
 =
 \iota_Y^{abc}\!\left(
 \frac{I_{\mathcal V_{ab}^c}}{N_{ab}^c}\right)
 \qquad(N_{ab}^c>0).
\]
\end{prop}
Appendix~\ref{app:merging-entropy} gives the proof and extends the proposition to the regions in Definition~\ref{def:two-hole-sandwich}.

{F}or the increasing two-hole regions $Y_N$, $N\ge0$, defined in
Section~\ref{sec:state-consistency}, fix the inner charges $a,b$.
Restriction from $Y_{N+1}$ to $Y_N$ is a bijection preserving
entropy differences on the information-convex state sets
with these inner charges.
It therefore preserves their unique maximum-entropy states
$\sigma_{Y_N}^{a\times b}$:
\[
 \tr_{Y_{N+1}\setminus Y_N}
 \sigma_{Y_{N+1}}^{a\times b}
 =
 \sigma_{Y_N}^{a\times b}.
\]
The fusion multiplicities and the weights in
\eqref{eq:abmergedweight} are independent of $N$.

\subsection{Merging three annuli}
\begin{setting}\label{mimisen}

Let $Z$ be a three-hole shell domain covered by regular
annuli $X_1,X_2,X_3$, surrounding the respective holes.
Suppose
\[
 X_1\cap X_3=\varnothing,
\]
and put
\[
 Y_{12}:=X_1\cup X_2,\qquad
 Y_{23}:=X_2\cup X_3,\qquad
 O_{12}:=X_1\cap X_2,\qquad
 O_{23}:=X_2\cap X_3.
\]
Assume that $Y_{12}$ and $Y_{23}$ are regular two-hole
regions.

For each ordered pair
\[
 (U,V)\in
 \bigl\{
 (X_1,X_2),\ (X_2,X_3),\
 (Y_{12},X_3),\ (X_1,Y_{23})
 \bigr\},
\]
write $L:=U\setminus V$ and $R:=V\setminus U$.
Suppose that the overlap admits partitions
\[
 U\cap V=M\sqcup N=M'\sqcup N'
\]
such that $M,N,M',N'$ and $U\cap V$ are disks, and
\[
 \begin{gathered}
 d(L,N),\ d(M,R),\ d(L,N'),\ d(M',R)>1,\\
 d(LM',NR)>2.
 \end{gathered}
\]
\end{setting}
\begin{lem}[Merging three annuli]
\label{lem:three-annulus-merging}
Consider Setting \ref{mimisen}.
Fix states $\sigma_i\in\Sigma(X_i)$, $i=1,2,3$.
Then both successive mergers are well defined and
\[
 \gamma
 :=
 (\sigma_1\bowtie\sigma_2)\bowtie\sigma_3
 =
 \sigma_1\bowtie(\sigma_2\bowtie\sigma_3)
 \in\Sigma(Z).
\]
Their pair marginals satisfy
\[
 \gamma_{Y_{12}}=\sigma_1\bowtie\sigma_2,
 \qquad
 \gamma_{Y_{23}}=\sigma_2\bowtie\sigma_3.
\]
Moreover, $\gamma$ is the unique entropy-maximizing state
on $Z$ subject to
\[
 \rho_{X_i}=\sigma_i\qquad(i=1,2,3),
\]
and
\[
 S(\gamma)
 =
 \sum_{i=1}^{3}S(\sigma_i)
 -S(\omega_{O_{12}})
 -S(\omega_{O_{23}}).
\]

For fixed $\sigma_3$, there is a fixed CPTP recovery map
$\mathcal R_3$ from $Y_{12}$ to $Z$ such that, for every
$\eta\in\Sigma(Y_{12})$ with $\eta_{X_2}=\sigma_2$,
\[
 \mathcal R_3(\eta)=\eta\bowtie\sigma_3\in\Sigma(Z),
\]
\[
 \bigl(\mathcal R_3(\eta)\bigr)_{Y_{12}}=\eta,
 \qquad
 \bigl(\mathcal R_3(\eta)\bigr)_{Y_{23}}
 =\sigma_2\bowtie\sigma_3.
\]
\end{lem}

\begin{proof}
By Lemma \ref{lem:localuniqueness},  the states agree on each overlap. A1 along the relative disk orderings shows that the four conditional mutual informations vanish for each cover.
Lemma~\ref{lem:merginglemma} and the argument using the two recovery maps in  \cite[Propositions~C.2 and C.5]{SKK} therefore construct
both successive mergers in $J(Z)$.
Since $Z$ is a shell domain, both merged states belong to $\Sigma(Z)$ by Lemma~\ref{lem:information-convex-bridge}.

For every state $\rho$ with the prescribed $X_i$
marginals, two applications of strong subadditivity provide
\[
 \begin{aligned}
 S(\rho)
 &\le S(\rho_{Y_{12}})+S(\sigma_3)
       -S(\omega_{O_{23}})\\
 &\le \sum_{i=1}^{3}S(\sigma_i)
       -S(\omega_{O_{12}})
       -S(\omega_{O_{23}}).
 \end{aligned}
\]
Both successive mergers attain this bound.
Strict concavity of entropy proves their equality and
uniqueness. The two asserted pair marginals follow because each merging preserves its input marginals.

Finally, take the Petz recovery map from $O_{23}$ to
$X_3$ determined by $\sigma_3$, extended by the identity
on $Y_{12}\setminus O_{23}$.
The same merging argument proves the assertions about $\caR_3$ for every $\eta$ satisfying the stated conditions. The restrictions of $\caR_3(\eta)$ to $Y_{23}$ is the merger of $\eta_{X_2}=\sigma_2$ with $\sigma_3$.
\end{proof}

The concrete three-hole configuration used in this paper satisfies these
hypotheses, as verified in Appendix~\ref{taiyou}.

\begin{figure}[ht]
 \centering
  {\includegraphics[width=0.4\linewidth]{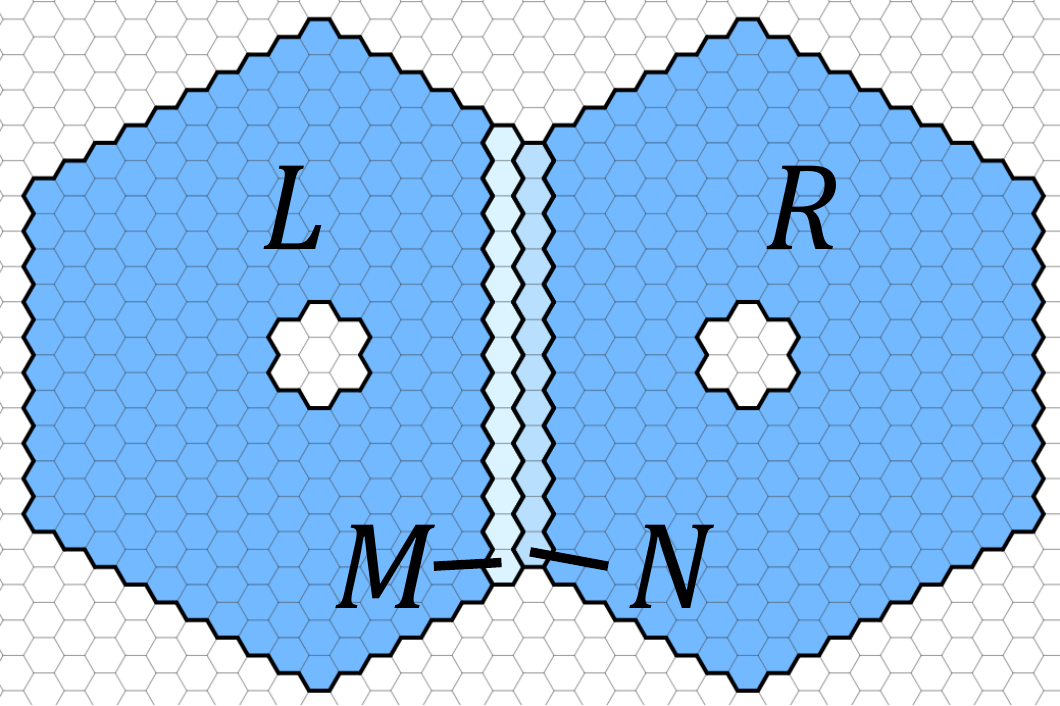}}
 \caption{Schematic partition of a two-hole disk $Y = LMNR$. The annuli $LMN$ and $MNR$ overlap on the disk $MN$.}
 \label{fig:Y=LMNR}
\end{figure}

\begin{setting}\label{mado}
Consider Setting \ref{mimisen}.
Assume that each of $Y_{12}$ and $Y_{23}$ satisfies either
Definition~\ref{def:two-hole-shell-family}
or Definition~\ref{def:two-hole-sandwich}, and that
both pairs $(X_1,X_2)$ and $(X_2,X_3)$ satisfy the
annular-partition, detector-containment, and separation
conditions of Setting \ref{setting:two-annulus-merging}. Theorem~\ref{thm:fusionspace} and Remark~\ref{lem:two-hole-sector-isomorphism}
give the orthogonal sector decomposition
\eqref{eq:two-hole-decomposition} for both regions.
Proposition~\ref{prop:two-annulus-merging} and
Corollary~\ref{cor:two-annulus-merging-sandwich}
provide the merger-weight formula \eqref{eq:abmergedweight}
for the corresponding canonical annular states of both pairs. 
For $(i,j)=(1,2),(2,3)$, choose a finite regular annulus
$C_{ij}\subset Y_{ij}$ whose hole contains both holes
of $Y_{ij}$ and a graph disk of radius at least $19$.
\end{setting}

The concrete annuli used later and the verification
of their charge-detection property are given in
Appendix~\ref{taiyou}.

\begin{prop}\label{concon}
Consider Setting \ref{mado}. For every $a\in\ceb$,
    \begin{equation}
        \tr\left[\left(\sigma^a_{X_1}\bowtie\sigma_{X_2}^{\bar a}\bowtie\sigma_{X_3}^a\right) {S_{C_{12}}^1}{S_{C_{23}}^1}\right]=e^{-4f(a)}>0.
    \end{equation}
\end{prop}
\begin{proof}
Let
\[
 Y_{12}:=X_1\cup X_2,\qquad
 Y_{23}:=X_2\cup X_3,\qquad
 Z:=X_1\cup X_2\cup X_3.
\]
Fix a charge label $a\in\ceb$, and let $\bar a$ be its antilabel in $\ceb$.
Write the two-annulus merger as
\[
 \sigma_{X_1}^{a}\bowtie\sigma_{X_2}^{\bar a}
 =
 \sum_c p_c\tau_{12}^{c},
\]
where the sum runs over the positive sector weights
$p_c$, and
$\tau_{12}^{c}\in\Sigma_{a\bar a}^{c}(Y_{12})$
is the normalized state in the sector with outer
charge $c$.
Since the fusion multiplicities satisfy
\[
 N_{a\bar a}^{1}=N_{\bar a a}^{1}=1
\]
and the annular entropy contribution satisfies
$f(1)=0$ and $f(\bar a)=f(a)$, the merger-weight
formula provides
\[
 p_1=e^{-2f(a)}>0.
\]
The merger
$\sigma_{X_2}^{\bar a}\bowtie\sigma_{X_3}^{a}$
has the same vacuum-sector weight.

We first identify the $X_2$ marginal of every
$\tau_{12}^{c}$.
Its second inner charge is $\bar a$, so it agrees
with $\sigma_{X_2}^{\bar a}$ on the inner detector
annulus defining that charge
and by injectivity of annular restriction in Lemma~\ref{lem:annular-restriction-injectivity}, we have
\[
 \tr_{Y_{12}\setminus X_2}
 \tau_{12}^{c}
 =
 \sigma_{X_2}^{\bar a}
 \qquad(p_c>0).
\]

Put
\[
 O:=X_2\cap X_3=Y_{12}\cap X_3,
\]
the overlap with the third input region.
Let
\[
 R^{\sigma_{X_3}^{a}}_{O\to X_3}:
 \cA_O\longrightarrow\cA_{X_3}
\]
be a fixed CPTP extension of the Petz recovery map
for restriction from $X_3$ to $O$, determined by
$\sigma_{X_3}^{a}$.
Define its extension to the pair region by
\[
 \mathcal R_3
 :=
 \operatorname{id}_{Y_{12}\setminus O}
 \otimes R^{\sigma_{X_3}^{a}}_{O\to X_3}
 :
 \cA_{Y_{12}}\longrightarrow\cA_Z.
\]

The final assertion of
Lemma~\ref{lem:three-annulus-merging} applies to each
$\tau_{12}^{c}$, since its $X_2$ marginal is
$\sigma_{X_2}^{\bar a}$.
Thus the state
\[
 \gamma^{c}:=\mathcal R_3(\tau_{12}^{c})
\]
is their Markov merger and satisfies
\[
 (\gamma^{c})_{Y_{12}}=\tau_{12}^{c}.
\]
Using the marginal identity established above, we obtain
\[
\begin{aligned}
 (\gamma^{c})_{Y_{23}}
 &=
 \left(
 \operatorname{id}_{X_2\setminus O}
 \otimes R^{\sigma_{X_3}^{a}}_{O\to X_3}
 \right)
 \left((\tau_{12}^{c})_{X_2}\right)\\
 &=
 \left(
 \operatorname{id}_{X_2\setminus O}
 \otimes R^{\sigma_{X_3}^{a}}_{O\to X_3}
 \right)
 \left(\sigma_{X_2}^{\bar a}\right)\\
 &=
 \sigma_{X_2}^{\bar a}\bowtie\sigma_{X_3}^{a},
\end{aligned}
\]
where the last equality is another application of
Lemma~\ref{lem:merginglemma}.
 
By linearity,
\[
 \bigl(
 \sigma_{X_1}^{a}\bowtie\sigma_{X_2}^{\bar a}
 \bigr)\bowtie\sigma_{X_3}^{a}
 =
 \sum_c p_c\gamma^{c}.
\]

We next verify the action of the vacuum detector.
Since $C_{12}\subset Y_{12}$, preservation of the
$Y_{12}$ marginal and the charge-detection property imply
\[
 \tr\!\left[
 \gamma^{c}S_{C_{12}}^1
 \right]
 =
 \tr\!\left[
 \tau_{12}^{c}S_{C_{12}}^1
 \right]
 =
 \delta_{c,1}.
\]
Since $\gamma^{c}$ is a state $S_{C_{12}}^1$ is a projection, 
\[
 \gamma^{c}S_{C_{12}}^1
 =
 S_{C_{12}}^1\gamma^{c}
 =
 \delta_{c,1}\gamma^{c}.
\]

Finally, using $C_{23}\subset Y_{23}$ and the
$Y_{23}$ marginal of $\gamma^{1}$, we obtain
\begin{align*}
 &\tr\!\left[
 \bigl(
   (\sigma_{X_1}^{a}\bowtie\sigma_{X_2}^{\bar a})
   \bowtie\sigma_{X_3}^{a}
 \bigr)
 S_{C_{12}}^1S_{C_{23}}^1
 \right]\\
 &\qquad=
 \sum_c p_c\tr\!\left[
 \gamma^{c}S_{C_{12}}^1S_{C_{23}}^1
 \right]\\
 &\qquad=
 p_1\tr\!\left[
 \gamma^{1}S_{C_{23}}^1
 \right]\\
 &\qquad=
 p_1\tr\!\left[
 \bigl(
 \sigma_{X_2}^{\bar a}\bowtie\sigma_{X_3}^{a}
 \bigr)S_{C_{23}}^1
 \right]\\
 &\qquad=p_1^2=e^{-4f(a)}>0.
\end{align*}
\end{proof}

\begin{figure}[ht]
 \centering
  {\includegraphics[width=0.5\linewidth]{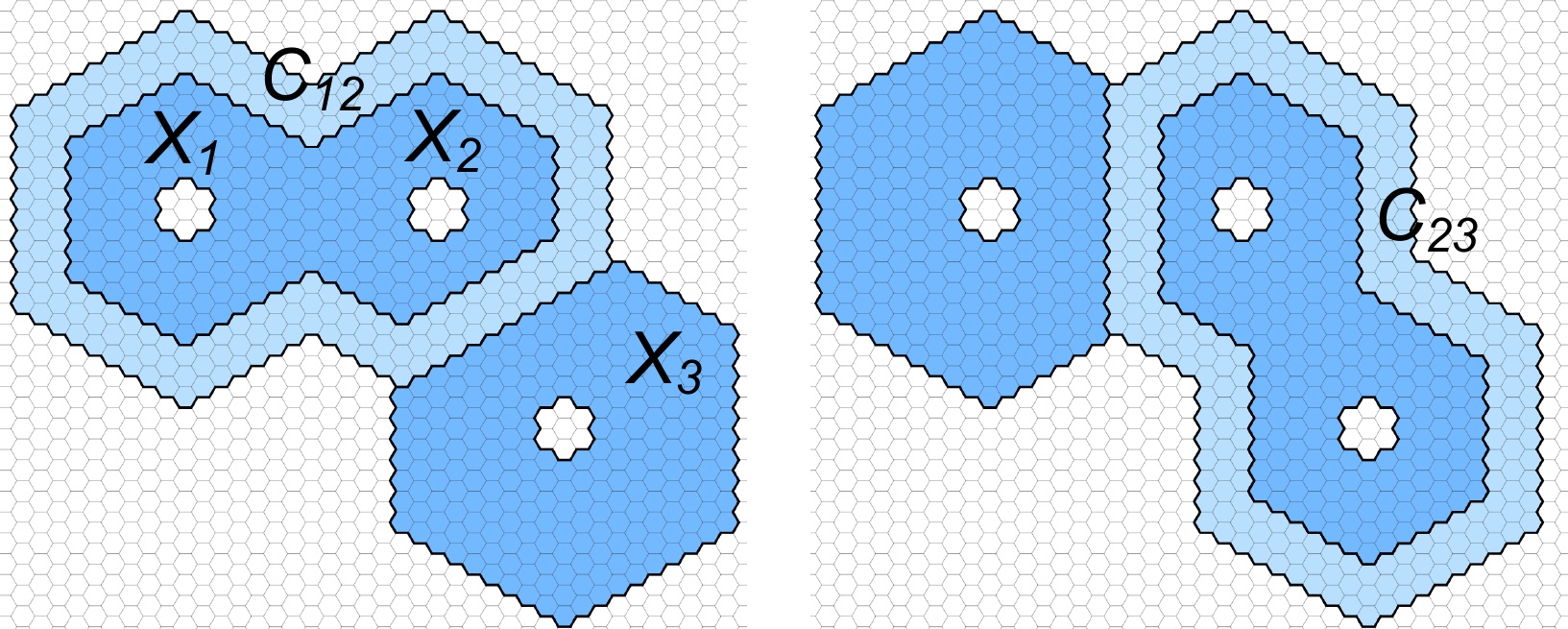}}
 \caption{The three-annulus configurations.}
 \label{fig:three-annuli}
\end{figure}

\subsection{A commuting projector parent Hamiltonian}

We summarize the commuting parent Hamiltonian construction of
Ref.~\cite{KLRS}.  This construction gives a local, frustration-free parent Hamiltonian with a stable gap. The information convex sets are defined directly
from the reference state and do not depend on the choice of parent Hamiltonian.

\subsubsection{Markov covers}

\begin{definition}[Markov cover]
A family $\cS$ of finite disks is a \emph{Markov cover} for $\omega$ if the
following conditions hold.
\begin{enumerate}[(i)]
 \item There are constants $R,\nu<\infty$ such that
 \begin{equation}
  \operatorname{diam}(X)\leq R
  \quad\text{for every }X\in\cS,
 \end{equation}
 and every face belongs to at most $\nu$ members of $\cS$.
 \item For every pair $X,Y\in\cS$ with $X\cap Y\neq\varnothing$,
 \begin{equation}
  I(X\setminus Y:Y\setminus X| X\cap Y)_\omega=0.
  \label{eq:markov-cover-markov}
 \end{equation}
 \item For every face $f$, there is an $X\in\cS$ such that
 \begin{equation}
  \{f\}^{+2}\subset X.
  \label{eq:markov-cover-cover}
 \end{equation}
\end{enumerate}
The range $r$ of a Markov cover is defined as
\begin{equation}
    r:=\sup_{X\in\cS}\sup_{f,g\in X}d(f,g)\,.
\end{equation}
\end{definition}

The first condition makes the resulting interaction uniformly local.  The
second condition is the Markov condition on overlaps, while the third gives the coverage used to prove local uniqueness. The extended A0 and A1
relations imply the existence of a Markov cover with constants independent of the position on the lattice \cite{KLRS}. However, in general, such a Markov cover contains a redundancy.
\begin{definition}[Irredundant Markov cover]
A Markov cover $\cS$ is said to be \emph{irredundant} if it is inclusion-minimal with respect to the cover condition; namely, for every $X\in\cS$, the subfamily $\cS\setminus\{X\}$ does not satisfy the cover condition.
\end{definition}
To see the existence of an irredundant Markov cover,
apply Zorn's lemma to the Markov subcovers ordered by reverse inclusion; 
finite multiplicity ensures that the intersection of every chain is again a Markov cover.

\begin{lem}
Let $\cS$ be an irredundant Markov cover. Then, for every $X\in\cS$, there exists a face $f_X$ such that
\begin{equation}
D_{f_X}^{+}\subset X,
\qquad
D_{f_X}^{+}\not\subset Y
\quad\text{for every }Y\in\cS\setminus\{X\}.
\end{equation}
In particular, every member of $\cS$ contains at least one thickened elementary disk.
\end{lem}

\begin{proof}
Fix $X\in\cS$. Conditions (i) and (ii) in the definition of a Markov
cover are inherited by every subfamily of $\cS$. Since $\cS$ is
irredundant, $\cS\setminus\{X\}$ is not a Markov cover and must therefore
fail condition (iii). Hence, there exists a face $f_X$ such that
\begin{equation}
 D_{f_X}^{+}\not\subset Y
 \qquad
 \text{for every }Y\in\cS\setminus\{X\}.
\end{equation}
Since $\cS$ itself satisfies condition (iii), $D_{f_X}^{+}$ must be
contained in some member of $\cS$. The only possibility is $X$, and hence
\begin{equation}
 D_{f_X}^{+}\subset X.
\end{equation}
\end{proof}

\begin{figure}[ht]
 \centering
 \IfFileExists{parentH.png}
  {\includegraphics[width=0.3\linewidth]{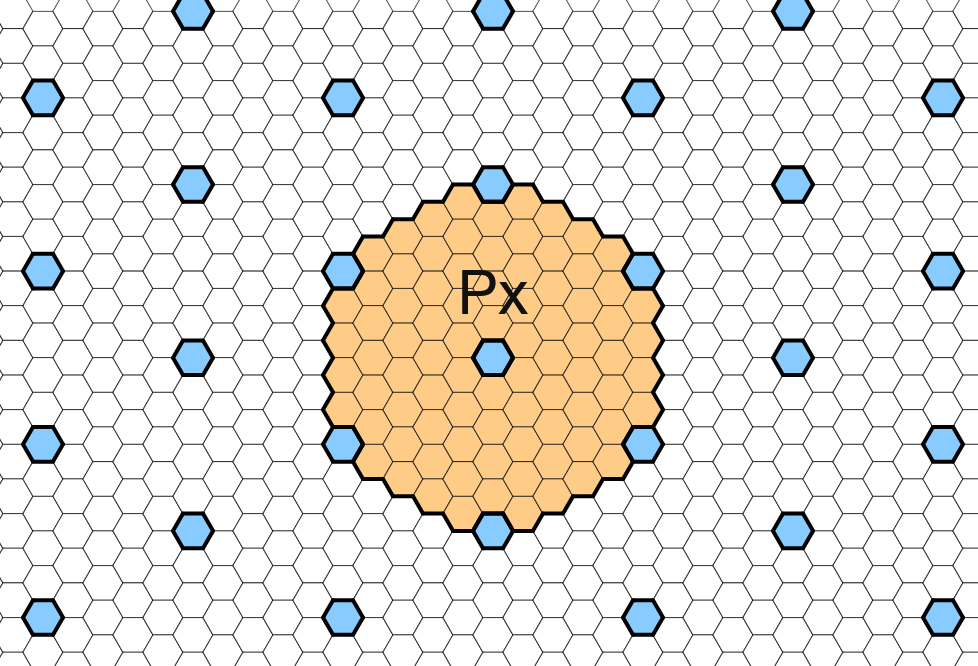}}
  {\fbox{\parbox[c][2.5cm][c]{0.32\linewidth}{\centering
     \texttt{parentH.png}}}}
 \caption{One possible choice of a Markov cover $\cS$ with range $r=10$. Each $X$ is associated to one blue face and the orange faces around it.}
 \label{fig:parentH}
\end{figure}

\subsubsection{The parent interaction}
For a finite region $X$, let
\begin{equation}
 Q_X:=\operatorname{proj}_{\operatorname{supp}(\omega_X)}
 \in\cA_X
 \label{eq:support-projector}
\end{equation}
be the orthogonal projection onto the support of $\omega_X$.
\begin{lem}\label{lem:qxneqI}
For any $X\in\cS$, where $\cS$ is an irredundant Markov cover, $Q_X\neq I$.
\end{lem}
\begin{proof}
For a finite region $Z$, write
$d_Z:=\dim\cH_Z=\prod_{g\in Z}q_g$, and abbreviate
$d_f:=d_{\{f\}}=q_f\ge2$.
    Let $f\in\mL$ be such that $D_f\subset X$. Consider a purification of $\omega_{D_f}$
    with ancilla $E$, $\Qsi_{D_fE}$. Then Axiom A0 implies $I(f:E)_\Qsi=0$; therefore, $\Qsi_{fE}=\omega_f\ot\Qsi_E$. Suppose $\omega_{D_f}$ is full-rank. We then have $\rank(\omega_f)=d_f$, $\rank(\omega_{\cN(f)})=d_{\cN(f)}$ and $\rank(\Qsi_E)=\rank(\omega_{D_f})=d_fd_{\cN(f)}$. Therefore, $\rank(\Qsi_{fE})=d_f^2d_{\cN(f)}$. However, this implies $\rank(\omega_{\cN(f)})=\rank(\Qsi_{fE})=d_f^2d_{\cN(f)}$, which only holds when $d_f=1$. This means $\omega_{D_f}$ is not full-rank, so $\omega_X$ is also not full-rank.   
\end{proof}
Fix an irredundant Markov cover $\cS$.  Define the local projective interaction
\begin{equation}
 h_X:=I_X-Q_X,\qquad X\in\cS,
 \label{eq:parent-local-term}
\end{equation}
and, for $\Lambda\Subset\mL$, the finite-volume Hamiltonian
\begin{equation}
 H_\Lambda:=\sum_{\substack{X\in\cS\wedge X\subset\Lambda}}h_X.
 \label{eq:finite-volume-parent-hamiltonian}
\end{equation}

\begin{thm}[Commuting parent Hamiltonian]\cite{KLRS}\label{thm:commutingparentH}
Let $\cS$ be a Markov cover for $\omega$.
Then the interaction in Eq.~\eqref{eq:parent-local-term} has the following
properties.
\begin{enumerate}[(i)]
 \item The local terms are pairwise commuting projections:
 \begin{equation}
  [h_X,h_Y]=0,\qquad X,Y\in\cS.
 \end{equation}
 \item The state $\omega$ is frustration free:
 \begin{equation}
  \omega(h_X)=0,\qquad X\in\cS.
 \end{equation}
 \item Every nonzero finite-volume Hamiltonian $H_\Lambda$ has a spectral gap at
 least one above its ground-state space.
 \item 

There is an integer $l_0\ge\max\{r,13\}$,
depending only on the cover, with the following
local uniqueness property.
 Let $D_0\subset D$ be finite disks
 satisfying
 \begin{equation}\label{eq:PXandinfoconvex}
  d(D_0,\mL\setminus D)\geq{l}_0.
 \end{equation}
 If $\eta_D\in\cD_D$ has zero energy for every term supported in $D$, then
 \begin{equation}
  \eta_{D_0}=\omega_{D_0}.
  \label{eq:parent-local-uniqueness}
 \end{equation}
 \item 
 The state $\omega$ is a gapped ground state of this Hamiltonian.
\end{enumerate}
Consequently, the interaction satisfies exact local
topological quantum order (LTQO).
Together with the uniform local gap, this implies
stability of the {{bulk}} spectral gap under
sufficiently weak {{short-range}} perturbations
{{\cite[Assumption~2.5 and Theorem~2.8]{NSYBulkGap}}}.
\end{thm}

\begin{remark}\label{rem:graphannulus}
    The bound $l_0\ge 13$ ensures that the annuli constructed in Lemma~\ref{lem:annular-limit} have inner radius at least $l_0+6\ge 19$ and width at least $2l_0\ge 25$, as required by our geometric conventions.
\end{remark}
\begin{defn}\label{standard}
For the rest of this paper, we fix $l_0$ satisfying the condition in Theorem \ref{thm:commutingparentH} (iv).
A disk of the form
\[
 D=\{f_0\}^{+s},
\]
where $f_0\in\mathbb L$ and $s$ is an integer satisfying
$s\ge 3(l_0+2)$, is called a standard disk of radius $s$.
For an integer $m$ with $0\le m\le s$, set
\[
 D^{-m}=\{f_0\}^{+(s-m)},
\]
the concentric disk obtained by removing $m$ layers from $D$.

Let $f$ be a face and let $0\le s<t$ be integers.
The concentric graph annulus
\[
 A_f(s,t):=K_f(t)\setminus K_f(s),
\]
where $K_f(r)$ is the graph disk of radius $r$ centered at $f$,
is called a \emph{standard annulus} if $K_f(s)$ is a
standard disk and $t-s\ge25$.
\end{defn}

\subsubsection{Ground subspace projections}
Let
\begin{equation}\label{eq:parentHamiltonian}
  P_A:=\prod_{\substack{X\in\cS\\ X\subseteq A}}Q_X,
\end{equation}
be the ground subspace projector on $A\Subset\bbL$. In particular, $P_X=Q_X$ for $X\in\cS$.
\begin{lem}\label{lem:GSprojection}
Let $A,B\Subset\mL$.
\begin{enumerate}[(i)]
 \item If $A\subset B$, then, after embedding $P_A$ in $\cA_B$,
 \begin{equation}
  P_B\leq P_A,
  \qquad
  P_AP_B=P_BP_A=P_B.
  \label{eq:ground-projector-monotonicity}
 \end{equation}
 Consequently, the marginal on $A$ of every ground state on $B$ is a
 ground state on $A$.

 \item Define the family of interaction supports crossing the overlap by
 \begin{equation}
  \cC(A,B)
  :=
  \left\{
  X\in\cS:
  X\subset A\cup B,\ 
  X\nsubseteq A,\ 
  X\nsubseteq B
  \right\}.
  \label{eq:crossing-interaction-supports}
 \end{equation}
 After embedding all projections in $\cA_{A\cup B}$,
 \begin{equation}
  P_{A\cup B}
  =
  P_AP_B
  \prod_{X\in\cC(A,B)}P_X.
  \label{eq:ground-projector-gluing-general}
 \end{equation}
 In particular,
 \begin{equation}
  P_{A\cup B}\leq P_AP_B.
  \label{eq:ground-projector-submultiplicativity}
 \end{equation}
 Equality holds if and only if
 \begin{equation}
  P_XP_AP_B=P_AP_B
  \qquad
  \text{for every }X\in\cC(A,B).
  \label{eq:ground-projector-gluing-criterion}
 \end{equation}

 \item A purely geometric sufficient condition for equality in
 Eq.~\eqref{eq:ground-projector-submultiplicativity} is
 \begin{equation}
  d(A\setminus B,B\setminus A)>r.
  \label{eq:ground-projector-buffer-condition}
 \end{equation}
 Under this condition, $\cC(A,B)=\varnothing$ and hence
 \begin{equation}
  P_AP_B=P_{A\cup B}.
  \label{eq:ground-projector-gluing}
 \end{equation}
\end{enumerate}
\end{lem}

\begin{proof}
If $A\subset B$, every interaction support contained in $A$ is also contained
in $B$.  Hence frustration-freeness implies $\operatorname{Im}P_B\subset\operatorname{Im}P_A$, which
provides Eq.~\eqref{eq:ground-projector-monotonicity}.  If
$\eta_B=P_B\eta_BP_B$, it follows that
\begin{equation}
 (P_A\otimes I_{B\setminus A})\eta_B
 (P_A\otimes I_{B\setminus A})
 =\eta_B.
\end{equation}
Taking the partial trace over $B\setminus A$ proves the last assertion in
(i). The supports of the factors defining $P_{A\cup B}$ are precisely the union of
the supports contained in $A$, those contained in $B$, and the crossing
supports in $\cC(A,B)$. Equation~\eqref{eq:ground-projector-gluing-general} follows from commutativity and idempotence of the factors. This implies (ii). Finally, if $X\in\cC(A,B)$, then $X$ contains some
$a\in A\setminus B$ and some $b\in B\setminus A$. Therefore,
\begin{equation}
 d(A\setminus B,B\setminus A)
 \leq d(a,b)
 \leq\operatorname{diam}(X)
 \leq r.
\end{equation}
Thus, Eq.~\eqref{eq:ground-projector-buffer-condition} excludes all crossing
supports and proves (iii).
\end{proof}

\begin{lem}\label{lem:ground-state-information-convex}
Let $X\Subset\mL$ be regular and set $Y:=X^{+{l}_0+1}$.
For every density matrix $\rho_Y$ satisfying
\begin{equation}
 \rho_Y=P_Y\rho_Y P_Y,
\end{equation}
we have $\rho_X\in\Sigma(X)$.
\end{lem}

\begin{proof}
Fix a face $f\in X$ and set $K_f:=D_f^{+{l}_0}$.
Both $D_f$ and $K_f$ are disks, and
\begin{equation}
 K_f\subset Y,
 \qquad
 d(D_f,\mL\setminus K_f)={l}_0+1\geq{l}_0.
\end{equation}
By Lemma~\ref{lem:GSprojection}(i), the marginal $\rho_{K_f}$ is a ground state
of $H_{K_f}$. Theorem~\ref{thm:commutingparentH}(iv), applied to
$D_0=D_f$ and $D=K_f$, therefore provides
\begin{equation}
 \rho_{D_f}=\omega_{D_f}.
\end{equation}
Since this holds for every $f\in X$ and $X^+\subset Y$,
the marginal $\rho_{X^+}$ belongs to $\widetilde\Sigma(X)$.
Its restriction to $X$ consequently belongs to $\Sigma(X)$.
\end{proof}

\begin{prop}
    Suppose a regular annulus $A'$ is an elementary enlargement of $A$, and 
    $X\Subset\mL$ satisfies $A'^{+l_0+1}\subset X$. 
    Assume that 
    all annuli satisfy (1) or (2) of
Theorem~\ref{thm:annlusconvex}. 
Then it holds that
    \begin{equation}
        {S_{A'}^a}P_X={S_A^a}P_X\,,\quad\forall a\in\cC^{\mathrm{EB}}\,.
    \end{equation}
\end{prop}
\begin{proof}
    The assumption $A'^{+l_0+1}\subset X$ guarantees that for any state $|\psi\rangle\in\supp(P_X)$, $\psi_{A'}\in\Sigma(A')$. Recall that $\sum_{b\in\cC^{\mathrm{EB}}}{S_{A'}^b}$ is the projection onto the support of $\Sigma(A')$. Thus it holds that
    \begin{align}
        \sum_{b\in\cC^{\mathrm{EB}}}{S_{A'}^b}P_X=P_X\,,
    \end{align}
    which completes the proof, since
    \begin{align}
        {S_A^a}P_X={S_A^a}\sum_{b\in\cC^{\mathrm{EB}}}{S_{A'}^b}P_X=\sum_{b\in\cC^{\mathrm{EB}}}\delta_{ab}{S_{A'}^b}P_X={S_{A'}^a}P_X\,.
    \end{align}
\end{proof}
\begin{cor}
    Consider $X\Subset\mL$ and two regular annuli $A_1,A_2\subset X$. Suppose that
     $A_1$ and $A_2$ are connected by a path of elementary deformations $A_1=B_0,\ldots,B_k,\ldots,B_n=A_2$ lying deep inside of $X$, i.e., $\left(\bigcup_kB_k\right)^{+l_0+1}\subset X$.
     Assume that every $B_k$ satisfies condition (1) or (2)
of Theorem~\ref{thm:annlusconvex}.
Then
    \begin{equation}
    {S_{A_1}^a}P_X={S_{A_2}^a}P_X, \quad\forall a\in\cC^{\mathrm{EB}}\,.
    \end{equation}
\end{cor}
\begin{proof}
Apply the preceding proposition to each step of the
deformation path.
\end{proof}

\endgroup

\section{Haag duality}\label{sec:haag-duality}
For a subset $\Lambda\subset\bbL$, we say that the GNS
representation $\pi$ of the reference state $\omega$
satisfies Haag duality with respect to $\Lambda$ if
\[
 \pi(\caA_{\Lambda^c})'=\pi(\caA_\Lambda)''.
\]
In most previous studies, $\Lambda$ was taken to be a planar cone.
However, since the system is defined on a lattice, care is needed
in specifying which lattice sites belong to the cone,
particularly along its boundary.
We prove Haag duality for a suitable lattice version
of cones, namely, lattice cones satisfying Condition M.
See below.

Our proof adapts the finite-volume argument in the proof
of \cite[Lemma~2.2.3]{OPR}.
For each number $M$, choose $\Omega_M\in\caH_{\Gamma_M}$ such that
\[
 \|\Omega_M\|=1,
 \qquad
 P_X\Omega_M=\Omega_M
 \quad
 \text{for every }X\in\mathcal S
 \text{ with }X\subset\Gamma_M.
\]
For $Y\subset\Gamma_M$, let $p_Y^{(M)}$ be the support
projection of the reduced density matrix of $\Omega_M$
on $Y$.
These projections satisfy the martingale property;
see \eqref{martingaele}.

The finite-volume criterion used below can be summarized
as follows.
 For a $*$-algebra $\mathcal B$, write
$\mathcal B^{\mathrm{sa}}:=\{b\in\mathcal B:b^*=b\}$
for its self-adjoint part.

\begin{lemnonum}
Suppose that, for each sufficiently large $\ell$, there
exist $k_\ell\ge\ell$, real-linear maps
\[
\begin{aligned}
 S_\ell^{(1)}&:
 \caA_{(\Lambda^c)_\ell}^{\mathrm{sa}}
 \longrightarrow
 \caA_{(\Lambda^c)_{k_\ell}}^{\mathrm{sa}},\\
 S_\ell^{(2)}&:
 \caA_{(\Lambda^c)_\ell}^{\mathrm{sa}}
 \longrightarrow
 \caA_{(\Lambda)_{k_\ell}}^{\mathrm{sa}},
\end{aligned}
\]
and a complex-linear map
\[
 \Theta_\ell:
 \caA_{\Gamma_\ell}
 \longrightarrow
 \caA_{(\Lambda^c)_{k_\ell}}
\]
such that
\[
\begin{aligned}
 i\,p_{(\Lambda^c)_M}^{(M)}A\Omega_M
 ={}&
 p_{(\Lambda^c)_M}^{(M)}S_\ell^{(1)}(A)\Omega_M\\
 &+
 i\,p_{\Lambda_M}^{(M)}S_\ell^{(2)}(A)\Omega_M
\end{aligned}
\]
for every $A\in\caA_{(\Lambda^c)_\ell}^{\mathrm{sa}}$, 
and
\[
 p_{\Lambda_M}^{(M)}B\Omega_M
 =
 \Theta_\ell(B)\Omega_M
\]
for every $B\in\caA_{\Gamma_\ell}$.
These identities are required to hold for all sufficiently
large $M$.
Then Haag duality holds for $\Lambda$.
\end{lemnonum}
The proof follows the argument of
\cite[Lemma~2.2.3]{OPR}.

 The second identity says that applying a local observable
to $\Omega_M$ and then compressing by the cone-side support
projection can be realized by a local operator in the
complement. 
The maps $S_\ell^{(1)}$ and $S_\ell^{(2)}$ provide the
corresponding control of self-adjoint parts.
The essential requirement is that the representing
operators have fixed finite supports (specified by $k_\ell$) and can be chosen
independently of the ambient volume $M$.

For each fixed finite volume, Tomita--Takesaki theory does give such maps; see Lemma~\ref{lem:haag-modular-decomposition}. But in general, we do not know if
these locality and uniformity requirements are satisfied.
The maps are given in terms of the modular automorphisms and modular conjugation
of
\[
 p_{\Lambda_M}^{(M)}
 \caA_{\Lambda_M}
 p_{\Lambda_M}^{(M)},
\]
for which $\Omega_M$ is cyclic and separating on the
cyclic subspace generated by this algebra.
Hence, it remains to prove locality of the modular
automorphisms and modular conjugation, uniformly in the
ambient volume, using the property of $\omega$.

In our setting, the Markov property of our state $\omega$ supplies the required
locality.
Lemma~\ref{lem:haag-markov-modular} shows that the modular
action on a local observable compressed by the support projection is
determined by a fixed enlarged region.
Lemma~\ref{lem:haag-selfadjoint-mirror} provides a self-adjoint operator in the complement whose action, after support compression, gives the vector obtained by modular conjugation.
The underlying Markov-state identity already appears in
Lemma~\ref{lem:petz-markov-modular}.
\subsection{Lattice cones and Haag duality}
We first define planar cones. 
For
each $\bm a\in \bbR^2$, $\theta\in\bbR$ and $\varphi\in (0,\pi)$,
we set 
\begin{align*}
\Lambda_{\bm a, \theta,\varphi}
:
=& \bm a+\left\{
t\bm e_{\beta}\mid t>0,\quad \beta\in (\theta-\varphi,\theta+\varphi)\right\}
\end{align*}
where  $\bm e_{\beta}=(\cos\beta,\sin\beta)$.
We call a set of this form a {(planar)} cone.
We set $\arg\Lambda_{\bm a, \theta,\varphi}:=(\theta-\varphi,\theta+\varphi)$, where the
 {right-hand} side should be understood $\mod 2\pi$.
We also set $|\arg\Lambda|=2\varphi$
{and} $\bm e_\Lambda:=\bm e_{\theta}$ for $\Lambda=\Lambda_{\bm a, \theta,\varphi}$.
For $0<\varepsilon<\pi-\varphi$, $t\in \bbR$ and $\Lambda=\Lambda_{\bm a, \theta,\varphi}$, {we set}
$\Lambda_\varepsilon=\Lambda_{\bm a, \theta,\varphi+\varepsilon}$ {and}
$\Lambda(t):=\ld +t {\bm e}_\ld$.

 We write
\begin{equation}\label{eq:strict-angular-inclusion}
 \Lambda\ll\widehat\Lambda
 \quad\Longleftrightarrow\quad
 \Lambda\subset\widehat\Lambda\ \text{ and }\ 
 \overline {\arg({\Lambda})}\subset \arg (\widehat\Lambda).
\end{equation}
Thus, the inclusion has a strict angular buffer; the apexes  need not
coincide. We say that two cones are angularly separated if their closed
intervals of directions are disjoint. For $r\ge0$, we write
\begin{equation}\label{eq:cone-separation}
 \Lambda_1\perp_r\Lambda_2
 \quad\Longleftrightarrow\quad
 \overline{\arg {\Lambda_1}}\cap\overline{\arg {\Lambda_2}}=\varnothing
 \ \text{ and }\ d(\Lambda_1,\Lambda_2):= d\bigl(\mathfrak G(\Lambda_1),\mathfrak G(\Lambda_2)\bigr)
\ge r,
\end{equation}

For
$\theta_0\in\bbR$ and $\varphi_0\in (0,\pi)$ we consider the following set of cones:
\begin{align}\label{kyuri}
\begin{split}
&\caC_{(\theta_0,\varphi_0)}:=\left\{
\Lambda : \text{cone}\mid \overline{\arg\Lambda}\cap{[\theta_0-\varphi_0,\theta_0+\varphi_0]}=\emptyset
\right\}.
\end{split}
\end{align}
Again, the equality in the parentheses should be understood $\mod 2\pi$.
Note that $\caC_{(\theta_0,\varphi_0)}$ is an upward-filtering set with respect to the inclusion relation.

{For a planar cone $\Lambda$, its associated lattice region is
${\mathfrak G}(\ld)$.
Whenever a planar cone occurs in an observable algebra,
a graph distance, a graph neighborhood, or a set operation
with a lattice region, we occasionally use the same symbol $\ld$
instead of ${\mathfrak G}(\ld)$.
 In such expressions, the complement of $\ld$ in $\bbL$ means
 ${\mathfrak G}(\ld)^c$.
(Angular conditions, translations, and the strict angular
inclusion relation $\ll$ defined above refer to the planar
cones themselves.)
}
The lattice region associated with a planar cone need not have the
geometric properties required below. We therefore introduce lattice
cones.

\begin{haagdefinition}[Lattice cones]\label{def:haag-lattice-cone}
A lattice region is called a \emph{lattice cone} if it has one of
the forms
\[
 \mathfrak F(C)^{+j},
 \qquad
 \bigl(\mathfrak F(C)^{+j}\bigr)^c,
\]
where $C\subset\mathbb R^2$ is a closed convex cone
of opening in $(0,\pi]$ and $j\ge0$ is an integer.
A lattice cone is called \emph{good} if it admits one of the
representations in Definition~\ref{def:haag-lattice-cone}
with $j=0$.
\end{haagdefinition}

\begin{defn}[Condition M]\label{conm}
We say that a lattice cone $\tilde \Lambda$ satisfies {\it Condition M} 
if there exist $l_{\tilde \ld}, r_{\tilde \ld}\in\bbN$ with $l_{\tilde \ld}\ge l_0\ge r$
and $ r_{\tilde \ld}\ge {l_0}$
satisfying the following.
Set $L:=\tilde\ld$ and $R:=\tilde\ld^c$,
and $L_k:=L\cap \Gamma_{k}$, 
$R_k:=R\cap \Gamma_k$ for each $k\in\bbN$.
Then  
for each $l, N\in\bbN$ with
\begin{align}
l_{\tilde \ld}\le l,\quad
 r_{\tilde \ld}\le N-l
\end{align}
we have
\begin{align}
\begin{split}
&\bbA_1^\omega\lmk  L_N\setminus L_l, \; L_l, \; R_N\rmk=0,\\
&\bbA_1^\omega\lmk  R_N\setminus R_l, \; R_l, \; L_N\rmk=0.
\end{split}
\end{align}
\end{defn}
Together with local uniqueness, Condition M ensures that
the finite-volume pure states introduced below satisfy
the quantum Markov property for the partitions in
Figure~\ref{fig:haag-setting};
see Lemma~\ref{lem:haag-cmi}.
\begin{defn}[Condition D]\label{notation:condition-d}
We say that a lattice cone
$\Lambda\subset\mathbb L$ satisfies Condition D if there exists a
strictly increasing sequence of positive integers $n_N\to\infty$
such that
\[
 K_N:=\Lambda\cap\Gamma_{n_N}
\]
is a finite disk for every $N\ge1$.

\end{defn}
{\begin{lemma}\label{oka}

Let $C\subset\mathbb R^2$ be a  closed convex cone
of opening in $(0,\pi]$, and set
\[
 L=\mathfrak F(C),\qquad R=L^c,\qquad
 L_n=L\cap\Gamma_n,\qquad R_n=R\cap\Gamma_n
 \quad(n\in\mathbb Z_{\ge0}).
\]
There exist a function
\[
 \ell_C:\mathbb Z_{\ge0}\longrightarrow\mathbb Z_{\ge1}
\]
and an integer $n_C\ge1$ such that the following hold.

\begin{enumerate}
\item
For every $j\in\mathbb Z_{\ge0}$, the regions
$L^{+j}$ and $(L^{+j})^c$ are lattice cones.
For all integers
\[
 N>l\ge\ell_C(j),
\]
the four regions
\[
\begin{gathered}
 L^{+j}\cap\Gamma_l,
 \qquad
 (L^{+j})^c\cap\Gamma_l,\\
 L^{+j}\cap(\Gamma_N\setminus\Gamma_l),
 \qquad
 (L^{+j})^c\cap(\Gamma_N\setminus\Gamma_l)
\end{gathered}
\]
are disks.

\item
For all integers $n\ge n_C$,
\[
\begin{aligned}
 &L_n^{+j}\text{ is a disk}
 &&(j\in\mathbb Z_{\ge0}),\\
 &(L_n^{+j})^{-1}\text{ is a disk}
 &&(j\in\mathbb Z_{\ge1}).
\end{aligned}
\]

\item
For every $j\in\mathbb Z_{\ge0}$, both
\[
 L^{+j}
 \qquad\text{and}\qquad
 (L^{+j})^c
\]
satisfy Conditions M and D.

\item
Let $\Lambda\subset\mathbb R^2$ be any open planar cone with
\[
 \Lambda\subset\operatorname{int}C,
 \qquad
 \overline{\arg\Lambda}
 \subset\arg(\operatorname{int}C),
\]
where $\arg$ denotes the interval of cone directions on the unit
circle. 
Then
\[
 \mathfrak G(\ld)\subset L,
 \qquad
 d(\mathfrak G(\ld)\setminus L_n,R)
 \longrightarrow\infty
 \quad(n\longrightarrow\infty).
\]
Equivalently, for every integer $q\ge0$, there is an integer
$n_0=n_0(C,\Lambda,q)\ge1$ such that
\[
 n\ge n_0
 \quad\Longrightarrow\quad
 d(\mathfrak G(\ld)\setminus L_n,R)>q
 \qquad(n\in\mathbb Z_{\ge0}).
\]
\end{enumerate}
\end{lemma}

\begin{proof}
    See Appendix~\ref{app:good-lattice-cones}\,.
\end{proof}
This lemma provides the lattice cones needed below, all satisfying Condition
M. 
\begin{lem}\label{rem:lattice-cone-approximation}
{For every $\varphi\in(0,\pi)$ and $\varepsilon>0$ satisfying
$\varphi+2\varepsilon<\pi$, there exists a constant
$R_{\varphi,\varepsilon}>0$ such that, for every cone
$\Lambda$ with {half-opening angle} $\varphi$,}
there are lattice cones
$\widetilde\Lambda_1,\widetilde\Lambda_2$
satisfying Condition M {and Condition D} such that
\[
  {|\widetilde\Lambda_1|}
  \subset \Lambda
  \subset {|\widetilde\Lambda_2|}
  {\subset
  \bigl(\Lambda-R_{\varphi,\varepsilon}e_\Lambda\bigr)_\varepsilon}.
\]
{The complements $\widetilde\Lambda_1^{\,c}$ and
$\widetilde\Lambda_2^{\,c}$, taken in the full face lattice,
also satisfy Conditions D and M.
The constant $R_{\varphi,\varepsilon}$ is independent of
the apex and the axial direction of $\Lambda$.
}
\end{lem}
\begin{proof}
{Indeed, take $t=2d_{\mathrm{hex}}/\sin\varphi$.
If $2\varphi\le\pi$, set
\[
 \widetilde\Lambda_1=\mathfrak F(\overline\Lambda+t e_\Lambda),
 \qquad \widetilde\Lambda_2=\mathfrak F(\overline\Lambda).
\]
If $2\varphi>\pi$, put $C=\mathbb R^2\setminus\Lambda$ and set
\[
 \widetilde\Lambda_1=\mathfrak F(C)^c,\qquad
 \widetilde\Lambda_2=\mathfrak F(C-t e_\Lambda)^c.
\]
The axial shifts leave a margin greater than one face diameter,
so in both cases
$|\widetilde\Lambda_1|\subset\Lambda\subset|\widetilde\Lambda_2|
\subset\Lambda-t e_\Lambda$.
Lemma~\ref{oka} supplies Conditions M and D,
also for the complements. Thus $R_{\varphi,\varepsilon}=t$ suffices.
}
\end{proof}

For lattice cones we consider 
Haag duality.
\begin{defn}
We say {that} $\omega$ satisfies Haag duality for a lattice cone $\tilde\ld$ {if}
\[
\pi(\caA_{\tilde \Lambda^c})'=
\pi\lmk\caA_{\tilde \Lambda}\rmk''
\]
holds.
\end{defn}
There is a weaker {notion} of {Haag} duality, called approximate Haag duality \cite{Ogata}.

{The main theorem} of this section {is} the following.
\begin{thm}\label{haag}
Consider the setting in Subsection~\ref{overallsetting}.
Then Haag duality holds for lattice cones
satisfying {\it Condition M}.
In particular, $\omega$ satisfies approximate Haag duality for any planar cones.
\end{thm}

Throughout this section $\tilde \ld$ denotes a given lattice cone
satisfying {\it Condition M}.
For each $j\in \bbN$, we set
\[
L_j:=\Gamma_j\cap\widetilde\Lambda,\qquad
 R_j:=\Gamma_j\cap\widetilde\Lambda^c.
 \]

We adapt \cite[Lemma~2.2.3]{OPR} using Tomita--Takesaki theory~\cite{BR1}. A proof of Haag duality using Tomita--Takesaki theory
in {the} Levin--Wen model was considered in \cite{NPW}.
{However}{, our use of Tomita}{--}{Takesaki theory is different from that of \cite{NPW}}{.}
Here we derive locality of modular data using the Markov property of our
$\omega$.

\subsection{Finite-volume states and modular objects}
First{,} we introduce the {data of} Tomita--Takesaki theory {for finite-dimensional systems}.
We use the {graph-distance disks} $\Gamma_M$
defined in Subsection~\ref{overallsetting}.
For each $M$, choose $\Omega_M\in \caH_{\Gamma_M}$
such that 
\begin{align}\label{OM}
\|\Omega_M\|=1,
 \ P_F\Omega_M=\Omega_M\text{ for every }F\in\mathcal S
 \text{ with }F\subset\Gamma_M.
\end{align}
Such a {vector} $\Omega_M$ exists because of the existence of $\omega$.
Set
\[
 \sigma^{(M)}:=|\Omega_M\rangle\langle\Omega_M|,
\]
and for every nonempty $Y\subset\Gamma_M$, set
\[
\begin{aligned}
 \sigma_Y^{(M)}&:=\tr_{\Gamma_M\setminus Y}\sigma^{(M)},&
 p_Y^{(M)}&:=\operatorname{proj}_{\operatorname{supp}(\sigma_Y^{(M)})},\\
 \mathcal M_Y^{(M)}&:=p_Y^{(M)}\A_Y p_Y^{(M)},&
 \mathcal K_Y^{(M)}&:=\mathcal M_Y^{(M)}\Omega_M.
\end{aligned}
\]
For $z\in\mathbb C$, $\sigma_Y^z$ denotes the complex power
of $\sigma_Y$ defined by spectral calculus on
$p_Y\mathcal H_Y$, extended by zero on its orthogonal
complement.

The vector $\Omega_M$ is cyclic and separating for $\mathcal M_Y^{(M)}$.
Let
\[
 \Delta_Y^{(M)},J_Y^{(M)}:
 \mathcal K_Y^{(M)}\longrightarrow\mathcal K_Y^{(M)}
\]
be the modular operator and conjugation for the pair
$(\mathcal M_Y^{(M)},\Omega_M)$. The former is positive and invertible;
the latter is antiunitary. By Tomita--Takesaki theory, define the conjugate-linear map
\[
 j_Y^{(M)}:\mathcal M_Y\longrightarrow
 (\mathcal M_Y)',\qquad
 j_Y^{(M)}(a):=J_Y^{(M)}aJ_Y^{(M)},
\]
where the commutant is taken in $B(\mathcal K_Y^{(M)})$.
Superscripts $(M)$ of $ \sigma_Y^{(M)}$,
$ p_Y^{(M)}$,
$ \mathcal M_Y^{(M)}$,
$ \mathcal K_Y^{(M)}$,$ j_Y^{(M)}$,$J_Y^{(M)}$, $\Delta_Y^{(M)}$
may be suppressed below.

\subsection{Locality of modular actions}\label{lma}

In this subsection, {we} fix $l, m, N, k,M\in\bbN$
so that
\begin{align}
\begin{split}
l\ge l_{\tilde \ld},\;
 m-l\ge r_{\tilde \ld}, \; N-m\ge r_{\tilde \ld}, \; k-N\ge r_{\tilde \ld},\;  M-k\ge r_{\tilde \ld}.
\end{split}
\end{align}
Here, $l_{\td\ld}$, $r_{\td\ld}$ are constants in Definition \ref{conm}.
We specify the regions $A,B,C,C',D$ as in
{Figures}~{\ref{fig:haag-setting} and~\ref{fig:haag-lemma7-regions}}.
The Condition M gurantees the following Markov property.
\begin{haaglemma}\label{lem:haag-cmi}
\[
 I(A:C'| B)_{\sigma}=I(A:C| B)_{\sigma}=0.
\]
\end{haaglemma}

\begin{proof}
Since $M-k\ge r_{\tilde\ld}\ge {l_0}$,
we have $\Gamma_k^{+l_0}\subset\Gamma_M$.
Hence, by Theorem~\ref{thm:commutingparentH}(iv),

\begin{equation}\label{eq:haag-local-uniqueness}
 \sigma_{\Gamma_k}=\omega_{\Gamma_k}.
\end{equation}
Because {the state defined by} $\Omega_M$ {is pure}, we have
\[
\begin{aligned}
 I(A:C'| B)_{\sigma}
 &=S(AB)_\sigma+S(BC')_\sigma-S(B)_\sigma-S(ABC')_\sigma\\
 &=S(AB)_\sigma+S(AD)_\sigma-S(B)_\sigma-S(D)_\sigma\\
  &=S(AB)_\omega+S(AD)_\omega-S(B)_\omega-S(D)_\omega\\
 &=\bbA_1^\omega\lmk B, A, D\rmk=0.
\end{aligned}
\]
The last equality follows from Condition M.

Finally, $C\subset C'$, so data processing and strong subadditivity imply
\[
 0\le I(A:C| B)_{\sigma}\le I(A:C'| B)_{\sigma}=0.
\]
This proves both identities.
\end{proof}

\begin{figure}[!htbp]
\centering
\resizebox{.48\textwidth}{!}{%
\begingroup
\definecolor{gdA}{HTML}{C4EB8C}
\definecolor{gdB}{HTML}{ADE4EA}
\definecolor{gdC}{HTML}{7DB473}
\definecolor{gdD}{HTML}{F2CCD8}
\definecolor{gdCprime}{HTML}{FFF579}
\definecolor{gdwhite}{HTML}{FFFFFF}
\definecolor{gdmesh}{HTML}{D9E4E6}
\definecolor{gdordinary}{HTML}{005F5F}
\definecolor{gdm}{HTML}{85B200}
\definecolor{gdk}{HTML}{C60000}
\definecolor{gdblack}{HTML}{000000}
\begin{tikzpicture}[x=3.6pt,y=3.6pt,line join=round,line cap=round]
\path[use as bounding box] (-26.0,-22.5) rectangle (26.0,26.1);
\foreach \gdi/\gdj in {-14/0,-14/1,-14/2,-14/3,-14/4,-14/5,-14/6,-14/7,-14/8,-14/9,-14/10,-14/11,-14/12,-14/13,-14/14,-13/-1,-13/0,-13/1,-13/2,-13/3,-13/4,-13/5,-13/6,-13/7,-13/8,-13/9,-13/10,-13/11,-13/12,-13/13,-13/14,-12/-2,-12/-1,-12/0,-12/1,-12/2,-12/3,-12/4,-12/5,-12/6,-12/7,-12/8,-12/9,-12/10,-12/11,-12/12,-12/13,-12/14,-11/-3,-11/-2,-11/-1,-11/0,-11/1,-11/2,-11/3,-11/4,-11/5,-11/6,-11/7,-11/8,-11/9,-11/10,-11/11,-11/12,-11/13,-11/14,-10/-4,-10/-3,-10/-2,-10/-1,-10/11,-10/12,-10/13,-10/14,-9/-5,-9/-4,-9/-3,-9/-2,-9/11,-9/12,-9/13,-9/14,-8/-6,-8/-5,-8/-4,-8/-3,-8/11,-8/12,-8/13,-8/14,-7/-7,-7/-6,-7/-5,-7/-4,-7/11,-7/12,-6/-8,-6/-7,-6/-6,-6/-5,-5/-9,-5/-8,-5/-7,-5/-6,-4/-10,-4/-9,-4/-8,-4/-7,-3/-11,-3/-10,-3/-9,-3/-8,-2/-12,-2/-11,-2/-10,-2/-9,-1/-13,-1/-12,-1/-11,-1/-10,0/-14,0/-13,0/-12,0/-11,1/-14,1/-13,1/-12,1/-11,1/10,1/11,1/12,1/13,2/-14,2/-13,2/-12,2/-11,2/9,2/10,2/11,2/12,3/-14,3/-13,3/-12,3/-11,3/8,3/9,3/10,3/11,4/-14,4/-13,4/-12,4/-11,4/7,4/8,4/9,4/10,5/-14,5/-13,5/-12,5/-11,5/6,5/7,5/8,5/9,6/-14,6/-13,6/-12,6/-11,6/5,6/6,6/7,6/8,7/-14,7/-13,7/-12,7/-11,7/4,7/5,7/6,7/7,8/-14,8/-13,8/-12,8/-11,8/3,8/4,8/5,8/6,9/-14,9/-13,9/-12,9/-11,9/2,9/3,9/4,9/5,10/-14,10/-13,10/-12,10/-11,10/1,10/2,10/3,10/4,11/-14,11/-13,11/-12,11/-11,11/-10,11/-9,11/-8,11/-7,11/-6,11/-5,11/-4,11/-3,11/-2,11/-1,11/0,11/1,11/2,11/3,12/-14,12/-13,12/-12,12/-11,12/-10,12/-9,12/-8,12/-7,12/-6,12/-5,12/-4,12/-3,12/-2,12/-1,12/0,12/1,12/2,13/-14,13/-13,13/-12,13/-11,13/-10,13/-9,13/-8,13/-7,13/-6,13/-5,13/-4,13/-3,13/-2,13/-1,13/0,13/1,14/-14,14/-13,14/-12,14/-11,14/-10,14/-9,14/-8,14/-7,14/-6,14/-5,14/-4,14/-3,14/-2,14/-1,14/0} {%
  \pgfmathsetmacro{\gdx}{sqrt(3)*(\gdi+\gdj/2)}%
  \pgfmathsetmacro{\gdy}{1.5*\gdj}%
  \begin{scope}[shift={(\gdx,\gdy)}]
    \path[fill=gdwhite,draw=gdmesh,line width=.12pt] (0.86603,0.5) -- (0,1) -- (-0.86603,0.5) -- (-0.86603,-0.5) -- (0,-1) -- (0.86603,-0.5) -- cycle;
  \end{scope}}
\foreach \gdi/\gdj in {-2/3,-2/4,-1/1,-1/2,-1/3,-1/4,0/0,0/1,0/2,0/3,0/4} {%
  \pgfmathsetmacro{\gdx}{sqrt(3)*(\gdi+\gdj/2)}%
  \pgfmathsetmacro{\gdy}{1.5*\gdj}%
  \begin{scope}[shift={(\gdx,\gdy)}]
    \path[fill=gdA,draw=gdmesh,line width=.12pt] (0.86603,0.5) -- (0,1) -- (-0.86603,0.5) -- (-0.86603,-0.5) -- (0,-1) -- (0.86603,-0.5) -- cycle;
  \end{scope}}
\foreach \gdi/\gdj in {-5/9,-5/10,-4/7,-4/8,-4/9,-4/10,-3/5,-3/6,-3/7,-3/8,-3/9,-3/10,-2/5,-2/6,-2/7,-2/8,-2/9,-2/10,-1/5,-1/6,-1/7,-1/8,-1/9,-1/10,0/5,0/6,0/7,0/8,0/9,0/10} {%
  \pgfmathsetmacro{\gdx}{sqrt(3)*(\gdi+\gdj/2)}%
  \pgfmathsetmacro{\gdy}{1.5*\gdj}%
  \begin{scope}[shift={(\gdx,\gdy)}]
    \path[fill=gdB,draw=gdmesh,line width=.12pt] (0.86603,0.5) -- (0,1) -- (-0.86603,0.5) -- (-0.86603,-0.5) -- (0,-1) -- (0.86603,-0.5) -- cycle;
  \end{scope}}
\foreach \gdi/\gdj in {-7/13,-7/14,-6/11,-6/12,-6/13,-6/14,-5/11,-5/12,-5/13,-5/14,-4/11,-4/12,-4/13,-4/14,-3/11,-3/12,-3/13,-3/14,-2/11,-2/12,-2/13,-2/14,-1/11,-1/12,-1/13,-1/14,0/11,0/12,0/13,0/14} {%
  \pgfmathsetmacro{\gdx}{sqrt(3)*(\gdi+\gdj/2)}%
  \pgfmathsetmacro{\gdy}{1.5*\gdj}%
  \begin{scope}[shift={(\gdx,\gdy)}]
    \path[fill=gdC,draw=gdmesh,line width=.12pt] (0.86603,0.5) -- (0,1) -- (-0.86603,0.5) -- (-0.86603,-0.5) -- (0,-1) -- (0.86603,-0.5) -- cycle;
  \end{scope}}
\foreach \gdi/\gdj in {-10/0,-10/1,-10/2,-10/3,-10/4,-10/5,-10/6,-10/7,-10/8,-10/9,-10/10,-9/-1,-9/0,-9/1,-9/2,-9/3,-9/4,-9/5,-9/6,-9/7,-9/8,-9/9,-9/10,-8/-2,-8/-1,-8/0,-8/1,-8/2,-8/3,-8/4,-8/5,-8/6,-8/7,-8/8,-8/9,-8/10,-7/-3,-7/-2,-7/-1,-7/0,-7/1,-7/2,-7/3,-7/4,-7/5,-7/6,-7/7,-7/8,-7/9,-7/10,-6/-4,-6/-3,-6/-2,-6/-1,-6/0,-6/1,-6/2,-6/3,-6/4,-6/5,-6/6,-6/7,-6/8,-6/9,-6/10,-5/-5,-5/-4,-5/-3,-5/-2,-5/-1,-5/0,-5/1,-5/2,-5/3,-5/4,-5/5,-5/6,-5/7,-5/8,-4/-6,-4/-5,-4/-4,-4/-3,-4/-2,-4/-1,-4/0,-4/1,-4/2,-4/3,-4/4,-4/5,-4/6,-3/-7,-3/-6,-3/-5,-3/-4,-3/-3,-3/-2,-3/-1,-3/0,-3/1,-3/2,-3/3,-3/4,-2/-8,-2/-7,-2/-6,-2/-5,-2/-4,-2/-3,-2/-2,-2/-1,-2/0,-2/1,-2/2,-1/-9,-1/-8,-1/-7,-1/-6,-1/-5,-1/-4,-1/-3,-1/-2,-1/-1,-1/0,0/-10,0/-9,0/-8,0/-7,0/-6,0/-5,0/-4,0/-3,0/-2,0/-1,1/-10,1/-9,1/-8,1/-7,1/-6,1/-5,1/-4,1/-3,1/-2,1/-1,1/0,1/1,1/2,1/3,1/4,1/5,1/6,1/7,1/8,1/9,2/-10,2/-9,2/-8,2/-7,2/-6,2/-5,2/-4,2/-3,2/-2,2/-1,2/0,2/1,2/2,2/3,2/4,2/5,2/6,2/7,2/8,3/-10,3/-9,3/-8,3/-7,3/-6,3/-5,3/-4,3/-3,3/-2,3/-1,3/0,3/1,3/2,3/3,3/4,3/5,3/6,3/7,4/-10,4/-9,4/-8,4/-7,4/-6,4/-5,4/-4,4/-3,4/-2,4/-1,4/0,4/1,4/2,4/3,4/4,4/5,4/6,5/-10,5/-9,5/-8,5/-7,5/-6,5/-5,5/-4,5/-3,5/-2,5/-1,5/0,5/1,5/2,5/3,5/4,5/5,6/-10,6/-9,6/-8,6/-7,6/-6,6/-5,6/-4,6/-3,6/-2,6/-1,6/0,6/1,6/2,6/3,6/4,7/-10,7/-9,7/-8,7/-7,7/-6,7/-5,7/-4,7/-3,7/-2,7/-1,7/0,7/1,7/2,7/3,8/-10,8/-9,8/-8,8/-7,8/-6,8/-5,8/-4,8/-3,8/-2,8/-1,8/0,8/1,8/2,9/-10,9/-9,9/-8,9/-7,9/-6,9/-5,9/-4,9/-3,9/-2,9/-1,9/0,9/1,10/-10,10/-9,10/-8,10/-7,10/-6,10/-5,10/-4,10/-3,10/-2,10/-1,10/0} {%
  \pgfmathsetmacro{\gdx}{sqrt(3)*(\gdi+\gdj/2)}%
  \pgfmathsetmacro{\gdy}{1.5*\gdj}%
  \begin{scope}[shift={(\gdx,\gdy)}]
    \path[fill=gdD,draw=gdmesh,line width=.12pt] (0.86603,0.5) -- (0,1) -- (-0.86603,0.5) -- (-0.86603,-0.5) -- (0,-1) -- (0.86603,-0.5) -- cycle;
  \end{scope}}
\path[draw=gdblack,line width=.75pt]
  (-1.73205,1) -- (-1.73205,2)
  (-1.73205,1) -- (-0.86603,0.5)
  (-1.73205,2) -- (-0.86603,2.5)
  (-1.73205,4) -- (-1.73205,5)
  (-1.73205,4) -- (-0.86603,3.5)
  (-1.73205,5) -- (-0.86603,5.5)
  (-1.73205,7) -- (-1.73205,8)
  (-1.73205,7) -- (-0.86603,6.5)
  (-1.73205,8) -- (-0.86603,8.5)
  (-1.73205,10) -- (-1.73205,11)
  (-1.73205,10) -- (-0.86603,9.5)
  (-1.73205,11) -- (-0.86603,11.5)
  (-1.73205,13) -- (-1.73205,14)
  (-1.73205,13) -- (-0.86603,12.5)
  (-1.73205,14) -- (-0.86603,14.5)
  (-1.73205,16) -- (-1.73205,17)
  (-1.73205,16) -- (-0.86603,15.5)
  (-1.73205,17) -- (-0.86603,17.5)
  (-1.73205,19) -- (-1.73205,20)
  (-1.73205,19) -- (-0.86603,18.5)
  (-1.73205,20) -- (-0.86603,20.5)
  (-0.86603,-0.5) -- (-0.86603,0.5)
  (-0.86603,-0.5) -- (0,-1)
  (-0.86603,2.5) -- (-0.86603,3.5)
  (-0.86603,5.5) -- (-0.86603,6.5)
  (-0.86603,8.5) -- (-0.86603,9.5)
  (-0.86603,11.5) -- (-0.86603,12.5)
  (-0.86603,14.5) -- (-0.86603,15.5)
  (-0.86603,17.5) -- (-0.86603,18.5)
  (-0.86603,20.5) -- (-0.86603,21.5)
  (0,-1) -- (0.86603,-0.5)
  (0.86603,-0.5) -- (0.86603,0.5)
  (0.86603,0.5) -- (1.73205,1)
  (1.73205,1) -- (1.73205,2)
  (1.73205,2) -- (2.59808,2.5)
  (2.59808,2.5) -- (2.59808,3.5)
  (2.59808,3.5) -- (3.4641,4)
  (3.4641,4) -- (3.4641,5)
  (3.4641,5) -- (4.33013,5.5)
  (4.33013,5.5) -- (4.33013,6.5)
  (4.33013,6.5) -- (5.19615,7)
  (5.19615,7) -- (5.19615,8)
  (5.19615,8) -- (6.06218,8.5)
  (6.06218,8.5) -- (6.06218,9.5)
  (6.06218,9.5) -- (6.9282,10)
  (6.9282,10) -- (6.9282,11)
  (6.9282,11) -- (7.79423,11.5)
  (7.79423,11.5) -- (7.79423,12.5)
  (7.79423,12.5) -- (8.66025,13)
  (8.66025,13) -- (8.66025,14)
  (8.66025,14) -- (9.52628,14.5)
  (9.52628,14.5) -- (9.52628,15.5)
  (9.52628,15.5) -- (10.3923,16)
  (10.3923,16) -- (10.3923,17)
  (10.3923,17) -- (11.25833,17.5)
  (11.25833,17.5) -- (11.25833,18.5)
  (11.25833,18.5) -- (12.12436,19)
  (12.12436,19) -- (12.12436,20)
  (12.12436,20) -- (12.99038,20.5);
\draw[draw=gdordinary,line width=.65pt] (-25.11474,-0.5) -- (-25.11474,0.5) -- (-24.24871,1) -- (-24.24871,2) -- (-23.38269,2.5) -- (-23.38269,3.5) -- (-22.51666,4) -- (-22.51666,5) -- (-21.65064,5.5) -- (-21.65064,6.5) -- (-20.78461,7) -- (-20.78461,8) -- (-19.91858,8.5) -- (-19.91858,9.5) -- (-19.05256,10) -- (-19.05256,11) -- (-18.18653,11.5) -- (-18.18653,12.5) -- (-17.32051,13) -- (-17.32051,14) -- (-16.45448,14.5) -- (-16.45448,15.5) -- (-15.58846,16) -- (-15.58846,17) -- (-14.72243,17.5) -- (-14.72243,18.5) -- (-13.85641,19) -- (-13.85641,20) -- (-12.99038,20.5) -- (-12.99038,21.5) -- (-12.12436,22) -- (-11.25833,21.5) -- (-10.3923,22) -- (-9.52628,21.5) -- (-8.66025,22) -- (-7.79423,21.5) -- (-6.9282,22) -- (-6.06218,21.5) -- (-5.19615,22) -- (-4.33013,21.5) -- (-3.4641,22) -- (-2.59808,21.5) -- (-1.73205,22) -- (-0.86603,21.5) -- (0,22) -- (0.86603,21.5) -- (1.73205,22) -- (2.59808,21.5) -- (3.4641,22) -- (4.33013,21.5) -- (5.19615,22) -- (6.06218,21.5) -- (6.9282,22) -- (7.79423,21.5) -- (8.66025,22) -- (9.52628,21.5) -- (10.3923,22) -- (11.25833,21.5) -- (12.12436,22) -- (12.99038,21.5) -- (12.99038,20.5) -- (13.85641,20) -- (13.85641,19) -- (14.72243,18.5) -- (14.72243,17.5) -- (15.58846,17) -- (15.58846,16) -- (16.45448,15.5) -- (16.45448,14.5) -- (17.32051,14) -- (17.32051,13) -- (18.18653,12.5) -- (18.18653,11.5) -- (19.05256,11) -- (19.05256,10) -- (19.91858,9.5) -- (19.91858,8.5) -- (20.78461,8) -- (20.78461,7) -- (21.65064,6.5) -- (21.65064,5.5) -- (22.51666,5) -- (22.51666,4) -- (23.38269,3.5) -- (23.38269,2.5) -- (24.24871,2) -- (24.24871,1) -- (25.11474,0.5) -- (25.11474,-0.5) -- (24.24871,-1) -- (24.24871,-2) -- (23.38269,-2.5) -- (23.38269,-3.5) -- (22.51666,-4) -- (22.51666,-5) -- (21.65064,-5.5) -- (21.65064,-6.5) -- (20.78461,-7) -- (20.78461,-8) -- (19.91858,-8.5) -- (19.91858,-9.5) -- (19.05256,-10) -- (19.05256,-11) -- (18.18653,-11.5) -- (18.18653,-12.5) -- (17.32051,-13) -- (17.32051,-14) -- (16.45448,-14.5) -- (16.45448,-15.5) -- (15.58846,-16) -- (15.58846,-17) -- (14.72243,-17.5) -- (14.72243,-18.5) -- (13.85641,-19) -- (13.85641,-20) -- (12.99038,-20.5) -- (12.99038,-21.5) -- (12.12436,-22) -- (11.25833,-21.5) -- (10.3923,-22) -- (9.52628,-21.5) -- (8.66025,-22) -- (7.79423,-21.5) -- (6.9282,-22) -- (6.06218,-21.5) -- (5.19615,-22) -- (4.33013,-21.5) -- (3.4641,-22) -- (2.59808,-21.5) -- (1.73205,-22) -- (0.86603,-21.5) -- (0,-22) -- (-0.86603,-21.5) -- (-1.73205,-22) -- (-2.59808,-21.5) -- (-3.4641,-22) -- (-4.33013,-21.5) -- (-5.19615,-22) -- (-6.06218,-21.5) -- (-6.9282,-22) -- (-7.79423,-21.5) -- (-8.66025,-22) -- (-9.52628,-21.5) -- (-10.3923,-22) -- (-11.25833,-21.5) -- (-12.12436,-22) -- (-12.99038,-21.5) -- (-12.99038,-20.5) -- (-13.85641,-20) -- (-13.85641,-19) -- (-14.72243,-18.5) -- (-14.72243,-17.5) -- (-15.58846,-17) -- (-15.58846,-16) -- (-16.45448,-15.5) -- (-16.45448,-14.5) -- (-17.32051,-14) -- (-17.32051,-13) -- (-18.18653,-12.5) -- (-18.18653,-11.5) -- (-19.05256,-11) -- (-19.05256,-10) -- (-19.91858,-9.5) -- (-19.91858,-8.5) -- (-20.78461,-8) -- (-20.78461,-7) -- (-21.65064,-6.5) -- (-21.65064,-5.5) -- (-22.51666,-5) -- (-22.51666,-4) -- (-23.38269,-3.5) -- (-23.38269,-2.5) -- (-24.24871,-2) -- (-24.24871,-1) -- cycle;
\draw[draw=gdordinary,line width=.65pt] (-18.18653,-0.5) -- (-18.18653,0.5) -- (-17.32051,1) -- (-17.32051,2) -- (-16.45448,2.5) -- (-16.45448,3.5) -- (-15.58846,4) -- (-15.58846,5) -- (-14.72243,5.5) -- (-14.72243,6.5) -- (-13.85641,7) -- (-13.85641,8) -- (-12.99038,8.5) -- (-12.99038,9.5) -- (-12.12436,10) -- (-12.12436,11) -- (-11.25833,11.5) -- (-11.25833,12.5) -- (-10.3923,13) -- (-10.3923,14) -- (-9.52628,14.5) -- (-9.52628,15.5) -- (-8.66025,16) -- (-7.79423,15.5) -- (-6.9282,16) -- (-6.06218,15.5) -- (-5.19615,16) -- (-4.33013,15.5) -- (-3.4641,16) -- (-2.59808,15.5) -- (-1.73205,16) -- (-0.86603,15.5) -- (0,16) -- (0.86603,15.5) -- (1.73205,16) -- (2.59808,15.5) -- (3.4641,16) -- (4.33013,15.5) -- (5.19615,16) -- (6.06218,15.5) -- (6.9282,16) -- (7.79423,15.5) -- (8.66025,16) -- (9.52628,15.5) -- (9.52628,14.5) -- (10.3923,14) -- (10.3923,13) -- (11.25833,12.5) -- (11.25833,11.5) -- (12.12436,11) -- (12.12436,10) -- (12.99038,9.5) -- (12.99038,8.5) -- (13.85641,8) -- (13.85641,7) -- (14.72243,6.5) -- (14.72243,5.5) -- (15.58846,5) -- (15.58846,4) -- (16.45448,3.5) -- (16.45448,2.5) -- (17.32051,2) -- (17.32051,1) -- (18.18653,0.5) -- (18.18653,-0.5) -- (17.32051,-1) -- (17.32051,-2) -- (16.45448,-2.5) -- (16.45448,-3.5) -- (15.58846,-4) -- (15.58846,-5) -- (14.72243,-5.5) -- (14.72243,-6.5) -- (13.85641,-7) -- (13.85641,-8) -- (12.99038,-8.5) -- (12.99038,-9.5) -- (12.12436,-10) -- (12.12436,-11) -- (11.25833,-11.5) -- (11.25833,-12.5) -- (10.3923,-13) -- (10.3923,-14) -- (9.52628,-14.5) -- (9.52628,-15.5) -- (8.66025,-16) -- (7.79423,-15.5) -- (6.9282,-16) -- (6.06218,-15.5) -- (5.19615,-16) -- (4.33013,-15.5) -- (3.4641,-16) -- (2.59808,-15.5) -- (1.73205,-16) -- (0.86603,-15.5) -- (0,-16) -- (-0.86603,-15.5) -- (-1.73205,-16) -- (-2.59808,-15.5) -- (-3.4641,-16) -- (-4.33013,-15.5) -- (-5.19615,-16) -- (-6.06218,-15.5) -- (-6.9282,-16) -- (-7.79423,-15.5) -- (-8.66025,-16) -- (-9.52628,-15.5) -- (-9.52628,-14.5) -- (-10.3923,-14) -- (-10.3923,-13) -- (-11.25833,-12.5) -- (-11.25833,-11.5) -- (-12.12436,-11) -- (-12.12436,-10) -- (-12.99038,-9.5) -- (-12.99038,-8.5) -- (-13.85641,-8) -- (-13.85641,-7) -- (-14.72243,-6.5) -- (-14.72243,-5.5) -- (-15.58846,-5) -- (-15.58846,-4) -- (-16.45448,-3.5) -- (-16.45448,-2.5) -- (-17.32051,-2) -- (-17.32051,-1) -- cycle;
\draw[draw=gdordinary,line width=.65pt] (-7.79423,-0.5) -- (-7.79423,0.5) -- (-6.9282,1) -- (-6.9282,2) -- (-6.06218,2.5) -- (-6.06218,3.5) -- (-5.19615,4) -- (-5.19615,5) -- (-4.33013,5.5) -- (-4.33013,6.5) -- (-3.4641,7) -- (-2.59808,6.5) -- (-1.73205,7) -- (-0.86603,6.5) -- (0,7) -- (0.86603,6.5) -- (1.73205,7) -- (2.59808,6.5) -- (3.4641,7) -- (4.33013,6.5) -- (4.33013,5.5) -- (5.19615,5) -- (5.19615,4) -- (6.06218,3.5) -- (6.06218,2.5) -- (6.9282,2) -- (6.9282,1) -- (7.79423,0.5) -- (7.79423,-0.5) -- (6.9282,-1) -- (6.9282,-2) -- (6.06218,-2.5) -- (6.06218,-3.5) -- (5.19615,-4) -- (5.19615,-5) -- (4.33013,-5.5) -- (4.33013,-6.5) -- (3.4641,-7) -- (2.59808,-6.5) -- (1.73205,-7) -- (0.86603,-6.5) -- (0,-7) -- (-0.86603,-6.5) -- (-1.73205,-7) -- (-2.59808,-6.5) -- (-3.4641,-7) -- (-4.33013,-6.5) -- (-4.33013,-5.5) -- (-5.19615,-5) -- (-5.19615,-4) -- (-6.06218,-3.5) -- (-6.06218,-2.5) -- (-6.9282,-2) -- (-6.9282,-1) -- cycle;
\node[text=gdordinary,fill=white,inner sep=.55pt,font=\fontsize{7}{8}\selectfont] at (-9.8,22.8) {$\Gamma_{M}$};
\node[text=gdordinary,fill=white,inner sep=.55pt,font=\fontsize{7}{8}\selectfont] at (-7,16.8) {$\Gamma_{N}$};
\node[text=gdordinary,fill=white,inner sep=.55pt,font=\fontsize{7}{8}\selectfont] at (-2.8,7.8) {$\Gamma_{l}$};
\node[text=gdblack,inner sep=0pt,font=\fontsize{10}{11}\selectfont] at (1.5,4.1) {{$A$}};
\node[text=gdblack,inner sep=0pt,font=\fontsize{10}{11}\selectfont] at (3.5,11.5) {{$B$}};
\node[text=gdblack,inner sep=0pt,font=\fontsize{10}{11}\selectfont] at (5.7,18.8) {{$C$}};
\node[text=gdblack,inner sep=0pt,font=\fontsize{10}{11}\selectfont] at (-10.3,-0.5) {{$D$}};
\node[text=gdblack,inner sep=0pt,font=\fontsize{9}{10}\selectfont] at (6.5,24.1) {{$\widetilde\Lambda$}};
\end{tikzpicture}
\endgroup
}\hfill
\resizebox{.48\textwidth}{!}{%
\begingroup
\definecolor{gdA}{HTML}{C4EB8C}
\definecolor{gdB}{HTML}{ADE4EA}
\definecolor{gdC}{HTML}{7DB473}
\definecolor{gdD}{HTML}{F2CCD8}
\definecolor{gdCprime}{HTML}{FFF579}
\definecolor{gdwhite}{HTML}{FFFFFF}
\definecolor{gdmesh}{HTML}{D9E4E6}
\definecolor{gdordinary}{HTML}{005F5F}
\definecolor{gdm}{HTML}{85B200}
\definecolor{gdk}{HTML}{C60000}
\definecolor{gdblack}{HTML}{000000}
\begin{tikzpicture}[x=3.6pt,y=3.6pt,line join=round,line cap=round]
\path[use as bounding box] (-26.0,-22.5) rectangle (26.0,26.1);
\foreach \gdi/\gdj in {-2/3,-2/4,-1/1,-1/2,-1/3,-1/4,0/0,0/1,0/2,0/3,0/4} {%
  \pgfmathsetmacro{\gdx}{sqrt(3)*(\gdi+\gdj/2)}%
  \pgfmathsetmacro{\gdy}{1.5*\gdj}%
  \begin{scope}[shift={(\gdx,\gdy)}]
    \path[fill=gdA,draw=gdmesh,line width=.12pt] (0.86603,0.5) -- (0,1) -- (-0.86603,0.5) -- (-0.86603,-0.5) -- (0,-1) -- (0.86603,-0.5) -- cycle;
  \end{scope}}
\foreach \gdi/\gdj in {-5/9,-5/10,-4/7,-4/8,-4/9,-4/10,-3/5,-3/6,-3/7,-3/8,-3/9,-3/10,-2/5,-2/6,-2/7,-2/8,-2/9,-2/10,-1/5,-1/6,-1/7,-1/8,-1/9,-1/10,0/5,0/6,0/7,0/8,0/9,0/10} {%
  \pgfmathsetmacro{\gdx}{sqrt(3)*(\gdi+\gdj/2)}%
  \pgfmathsetmacro{\gdy}{1.5*\gdj}%
  \begin{scope}[shift={(\gdx,\gdy)}]
    \path[fill=gdB,draw=gdmesh,line width=.12pt] (0.86603,0.5) -- (0,1) -- (-0.86603,0.5) -- (-0.86603,-0.5) -- (0,-1) -- (0.86603,-0.5) -- cycle;
  \end{scope}}
\foreach \gdi/\gdj in {-10/0,-10/1,-10/2,-10/3,-10/4,-10/5,-10/6,-10/7,-10/8,-10/9,-10/10,-9/-1,-9/0,-9/1,-9/2,-9/3,-9/4,-9/5,-9/6,-9/7,-9/8,-9/9,-9/10,-8/-2,-8/-1,-8/0,-8/1,-8/2,-8/3,-8/4,-8/5,-8/6,-8/7,-8/8,-8/9,-8/10,-7/-3,-7/-2,-7/-1,-7/0,-7/1,-7/2,-7/3,-7/4,-7/5,-7/6,-7/7,-7/8,-7/9,-7/10,-6/-4,-6/-3,-6/-2,-6/-1,-6/0,-6/1,-6/2,-6/3,-6/4,-6/5,-6/6,-6/7,-6/8,-6/9,-6/10,-5/-5,-5/-4,-5/-3,-5/-2,-5/-1,-5/0,-5/1,-5/2,-5/3,-5/4,-5/5,-5/6,-5/7,-5/8,-4/-6,-4/-5,-4/-4,-4/-3,-4/-2,-4/-1,-4/0,-4/1,-4/2,-4/3,-4/4,-4/5,-4/6,-3/-7,-3/-6,-3/-5,-3/-4,-3/-3,-3/-2,-3/-1,-3/0,-3/1,-3/2,-3/3,-3/4,-2/-8,-2/-7,-2/-6,-2/-5,-2/-4,-2/-3,-2/-2,-2/-1,-2/0,-2/1,-2/2,-1/-9,-1/-8,-1/-7,-1/-6,-1/-5,-1/-4,-1/-3,-1/-2,-1/-1,-1/0,0/-10,0/-9,0/-8,0/-7,0/-6,0/-5,0/-4,0/-3,0/-2,0/-1,1/-10,1/-9,1/-8,1/-7,1/-6,1/-5,1/-4,1/-3,1/-2,1/-1,1/0,1/1,1/2,1/3,1/4,1/5,1/6,1/7,1/8,1/9,2/-10,2/-9,2/-8,2/-7,2/-6,2/-5,2/-4,2/-3,2/-2,2/-1,2/0,2/1,2/2,2/3,2/4,2/5,2/6,2/7,2/8,3/-10,3/-9,3/-8,3/-7,3/-6,3/-5,3/-4,3/-3,3/-2,3/-1,3/0,3/1,3/2,3/3,3/4,3/5,3/6,3/7,4/-10,4/-9,4/-8,4/-7,4/-6,4/-5,4/-4,4/-3,4/-2,4/-1,4/0,4/1,4/2,4/3,4/4,4/5,4/6,5/-10,5/-9,5/-8,5/-7,5/-6,5/-5,5/-4,5/-3,5/-2,5/-1,5/0,5/1,5/2,5/3,5/4,5/5,6/-10,6/-9,6/-8,6/-7,6/-6,6/-5,6/-4,6/-3,6/-2,6/-1,6/0,6/1,6/2,6/3,6/4,7/-10,7/-9,7/-8,7/-7,7/-6,7/-5,7/-4,7/-3,7/-2,7/-1,7/0,7/1,7/2,7/3,8/-10,8/-9,8/-8,8/-7,8/-6,8/-5,8/-4,8/-3,8/-2,8/-1,8/0,8/1,8/2,9/-10,9/-9,9/-8,9/-7,9/-6,9/-5,9/-4,9/-3,9/-2,9/-1,9/0,9/1,10/-10,10/-9,10/-8,10/-7,10/-6,10/-5,10/-4,10/-3,10/-2,10/-1,10/0} {%
  \pgfmathsetmacro{\gdx}{sqrt(3)*(\gdi+\gdj/2)}%
  \pgfmathsetmacro{\gdy}{1.5*\gdj}%
  \begin{scope}[shift={(\gdx,\gdy)}]
    \path[fill=gdD,draw=gdmesh,line width=.12pt] (0.86603,0.5) -- (0,1) -- (-0.86603,0.5) -- (-0.86603,-0.5) -- (0,-1) -- (0.86603,-0.5) -- cycle;
  \end{scope}}
\foreach \gdi/\gdj in {-14/0,-14/1,-14/2,-14/3,-14/4,-14/5,-14/6,-14/7,-14/8,-14/9,-14/10,-14/11,-14/12,-14/13,-14/14,-13/-1,-13/0,-13/1,-13/2,-13/3,-13/4,-13/5,-13/6,-13/7,-13/8,-13/9,-13/10,-13/11,-13/12,-13/13,-13/14,-12/-2,-12/-1,-12/0,-12/1,-12/2,-12/3,-12/4,-12/5,-12/6,-12/7,-12/8,-12/9,-12/10,-12/11,-12/12,-12/13,-12/14,-11/-3,-11/-2,-11/-1,-11/0,-11/1,-11/2,-11/3,-11/4,-11/5,-11/6,-11/7,-11/8,-11/9,-11/10,-11/11,-11/12,-11/13,-11/14,-10/-4,-10/-3,-10/-2,-10/-1,-10/11,-10/12,-10/13,-10/14,-9/-5,-9/-4,-9/-3,-9/-2,-9/11,-9/12,-9/13,-9/14,-8/-6,-8/-5,-8/-4,-8/-3,-8/11,-8/12,-8/13,-8/14,-7/-7,-7/-6,-7/-5,-7/-4,-7/11,-7/12,-7/13,-7/14,-6/-8,-6/-7,-6/-6,-6/-5,-6/11,-6/12,-6/13,-6/14,-5/-9,-5/-8,-5/-7,-5/-6,-5/11,-5/12,-5/13,-5/14,-4/-10,-4/-9,-4/-8,-4/-7,-4/11,-4/12,-4/13,-4/14,-3/-11,-3/-10,-3/-9,-3/-8,-3/11,-3/12,-3/13,-3/14,-2/-12,-2/-11,-2/-10,-2/-9,-2/11,-2/12,-2/13,-2/14,-1/-13,-1/-12,-1/-11,-1/-10,-1/11,-1/12,-1/13,-1/14,0/-14,0/-13,0/-12,0/-11,0/11,0/12,0/13,0/14,1/-14,1/-13,1/-12,1/-11,1/10,1/11,1/12,1/13,2/-14,2/-13,2/-12,2/-11,2/9,2/10,2/11,2/12,3/-14,3/-13,3/-12,3/-11,3/8,3/9,3/10,3/11,4/-14,4/-13,4/-12,4/-11,4/7,4/8,4/9,4/10,5/-14,5/-13,5/-12,5/-11,5/6,5/7,5/8,5/9,6/-14,6/-13,6/-12,6/-11,6/5,6/6,6/7,6/8,7/-14,7/-13,7/-12,7/-11,7/4,7/5,7/6,7/7,8/-14,8/-13,8/-12,8/-11,8/3,8/4,8/5,8/6,9/-14,9/-13,9/-12,9/-11,9/2,9/3,9/4,9/5,10/-14,10/-13,10/-12,10/-11,10/1,10/2,10/3,10/4,11/-14,11/-13,11/-12,11/-11,11/-10,11/-9,11/-8,11/-7,11/-6,11/-5,11/-4,11/-3,11/-2,11/-1,11/0,11/1,11/2,11/3,12/-14,12/-13,12/-12,12/-11,12/-10,12/-9,12/-8,12/-7,12/-6,12/-5,12/-4,12/-3,12/-2,12/-1,12/0,12/1,12/2,13/-14,13/-13,13/-12,13/-11,13/-10,13/-9,13/-8,13/-7,13/-6,13/-5,13/-4,13/-3,13/-2,13/-1,13/0,13/1,14/-14,14/-13,14/-12,14/-11,14/-10,14/-9,14/-8,14/-7,14/-6,14/-5,14/-4,14/-3,14/-2,14/-1,14/0} {%
  \pgfmathsetmacro{\gdx}{sqrt(3)*(\gdi+\gdj/2)}%
  \pgfmathsetmacro{\gdy}{1.5*\gdj}%
  \begin{scope}[shift={(\gdx,\gdy)}]
    \path[fill=gdCprime,draw=gdmesh,line width=.12pt] (0.86603,0.5) -- (0,1) -- (-0.86603,0.5) -- (-0.86603,-0.5) -- (0,-1) -- (0.86603,-0.5) -- cycle;
  \end{scope}}
\path[draw=gdblack,line width=.75pt]
  (-1.73205,1) -- (-1.73205,2)
  (-1.73205,1) -- (-0.86603,0.5)
  (-1.73205,2) -- (-0.86603,2.5)
  (-1.73205,4) -- (-1.73205,5)
  (-1.73205,4) -- (-0.86603,3.5)
  (-1.73205,5) -- (-0.86603,5.5)
  (-1.73205,7) -- (-1.73205,8)
  (-1.73205,7) -- (-0.86603,6.5)
  (-1.73205,8) -- (-0.86603,8.5)
  (-1.73205,10) -- (-1.73205,11)
  (-1.73205,10) -- (-0.86603,9.5)
  (-1.73205,11) -- (-0.86603,11.5)
  (-1.73205,13) -- (-1.73205,14)
  (-1.73205,13) -- (-0.86603,12.5)
  (-1.73205,14) -- (-0.86603,14.5)
  (-1.73205,16) -- (-1.73205,17)
  (-1.73205,16) -- (-0.86603,15.5)
  (-1.73205,17) -- (-0.86603,17.5)
  (-1.73205,19) -- (-1.73205,20)
  (-1.73205,19) -- (-0.86603,18.5)
  (-1.73205,20) -- (-0.86603,20.5)
  (-0.86603,-0.5) -- (-0.86603,0.5)
  (-0.86603,-0.5) -- (0,-1)
  (-0.86603,2.5) -- (-0.86603,3.5)
  (-0.86603,5.5) -- (-0.86603,6.5)
  (-0.86603,8.5) -- (-0.86603,9.5)
  (-0.86603,11.5) -- (-0.86603,12.5)
  (-0.86603,14.5) -- (-0.86603,15.5)
  (-0.86603,17.5) -- (-0.86603,18.5)
  (-0.86603,20.5) -- (-0.86603,21.5)
  (0,-1) -- (0.86603,-0.5)
  (0.86603,-0.5) -- (0.86603,0.5)
  (0.86603,0.5) -- (1.73205,1)
  (1.73205,1) -- (1.73205,2)
  (1.73205,2) -- (2.59808,2.5)
  (2.59808,2.5) -- (2.59808,3.5)
  (2.59808,3.5) -- (3.4641,4)
  (3.4641,4) -- (3.4641,5)
  (3.4641,5) -- (4.33013,5.5)
  (4.33013,5.5) -- (4.33013,6.5)
  (4.33013,6.5) -- (5.19615,7)
  (5.19615,7) -- (5.19615,8)
  (5.19615,8) -- (6.06218,8.5)
  (6.06218,8.5) -- (6.06218,9.5)
  (6.06218,9.5) -- (6.9282,10)
  (6.9282,10) -- (6.9282,11)
  (6.9282,11) -- (7.79423,11.5)
  (7.79423,11.5) -- (7.79423,12.5)
  (7.79423,12.5) -- (8.66025,13)
  (8.66025,13) -- (8.66025,14)
  (8.66025,14) -- (9.52628,14.5)
  (9.52628,14.5) -- (9.52628,15.5)
  (9.52628,15.5) -- (10.3923,16)
  (10.3923,16) -- (10.3923,17)
  (10.3923,17) -- (11.25833,17.5)
  (11.25833,17.5) -- (11.25833,18.5)
  (11.25833,18.5) -- (12.12436,19)
  (12.12436,19) -- (12.12436,20)
  (12.12436,20) -- (12.99038,20.5);
\draw[draw=gdordinary,line width=.65pt] (-25.11474,-0.5) -- (-25.11474,0.5) -- (-24.24871,1) -- (-24.24871,2) -- (-23.38269,2.5) -- (-23.38269,3.5) -- (-22.51666,4) -- (-22.51666,5) -- (-21.65064,5.5) -- (-21.65064,6.5) -- (-20.78461,7) -- (-20.78461,8) -- (-19.91858,8.5) -- (-19.91858,9.5) -- (-19.05256,10) -- (-19.05256,11) -- (-18.18653,11.5) -- (-18.18653,12.5) -- (-17.32051,13) -- (-17.32051,14) -- (-16.45448,14.5) -- (-16.45448,15.5) -- (-15.58846,16) -- (-15.58846,17) -- (-14.72243,17.5) -- (-14.72243,18.5) -- (-13.85641,19) -- (-13.85641,20) -- (-12.99038,20.5) -- (-12.99038,21.5) -- (-12.12436,22) -- (-11.25833,21.5) -- (-10.3923,22) -- (-9.52628,21.5) -- (-8.66025,22) -- (-7.79423,21.5) -- (-6.9282,22) -- (-6.06218,21.5) -- (-5.19615,22) -- (-4.33013,21.5) -- (-3.4641,22) -- (-2.59808,21.5) -- (-1.73205,22) -- (-0.86603,21.5) -- (0,22) -- (0.86603,21.5) -- (1.73205,22) -- (2.59808,21.5) -- (3.4641,22) -- (4.33013,21.5) -- (5.19615,22) -- (6.06218,21.5) -- (6.9282,22) -- (7.79423,21.5) -- (8.66025,22) -- (9.52628,21.5) -- (10.3923,22) -- (11.25833,21.5) -- (12.12436,22) -- (12.99038,21.5) -- (12.99038,20.5) -- (13.85641,20) -- (13.85641,19) -- (14.72243,18.5) -- (14.72243,17.5) -- (15.58846,17) -- (15.58846,16) -- (16.45448,15.5) -- (16.45448,14.5) -- (17.32051,14) -- (17.32051,13) -- (18.18653,12.5) -- (18.18653,11.5) -- (19.05256,11) -- (19.05256,10) -- (19.91858,9.5) -- (19.91858,8.5) -- (20.78461,8) -- (20.78461,7) -- (21.65064,6.5) -- (21.65064,5.5) -- (22.51666,5) -- (22.51666,4) -- (23.38269,3.5) -- (23.38269,2.5) -- (24.24871,2) -- (24.24871,1) -- (25.11474,0.5) -- (25.11474,-0.5) -- (24.24871,-1) -- (24.24871,-2) -- (23.38269,-2.5) -- (23.38269,-3.5) -- (22.51666,-4) -- (22.51666,-5) -- (21.65064,-5.5) -- (21.65064,-6.5) -- (20.78461,-7) -- (20.78461,-8) -- (19.91858,-8.5) -- (19.91858,-9.5) -- (19.05256,-10) -- (19.05256,-11) -- (18.18653,-11.5) -- (18.18653,-12.5) -- (17.32051,-13) -- (17.32051,-14) -- (16.45448,-14.5) -- (16.45448,-15.5) -- (15.58846,-16) -- (15.58846,-17) -- (14.72243,-17.5) -- (14.72243,-18.5) -- (13.85641,-19) -- (13.85641,-20) -- (12.99038,-20.5) -- (12.99038,-21.5) -- (12.12436,-22) -- (11.25833,-21.5) -- (10.3923,-22) -- (9.52628,-21.5) -- (8.66025,-22) -- (7.79423,-21.5) -- (6.9282,-22) -- (6.06218,-21.5) -- (5.19615,-22) -- (4.33013,-21.5) -- (3.4641,-22) -- (2.59808,-21.5) -- (1.73205,-22) -- (0.86603,-21.5) -- (0,-22) -- (-0.86603,-21.5) -- (-1.73205,-22) -- (-2.59808,-21.5) -- (-3.4641,-22) -- (-4.33013,-21.5) -- (-5.19615,-22) -- (-6.06218,-21.5) -- (-6.9282,-22) -- (-7.79423,-21.5) -- (-8.66025,-22) -- (-9.52628,-21.5) -- (-10.3923,-22) -- (-11.25833,-21.5) -- (-12.12436,-22) -- (-12.99038,-21.5) -- (-12.99038,-20.5) -- (-13.85641,-20) -- (-13.85641,-19) -- (-14.72243,-18.5) -- (-14.72243,-17.5) -- (-15.58846,-17) -- (-15.58846,-16) -- (-16.45448,-15.5) -- (-16.45448,-14.5) -- (-17.32051,-14) -- (-17.32051,-13) -- (-18.18653,-12.5) -- (-18.18653,-11.5) -- (-19.05256,-11) -- (-19.05256,-10) -- (-19.91858,-9.5) -- (-19.91858,-8.5) -- (-20.78461,-8) -- (-20.78461,-7) -- (-21.65064,-6.5) -- (-21.65064,-5.5) -- (-22.51666,-5) -- (-22.51666,-4) -- (-23.38269,-3.5) -- (-23.38269,-2.5) -- (-24.24871,-2) -- (-24.24871,-1) -- cycle;
\draw[draw=gdordinary,line width=.65pt] (-18.18653,-0.5) -- (-18.18653,0.5) -- (-17.32051,1) -- (-17.32051,2) -- (-16.45448,2.5) -- (-16.45448,3.5) -- (-15.58846,4) -- (-15.58846,5) -- (-14.72243,5.5) -- (-14.72243,6.5) -- (-13.85641,7) -- (-13.85641,8) -- (-12.99038,8.5) -- (-12.99038,9.5) -- (-12.12436,10) -- (-12.12436,11) -- (-11.25833,11.5) -- (-11.25833,12.5) -- (-10.3923,13) -- (-10.3923,14) -- (-9.52628,14.5) -- (-9.52628,15.5) -- (-8.66025,16) -- (-7.79423,15.5) -- (-6.9282,16) -- (-6.06218,15.5) -- (-5.19615,16) -- (-4.33013,15.5) -- (-3.4641,16) -- (-2.59808,15.5) -- (-1.73205,16) -- (-0.86603,15.5) -- (0,16) -- (0.86603,15.5) -- (1.73205,16) -- (2.59808,15.5) -- (3.4641,16) -- (4.33013,15.5) -- (5.19615,16) -- (6.06218,15.5) -- (6.9282,16) -- (7.79423,15.5) -- (8.66025,16) -- (9.52628,15.5) -- (9.52628,14.5) -- (10.3923,14) -- (10.3923,13) -- (11.25833,12.5) -- (11.25833,11.5) -- (12.12436,11) -- (12.12436,10) -- (12.99038,9.5) -- (12.99038,8.5) -- (13.85641,8) -- (13.85641,7) -- (14.72243,6.5) -- (14.72243,5.5) -- (15.58846,5) -- (15.58846,4) -- (16.45448,3.5) -- (16.45448,2.5) -- (17.32051,2) -- (17.32051,1) -- (18.18653,0.5) -- (18.18653,-0.5) -- (17.32051,-1) -- (17.32051,-2) -- (16.45448,-2.5) -- (16.45448,-3.5) -- (15.58846,-4) -- (15.58846,-5) -- (14.72243,-5.5) -- (14.72243,-6.5) -- (13.85641,-7) -- (13.85641,-8) -- (12.99038,-8.5) -- (12.99038,-9.5) -- (12.12436,-10) -- (12.12436,-11) -- (11.25833,-11.5) -- (11.25833,-12.5) -- (10.3923,-13) -- (10.3923,-14) -- (9.52628,-14.5) -- (9.52628,-15.5) -- (8.66025,-16) -- (7.79423,-15.5) -- (6.9282,-16) -- (6.06218,-15.5) -- (5.19615,-16) -- (4.33013,-15.5) -- (3.4641,-16) -- (2.59808,-15.5) -- (1.73205,-16) -- (0.86603,-15.5) -- (0,-16) -- (-0.86603,-15.5) -- (-1.73205,-16) -- (-2.59808,-15.5) -- (-3.4641,-16) -- (-4.33013,-15.5) -- (-5.19615,-16) -- (-6.06218,-15.5) -- (-6.9282,-16) -- (-7.79423,-15.5) -- (-8.66025,-16) -- (-9.52628,-15.5) -- (-9.52628,-14.5) -- (-10.3923,-14) -- (-10.3923,-13) -- (-11.25833,-12.5) -- (-11.25833,-11.5) -- (-12.12436,-11) -- (-12.12436,-10) -- (-12.99038,-9.5) -- (-12.99038,-8.5) -- (-13.85641,-8) -- (-13.85641,-7) -- (-14.72243,-6.5) -- (-14.72243,-5.5) -- (-15.58846,-5) -- (-15.58846,-4) -- (-16.45448,-3.5) -- (-16.45448,-2.5) -- (-17.32051,-2) -- (-17.32051,-1) -- cycle;
\draw[draw=gdordinary,line width=.65pt] (-7.79423,-0.5) -- (-7.79423,0.5) -- (-6.9282,1) -- (-6.9282,2) -- (-6.06218,2.5) -- (-6.06218,3.5) -- (-5.19615,4) -- (-5.19615,5) -- (-4.33013,5.5) -- (-4.33013,6.5) -- (-3.4641,7) -- (-2.59808,6.5) -- (-1.73205,7) -- (-0.86603,6.5) -- (0,7) -- (0.86603,6.5) -- (1.73205,7) -- (2.59808,6.5) -- (3.4641,7) -- (4.33013,6.5) -- (4.33013,5.5) -- (5.19615,5) -- (5.19615,4) -- (6.06218,3.5) -- (6.06218,2.5) -- (6.9282,2) -- (6.9282,1) -- (7.79423,0.5) -- (7.79423,-0.5) -- (6.9282,-1) -- (6.9282,-2) -- (6.06218,-2.5) -- (6.06218,-3.5) -- (5.19615,-4) -- (5.19615,-5) -- (4.33013,-5.5) -- (4.33013,-6.5) -- (3.4641,-7) -- (2.59808,-6.5) -- (1.73205,-7) -- (0.86603,-6.5) -- (0,-7) -- (-0.86603,-6.5) -- (-1.73205,-7) -- (-2.59808,-6.5) -- (-3.4641,-7) -- (-4.33013,-6.5) -- (-4.33013,-5.5) -- (-5.19615,-5) -- (-5.19615,-4) -- (-6.06218,-3.5) -- (-6.06218,-2.5) -- (-6.9282,-2) -- (-6.9282,-1) -- cycle;
\node[text=gdordinary,fill=white,inner sep=.55pt,font=\fontsize{7}{8}\selectfont] at (-9.8,22.8) {$\Gamma_{M}$};
\node[text=gdordinary,fill=white,inner sep=.55pt,font=\fontsize{7}{8}\selectfont] at (-7,16.8) {$\Gamma_{N}$};
\node[text=gdordinary,fill=white,inner sep=.55pt,font=\fontsize{7}{8}\selectfont] at (-2.8,7.8) {$\Gamma_{l}$};
\node[text=gdblack,inner sep=0pt,font=\fontsize{10}{11}\selectfont] at (1.5,4.1) {{$A$}};
\node[text=gdblack,inner sep=0pt,font=\fontsize{10}{11}\selectfont] at (3.5,11.5) {{$B$}};
\node[text=gdblack,inner sep=0pt,font=\fontsize{10}{11}\selectfont] at (5.7,18.8) {{$C'$}};
\node[text=gdblack,inner sep=0pt,font=\fontsize{10}{11}\selectfont] at (-10.3,-0.5) {{$D$}};
\node[text=gdblack,inner sep=0pt,font=\fontsize{9}{10}\selectfont] at (6.5,24.1) {{$\widetilde\Lambda$}};
\end{tikzpicture}
\endgroup
}
\caption{Two views of the partition of nested {concentric graph-distance disks}
$\Gamma_{l}\subset\Gamma_N\subset\Gamma_M$ in the hexagonal lattice
$\LL$, for ${l}<N<M$. The lattice cone is $\widetilde\Lambda\subset\LL$,
and $\widetilde\Lambda^c=\LL\setminus\widetilde\Lambda$.
All outlines follow hexagonal edges{.} {Layer counts are illustrative.} Left: $A$ is yellow-green,
$B$ light blue, $C$ green, and $D$ pink.
Right: the full outer shell $C'$ is yellow.}
\label{fig:haag-setting}
\label{haag23}
\par\medskip
\begin{tabular}{@{}ll@{}}
$A=\Gamma_{l}\cap\widetilde\Lambda$
 & $B=(\Gamma_N\setminus\Gamma_{l})\cap\widetilde\Lambda$\\[3pt]
$C=(\Gamma_M\setminus\Gamma_N)\cap\widetilde\Lambda$
 & $D=\Gamma_N\cap\widetilde\Lambda^c$\\[3pt]
$C'=\Gamma_M\setminus\Gamma_N$
 & $V=\Gamma_M\setminus D$\\[3pt]
$L=\Gamma_M\cap\widetilde\Lambda$
 & $R=\Gamma_M\cap\widetilde\Lambda^c$
\end{tabular}
\par\medskip
\begin{minipage}{\textwidth}
\small
Here juxtaposition of disjoint regions denotes their union.
Thus $AB=\Gamma_N\cap\widetilde\Lambda$ and
$AB\sqcup D=\Gamma_N$:
the cone part $AB$ and its complement part $D$ reach the same
{outer boundary}.
Also $L=A\sqcup B\sqcup C$ and $V=A\sqcup B\sqcup C'$.
\end{minipage}
\end{figure}
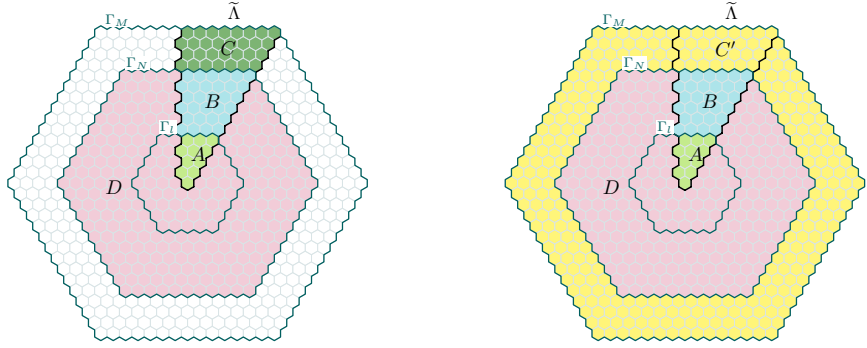

\begin{figure}[!htbp]
\centering
\resizebox{.64\textwidth}{!}{%
\begingroup
\definecolor{gdA}{HTML}{C4EB8C}
\definecolor{gdB}{HTML}{ADE4EA}
\definecolor{gdC}{HTML}{7DB473}
\definecolor{gdD}{HTML}{F2CCD8}
\definecolor{gdCprime}{HTML}{FFF579}
\definecolor{gdwhite}{HTML}{FFFFFF}
\definecolor{gdmesh}{HTML}{D9E4E6}
\definecolor{gdordinary}{HTML}{005F5F}
\definecolor{gdm}{HTML}{85B200}
\definecolor{gdk}{HTML}{C60000}
\definecolor{gdblack}{HTML}{000000}
\begin{tikzpicture}[x=3.6pt,y=3.6pt,line join=round,line cap=round]
\path[use as bounding box] (-26.0,-22.5) rectangle (26.0,26.1);
\foreach \gdi/\gdj in {-14/0,-14/1,-14/2,-14/3,-14/4,-14/5,-14/6,-14/7,-14/8,-14/9,-14/10,-14/11,-14/12,-14/13,-14/14,-13/-1,-13/0,-13/1,-13/2,-13/3,-13/4,-13/5,-13/6,-13/7,-13/8,-13/9,-13/10,-13/11,-13/12,-13/13,-13/14,-12/-2,-12/-1,-12/0,-12/1,-12/2,-12/3,-12/4,-12/5,-12/6,-12/7,-12/8,-12/9,-12/10,-12/11,-12/12,-12/13,-12/14,-11/-3,-11/-2,-11/-1,-11/0,-11/1,-11/2,-11/3,-11/4,-11/5,-11/6,-11/7,-11/8,-11/9,-11/10,-11/11,-11/12,-11/13,-11/14,-10/-4,-10/-3,-10/-2,-10/-1,-10/11,-10/12,-10/13,-10/14,-9/-5,-9/-4,-9/-3,-9/-2,-9/11,-9/12,-9/13,-9/14,-8/-6,-8/-5,-8/-4,-8/-3,-8/11,-8/12,-8/13,-8/14,-7/-7,-7/-6,-7/-5,-7/-4,-7/11,-7/12,-6/-8,-6/-7,-6/-6,-6/-5,-5/-9,-5/-8,-5/-7,-5/-6,-4/-10,-4/-9,-4/-8,-4/-7,-3/-11,-3/-10,-3/-9,-3/-8,-2/-12,-2/-11,-2/-10,-2/-9,-1/-13,-1/-12,-1/-11,-1/-10,0/-14,0/-13,0/-12,0/-11,1/-14,1/-13,1/-12,1/-11,1/10,1/11,1/12,1/13,2/-14,2/-13,2/-12,2/-11,2/9,2/10,2/11,2/12,3/-14,3/-13,3/-12,3/-11,3/8,3/9,3/10,3/11,4/-14,4/-13,4/-12,4/-11,4/7,4/8,4/9,4/10,5/-14,5/-13,5/-12,5/-11,5/6,5/7,5/8,5/9,6/-14,6/-13,6/-12,6/-11,6/5,6/6,6/7,6/8,7/-14,7/-13,7/-12,7/-11,7/4,7/5,7/6,7/7,8/-14,8/-13,8/-12,8/-11,8/3,8/4,8/5,8/6,9/-14,9/-13,9/-12,9/-11,9/2,9/3,9/4,9/5,10/-14,10/-13,10/-12,10/-11,10/1,10/2,10/3,10/4,11/-14,11/-13,11/-12,11/-11,11/-10,11/-9,11/-8,11/-7,11/-6,11/-5,11/-4,11/-3,11/-2,11/-1,11/0,11/1,11/2,11/3,12/-14,12/-13,12/-12,12/-11,12/-10,12/-9,12/-8,12/-7,12/-6,12/-5,12/-4,12/-3,12/-2,12/-1,12/0,12/1,12/2,13/-14,13/-13,13/-12,13/-11,13/-10,13/-9,13/-8,13/-7,13/-6,13/-5,13/-4,13/-3,13/-2,13/-1,13/0,13/1,14/-14,14/-13,14/-12,14/-11,14/-10,14/-9,14/-8,14/-7,14/-6,14/-5,14/-4,14/-3,14/-2,14/-1,14/0} {%
  \pgfmathsetmacro{\gdx}{sqrt(3)*(\gdi+\gdj/2)}%
  \pgfmathsetmacro{\gdy}{1.5*\gdj}%
  \begin{scope}[shift={(\gdx,\gdy)}]
    \path[fill=gdwhite,draw=gdmesh,line width=.12pt] (0.86603,0.5) -- (0,1) -- (-0.86603,0.5) -- (-0.86603,-0.5) -- (0,-1) -- (0.86603,-0.5) -- cycle;
  \end{scope}}
\foreach \gdi/\gdj in {-2/3,-2/4,-1/1,-1/2,-1/3,-1/4,0/0,0/1,0/2,0/3,0/4} {%
  \pgfmathsetmacro{\gdx}{sqrt(3)*(\gdi+\gdj/2)}%
  \pgfmathsetmacro{\gdy}{1.5*\gdj}%
  \begin{scope}[shift={(\gdx,\gdy)}]
    \path[fill=gdA,draw=gdmesh,line width=.12pt] (0.86603,0.5) -- (0,1) -- (-0.86603,0.5) -- (-0.86603,-0.5) -- (0,-1) -- (0.86603,-0.5) -- cycle;
  \end{scope}}
\foreach \gdi/\gdj in {-5/9,-5/10,-4/7,-4/8,-4/9,-4/10,-3/5,-3/6,-3/7,-3/8,-3/9,-3/10,-2/5,-2/6,-2/7,-2/8,-2/9,-2/10,-1/5,-1/6,-1/7,-1/8,-1/9,-1/10,0/5,0/6,0/7,0/8,0/9,0/10} {%
  \pgfmathsetmacro{\gdx}{sqrt(3)*(\gdi+\gdj/2)}%
  \pgfmathsetmacro{\gdy}{1.5*\gdj}%
  \begin{scope}[shift={(\gdx,\gdy)}]
    \path[fill=gdB,draw=gdmesh,line width=.12pt] (0.86603,0.5) -- (0,1) -- (-0.86603,0.5) -- (-0.86603,-0.5) -- (0,-1) -- (0.86603,-0.5) -- cycle;
  \end{scope}}
\foreach \gdi/\gdj in {-7/13,-7/14,-6/11,-6/12,-6/13,-6/14,-5/11,-5/12,-5/13,-5/14,-4/11,-4/12,-4/13,-4/14,-3/11,-3/12,-3/13,-3/14,-2/11,-2/12,-2/13,-2/14,-1/11,-1/12,-1/13,-1/14,0/11,0/12,0/13,0/14} {%
  \pgfmathsetmacro{\gdx}{sqrt(3)*(\gdi+\gdj/2)}%
  \pgfmathsetmacro{\gdy}{1.5*\gdj}%
  \begin{scope}[shift={(\gdx,\gdy)}]
    \path[fill=gdC,draw=gdmesh,line width=.12pt] (0.86603,0.5) -- (0,1) -- (-0.86603,0.5) -- (-0.86603,-0.5) -- (0,-1) -- (0.86603,-0.5) -- cycle;
  \end{scope}}
\foreach \gdi/\gdj in {-10/0,-10/1,-10/2,-10/3,-10/4,-10/5,-10/6,-10/7,-10/8,-10/9,-10/10,-9/-1,-9/0,-9/1,-9/2,-9/3,-9/4,-9/5,-9/6,-9/7,-9/8,-9/9,-9/10,-8/-2,-8/-1,-8/0,-8/1,-8/2,-8/3,-8/4,-8/5,-8/6,-8/7,-8/8,-8/9,-8/10,-7/-3,-7/-2,-7/-1,-7/0,-7/1,-7/2,-7/3,-7/4,-7/5,-7/6,-7/7,-7/8,-7/9,-7/10,-6/-4,-6/-3,-6/-2,-6/-1,-6/0,-6/1,-6/2,-6/3,-6/4,-6/5,-6/6,-6/7,-6/8,-6/9,-6/10,-5/-5,-5/-4,-5/-3,-5/-2,-5/-1,-5/0,-5/1,-5/2,-5/3,-5/4,-5/5,-5/6,-5/7,-5/8,-4/-6,-4/-5,-4/-4,-4/-3,-4/-2,-4/-1,-4/0,-4/1,-4/2,-4/3,-4/4,-4/5,-4/6,-3/-7,-3/-6,-3/-5,-3/-4,-3/-3,-3/-2,-3/-1,-3/0,-3/1,-3/2,-3/3,-3/4,-2/-8,-2/-7,-2/-6,-2/-5,-2/-4,-2/-3,-2/-2,-2/-1,-2/0,-2/1,-2/2,-1/-9,-1/-8,-1/-7,-1/-6,-1/-5,-1/-4,-1/-3,-1/-2,-1/-1,-1/0,0/-10,0/-9,0/-8,0/-7,0/-6,0/-5,0/-4,0/-3,0/-2,0/-1,1/-10,1/-9,1/-8,1/-7,1/-6,1/-5,1/-4,1/-3,1/-2,1/-1,1/0,1/1,1/2,1/3,1/4,1/5,1/6,1/7,1/8,1/9,2/-10,2/-9,2/-8,2/-7,2/-6,2/-5,2/-4,2/-3,2/-2,2/-1,2/0,2/1,2/2,2/3,2/4,2/5,2/6,2/7,2/8,3/-10,3/-9,3/-8,3/-7,3/-6,3/-5,3/-4,3/-3,3/-2,3/-1,3/0,3/1,3/2,3/3,3/4,3/5,3/6,3/7,4/-10,4/-9,4/-8,4/-7,4/-6,4/-5,4/-4,4/-3,4/-2,4/-1,4/0,4/1,4/2,4/3,4/4,4/5,4/6,5/-10,5/-9,5/-8,5/-7,5/-6,5/-5,5/-4,5/-3,5/-2,5/-1,5/0,5/1,5/2,5/3,5/4,5/5,6/-10,6/-9,6/-8,6/-7,6/-6,6/-5,6/-4,6/-3,6/-2,6/-1,6/0,6/1,6/2,6/3,6/4,7/-10,7/-9,7/-8,7/-7,7/-6,7/-5,7/-4,7/-3,7/-2,7/-1,7/0,7/1,7/2,7/3,8/-10,8/-9,8/-8,8/-7,8/-6,8/-5,8/-4,8/-3,8/-2,8/-1,8/0,8/1,8/2,9/-10,9/-9,9/-8,9/-7,9/-6,9/-5,9/-4,9/-3,9/-2,9/-1,9/0,9/1,10/-10,10/-9,10/-8,10/-7,10/-6,10/-5,10/-4,10/-3,10/-2,10/-1,10/0} {%
  \pgfmathsetmacro{\gdx}{sqrt(3)*(\gdi+\gdj/2)}%
  \pgfmathsetmacro{\gdy}{1.5*\gdj}%
  \begin{scope}[shift={(\gdx,\gdy)}]
    \path[fill=gdD,draw=gdmesh,line width=.12pt] (0.86603,0.5) -- (0,1) -- (-0.86603,0.5) -- (-0.86603,-0.5) -- (0,-1) -- (0.86603,-0.5) -- cycle;
  \end{scope}}
\path[draw=gdblack,line width=.75pt]
  (-1.73205,1) -- (-1.73205,2)
  (-1.73205,1) -- (-0.86603,0.5)
  (-1.73205,2) -- (-0.86603,2.5)
  (-1.73205,4) -- (-1.73205,5)
  (-1.73205,4) -- (-0.86603,3.5)
  (-1.73205,5) -- (-0.86603,5.5)
  (-1.73205,7) -- (-1.73205,8)
  (-1.73205,7) -- (-0.86603,6.5)
  (-1.73205,8) -- (-0.86603,8.5)
  (-1.73205,10) -- (-1.73205,11)
  (-1.73205,10) -- (-0.86603,9.5)
  (-1.73205,11) -- (-0.86603,11.5)
  (-1.73205,13) -- (-1.73205,14)
  (-1.73205,13) -- (-0.86603,12.5)
  (-1.73205,14) -- (-0.86603,14.5)
  (-1.73205,16) -- (-1.73205,17)
  (-1.73205,16) -- (-0.86603,15.5)
  (-1.73205,17) -- (-0.86603,17.5)
  (-1.73205,19) -- (-1.73205,20)
  (-1.73205,19) -- (-0.86603,18.5)
  (-1.73205,20) -- (-0.86603,20.5)
  (-0.86603,-0.5) -- (-0.86603,0.5)
  (-0.86603,-0.5) -- (0,-1)
  (-0.86603,2.5) -- (-0.86603,3.5)
  (-0.86603,5.5) -- (-0.86603,6.5)
  (-0.86603,8.5) -- (-0.86603,9.5)
  (-0.86603,11.5) -- (-0.86603,12.5)
  (-0.86603,14.5) -- (-0.86603,15.5)
  (-0.86603,17.5) -- (-0.86603,18.5)
  (-0.86603,20.5) -- (-0.86603,21.5)
  (0,-1) -- (0.86603,-0.5)
  (0.86603,-0.5) -- (0.86603,0.5)
  (0.86603,0.5) -- (1.73205,1)
  (1.73205,1) -- (1.73205,2)
  (1.73205,2) -- (2.59808,2.5)
  (2.59808,2.5) -- (2.59808,3.5)
  (2.59808,3.5) -- (3.4641,4)
  (3.4641,4) -- (3.4641,5)
  (3.4641,5) -- (4.33013,5.5)
  (4.33013,5.5) -- (4.33013,6.5)
  (4.33013,6.5) -- (5.19615,7)
  (5.19615,7) -- (5.19615,8)
  (5.19615,8) -- (6.06218,8.5)
  (6.06218,8.5) -- (6.06218,9.5)
  (6.06218,9.5) -- (6.9282,10)
  (6.9282,10) -- (6.9282,11)
  (6.9282,11) -- (7.79423,11.5)
  (7.79423,11.5) -- (7.79423,12.5)
  (7.79423,12.5) -- (8.66025,13)
  (8.66025,13) -- (8.66025,14)
  (8.66025,14) -- (9.52628,14.5)
  (9.52628,14.5) -- (9.52628,15.5)
  (9.52628,15.5) -- (10.3923,16)
  (10.3923,16) -- (10.3923,17)
  (10.3923,17) -- (11.25833,17.5)
  (11.25833,17.5) -- (11.25833,18.5)
  (11.25833,18.5) -- (12.12436,19)
  (12.12436,19) -- (12.12436,20)
  (12.12436,20) -- (12.99038,20.5);
\draw[draw=gdordinary,line width=.65pt] (-25.11474,-0.5) -- (-25.11474,0.5) -- (-24.24871,1) -- (-24.24871,2) -- (-23.38269,2.5) -- (-23.38269,3.5) -- (-22.51666,4) -- (-22.51666,5) -- (-21.65064,5.5) -- (-21.65064,6.5) -- (-20.78461,7) -- (-20.78461,8) -- (-19.91858,8.5) -- (-19.91858,9.5) -- (-19.05256,10) -- (-19.05256,11) -- (-18.18653,11.5) -- (-18.18653,12.5) -- (-17.32051,13) -- (-17.32051,14) -- (-16.45448,14.5) -- (-16.45448,15.5) -- (-15.58846,16) -- (-15.58846,17) -- (-14.72243,17.5) -- (-14.72243,18.5) -- (-13.85641,19) -- (-13.85641,20) -- (-12.99038,20.5) -- (-12.99038,21.5) -- (-12.12436,22) -- (-11.25833,21.5) -- (-10.3923,22) -- (-9.52628,21.5) -- (-8.66025,22) -- (-7.79423,21.5) -- (-6.9282,22) -- (-6.06218,21.5) -- (-5.19615,22) -- (-4.33013,21.5) -- (-3.4641,22) -- (-2.59808,21.5) -- (-1.73205,22) -- (-0.86603,21.5) -- (0,22) -- (0.86603,21.5) -- (1.73205,22) -- (2.59808,21.5) -- (3.4641,22) -- (4.33013,21.5) -- (5.19615,22) -- (6.06218,21.5) -- (6.9282,22) -- (7.79423,21.5) -- (8.66025,22) -- (9.52628,21.5) -- (10.3923,22) -- (11.25833,21.5) -- (12.12436,22) -- (12.99038,21.5) -- (12.99038,20.5) -- (13.85641,20) -- (13.85641,19) -- (14.72243,18.5) -- (14.72243,17.5) -- (15.58846,17) -- (15.58846,16) -- (16.45448,15.5) -- (16.45448,14.5) -- (17.32051,14) -- (17.32051,13) -- (18.18653,12.5) -- (18.18653,11.5) -- (19.05256,11) -- (19.05256,10) -- (19.91858,9.5) -- (19.91858,8.5) -- (20.78461,8) -- (20.78461,7) -- (21.65064,6.5) -- (21.65064,5.5) -- (22.51666,5) -- (22.51666,4) -- (23.38269,3.5) -- (23.38269,2.5) -- (24.24871,2) -- (24.24871,1) -- (25.11474,0.5) -- (25.11474,-0.5) -- (24.24871,-1) -- (24.24871,-2) -- (23.38269,-2.5) -- (23.38269,-3.5) -- (22.51666,-4) -- (22.51666,-5) -- (21.65064,-5.5) -- (21.65064,-6.5) -- (20.78461,-7) -- (20.78461,-8) -- (19.91858,-8.5) -- (19.91858,-9.5) -- (19.05256,-10) -- (19.05256,-11) -- (18.18653,-11.5) -- (18.18653,-12.5) -- (17.32051,-13) -- (17.32051,-14) -- (16.45448,-14.5) -- (16.45448,-15.5) -- (15.58846,-16) -- (15.58846,-17) -- (14.72243,-17.5) -- (14.72243,-18.5) -- (13.85641,-19) -- (13.85641,-20) -- (12.99038,-20.5) -- (12.99038,-21.5) -- (12.12436,-22) -- (11.25833,-21.5) -- (10.3923,-22) -- (9.52628,-21.5) -- (8.66025,-22) -- (7.79423,-21.5) -- (6.9282,-22) -- (6.06218,-21.5) -- (5.19615,-22) -- (4.33013,-21.5) -- (3.4641,-22) -- (2.59808,-21.5) -- (1.73205,-22) -- (0.86603,-21.5) -- (0,-22) -- (-0.86603,-21.5) -- (-1.73205,-22) -- (-2.59808,-21.5) -- (-3.4641,-22) -- (-4.33013,-21.5) -- (-5.19615,-22) -- (-6.06218,-21.5) -- (-6.9282,-22) -- (-7.79423,-21.5) -- (-8.66025,-22) -- (-9.52628,-21.5) -- (-10.3923,-22) -- (-11.25833,-21.5) -- (-12.12436,-22) -- (-12.99038,-21.5) -- (-12.99038,-20.5) -- (-13.85641,-20) -- (-13.85641,-19) -- (-14.72243,-18.5) -- (-14.72243,-17.5) -- (-15.58846,-17) -- (-15.58846,-16) -- (-16.45448,-15.5) -- (-16.45448,-14.5) -- (-17.32051,-14) -- (-17.32051,-13) -- (-18.18653,-12.5) -- (-18.18653,-11.5) -- (-19.05256,-11) -- (-19.05256,-10) -- (-19.91858,-9.5) -- (-19.91858,-8.5) -- (-20.78461,-8) -- (-20.78461,-7) -- (-21.65064,-6.5) -- (-21.65064,-5.5) -- (-22.51666,-5) -- (-22.51666,-4) -- (-23.38269,-3.5) -- (-23.38269,-2.5) -- (-24.24871,-2) -- (-24.24871,-1) -- cycle;
\draw[draw=gdk,line width=.65pt] (-21.65064,-0.5) -- (-21.65064,0.5) -- (-20.78461,1) -- (-20.78461,2) -- (-19.91858,2.5) -- (-19.91858,3.5) -- (-19.05256,4) -- (-19.05256,5) -- (-18.18653,5.5) -- (-18.18653,6.5) -- (-17.32051,7) -- (-17.32051,8) -- (-16.45448,8.5) -- (-16.45448,9.5) -- (-15.58846,10) -- (-15.58846,11) -- (-14.72243,11.5) -- (-14.72243,12.5) -- (-13.85641,13) -- (-13.85641,14) -- (-12.99038,14.5) -- (-12.99038,15.5) -- (-12.12436,16) -- (-12.12436,17) -- (-11.25833,17.5) -- (-11.25833,18.5) -- (-10.3923,19) -- (-9.52628,18.5) -- (-8.66025,19) -- (-7.79423,18.5) -- (-6.9282,19) -- (-6.06218,18.5) -- (-5.19615,19) -- (-4.33013,18.5) -- (-3.4641,19) -- (-2.59808,18.5) -- (-1.73205,19) -- (-0.86603,18.5) -- (0,19) -- (0.86603,18.5) -- (1.73205,19) -- (2.59808,18.5) -- (3.4641,19) -- (4.33013,18.5) -- (5.19615,19) -- (6.06218,18.5) -- (6.9282,19) -- (7.79423,18.5) -- (8.66025,19) -- (9.52628,18.5) -- (10.3923,19) -- (11.25833,18.5) -- (11.25833,17.5) -- (12.12436,17) -- (12.12436,16) -- (12.99038,15.5) -- (12.99038,14.5) -- (13.85641,14) -- (13.85641,13) -- (14.72243,12.5) -- (14.72243,11.5) -- (15.58846,11) -- (15.58846,10) -- (16.45448,9.5) -- (16.45448,8.5) -- (17.32051,8) -- (17.32051,7) -- (18.18653,6.5) -- (18.18653,5.5) -- (19.05256,5) -- (19.05256,4) -- (19.91858,3.5) -- (19.91858,2.5) -- (20.78461,2) -- (20.78461,1) -- (21.65064,0.5) -- (21.65064,-0.5) -- (20.78461,-1) -- (20.78461,-2) -- (19.91858,-2.5) -- (19.91858,-3.5) -- (19.05256,-4) -- (19.05256,-5) -- (18.18653,-5.5) -- (18.18653,-6.5) -- (17.32051,-7) -- (17.32051,-8) -- (16.45448,-8.5) -- (16.45448,-9.5) -- (15.58846,-10) -- (15.58846,-11) -- (14.72243,-11.5) -- (14.72243,-12.5) -- (13.85641,-13) -- (13.85641,-14) -- (12.99038,-14.5) -- (12.99038,-15.5) -- (12.12436,-16) -- (12.12436,-17) -- (11.25833,-17.5) -- (11.25833,-18.5) -- (10.3923,-19) -- (9.52628,-18.5) -- (8.66025,-19) -- (7.79423,-18.5) -- (6.9282,-19) -- (6.06218,-18.5) -- (5.19615,-19) -- (4.33013,-18.5) -- (3.4641,-19) -- (2.59808,-18.5) -- (1.73205,-19) -- (0.86603,-18.5) -- (0,-19) -- (-0.86603,-18.5) -- (-1.73205,-19) -- (-2.59808,-18.5) -- (-3.4641,-19) -- (-4.33013,-18.5) -- (-5.19615,-19) -- (-6.06218,-18.5) -- (-6.9282,-19) -- (-7.79423,-18.5) -- (-8.66025,-19) -- (-9.52628,-18.5) -- (-10.3923,-19) -- (-11.25833,-18.5) -- (-11.25833,-17.5) -- (-12.12436,-17) -- (-12.12436,-16) -- (-12.99038,-15.5) -- (-12.99038,-14.5) -- (-13.85641,-14) -- (-13.85641,-13) -- (-14.72243,-12.5) -- (-14.72243,-11.5) -- (-15.58846,-11) -- (-15.58846,-10) -- (-16.45448,-9.5) -- (-16.45448,-8.5) -- (-17.32051,-8) -- (-17.32051,-7) -- (-18.18653,-6.5) -- (-18.18653,-5.5) -- (-19.05256,-5) -- (-19.05256,-4) -- (-19.91858,-3.5) -- (-19.91858,-2.5) -- (-20.78461,-2) -- (-20.78461,-1) -- cycle;
\draw[draw=gdordinary,line width=.65pt] (-18.18653,-0.5) -- (-18.18653,0.5) -- (-17.32051,1) -- (-17.32051,2) -- (-16.45448,2.5) -- (-16.45448,3.5) -- (-15.58846,4) -- (-15.58846,5) -- (-14.72243,5.5) -- (-14.72243,6.5) -- (-13.85641,7) -- (-13.85641,8) -- (-12.99038,8.5) -- (-12.99038,9.5) -- (-12.12436,10) -- (-12.12436,11) -- (-11.25833,11.5) -- (-11.25833,12.5) -- (-10.3923,13) -- (-10.3923,14) -- (-9.52628,14.5) -- (-9.52628,15.5) -- (-8.66025,16) -- (-7.79423,15.5) -- (-6.9282,16) -- (-6.06218,15.5) -- (-5.19615,16) -- (-4.33013,15.5) -- (-3.4641,16) -- (-2.59808,15.5) -- (-1.73205,16) -- (-0.86603,15.5) -- (0,16) -- (0.86603,15.5) -- (1.73205,16) -- (2.59808,15.5) -- (3.4641,16) -- (4.33013,15.5) -- (5.19615,16) -- (6.06218,15.5) -- (6.9282,16) -- (7.79423,15.5) -- (8.66025,16) -- (9.52628,15.5) -- (9.52628,14.5) -- (10.3923,14) -- (10.3923,13) -- (11.25833,12.5) -- (11.25833,11.5) -- (12.12436,11) -- (12.12436,10) -- (12.99038,9.5) -- (12.99038,8.5) -- (13.85641,8) -- (13.85641,7) -- (14.72243,6.5) -- (14.72243,5.5) -- (15.58846,5) -- (15.58846,4) -- (16.45448,3.5) -- (16.45448,2.5) -- (17.32051,2) -- (17.32051,1) -- (18.18653,0.5) -- (18.18653,-0.5) -- (17.32051,-1) -- (17.32051,-2) -- (16.45448,-2.5) -- (16.45448,-3.5) -- (15.58846,-4) -- (15.58846,-5) -- (14.72243,-5.5) -- (14.72243,-6.5) -- (13.85641,-7) -- (13.85641,-8) -- (12.99038,-8.5) -- (12.99038,-9.5) -- (12.12436,-10) -- (12.12436,-11) -- (11.25833,-11.5) -- (11.25833,-12.5) -- (10.3923,-13) -- (10.3923,-14) -- (9.52628,-14.5) -- (9.52628,-15.5) -- (8.66025,-16) -- (7.79423,-15.5) -- (6.9282,-16) -- (6.06218,-15.5) -- (5.19615,-16) -- (4.33013,-15.5) -- (3.4641,-16) -- (2.59808,-15.5) -- (1.73205,-16) -- (0.86603,-15.5) -- (0,-16) -- (-0.86603,-15.5) -- (-1.73205,-16) -- (-2.59808,-15.5) -- (-3.4641,-16) -- (-4.33013,-15.5) -- (-5.19615,-16) -- (-6.06218,-15.5) -- (-6.9282,-16) -- (-7.79423,-15.5) -- (-8.66025,-16) -- (-9.52628,-15.5) -- (-9.52628,-14.5) -- (-10.3923,-14) -- (-10.3923,-13) -- (-11.25833,-12.5) -- (-11.25833,-11.5) -- (-12.12436,-11) -- (-12.12436,-10) -- (-12.99038,-9.5) -- (-12.99038,-8.5) -- (-13.85641,-8) -- (-13.85641,-7) -- (-14.72243,-6.5) -- (-14.72243,-5.5) -- (-15.58846,-5) -- (-15.58846,-4) -- (-16.45448,-3.5) -- (-16.45448,-2.5) -- (-17.32051,-2) -- (-17.32051,-1) -- cycle;
\draw[draw=gdm,line width=.65pt] (-12.99038,-0.5) -- (-12.99038,0.5) -- (-12.12436,1) -- (-12.12436,2) -- (-11.25833,2.5) -- (-11.25833,3.5) -- (-10.3923,4) -- (-10.3923,5) -- (-9.52628,5.5) -- (-9.52628,6.5) -- (-8.66025,7) -- (-8.66025,8) -- (-7.79423,8.5) -- (-7.79423,9.5) -- (-6.9282,10) -- (-6.9282,11) -- (-6.06218,11.5) -- (-5.19615,11) -- (-4.33013,11.5) -- (-3.4641,11) -- (-2.59808,11.5) -- (-1.73205,11) -- (-0.86603,11.5) -- (0,11) -- (0.86603,11.5) -- (1.73205,11) -- (2.59808,11.5) -- (3.4641,11) -- (4.33013,11.5) -- (5.19615,11) -- (6.06218,11.5) -- (6.9282,11) -- (6.9282,10) -- (7.79423,9.5) -- (7.79423,8.5) -- (8.66025,8) -- (8.66025,7) -- (9.52628,6.5) -- (9.52628,5.5) -- (10.3923,5) -- (10.3923,4) -- (11.25833,3.5) -- (11.25833,2.5) -- (12.12436,2) -- (12.12436,1) -- (12.99038,0.5) -- (12.99038,-0.5) -- (12.12436,-1) -- (12.12436,-2) -- (11.25833,-2.5) -- (11.25833,-3.5) -- (10.3923,-4) -- (10.3923,-5) -- (9.52628,-5.5) -- (9.52628,-6.5) -- (8.66025,-7) -- (8.66025,-8) -- (7.79423,-8.5) -- (7.79423,-9.5) -- (6.9282,-10) -- (6.9282,-11) -- (6.06218,-11.5) -- (5.19615,-11) -- (4.33013,-11.5) -- (3.4641,-11) -- (2.59808,-11.5) -- (1.73205,-11) -- (0.86603,-11.5) -- (0,-11) -- (-0.86603,-11.5) -- (-1.73205,-11) -- (-2.59808,-11.5) -- (-3.4641,-11) -- (-4.33013,-11.5) -- (-5.19615,-11) -- (-6.06218,-11.5) -- (-6.9282,-11) -- (-6.9282,-10) -- (-7.79423,-9.5) -- (-7.79423,-8.5) -- (-8.66025,-8) -- (-8.66025,-7) -- (-9.52628,-6.5) -- (-9.52628,-5.5) -- (-10.3923,-5) -- (-10.3923,-4) -- (-11.25833,-3.5) -- (-11.25833,-2.5) -- (-12.12436,-2) -- (-12.12436,-1) -- cycle;
\draw[draw=gdordinary,line width=.65pt] (-7.79423,-0.5) -- (-7.79423,0.5) -- (-6.9282,1) -- (-6.9282,2) -- (-6.06218,2.5) -- (-6.06218,3.5) -- (-5.19615,4) -- (-5.19615,5) -- (-4.33013,5.5) -- (-4.33013,6.5) -- (-3.4641,7) -- (-2.59808,6.5) -- (-1.73205,7) -- (-0.86603,6.5) -- (0,7) -- (0.86603,6.5) -- (1.73205,7) -- (2.59808,6.5) -- (3.4641,7) -- (4.33013,6.5) -- (4.33013,5.5) -- (5.19615,5) -- (5.19615,4) -- (6.06218,3.5) -- (6.06218,2.5) -- (6.9282,2) -- (6.9282,1) -- (7.79423,0.5) -- (7.79423,-0.5) -- (6.9282,-1) -- (6.9282,-2) -- (6.06218,-2.5) -- (6.06218,-3.5) -- (5.19615,-4) -- (5.19615,-5) -- (4.33013,-5.5) -- (4.33013,-6.5) -- (3.4641,-7) -- (2.59808,-6.5) -- (1.73205,-7) -- (0.86603,-6.5) -- (0,-7) -- (-0.86603,-6.5) -- (-1.73205,-7) -- (-2.59808,-6.5) -- (-3.4641,-7) -- (-4.33013,-6.5) -- (-4.33013,-5.5) -- (-5.19615,-5) -- (-5.19615,-4) -- (-6.06218,-3.5) -- (-6.06218,-2.5) -- (-6.9282,-2) -- (-6.9282,-1) -- cycle;
\node[text=gdordinary,fill=white,inner sep=.55pt,font=\fontsize{7}{8}\selectfont] at (-9.8,22.8) {$\Gamma_{M}$};
\node[text=gdk,fill=white,inner sep=.55pt,font=\fontsize{7}{8}\selectfont] at (-8.4,19.8) {$\Gamma_{k}$};
\node[text=gdordinary,fill=white,inner sep=.55pt,font=\fontsize{7}{8}\selectfont] at (-7,16.8) {$\Gamma_{N}$};
\node[text=gdm,fill=white,inner sep=.55pt,font=\fontsize{7}{8}\selectfont] at (-4.9,12.3) {$\Gamma_{m}$};
\node[text=gdordinary,fill=white,inner sep=.55pt,font=\fontsize{7}{8}\selectfont] at (-2.8,7.8) {$\Gamma_{l}$};
\node[text=gdblack,inner sep=0pt,font=\fontsize{10}{11}\selectfont] at (1.5,4.1) {{$A$}};
\node[text=gdblack,inner sep=0pt,font=\fontsize{10}{11}\selectfont] at (3.5,13.1) {{$B$}};
\node[text=gdblack,inner sep=0pt,font=\fontsize{10}{11}\selectfont] at (5.7,20.25) {{$C$}};
\node[text=gdblack,inner sep=0pt,font=\fontsize{10}{11}\selectfont] at (-10.3,-0.5) {{$D$}};
\node[text=gdblack,inner sep=0pt,font=\fontsize{9}{10}\selectfont] at (6.5,24.1) {{$\widetilde\Lambda$}};
\end{tikzpicture}
\endgroup
}
\caption{The same partition with an intermediate {concentric graph-distance
disk} $\Gamma_m$, where
$\Gamma_{l}\subset\Gamma_m\subset\Gamma_N\subset\Gamma_k\subset\Gamma_M$
and ${l}<m<N<k<M$. The lattice cone is $\widetilde\Lambda$.
The yellow-green outline marks the boundary of $\Gamma_m$ within the light-blue
region $B$.}
\label{fig:haag-lemma7-regions}
\label{haag44}
\par\medskip
\begin{tabular}{@{}ll@{}}
$A=\Gamma_{l}\cap\widetilde\Lambda$
 & $B=(\Gamma_N\setminus\Gamma_{l})\cap\widetilde\Lambda$\\[3pt]
$C=(\Gamma_M\setminus\Gamma_N)\cap\widetilde\Lambda$
 & $D=\Gamma_N\cap\widetilde\Lambda^c$\\[3pt]
$B_1=(\Gamma_m\setminus\Gamma_{l})\cap\widetilde\Lambda$
 & $B_2=(\Gamma_N\setminus\Gamma_m)\cap\widetilde\Lambda$
\end{tabular}
\par\smallskip
Here $B=B_1\sqcup B_2$; the two parts are separated by the
intermediate outline. 

\end{figure}
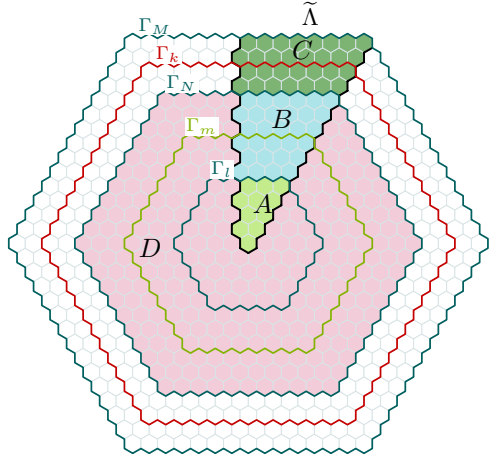

The {vanishing of the conditional mutual information} implies locality of modular automorphisms.
\begin{haaglemma}\label{lem:haag-markov-modular}
For every $X_A\in\A_A$ and $t\in\mathbb R$,
\[
 \sigma_L^{it}p_LX_Ap_L\sigma_L^{-it}
 =\sigma_{AB}^{it}p_{AB}X_Ap_{AB}\sigma_{AB}^{-it}p_L,
\]
and
\[
 \sigma_{ABC'}^{it}p_{ABC'}X_Ap_{ABC'}\sigma_{ABC'}^{-it}
 =\sigma_{AB}^{it}p_{AB}X_Ap_{AB}\sigma_{AB}^{-it}p_{ABC'}.
\]
Moreover,
\begin{align}\label{martingaele}
[p_{AB},p_{BC}]=0,\qquad p_L=p_{AB}p_{BC}.
\end{align}

\end{haaglemma}
\begin{proof}
By Lemma~{\ref{lem:haag-cmi}}, both $\sigma_{ABC}$ and $\sigma_{ABC'}$ are quantum
Markov states for the tripartitions $A$--$B$--$C$ and
$A$--$B$--$C'$, respectively. 
Hence, applying Lemma~\ref{lem:supportprojection} to $\sigma_{ABZ}$ with $Z=C,C'$ provides
\begin{equation}
[p_{AB},p_{BZ}]=0,
\qquad
p_{ABZ}=p_{AB}p_{BZ}.
\end{equation}
Combining this with Lemma~\ref{lem:petz-markov-modular} provides
\begin{equation}
\sigma_{ABZ}^{it}
p_{ABZ}X_Ap_{ABZ}
\sigma_{ABZ}^{-it}
=
\sigma_{AB}^{it}
p_{AB}X_Ap_{AB}
\sigma_{AB}^{-it}
p_{ABZ}.
\end{equation}
Taking $Z=C$ and using $L=ABC$ provides the first identity, while
$Z=C'$ provides the second.

\end{proof}

The same argument proves the next lemma.
\begin{haaglemma}\label{lem:haag-complement-compression}
Every $Y\in\A_D$ satisfies
\[
 p_{R_k}Yp_{R_k}p_{R_M}
 =p_{R_M}Yp_{R_M}
\]
as an operator on $\mathcal H_{\Gamma_M}$.
\end{haaglemma}
We also have {the following} intersection property.
\begin{haaglemma}\label{shika}
\begin{align}
p_{L_{N}}p_{L_{j}\setminus L_{l}}
=p_{L_j},\quad
N\le j\le M.
\end{align}
\end{haaglemma}
\begin{proof}
By data processing and Lemma~\ref{lem:haag-cmi},
\begin{align}
0&\le I\bigl(L_l:L_j\setminus L_N| L_N\setminus L_l\bigr)_{\sigma}
\nonumber\\
&\le I\bigl(L_l:L_M\setminus L_N| L_N\setminus L_l\bigr)_{\sigma}=I(A:C| B)_{\sigma}=0.
\end{align}
Hence $I\bigl(L_l:L_j\setminus L_N| L_N\setminus L_l\bigr)_{\sigma}=0$. 
Applying Lemma~\ref{lem:supportprojection} to the tripartition 
$L_l\cup (L_N\setminus L_l)\cup(L_j\setminus L_N)$ we obtain $p_{L_N}p_{L_j\setminus L_l}=p_{L_j}$.
\end{proof}
We recall the following standard fact.
\begin{haaglemma}\label{lem:haag-finite-modular}
For every nonempty $Y\subset\Gamma_M$, $X_Y\in\A_Y$, and
$t\in\mathbb R$,
\[
 \Delta_Y^{it}p_YX_Yp_Y\Delta_Y^{-it}
 =\sigma_Y^{it}p_YX_Yp_Y\sigma_Y^{-it}
\]
as operators on $\mathcal K_Y$.
\end{haaglemma}

\begin{proof}
This is a basic fact of finite-dimensional modular theory; see, for
example, \cite[Section~4]{DPModular}.
\end{proof}

\begin{haaglemma}\label{lem:haag-selfadjoint-mirror}
For every self-adjoint $X_A\in\A_A$, there exists a self-adjoint
$Y_D\in\A_D$ such that
\[
 j_L(p_LX_Ap_L)\Omega_M=p_RY_Dp_R\Omega_M.
\]
\end{haaglemma}

\begin{proof}
{Set $V:=ABC'$.
We fix} a self-adjoint $X_A\in\A_A$. For $Y=L,V$, set
\[
 a_Y:=p_YX_Ap_Y.
\]
Since $a_Y=a_Y^*$,  using Lemma \ref{lem:haag-finite-modular}, we have
\[
 j_Y(a_Y)\Omega_M
 =\Delta_Y^{1/2}a_Y\Omega_M
 =\sigma_Y^{1/2}a_Y\sigma_Y^{-1/2}\Omega_M.
\]

We next apply Lemma~\ref{lem:haag-markov-modular} at exponent $1/2$.
The density-matrix powers defined above are entire matrix-valued functions.
The identities of Lemma~\ref{lem:haag-markov-modular}, initially
stated for $z=it$, therefore extend to all $z\in\mathbb C$ by the
identity theorem. Taking $z=1/2$ and using $p_Y\Omega_M=\Omega_M$,
we obtain
\[
 \sigma_Y^{1/2}a_Y\sigma_Y^{-1/2}\Omega_M
 =\sigma_{AB}^{1/2}p_{AB}X_Ap_{AB}
 \sigma_{AB}^{-1/2}\Omega_M,
 \qquad Y=L,V.
\]
Consequently,
\begin{equation}\label{eq:haag-j-comparison}
 j_L(a_L)\Omega_M=j_V(a_V)\Omega_M.
\end{equation}

The Schmidt decomposition of $\Omega_M$ across $V\sqcup D$ provides
\[
 \mathcal K_V=p_V\mathcal H_V\otimes p_D\mathcal H_D.
\]
On this Hilbert space, the algebra $p_V\A_Vp_V$ acts as
\[
 B(p_V\mathcal H_V)\otimes I_{p_D\mathcal H_D},
\]
whose commutant is
\[
 I_{p_V\mathcal H_V}\otimes B(p_D\mathcal H_D).
\]
By the Tomita--Takesaki theorem, there is therefore a unique
$b_D\in B(p_D\mathcal H_D)$ with  $b_D=b_D^*$ such that
\[
 j_V(a_V)=I_{p_V\mathcal H_V}\otimes b_D
 \quad\text{on }\mathcal K_V.
\]

Extend $b_D$ by zero on $(p_D\mathcal H_D)^\perp$, and denote the
resulting self-adjoint element of $\A_D$ by $Y_D$.
Equation~\eqref{eq:haag-j-comparison} implies
\[
 \xi:=j_L(a_L)\Omega_M=Y_D\Omega_M.
\]
Since $p_R\Omega_M=\Omega_M$ and $p_R$ commutes with $p_L\A_L$,
it acts as the identity on $\mathcal K_L=p_L\A_L\Omega_M$.
In particular, $p_R\xi=\xi$, since $\xi\in\mathcal K_L$. Hence
\[
 j_L(a_L)\Omega_M
 =\xi=p_RY_D\Omega_M=p_RY_Dp_R\Omega_M,
\]
as required.
\end{proof}

\subsection{An \texorpdfstring{$\bbR$}{R}-linear map given by modular conjugation}\label{tl}
In this subsection, {we} fix $l, m, N, k,M_0\in\bbN$
so that
\begin{align}
\begin{split}
l\ge l_{\tilde \ld},\;
 m-l\ge r_{\tilde \ld}, \; N-m\ge r_{\tilde \ld}, \; k-N\ge r_{\tilde \ld},\;  M_0-k\ge r_{\tilde \ld}.
\end{split}
\end{align}
We specify the regions $A,B,C,C',D$ as in
{Figures}~{\ref{fig:haag-setting} and~\ref{fig:haag-lemma7-regions}}.

\begin{haaglemma}\label{lem:haag-volume-independent-mirrors}
There exists a real-linear map
\[
 T_{m,N}:\A_{L_m}^{\mathrm{sa}}\longrightarrow\A_{R_{N}}^{\mathrm{sa}}
\]
such that the following {holds: for every}
$M\ge M_0$, {every} $\Omega_M$ satisfying
{\eqref{OM}}, {and every}
{$X\in\A_{L_m}^{\mathrm{sa}}$}{,}
\[
 j_{L_M}^{(M)}\!\left(p_{L_M}^{(M)}X p_{L_M}^{(M)}\right)\Omega_M
 =p_{R_M}^{(M)}T_{m,N}(X)p_{R_M}^{(M)}\Omega_M.
\]

\end{haaglemma}
\begin{rem}
In particular, the map $T_{m,N}$ is independent of $M$ and
$\Omega_M$.
\end{rem}
\begin{proof}
Set $A':=L_m$, $B':=L_N\setminus L_m$, and $D:=R_N$.
These regions do not depend on $M$.
Choose a real basis $X_1,\ldots,X_s$ of $\A_{A'}^{\mathrm{sa}}$,
and choose a unit vector $\Omega_{M_0}\in\mathcal H_{\Gamma_{M_0}}$
satisfying (\ref{OM}).

By the same argument as in the proof of
Lemma~\ref{lem:haag-selfadjoint-mirror},
for every $i=1,\ldots,s$, there exists
$Y_i\in\A_D^{\mathrm{sa}}$ such that
\begin{equation}\label{eq:lem39a}
j_{L_{M_0}}^{(M_0)}
\left(
p_{L_{M_0}}^{(M_0)}
X_i
p_{L_{M_0}}^{(M_0)}
\right)
\Omega_{M_0}
=
p_{R_{M_0}}^{(M_0)}
Y_i
p_{R_{M_0}}^{(M_0)}
\Omega_{M_0}.
\end{equation}
Define $T_{m,N}(X_i):=Y_i$ and extend $T_{m,N}$ by real linearity to
$\A_{L_m}^{\mathrm{sa}}$.

We show that the same operators $Y_i$ work for every $M\ge M_0$
and every admissible $\Omega_M$.
By Theorem~\ref{thm:commutingparentH}(iv),
\begin{equation}\label{eq:SgOgSg}
\sigma_{\Gamma_k}^{(M)}
=
\omega_{\Gamma_k}
=
\sigma_{\Gamma_k}^{(M_0)}.
\end{equation}
Consequently, all reduced density matrices and their support
projections on subsets of $\Gamma_k$ are independent of $M$ and
of the choice of $\Omega_M$.

Set
\begin{equation}
W_i
:=
\sigma_{A'B'}^{1/2}
p_{A'B'}X_ip_{A'B'}
\sigma_{A'B'}^{-1/2},
\qquad
Z_i
:=
p_{R_k}Y_ip_{R_k}.
\end{equation}
Both $W_i$ and $Z_i$ are independent of $M$ and $\Omega_M$.
By Lemmas~\ref{lem:haag-markov-modular} and~\ref{lem:haag-finite-modular},
\begin{equation}\label{eq:lem39b}
j_{L_M}^{(M)}
\left(
p_{L_M}^{(M)}X_ip_{L_M}^{(M)}
\right)
\Omega_M
=
W_i\Omega_M.
\end{equation}
By Lemma~\ref{lem:haag-complement-compression},
\begin{equation}\label{eq:lem39c}
p_{R_M}^{(M)}
Y_i
p_{R_M}^{(M)}
\Omega_M
=
Z_i\Omega_M.
\end{equation}

Since $W_i-Z_i\in\A_{\Gamma_k}$ and
$\sigma_{\Gamma_k}^{(M)}=\sigma_{\Gamma_k}^{(M_0)}$, we have
\begin{equation}
\|(W_i-Z_i)\Omega_M\|^2
=
\|(W_i-Z_i)\Omega_{M_0}\|^2.
\end{equation}
By Eqs.~{{\eqref{eq:lem39a}, \eqref{eq:lem39b},
and~\eqref{eq:lem39c}}}, the right-hand side is zero.
Therefore
\begin{equation}
j_{L_M}^{(M)}
\left(
p_{L_M}^{(M)}X_ip_{L_M}^{(M)}
\right)
\Omega_M
=
p_{R_M}^{(M)}
Y_i
p_{R_M}^{(M)}
\Omega_M.
\end{equation}

Every $X\in\A_{L_m}^{\mathrm{sa}}$ has a unique expansion
\begin{equation}
X=\sum_{i=1}^s r_iX_i,
\qquad r_i\in\rr.
\end{equation}
Since $j_{L_M}^{(M)}$ is real-linear on the self-adjoint part, and
\begin{equation}
T_{m,N}(X)=\sum_{i=1}^s r_iY_i,
\end{equation}
the required identity follows.
\end{proof}

\subsection{Proof of Theorem \ref{haag}}
Now we prove the main theorem of this section.
Let $\tilde \ld$ be the given lattice cone {satisfying} {\it Condition M}. Set $L:=\tilde\ld$ and $R:=\tilde\ld^c$.

\begin{haaglemma}\label{lem:haag-modular-decomposition}
For $M\in\bbN$ {and}
every self-adjoint $x\in\mathcal M_{L_M}$, set
\[
 h:=\int_{\mathbb R}
 \frac{\sigma_{L_M}^{it}x\sigma_{L_M}^{-it}}
 {e^{\pi t}+e^{-\pi t}}\,dt,
\]
and define
\[
 \begin{aligned}
 y&:=i\left(
 \sigma_{L_M}^{-1/2}h\sigma_{L_M}^{1/2}
 -\sigma_{L_M}^{1/2}h\sigma_{L_M}^{-1/2}
 \right),\\
 z&:=2j_{L_M}(h).
 \end{aligned}
\]
Then the integral converges in operator norm,
\[
 y=y^*\in\mathcal M_{L_M},\qquad
 z=z^*\in\mathcal M_{L_M}',
\]
and
\begin{align}\label{saka}
    ix\Omega_M=y\Omega_M+iz\Omega_M
\end{align}
as an equality in $\mathcal K_{L_M}$.
\end{haaglemma}
\begin{proof}The first statement is trivial.
We show (\ref{saka}).

Define the linear map
\[
 \begin{aligned}
 I_1&:B(\mathcal K_{L_M})\longrightarrow B(\mathcal K_{L_M}),\\
 I_1(T)&:=\int_{\mathbb R}
 \frac{\Delta_{L_M}^{it}T\Delta_{L_M}^{-it}}
 {e^{\pi t}+e^{-\pi t}}\,dt.
 \end{aligned}
\]
By Lemma~\ref{lem:haag-finite-modular}, $I_1(x)=h$.
By the integral identity in
\cite[proof of Theorem~2.5.14, pp.~94--95]{BR1},
\[
 \Delta_{L_M}^{-1/2}I_1(x)\Delta_{L_M}^{1/2}
 +\Delta_{L_M}^{1/2}I_1(x)\Delta_{L_M}^{-1/2}=x,
\]
hence
\begin{equation}\label{eq:haag-modular-integral-inverse}
 \sigma_{L_M}^{-1/2}h\sigma_{L_M}^{1/2}
 +\sigma_{L_M}^{1/2}h\sigma_{L_M}^{-1/2}=x.
\end{equation}

Using $J_{L_M}\Omega_M=\Omega_M$ and $h=h^*$,
the polar decomposition of the Tomita operator provides
\[
 j_{L_M}(h)\Omega_M
 =\Delta_{L_M}^{1/2}h\Omega_M
 =\sigma_{L_M}^{1/2}h\sigma_{L_M}^{-1/2}\Omega_M.
\]
Therefore, by \eqref{eq:haag-modular-integral-inverse},
\[
 \begin{aligned}
 y\Omega_M+iz\Omega_M
 &=i\left(
 \sigma_{L_M}^{-1/2}h\sigma_{L_M}^{1/2}
 -\sigma_{L_M}^{1/2}h\sigma_{L_M}^{-1/2}
 \right)\Omega_M\\
 &\quad+2i\sigma_{L_M}^{1/2}h\sigma_{L_M}^{-1/2}\Omega_M\\
 &=ix\Omega_M.
 \end{aligned}
\]
This proves (\ref{saka}).
\end{proof}

Let ${l}_{\widetilde\Lambda}$ and $r_{\widetilde\Lambda}$ be the
constants in Condition M. 
For the rest of this section,
for each $l_{\tilde \ld}\le l\in\bbN$,
we set
\begin{align}
\begin{split}
 m_l:=l+ r_{\tilde \ld}, \; N_l:=m_l+  r_{\tilde \ld}, \; k_l:=N_l+ r_{\tilde \ld},\;  M_l^{(0)}:=k_l+ r_{\tilde \ld}.
\end{split}
\end{align}
Thus, for $M\ge M_{l}^{(0)}$,
\[
 \Gamma_{l}\subset\Gamma_{m_{l}}\subset\Gamma_{N_{l}}
 \subset\Gamma_{k_{l}}\subset\Gamma_M.
\]

We obtain uniform local maps for the vector decomposition.

\begin{haaglemma}\label{lem:haag-uniform-local-decomposition}
 For each $l_{\tilde \ld}\le l\in\bbN$,
there exist real-linear maps
\[
 S_{l}^{(1)}:\A_{L_{l}}^{\mathrm{sa}}\longrightarrow\A_{L_{N_l}}^{\mathrm{sa}},
 \qquad
 S_{l}^{(2)}:\A_{L_{l}}^{\mathrm{sa}}\longrightarrow\A_{R_{{N_l}}}^{\mathrm{sa}}
\]
such that, for every $X\in\A_{L_{l}}^{\mathrm{sa}}$ and $M\ge M_l^{(0)}$
and $\Omega_M$ satisfying (\ref{OM}),
\[
 i\,p_{L_M}^{(M)}X\Omega_M
 =p_{L_M}^{(M)}S_{l}^{(1)}(X)\Omega_M
 +i\,p_{R_M}^{(M)}S_{l}^{(2)}(X)\Omega_M.
\]
\end{haaglemma}

\begin{proof}
For each $M$ and each choice of $\Omega_M$ satisfying \eqref{OM},
define, for $j\le M$, the real-linear map
\[
 H_j^{(M)}:\A_{L_j}^{\mathrm{sa}}
 \longrightarrow\A_{L_j}^{\mathrm{sa}}
\]
by
\[
 H_j^{(M)}(X):=\int_{\mathbb R}
 \frac{\sigma_{L_{j}}^{it}p_{L_{j}}Xp_{L_{j}}\sigma_{L_{j}}^{-it}}
 {e^{\pi t}+e^{-\pi t}}\,dt.
\]
\eqref{eq:SgOgSg} implies that 
these maps are independent of $M$ and $\Omega_M$ for $M\ge M_l^{(0)}$ and $j\le k_l$;
we denote them by $H_j$.

Let  
$T_{m_l,N_l}:\A_{L_{m_l}}^{\mathrm{sa}}\longrightarrow\A_{R_{N_l}}^{\mathrm{sa}}$
 be the map in Lemma \ref{lem:haag-volume-independent-mirrors}. 
For $X\in\A_{L_{l}}^{\mathrm{sa}}$, set
\[
 \begin{aligned}
 S_{l}^{(1)}(X)&:=i\left(
 \sigma_{L_{N_l}}^{-1/2}H_{N_l}(X)\sigma_{L_{N_l}}^{1/2}
 -\sigma_{L_{N_l}}^{1/2}H_{N_l}(X)\sigma_{L_{N_l}}^{-1/2}\right),\\
 S_{l}^{(2)}(X)&:=2T_{m_l, N_l}\!\left(H_{m_l}(X)\right).
 \end{aligned}
\]
These define real-linear maps
 $S_{l}^{(1)}:\A_{L_{l}}^{\mathrm{sa}}\longrightarrow\A_{L_{N_l}}^{\mathrm{sa}}$
 and
 $ S_{l}^{(2)}:\A_{L_{l}}^{\mathrm{sa}}\longrightarrow\A_{R_{N_l}}^{\mathrm{sa}}$.

Let $M\ge {M_l^{(0)}}$ and {choose}
$\Omega_M$ satisfying \eqref{OM}.
For the calculation, abbreviate
\[
 \rho:=\sigma_{L_M}^{(M)},\quad p:=p_{L_M}^{(M)},\quad
 \mathcal M:=\mathcal M_{L_M}^{(M)},\quad
 j:=j_{L_M}^{(M)}.
\]

Let $X\in\A_{{L_l}}^{\mathrm{sa}}$.
From Lemma~\ref{lem:haag-modular-decomposition},
we have
\begin{align}
i pXp\Omega_M=
i\lmk \rho^{-\frac 12} H_M(X) \rho^{\frac 12}- \rho^{\frac 12} H_M(X) \rho^{-\frac 12}\rmk\Omega_M
+2ij\lmk H_M(X)\rmk\Omega_M.
\end{align}
From Lemma \ref{lem:haag-markov-modular}, we have
\begin{align}
i\lmk \rho^{-\frac 12} H_M(X) \rho^{\frac 12}- \rho^{\frac 12} H_M(X) \rho^{-\frac 12}\rmk\Omega_M
=pS_l^{(1)}(X)\Omega_M.
\end{align}
On the other hand, {Lemma}~\ref{lem:haag-volume-independent-mirrors} {implies that}
\begin{align}\label{neko}
j\!\left(pY p\right)\Omega_M
 =p_{R_M}^{(M)}T_{m_l,N_l}(Y)p_{R_M}^{(M)}\Omega_M
\end{align}
{holds} for every $Y\in\A_{L_{m_l}}^{\mathrm{sa}}$.
From Lemma~{\ref{lem:haag-markov-modular}}
and \eqref{neko},
\begin{align}
\begin{split}
&j\lmk H_M(X)\rmk\Omega_M
=j\lmk {pH_{m_l}(X)p}\rmk\Omega_M\\
&=p_{R_M}^{(M)}T_{m_l,N_l}\lmk H_{m_l}(X)\rmk p_{R_M}^{(M)}\Omega_M
=\frac 12p_{R_M}^{(M)} S_{l}^{(2)}\lmk X \rmk\Omega_M.
\end{split}
\end{align}
This completes the proof.
\end{proof}

Exchanging the {roles} of {$R$} and {$L$}, we obtain the following{.}
\begin{haaglemma}\label{lem:haag-uniform-local-decomposition-complement}
For each $l_{\tilde \ld}\le l\in\bbN$,
there exist real-linear maps
\[
 \tilde S_{l}^{(1)}:\A_{R_{l}}^{\mathrm{sa}}\longrightarrow\A_{R_{N_l}}^{\mathrm{sa}},
 \qquad
 \tilde S_{l}^{(2)}:\A_{R_{l}}^{\mathrm{sa}}\longrightarrow\A_{L_{{N_l}}}^{\mathrm{sa}}
\]
such that, for every $X\in\A_{R_{l}}^{\mathrm{sa}}$ and $M\ge M_l^{(0)}$
and $\Omega_M$ satisfying (\ref{OM}),
\[
 i\,p_{R_M}^{(M)}X\Omega_M
 =p_{R_M}^{(M)}\tilde S_{l}^{(1)}(X)\Omega_M
 +i\,p_{L_M}^{(M)}\tilde S_{l}^{(2)}(X)\Omega_M.
\]
\end{haaglemma}

The next lemma represents the cone-side support compression of any
local vector by an operator in the fixed complement region $R_{k_{l}}$.
\begin{haaglemma}\label{lem:haag-local-complement-map}
For each $l_{\tilde \ld}\le l\in\bbN$,
there exists a complex-linear map
\[
 \Theta_{l}:\A_{\Gamma_{l}}\longrightarrow\A_{R_{k_l}}
\]
such that
\begin{align}\label{inu}
 p_{L_M}^{(M)}B\Omega_M=\Theta_{l}(B)\Omega_M
 \qquad B\in\A_{\Gamma_{l}},
\end{align}
for all $M\ge  M_l^{(0)}${.}
\end{haaglemma}

\begin{proof}
Let $\widehat T_{m_l,N_l}:\A_{L_{m_l}}\longrightarrow\A_{R_{N_l}}$
be the complex-linear extension of the map
$T_{m_l,N_l}$ in
Lemma~\ref{lem:haag-volume-independent-mirrors}.
By Lemma~\ref{lem:haag-volume-independent-mirrors}, {for every}
$W\in \A_{L_{m_l}}$, {we have}
\begin{equation}\label{eq:haag-complex-mirror}
 p_{R_M}^{(M)}\widehat T_{m_l,N_l}(W)\Omega_M
 =j_{L_M}^{(M)}\!\left(p_{L_M}^{(M)}W^*p_{L_M}^{(M)}\right)\Omega_M.
\end{equation}

Define the complex-linear map
\[
\begin{aligned}
 F_{l}&:\A_{L_l}\longrightarrow\A_{L_{m_l}},\\
 F_{l}(X)&:=\sigma_{L_{m_l}}^{-1/2}p_{L_{m_l}}X p_{L_{m_l}}
                 \sigma_{L_{m_l}}^{1/2}.
\end{aligned}
\]
By \eqref{eq:SgOgSg}, for every $M\ge M_l^{(0)}$,
\[
p_{R_{k_l}}^{(M)}
=\operatorname{proj}_{\operatorname{supp}\omega_{R_{k_l}}}
=:q_{R_{k_l}}.
\]
We may define a {complex-linear} map $ \Theta_{l}:\A_{\Gamma_{l}}\longrightarrow\A_{R_{k_l}}$ by
\begin{equation}\label{eq:haag-theta-definition}
\Theta_l(X\otimes Z)
:=Zq_{R_{k_l}}\widehat T_{m_l,N_l}(F_l(X)),
\qquad X\in\A_{L_l},\quad Z\in\A_{R_l}.
\end{equation}

We verify (\ref{inu}). 
By (\ref{eq:haag-complex-mirror}) and Lemma \ref{lem:haag-markov-modular},
for any $X\in\A_{L_{l}}$
\begin{equation}\label{eq:haag-theta-cone-input}
\begin{aligned}
 p_{R_M}^{(M)}\widehat T_{m_l, N_l}\!\left(F_{l}(X)\right)\Omega_M
 &=j_{L_M}^{(M)}\!\left(p_{L_M}^{(M)}\!\left(F_{l}(X)\right)^*p_{L_M}^{(M)}\right)\Omega_M\\
 &=\sigma_{L_{M}}^{1/2}F_{l}(X) \sigma_{L_{M}}^{-1/2}\Omega_M\\
 &=\sigma_{L_{M}}^{1/2}\sigma_{L_{m_l}}^{-1/2}p_{L_{m_l}}X p_{L_{m_l}}
                 \sigma_{L_{m_l}}^{1/2}\sigma_{L_{M}}^{-1/2}\Omega_M\\
 &= \sigma_{L_{M}}^{1/2}\sigma_{L_{M}}^{-1/2}p_{L_{M}}X p_{L_{M}}
                 \sigma_{L_{M}}^{1/2}\sigma_{L_{M}}^{-1/2}\Omega_M            \\
  &=               p_{L_{M}}^{(M)}X \Omega_M.
  \end{aligned}
\end{equation}

Finally, let us consider {arbitrary} $X\in\A_{L_{l}}$ {and} $Z\in\A_{R_{{l}}}$.
Because
$Z$ commutes {with} $p_{L_{M}}$, using Lemma \ref{lem:haag-complement-compression},
we have
\[
\begin{aligned}
  p_{L_{M}}^{(M)}(X\otimes Z)\Omega_M
 &=Z p_{L_{M}}^{(M)}X\Omega_M\\
 &=Zp_{R_M}^{(M)}\widehat T_{m_l, N_l}\!\left(F_{l}(X)\right)\Omega_M\\
  &=Zp_{R_{k_l}}^{(M)}\widehat T_{m_l, N_l}\!\left(F_{l}(X)\right)\Omega_M\\
 &=\Theta_{l}(X\otimes Z)\Omega_M.
\end{aligned}
\]
\end{proof}
\begin{proofof}[Theorem \ref{haag}]
We prove Haag duality using the support projections of the finite-volume
states. The argument follows the proof of \cite[Lemma~2.2.3]{OPR} in the
case of vanishing error, with our notation and length scales.

Recall $Q_X$ from \eqref{eq:support-projector}.
By local uniqueness
(Theorem~\ref{thm:commutingparentH}(iv)), 
\begin{equation}\label{eq:haag-support-local-uniqueness}
 \sigma_{\Gamma_n}^{(M)}=\omega_{\Gamma_n}
\end{equation}
holds for each fixed $n$ and all sufficiently large $M$. 
In particular, $p_X^{(M)}=Q_X$ for $X\subset\Gamma_n$ and such $M$.

By Lemma~\ref{lem:haag-uniform-local-decomposition-complement},
 there exist real-linear maps
\[
 \widetilde S_{l}^{(1)}:
 \mathcal A_{R_{l}}^{\rm sa}\longrightarrow
 \mathcal A_{R_{N_{l}}}^{\rm sa},\qquad
 \widetilde S_{l}^{(2)}:
 \mathcal A_{R_{l}}^{\rm sa}\longrightarrow
 \mathcal A_{L_{N_{l}}}^{\rm sa}
\]
such that, for $M\ge M_{l}^{(0)}$,
\begin{equation}\label{eq:haag-support-finite-decomposition}
\begin{aligned}
 i\,p_{R_M}^{(M)}A\Omega_M
 ={}&p_{R_M}^{(M)}\widetilde S_{l}^{(1)}(A)\Omega_M\\
 &+i\,p_{L_M}^{(M)}\widetilde S_{l}^{(2)}(A)\Omega_M,
 \qquad A\in\mathcal A_{R_{l}}^{\rm sa}.
\end{aligned}
\end{equation}
By Lemma~\ref{lem:haag-local-complement-map}, there exists a complex-linear
map
\[
 \Theta_{l}:\mathcal A_{\Gamma_{l}}
 \longrightarrow\mathcal A_{R_{k_{l}}}
\]
such that, for $M\ge M_l^{(0)}$,
\begin{equation}\label{eq:haag-support-finite-theta}
 p_{L_M}^{(M)}B\Omega_M=\Theta_{l}(B)\Omega_M,
 \qquad B\in\mathcal A_{\Gamma_{l}}.
\end{equation}
These maps are independent of $M$ and of the admissible choice of
$\Omega_M$.

By the same argument as in Lemma~\ref{shika},
applied to both $L$ and $R$, we obtain
\begin{equation}\label{eq:haag-support-finite-factorization}
\begin{aligned}
 p_{R_n}^{(M)}&=p_{R_{k_{l}}}^{(M)}
 p_{R_n\setminus R_{N_{l}}}^{(M)},\\
 p_{L_n}^{(M)}&=p_{L_{k_{l}}}^{(M)}
 p_{L_n\setminus L_{N_{l}}}^{(M)},
 \qquad k_{l}\le n\le M,\quad M\ge M_{l}^{(0)}.
\end{aligned}
\end{equation}
The inner and outer regions overlap between the cutoffs $N_{l}$ and
$k_{l}$. Equations~\eqref{eq:haag-support-finite-decomposition}--\eqref{eq:haag-support-finite-factorization}
replace the corresponding finite-volume conditions in \cite{OPR}, with
zero error.

For finite regions $X\subset Y$, the support projections satisfy
\[
 Q_Y\le Q_X\otimes I_{Y\setminus X}.
\]
Indeed, $\omega(Q_X)=1$, so the support of $\omega_Y$ is contained in the
range of $Q_X\otimes I_{Y\setminus X}$. Hence, for $Y=L,R$, the
decreasing sequence $\pi(Q_{Y_n})$ has a strong limit. Set
\[
 Q_Y:=\operatorname*{s-lim}_{n\to\infty}\pi(Q_{Y_n}).
\]
These projections satisfy
\begin{equation}\label{eq:haag-support-infinite-projections}
 Q_L\in\pi(\mathcal A_L)'',\qquad
 Q_R\in\pi(\mathcal A_R)'',\qquad
 Q_L\Omega=Q_R\Omega=\Omega.
\end{equation}
Since the two represented algebras commute, $Q_L$ and $Q_R$ commute.

Fix ${l}\ge{l}_{\widetilde\Lambda}$ and $n\ge k_{l}$. Choose $M$
sufficiently large that \eqref{eq:haag-support-finite-factorization}
holds and \eqref{eq:haag-support-local-uniqueness} applies on $\Gamma_n$.
Then
\begin{equation}\label{eq:haag-support-reference-factorization}
 Q_{Y_n}=Q_{Y_{k_{l}}}Q_{Y_n\setminus Y_{N_{l}}},
 \qquad Y=L,R.
\end{equation}
For $Z\in\mathcal A_{\Gamma_{N_{l}}}$, the outer projection commutes
with $Z$, and its image under $\pi$ fixes $\Omega$. Therefore
\[
 \pi(Q_{Y_n})\pi(Z)\Omega
 =\pi(Q_{Y_{k_{l}}})\pi(Z)\Omega.
\]
Taking $n\to\infty$, we obtain
\begin{equation}\label{eq:haag-support-localization}
 Q_Y\pi(Z)\Omega=\pi(Q_{Y_{k_{l}}})\pi(Z)\Omega,
 \qquad Z\in\mathcal A_{\Gamma_{N_{l}}},\quad Y=L,R.
\end{equation}

We first prove the counterpart of \cite[Equation~(12)]{OPR}:
\begin{equation}\label{eq:haag-support-cyclicity}
 Q_L\mathcal H=\overline{\pi(\mathcal A_R)\Omega}.
\end{equation}
Fix $B\in\mathcal A_{\Gamma_{l}}$.
Equations~\eqref{eq:haag-support-finite-theta} and
\eqref{eq:haag-support-finite-factorization}, with $n=M$, provide
\begin{equation}\label{eq:haag-support-localized-theta}
\begin{aligned}
 \Theta_{l}(B)\Omega_M
 &=p_{L_M}^{(M)}B\Omega_M\\
 &=p_{L_{k_{l}}}^{(M)}
   p_{L_M\setminus L_{N_{l}}}^{(M)}B\Omega_M\\
 &=p_{L_{k_{l}}}^{(M)}B\Omega_M.
\end{aligned}
\end{equation}
The last equality follows because the outer projection commutes with $B$
and fixes $\Omega_M$.

Choose $M$ still larger, if necessary, so that
\eqref{eq:haag-support-local-uniqueness} applies on $\Gamma_{k_{l}}$.
Set
\[
 D_{{l},B}:=Q_{L_{k_{l}}}B-\Theta_{l}(B)
 \in\mathcal A_{\Gamma_{k_{l}}}.
\]
Equation~\eqref{eq:haag-support-localized-theta} implies $D_{{l},B}\Omega_M=0$. Since $D_{{l},B}$ is a fixed local operator,
local uniqueness yields
\[
\begin{aligned}
 \|\pi(D_{{l},B})\Omega\|^2
 &=\omega(D_{{l},B}^*D_{{l},B})\\
 &=\langle\Omega_M,D_{{l},B}^*D_{{l},B}\Omega_M\rangle=0.
\end{aligned}
\]
Together with \eqref{eq:haag-support-localization}, this implies
\begin{equation}\label{eq:haag-support-infinite-theta}
\begin{aligned}
 Q_L\pi(B)\Omega
 &=\pi(Q_{L_{k_{l}}})\pi(B)\Omega\\
 &=\pi(\Theta_{l}(B))\Omega\in\pi(\mathcal A_R)\Omega.
\end{aligned}
\end{equation}
Every local observable belongs to $\mathcal A_{\Gamma_{l}}$ for some
sufficiently large ${l}$. Since local observables applied to $\Omega$
form a dense subspace of $\mathcal H$,
Equation~\eqref{eq:haag-support-infinite-theta} implies 
\[
 Q_L\mathcal H\subset\overline{\pi(\mathcal A_R)\Omega}.
\]
The reverse inclusion follows from the facts that $Q_L$ commutes with
$\pi(\mathcal A_R)$ and fixes $\Omega$. This proves
\eqref{eq:haag-support-cyclicity}.

Next, set
\[
 \mathcal M:=\pi(\mathcal A_R)'',\qquad
 \mathcal N:=\pi(\mathcal A_L)'',\qquad E:=Q_RQ_L.
\]
Thus $\mathcal M,\mathcal N$ are commuting von Neumann algebras on
$\mathcal H$, and $E$ is the projection onto the intersection of the
ranges of $Q_R$ and $Q_L$. We prove the counterpart of
\cite[Equation~(11)]{OPR}:
\begin{equation}\label{eq:haag-support-real-density}
 \overline{\bigl(E\mathcal M^{\rm sa}E
 +iE\mathcal N^{\rm sa}E\bigr)\Omega}=E\mathcal H.
\end{equation}
Fix $A\in\mathcal A_{R_{l}}^{\rm sa}$. Both
$\widetilde S_{l}^{(1)}(A)$ and $\widetilde S_{l}^{(2)}(A)$ are
supported in $\Gamma_{N_{l}}$. Applying
\eqref{eq:haag-support-finite-factorization} to each term in
\eqref{eq:haag-support-finite-decomposition} therefore gives 
\begin{equation}\label{eq:haag-support-localized-decomposition}
\begin{aligned}
 i\,p_{R_{k_{l}}}^{(M)}A\Omega_M
 ={}&p_{R_{k_{l}}}^{(M)}\widetilde S_{l}^{(1)}(A)\Omega_M\\
 &+i\,p_{L_{k_{l}}}^{(M)}\widetilde S_{l}^{(2)}(A)\Omega_M.
\end{aligned}
\end{equation}
Choose $M$ sufficiently large that
\eqref{eq:haag-support-local-uniqueness} applies on $\Gamma_{k_{l}}$,
and set
\[
\begin{aligned}
 D_{{l},A}:={}&iQ_{R_{k_{l}}}A
 -Q_{R_{k_{l}}}\widetilde S_{l}^{(1)}(A)\\
 &-iQ_{L_{k_{l}}}\widetilde S_{l}^{(2)}(A)
 \in\mathcal A_{\Gamma_{k_{l}}}.
\end{aligned}
\]
Equation~\eqref{eq:haag-support-localized-decomposition} and local
uniqueness imply
\[
 \|\pi(D_{{l},A})\Omega\|^2
 =\langle\Omega_M,D_{{l},A}^*D_{{l},A}\Omega_M\rangle=0.
\]
Using \eqref{eq:haag-support-localization}, we obtain
\begin{equation}\label{eq:haag-support-infinite-decomposition}
\begin{aligned}
 iQ_R\pi(A)\Omega
 ={}&Q_R\pi\!\left(\widetilde S_{l}^{(1)}(A)\right)\Omega\\
 &+iQ_L\pi\!\left(\widetilde S_{l}^{(2)}(A)\right)\Omega.
\end{aligned}
\end{equation}
Denote the left-hand side of \eqref{eq:haag-support-real-density} by
$\mathcal D$. It is a closed real-linear subspace of $E\mathcal H$.
For every self-adjoint local observable $B\in\mathcal A_R$,
\[
 Q_R\pi(B)\Omega=E\pi(B)E\Omega\in\mathcal D.
\]
Multiplying \eqref{eq:haag-support-infinite-decomposition} by $E$ from
the left also provides

\begin{align}\label{oil}
\begin{split}
 iQ_R\pi(A)\Omega
 ={}&E\pi\!\left(\widetilde S_{l}^{(1)}(A)\right)E\Omega\\
 &+iE\pi\!\left(\widetilde S_{l}^{(2)}(A)\right)E\Omega
 \in\mathcal D.
 \end{split}
\end{align}

Here the two terms belong to the required real-linear spaces because the
values of the two maps are self-adjoint and supported in $R$ and $L$,
respectively.

Every local observable is the sum of a self-adjoint observable and $i$
times a self-adjoint observable. Consequently, since \eqref{oil} holds for every sufficiently
large $l$ and every $A\in\mathcal A_{R_l}^{\rm sa}$,
\[
 Q_R\pi(\mathcal A_{R,\mathrm{loc}})\Omega\subset\mathcal D,
\]
where $\mathcal A_{R,\mathrm{loc}}$ denotes the algebra of observables
with finite support in $R$. By \eqref{eq:haag-support-cyclicity},
\[
 E\mathcal H
 =\overline{Q_R\pi(\mathcal A_{R,\mathrm{loc}})\Omega}
 \subset\mathcal D.
\]
The reverse inclusion follows from the definition of $\mathcal D$.
This proves \eqref{eq:haag-support-real-density}.

We now repeat the proof of \cite[Lemma~2.1.4]{OPR}, with its region
$\Gamma$ replaced by $R$ and its projections $P_\Gamma,P_{\Gamma^c}$
replaced by $Q_R,Q_L$. The required projection properties and density
conditions are \eqref{eq:haag-support-infinite-projections},
\eqref{eq:haag-support-real-density}, and
\eqref{eq:haag-support-cyclicity}, and $\omega$ is pure. Hence
\[
 \pi(\mathcal A_L)'=\pi(\mathcal A_R)''.
\]
Taking commutants yields
\[
 \pi(\mathcal A_R)'=\pi(\mathcal A_L)''.
\]

The required inclusion for every planar cone follows from Lemma~\ref{rem:lattice-cone-approximation} and lattice Haag duality.
Thus approximate Haag duality holds with all unitaries in \cite[Definition~1.1]{Ogata} chosen to be the identity.
\end{proofof}

\section{DHR theory}\label{sec:superselection-framework}
We briefly recall DHR theory for two-dimensional quantum spin systems.
Objects of the DHR category will be given by representations satisfying the superselection criterion.

\begin{defn}
We say a representation $\rho$ of $\caA$ on $\caH$
satisfies the superselection criterion for $\pi$ if 
$\rho\vert_{\caA_{\Lambda^c}}$ and
$\pi\vert_{{\caA_{\Lambda^c}}}$ are
unitarily equivalent
for any planar cone $\Lambda$ in $\bbR^2$.
\end{defn}
We extend such representations to endomorphisms of a larger algebra as follows.
For a cone $\ld_0$, we denote by $\caO_{\ld_0}$ the set of all representations of $\caA$ on $\caH$
satisfying the superselection criterion for $\pi$
and 
\begin{align}
\rho\vert_{\caA_{\Lambda_0^c}}=
 \pi\vert_{{\caA_{\Lambda_0^c}}}.
\end{align}
In this case, we say $\rho$ is localized in $\ld_0$.

{{
For each $\theta\in\mathbb R$ and $\varphi\in(0,\pi)$,
set
\begin{equation}\label{kiiro}
 \mathcal B_{(\theta,\varphi)}
 :=
 \overline{
  \bigcup_{\Lambda\in\mathcal C_{(\theta,\varphi)}}
  \pi(\mathcal A_{\Lambda^c})'
 }^{\|\cdot\|}.
\end{equation}
Recall $\mathcal C_{(\theta,\varphi)}$
from \eqref{kyuri}.
{Throughout the rest of this paper, we fix
$\theta_0=3\pi/4$ and $\varphi_0=\pi/36$, and write
$\mathcal B:=\mathcal B_{(\theta_0,\varphi_0)}$.
}

}}
{{
Whenever we use DHR tensor products or representation
extensions below, the specified localization cones
of the representations involved are chosen from
$\mathcal C_{(\theta_0,\varphi_0)}$.
}}
A representation satisfying the superselection criterion can be extended to an endomorphism on this algebra.
\begin{lem}
{{
Let
$\Lambda_0\in\mathcal C_{(\theta_0,\varphi_0)}$ and $\rho\in \caO_{\ld_0}$.
There is a unique endomorphism
$ T_\rho$ on $\mathcal B$
such that
\[
 T_\rho\circ\pi=\rho
\]
and $T_\rho$ is normal on
$\pi(\mathcal A_\Lambda)''$ and $\pi(\caA_{\ld^c})'$
for every
$\Lambda\in\mathcal C_{(\theta_0,\varphi_0)}$.
For any two representations $\rho,\sigma$
satisfying these conditions,
$T_\rho\circ T_\sigma$ is also normal on
each $\pi(\caA_{\ld^c})'$.
}}
\end{lem}

\begin{proof}
By Theorem \ref{haag}, $\omega$ satisfies
approximate Haag duality for any planar cones. 
{{
Apply \cite[Lemmas~2.14, 2.17, and~3.5]{Ogata},
taking the localizing unitary to be the identity.
}}
\end{proof}

Under the hypotheses below, these representations form a braided $C^*$-tensor category describing anyons.
\begin{ithm}{\cite[proof of Theorem~5.2]{Ogata}}
Let  $\omega_{\Phi}$ be a  gapped ground state of a uniformly bounded {finite-range} interaction $\Phi$
on $\caA$. Suppose that its
GNS representation satisfies
 {approximate} Haag duality.
 Then its superselection sectors 
 form a braided $C^*$-tensor category $\caC_\omega^{\ld_0}$ with {nonzero} objects
 $\caO_{\ld_0}$.
 \end{ithm}

See \cite{NT} for the definition of
 {a} braided $C^*$-tensor category.
 In particular, 
for objects $\rho,\sigma\in \caO_{\ld_0}$,
the homomorphism space from $\rho$ to $\sigma$
 is the space of intertwiners
 \begin{align}
     (\rho,\sigma):=
     \left\{
R\in\caB(\caH) \mid \sigma(A)R=R\rho(A),\;\;A\in\caA
     \right\}.
 \end{align}
 An object $\rho$ is called simple if
\[
\mathrm{End}(\rho):=(\rho,\rho)
=\mathbb C\,\mathrm{id}_{\caH}.
\]
 For objects $\rho_1,\rho_2\in \caO_{\ld_0}$,
  their tensor {product}
 $\rho_1\otimes \rho_2\in\caO_{\ld_0}$
 is defined by
 \begin{align}
     \rho_1\otimes \rho_2
     :=T_{\rho_1}T_{\rho_2}\pi
 \end{align}
 The tensor unit is $\pi$. For $\rho_i,\sigma_i\in\caO_{\ld_0}$, $i=1,2$,
if $R\in (\rho_1,\sigma_1)$ and $S\in (\rho_2,\sigma_2)$,
we have $R,S\in\caB$ and
their tensor product is defined by
\[
R\otimes S:=RT_{\rho_1}(S).
\]See \cite[proofs of Lemmas~3.6, 4.3, and~5.6]{Ogata}.
If cones $\ld_1,\ld_2\in  C_{(\theta_0,\varphi_0)}$  are disjoint and
 both are included in {another} cone $\ld_3\in  C_{(\theta_0,\varphi_0)}
 $, for any 
 $\rho_1^{\ld_1}\in \caO_{\ld_1}$ and $\rho_2^{\ld_2}\in\caO_{\ld_2}$,
we can define
$\rho_1^{\ld_1}\otimes \rho_2^{\ld_2}\in \caO_{\ld_3}$, satisfying
\begin{align}
\left(\rho_1^{\ld_1}\otimes \rho_2^{\ld_2}\right)\lmk AB\rmk
=\rho_1^{\ld_1}(A)\rho_2^{\ld_2}(B),
\quad A\in \caA_{\ld_1},\quad B\in \caA_{\ld_2}.
\end{align}

\section{Infinite-volume states and superselection sectors}\label{sec:infinite-volume}
In this section we show the following theorem.
\begin{thm}\label{anyontheorem}
In the setting of Subsection~\ref{overallsetting},
there is a one-to-one correspondence between $\ceb$
and  equivalence classes of simple objects in $\caC_\omega^{\ld_0}$. 
\end{thm}

For an entanglement-bootstrap charge $a\in\mathcal C^{\mathrm{EB}}$
and a thick enough disk $D$, we take the
thermodynamic limit of the charge-$a$ annular states
 to obtain a state
$\widetilde\varphi_{aD}$, on an infinite annulus
(Lemma~\ref{lem:annular-limit}).

Using Theorem \ref{thm:commutingparentH},
$\widetilde\varphi_{aD}$ can be seen as a frustration free ground state with respect to the interaction terms outside of a disk $D'$. The set of states on $\caA$
that are frustration free on the complement of $D'$ forms a convex weak-$^*$closed face (Lemma~\ref{lem:exterior-ground-states}), hence all its extreme points are pure.
And
any extension of $\widetilde\varphi_{aD}$ to $\caA$ belongs to this face (Lemma \ref{lem:state-extensions}).
From this fact and the annulus simplex theorem (Theorem~\ref{thm:annlusconvex})
we show the existence of a pure state extension
$\varphi_{aD}$ of $\widetilde\varphi_{aD}$ to $\caA$
(Lemma~\ref{lem:pure-charge-states}).
The GNS representations of any two such pure extensions
with the same charge are unitarily equivalent, since
the states agree outside a finite disk
(Lemma~\ref{lem:state-extensions}).

Let $(\mathcal H,\pi,\Omega)$ be the GNS triple of $\omega$,
where $\pi:\caA\to B(\mathcal H)$ is its representation
and $\Omega$ its cyclic unit vector.
Since $\varphi_{aD}$ agrees with $\omega$ on every
lattice cone satisfying Condition D away from $D$,
any GNS representation of $\varphi_{aD}$ on $\mathcal H$
satisfies the superselection criterion, by
Lemma~\ref{lem:superselection-criterion}.
Although the choice of a GNS representation of
$\varphi_{aD}$ is not unique, all such choices are
unitarily equivalent.
Since $\varphi_{aD}$ is pure, these representations
are irreducible.
We therefore obtain a well-defined map from the charge
label set $\mathcal C^{\mathrm{EB}}$ to the unitary
equivalence classes of irreducible representations of
$\caA$ satisfying the superselection criterion with
respect to $\pi$.

For the converse, let $\rho:\caA\to B(\mathcal H)$ be
an irreducible representation satisfying the superselection
criterion, and let $\xi\in\mathcal H$ be any unit vector.
Adapting the argument of Bols and Vadnerkar
\cite[Proposition~5.2]{BV} to our commuting projector
parent Hamiltonian, we show that the vector state
$\langle\xi,\rho(\,\cdot\,)\xi\rangle$ is unitarily
equivalent to
a state on $\caA$ that is frustration free outside
a finite disk; see the proof of Lemma~\ref{lem:ssc-normality}.
By LTQO, the restriction of this state to the exterior
of a sufficiently enlarged disk belongs to the
corresponding information convex set.
The exterior decomposition, the irreducibility of $\rho$,
and the disjointness of distinct charges then show that
$\rho$ corresponds to a unique charge.
This proves Theorem~\ref{anyontheorem}.

Furthermore, when $D$ is deep inside a prescribed cone
$\Lambda$, we can choose a GNS representation
$\rho_a^{D\Lambda}:\caA\to B(\mathcal H)$ of
$\varphi_{aD}$ with cyclic vector $\Omega$ that is
localized in $\Lambda$, meaning that it agrees with
$\pi$ on $\caA_{\Lambda^c}$
(Proposition~\ref{sushi}).
Under the geometric hypotheses of
Lemma~\ref{lem:cone-local-approximation}, vectors in
$\pi(\caA_\Lambda)''\Omega$ can be approximated in norm
by vectors in $\rho_a^{D\Lambda}(\caA_{\Lambda'})\Omega$
for a suitably enlarged cone $\Lambda'\supset\Lambda$.
This choice will be useful later.

\subsection{States in the information convex on an infinite annulus}

Let 
${\mathfrak S}(\mathcal A_{\Gamma})$ denote the state space of the observable
algebra on $\Gamma\subset \bbL$.
We introduce
information convex on
infinite sets:
\begin{notation}\label{notation:regions}
    For an infinite lattice region $X$, set
\[
 \widetilde\Sigma_\omega(X)
 :=\left\{\varphi\in {\mathfrak S}(\mathcal A_{X^+}):
 \varphi|_{\mathcal A_{D_f}}=\omega|_{\mathcal A_{D_f}}
 \text{ for every }f\in X\right\}.
\]
Then information convex on $X$ is
\[
\Sigma_\omega(X)
:=\left\{
\varphi|_{\mathcal A_X}:
\varphi\in\widetilde\Sigma_\omega(X)
\right\}
\subset {\mathfrak S}(\mathcal A_X).
\]
We occasionally omit $\omega$ and write 
$\widetilde\Sigma(X)$ and $\Sigma(X)$.
\end{notation}

\begin{lemma}\label{lem:cone-disk-uniqueness}
{Let $\td \Lambda$ be a lattice cone satisfying Condition D.
Then
\[
 \Sigma_\omega(\td \Lambda)
 =\{\omega|_{\mathcal A_{\td \Lambda}}\}.
\]
}

\end{lemma}

\begin{proof}
This {follows immediately} from the {definition} of Condition D and Lemma~\ref{lem:finite-disk-local-uniqueness}.
\end{proof}

{
{%
\long\gdef\DiskBlueStrikeBox#1{%
  \leavevmode\begingroup
  \setbox0=\hbox{#1}%
  \rlap{\raisebox{.5ex}{{\rule{\wd0}{.45pt}}}}%
  \box0\endgroup}
\long\gdef\DiskBlueDelete#1{%
  \begingroup\DiskBlueDeleteWords#1 {\DiskBlueDeleteEnd} \endgroup}
\long\gdef\DiskBlueDeleteWords#1 {%
  \def\DiskBlueWord{#1}\def\DiskBlueStop{\DiskBlueDeleteEnd}%
  \ifx\DiskBlueWord\DiskBlueStop
  \else
    \DiskBlueStrikeBox{#1}\space
    \expandafter\DiskBlueDeleteWords
  \fi}
\long\gdef\DiskBlueDeleteMath#1{%
  \mbox{\DiskBlueStrikeBox{$\displaystyle #1$}}}

}}
For standard disks (Definition \ref{standard}), the following lemma constructs the thermodynamic limit of the charge-$a$ annular states.
\begin{lemma}\label{lem:annular-limit}
Let $D$ be a standard disk and {set}

\begin{align}\label{hikooki}
 A_{l}:=D^{+{l}}\setminus D^{-2l_0},
 \quad {l}\ge0.
\end{align}
For $a\in\mathcal C^{\mathrm{EB}}$, let
\[
 \sigma_a^{{l} D}\in\operatorname{ext}\Sigma(A_{l})
\]
denote the corresponding extreme {state} given in Eq.~\eqref{eq:annulus-simplex}.
 Then there exists a unique state
$\widetilde\varphi_{aD}$ on $\A_{(D^{-l_0})^c}$ such that
\[
 \widetilde\varphi_{aD}|_{\A_{A_{l}\cap (D^{-l_0})^c}}
 =\sigma_a^{{l} D}|_{\A_{A_{l}\cap (D^{-l_0})^c}}\qquad({l}\ge0).
\]

\end{lemma}
{\begin{proof}
Since $D$ is a standard disk and $l_0\ge13$,
each $A_l$ is a regular graph annulus (see Remark~\ref{rem:graphannulus}).
The nested annuli $A_l\subset A_{l+1}$ have
the same hole. Hence ~\eqref{hebi}
implies
\begin{align}\label{sl}
 \sigma_a^{(l+1)D}|_{\mathcal A_{A_l}}
 =\sigma_a^{lD}.
\end{align}
The regions $A_l\cap(D^{-l_0})^c$ increase to
$(D^{-l_0})^c$.
The compatible restrictions, therefore, determine
a unique state on $\mathcal A_{(D^{-l_0})^c}$.
\end{proof}
}

\begin{nota}
For the rest of this paper{,} for each standard disk 
$ D$ and $a\in\mathcal C^{\mathrm{EB}}$, the state 
$\widetilde\varphi_{aD}$
on $(D^{-l_0})^c$
will denote the unique state given above.
\end{nota}

\begin{lem}\label{tomato}
Let $ D=\{f_0\}^{+s}$ be a standard disk.
Then each $\widetilde\varphi_{aD}$, $a \in\mathcal C^{\mathrm{EB}}${,} is a factor state on
$\A_{(D^+)^c}$.
\end{lem}
\begin{proof}
By \cite[Corollary~D.5.1]{SKK} and the definition of $\widetilde\varphi_{aD}$,
for any finite $F\subset (D^+)^c$, there exists
an $N\in\bbN$ such that
$\widetilde\varphi_{aD}(xy)=\widetilde\varphi_{aD}(x)\widetilde\varphi_{aD}(y)$
for any $x\in\caA_{F}$ and $y\in \caA_{\Gamma_N^c}$.
This, together with \cite[Theorem~2.6.10 and Example~2.6.12]{BR1}, proves factoriality.

\end{proof}
\begin{lemma}\label{lem:annular-separation}
Let $ D$ be a standard disk.
Let $D'$, $D''$ be disks  such that $D^{+2}\subset {D{'}}$, ${D'}^{+2}\subset D''$ 
and set $A:={D''}\setminus D'$.
Assume that $A$ is a regular annulus of thickness at least $12$.
Then for any $a,b\in\mathcal C^{\mathrm{EB}}$, 
we have
\[
 \widetilde\varphi_{aD}({S_A^b})=\delta_{ab},
 \qquad a,b\in\mathcal C^{\mathrm{EB}}.
\]
In particular, the states $\widetilde\varphi_{aD}$,
$a\in\mathcal C^{\mathrm{EB}}$, are mutually singular and have
pairwise disjoint GNS representations.

\end{lemma}
\begin{proof}
This follows from Lemma \ref{hashi} and the definition of
$\widetilde\varphi_{aD}$ and  \cite[Corollary~2.6.11]{BR1}.
\end{proof}

Now we would like to extend $\tilde\varphi_{aD}$ to a \emph{pure} state on $\caA$.
In order to do so, we introduce the following.
\begin{notation}

For $S\Subset\LL$, set
\[
 \mathcal F_S:=\bigl\{\psi\in {\mathfrak S}(\A):
 \psi(I-Q_X)=0\ \text{for every }X\in\mathcal S
 \text{ with }X\subset S^c\bigr\}.
\]
\end{notation}
\begin{lemma}\label{lem:exterior-ground-states}
Let $S\Subset\LL$. Then $\mathcal F_S$ is a nonempty weak-$^*$ closed
face of ${\mathfrak S}(\A)$.
\end{lemma}

\begin{proof}
Since $\omega$ is frustration free, $\omega\in\mathcal F_S$.
Each defining condition is a weak-$^*$ continuous affine equality, so
$\mathcal F_S$ is convex and weak-$^*$ closed. If
\[
 t\psi_1+(1-t)\psi_2\in\mathcal F_S,\qquad 0<t<1,
\]
then positivity of $I-Q_X$ implies
\[
 \psi_1(I-Q_X)=\psi_2(I-Q_X)=0
\]
for every $X\in\mathcal S$ contained in $S^c$. Hence
$\psi_1,\psi_2\in\mathcal F_S$, proving the face property.

\end{proof}
\begin{lemma}\label{lem:exterior-cone-agreement}
Let $S\Subset\LL$.
Let $\ld\subset\LL$ be a cone {satisfying} {\it Condition D}
such that $\ld\subset \lmk S^{+l_0}\rmk^c$.
Then every $\psi\in\mathcal F_S$ satisfies
\[
 \psi|_{\A_\ld}=\omega|_{\A_\ld}.
\]
\end{lemma}

\begin{proof}

Let $\psi\in\mathcal F_S$. Consider $K_N$ in {\it Condition D}.
For every $N$,
\[
 K_N^{+{l}_0}\subset\ld^{+{l}_0}\subset S^c.
\]
Thus $\psi_{K_N^{+{l}_0}}$ has zero energy for every interaction term
supported in $K_N^{+{l}_0}$. Moreover,
\[
 d\!\left(K_N,\LL\setminus K_N^{+{l}_0}\right)\ge{l}_0+1.
\]
Theorem~\ref{thm:commutingparentH}(iv){,
applied to the disks $D_f=\{f\}^{+1}$ and
$\{f\}^{+l_0}\subset K_N^{+l_0}$
for each $f\in K_N$,}
therefore provides
{$\psi_{D_f}=\omega_{D_f}$.
Lemma~\ref{lem:finite-disk-local-uniqueness} then implies
}
$\psi_{K_N}=\omega_{K_N}$.
Taking $N\to\infty$ completes the proof.
\end{proof}
Now we relate $\td\varphi_{aD}$ with $\caF_S$.
\begin{lemma}\label{lem:state-extensions}
Let $f_0\in\LL$ and $s\in\mathbb Z_{\ge0}$ and let
 $D:=\{f_0\}^{+s}$ be a standard disk{, and let}
  $a\in\mathcal C^{\mathrm{EB}}$.
Then the following hold.
\begin{enumerate}
\item[(i)] The state $\widetilde\varphi_{aD}$ in Lemma \ref{lem:annular-limit} has a state extension
$\varphi_{aD}$ to $\A$.
\item[(ii)] Every such extension belongs to $\mathcal F_{D^+}$.
\item[(iii)] Any two such extensions have quasi-equivalent GNS
representations.
\end{enumerate}
\end{lemma}

\begin{proof}
(i) is trivial.
For (ii), let $\varphi$ be any such extension, and let
$X\in\mathcal S$ satisfy $X\subset(D^+)^c$.
We have
\[
 \varphi(I-Q_X)
 =\widetilde\varphi_{aD}(I-Q_X)
 =\omega_X(I-Q_X)=0.
\]
Thus $\varphi\in\mathcal F_{D^+}$.

For (iii), since $\A_{D^+}$ is a full matrix algebra, any two state
extensions of $\widetilde\varphi_{aD}$ to
$\A_{D^+}\otimes\A_{(D^+)^c}$ have quasi-equivalent GNS
representations.
\end{proof}
By LTQO, we obtain the following.
\begin{lemma}\label{lem:exterior-decomposition}
Let $D$ be a standard disk. For every $\psi\in\mathcal F_{D^{-l_0}}$, there
are unique coefficients $c_a^\psi\ge0$, $a\in\mathcal C^{\mathrm{EB}}$ with
$\sum_{a\in\mathcal C^{\mathrm{EB}}}c_a^\psi=1$ such that

\[
 \psi|_{\A_{(D^+)^c}}
 =\sum_{a\in\mathcal C^{\mathrm{EB}}}
 c_a^\psi\,\widetilde\varphi_{aD}
 \leavevmode{\big|_{\A_{(D^+)^c}}}.
\]
\end{lemma}

\begin{proof}
Let  $\psi\in\mathcal F_{D^{-l_0}}$. 
By Lemma~\ref{lem:ground-state-information-convex},
\[
\left.\psi\right|_{
  \caA_{D^{+(l+1)}\setminus D^{+}}}
\in \Sigma_\omega\lmk
  \leavevmode{D^{+(l+1)}\setminus D^{+}
  }
\rmk,\quad
\text{for all}\quad l\ge1.
\]
Because $D$ is standard, for $l\geq26$, 
Theorem~\ref{thm:annlusconvex} yields
a unique convex decomposition
\[
 \left.\psi\right|_{\caA_{D^{+l}\setminus D^+}}
 =
 \sum_a c_a^\psi\,
 \left.\sigma_a^{lD}\right|_{\caA_{D^{+l}\setminus D^+}}.
\]
Note that because of the uniqueness in
\leavevmode{Theorem~\ref{thm:annlusconvex}}
and (\ref{sl}),
$c_a^\psi$ does not depend on $l$. Taking the weak$*$ limit completes the proof.
\end{proof}

In particular, for extremal points, we have the following.
\begin{lemma}\label{lem:extreme-ground-states}
Let $D=\{f_0\}^{+s}$ be a standard disk.
Then the set $ \operatorname{ext}\mathcal F_{D^{-l_0}}$ of all
extremal points of $\mathcal F_{D^{-l_0}}$
is non-empty.
Every $\psi\in\operatorname{ext}\mathcal F_{D^{-l_0}}$ is a pure state,
and there exists a unique $a_\psi\in\mathcal C^{\mathrm{EB}}$ such that

\[
 \psi|_{\A_{(D^+)^c}}
 =\widetilde\varphi_{a_\psi D}
 \leavevmode{\big|_{\A_{(D^+)^c}}}.
\]
\end{lemma}
\begin{proof}
This follows from  Lemma~\ref{lem:exterior-ground-states},
 Lemma~\ref{lem:exterior-decomposition} and
Lemma~\ref{lem:annular-separation}.
\end{proof}
These lemmas yield a pure extension of $\td\varphi_{aD}$.
\begin{lemma}\label{lem:pure-charge-states}
Let $D=\{f_0\}^{+s}$ be a
standard disk of radius $s$, 
 with
 \[
 s\ge 6(l_0+2).
 \]
Then for every charge $a\in\mathcal C^{\mathrm{EB}}$, there exists
a pure state 
\[
 \varphi_{aD}\in\mathcal F_{D^{-3{l}_0}}
\]
that satisfies 
\begin{align}\label{koke}
 \varphi_{aD}|_{\A_{(D^{-2l_0+1})^c}}
 =\widetilde\varphi_{a\,D^{-2l_0}}
   |_{\A_{(D^{-2l_0+1})^c}}.
\end{align}

\end{lemma}

\begin{proof}
Fix $a\in\mathcal C^{\mathrm{EB}}$. By
Lemma~\ref{lem:state-extensions}, choose a state extension $\varphi$
of $\widetilde\varphi_{a{D^{-(3l_0+1)}}}$ to $\A$. Then
$\varphi\in\mathcal F_{{D^{-(3l_0)}}}$.

By Lemma~\ref{lem:exterior-ground-states} and the Krein--Milman theorem,
\[
 \mathcal F_{D^{-3l_0}}
 =\overline{\operatorname{co}
 \bigl(\operatorname{ext}\mathcal F_{D^{-(3{l}_0)}}\bigr)}^{\,{wk}^*}.
\]
Hence there exists a net of finite convex combinations
\[
 \varphi_\alpha
 =\sum_{k=1}^{N_\alpha}t_{\alpha k}\psi_{\alpha k}
 \xrightarrow{w^*}\varphi,
\]
where
\[
 \psi_{\alpha k}\in\operatorname{ext}\mathcal F_{D^{-3l_0}},
 \qquad t_{\alpha k}\ge0,
 \qquad \sum_{k=1}^{N_\alpha}t_{\alpha k}=1.
\]
By Lemma~\ref{lem:extreme-ground-states},
applied with $D$ replaced by $D^{-2l_0}$,
each $\psi_{\alpha k}$ is
pure and has a label $a_{\alpha k}\in\mathcal C^{\mathrm{EB}}$ such that
\[
 \psi_{\alpha k}|_{\A_{(D^{-2l_0+1})^c}}
{=\widetilde\varphi_{a_{\alpha k}D^{-(2l_0)}}|_{\A_{(D^{-2l_0+1})^c}}.}
\]
Set
\[
 C_n:=D^{+(n+25)}\setminus D^{+n},
 \qquad n\in\mathbb N.
\]
Fix $n$ sufficiently large that $C_n\subset(D)^c$
and Lemma~\ref{lem:annular-separation} applies to
$\widetilde\varphi_{aD}$ and all
$\widetilde\varphi_{b\leavevmode{D^{-2l_0}}}$.
Then
\[
\begin{aligned}
 \sum_{k=1}^{N_\alpha}t_{\alpha k}\delta_{a_{\alpha k},b}
 &=\sum_{k=1}^{N_\alpha}t_{\alpha k}\psi_{\alpha k}({S_{C_n}^b})\\
 &=\varphi_\alpha({S_{C_n}^b})\\
 &\longrightarrow\varphi({S_{C_n}^b})
 =\widetilde\varphi_{aD}({S_{C_n}^b})=\delta_{ab}.
\end{aligned}
\]
Taking $b=a$, we obtain
\[
 \sum_{k=1}^{N_\alpha}t_{\alpha k}\delta_{a_{\alpha k},a}
 \longrightarrow1.
\]
Consequently, for some $\alpha$ and $k$,
\[
 t_{\alpha k}>0,
 \qquad
 a_{\alpha k}=a.
\]
Set $\varphi_{aD}:=\psi_{\alpha k}$.
This is a pure state in $\mathcal F_{D^{-3l_0}}$ satisfying
the requirement.
\end{proof}

\begin{lem}\label{lemon}
    Let $a\in\ceb$.
For $i=1,2$, let $D_i$ be a standard disk
of radius $s_i\ge 6(l_0+2)$
and $\varphi_i$ be a pure state 
such that
\[
\td\varphi_{aD_i}\vert_{\caA_{D_i^c}}=\varphi_{i}\vert_{\caA_{D_i^c}},\quad i=1,2.
\]
Then $\varphi_1$ and $\varphi_2$ are unitarily equivalent.
\end{lem}
\begin{proof}
    Let $D$ be a standard disk including both $D_1$, $D_2$.
    Then 
    \[
\td\varphi_{aD_i}\vert_{\caA_{D^c}}
=\td\varphi_{aD}\vert_{\caA_{D^c}},\quad i=1,2
\]
by \eqref{hebi} and the construction of $\td\varphi_{aD_i}$.
This proves the Lemma because
$D$ is finite.
\end{proof}

{{
\begin{nota}\label{vad}
Throughout the rest of this paper, for each standard disk {$D$}
of radius $s\ge 6(l_0+2)$, and  for each $a\in\mathcal C^{\mathrm{EB}}$
we assign a pure state $\varphi_{aD}\in \mathcal F_{D^{-3{l}_0}}$ satisfying (\ref{koke}).
{This radius condition is understood whenever
$\varphi_{aD}$ is used in the remainder of this paper.
}
\end{nota}
}}

\subsection{Proof of Theorem \ref{anyontheorem}}
In this subsection, we prove Theorem \ref{anyontheorem}.
First we consider a map from EB charge labels to equivalence classes of simple DHR objects. Because of Lemma \ref{lem:cone-disk-uniqueness}, any
GNS representation of $\varphi_{aD}$ satisfies the superselection criterion.
\begin{lemma}\label{lem:superselection-criterion}
Let $\varphi_{aD}$
be the state given in Notation \ref{vad} for 
a standard disk $D$ and $a\in\mathcal C^{\mathrm{EB}}$.
Then for any GNS triple
$ (\mathcal H,\rho_D^{a\prime},\xi_a')$
of $\varphi_{aD}$ acting on $\caH$,
$\rho_D^{a\prime}$ satisfies the superselection criterion with respect
to $\pi$.
\end{lemma}

\begin{proof}
Let $\ld$ be an arbitrary planar cone.
We can find a finite set $F\subset \ld^c$ {and}
 a lattice cone $\tilde \ld$ satisfying Condition D
 such that $\ld^c\setminus F\subset \tilde\ld \subset (D^{+2+l_0})^c$.
By Lemma \ref{lem:cone-disk-uniqueness} {and} Lemma \ref{lem:annular-limit},
\[
 \varphi_{aD}|_{\mathcal A_{\tilde \ld}}=\omega|_{\mathcal A_{\tilde \ld}}.
\]
Because both $\varphi_{aD}$ {and} $\omega$ are factor states,
we see that 
$\rho_D^{a\prime}|_{\mathcal A_{\tilde \ld}}$
 and $\pi|_{\mathcal A_{\tilde \ld}}$ are quasi-equivalent.
Since  $\ld^c\setminus F\subset \tilde\ld \subset (D^{+2})^c$ and $F$ is finite,
$\rho_D^{a\prime}|_{\mathcal A_{\ld^c}}$
 and $\pi|_{\mathcal A_{\ld^c}}$ are quasi-equivalent.
By the {proper} infiniteness of 
$\rho_D^{a\prime}({\mathcal A_{\ld}})''$
 and $\pi({\mathcal A_{\ld}})''$ 
they are unitarily equivalent \cite{Ogata}.
This completes the proof.
\end{proof}
Now we connect representations satisfying the superselection criterion to the objects in $\ceb$. We introduce the following notation for the proof.
 \begin{notation}\label{notation:ground-projections}
Let $(\mathcal K,\rho)$ be a unital representation of $\A$, and let
$\Lambda\subset\LL$. Denote by $P_{\Lambda,\rho}$ the orthogonal
projection onto
\[
 \bigcap_{X\Subset\Lambda, X\in\caS}\ker\rho(I-P_X),
\]
where $P_X$ is the finite-volume ground-space projection in Eq.~\eqref{eq:parentHamiltonian}.
Equivalently,
\[
 P_{\Lambda,\rho}=\bigwedge_{X\Subset\Lambda, X\in \caS}\rho(P_X).
\]
\end{notation}

\begin{lemma}\label{lem:ssc-normality}
Let $D$ be a standard disk.
Let $\rho:\A\to B(\mathcal H)$ be a representation satisfying the
superselection criterion. Then, for every $\eta\in\mathcal H$, the
positive functional
\[
 \psi_\eta(A):=\langle\eta,\rho(A)\eta\rangle,\qquad A\in\A,
\]
is $\sum_{a\in\mathcal C^{\mathrm{EB}} }\varphi_{aD}$-normal. 
\end{lemma}
\begin{rem}
Note that $\rho$ in this {lemma} does not need to be irreducible.
\end{rem}
\begin{proof}
{\it 1.} By the superselection criterion, we may assume that $\rho$ is localized in a cone $\Lambda$. By irreducibility of $\pi$ and norm-closedness of normal positive functionals, it suffices to consider $\eta=\pi(A)\Omega$ with $A\in\caA$ local.

{Choose planar cones
$\widehat\Lambda,\widehat{\widehat\Lambda},\Lambda'$
as in Figure~\ref{fig:normality-fixed-cones}, such that
\begin{align}
\begin{gathered}
\Lambda\ll\widehat\Lambda,
\qquad
\widehat\Lambda\ll\widehat{\widehat\Lambda},
\\
\Lambda^{+2(l_0+2)}\subset\widehat\Lambda,
\qquad
\widehat\Lambda^{+2(l_0+2)}
\subset\widehat{\widehat\Lambda},
\\
(\supp A)^{+2(l_0+2)}\subset\widehat\Lambda,
\qquad
\Lambda'\cap\widehat{\widehat\Lambda}=\varnothing.
\end{gathered}
\end{align}
}

\begin{figure}[htbp]
\centering
\resizebox{\textwidth}{!}{%
\begin{tikzpicture}[x=.74cm,y=.74cm,font=\large,>=stealth]
 \definecolor{figFourConeFill}{HTML}{D6E9E6}
 \definecolor{figFourConeLine}{HTML}{287D73}
 \clip (-6.1,-3.15) rectangle (6.15,3.15);
 \filldraw[fill=red!9,draw=red!65!black,line width=1.2pt]
   (-3.7,0)--(-6.4,2.7)--(-6.4,-2.7)--cycle;
 \filldraw[fill=blue!7,draw=blue!68!black,line width=1.25pt]
   (-2.3,0)--(6.55,4.3)--(6.55,-4.3)--cycle;
 \filldraw[fill=orange!8,draw=orange!78!black,line width=1.2pt,dashed]
   (-1.05,0)--(6.55,3.42)--(6.55,-3.42)--cycle;
 \filldraw[fill=figFourConeFill,draw=figFourConeLine,line width=1.5pt]
   (.55,0)--(6.55,2.5)--(6.55,-2.5)--cycle;
 \filldraw[fill=violet!26,draw=violet!77!black,line width=1pt]
   (.64,.32)--(.99,.43)--(1.27,.16)--(1.14,-.14)--(.72,-.13)--cycle;
 \node[fill=white,inner sep=1pt] at (-4.82,.55) {$\Lambda'$};
 \node[fill=white,inner sep=1pt] (outer) at (-.62,2.7)
    {$\widehat{\widehat\Lambda}$};
 \draw[->,blue!68!black] (outer.east) to[bend left=10] (1.06,1.63);
 \node[fill=white,inner sep=1pt] (middle) at (3.45,2.53)
    {$\widehat\Lambda$};
 \draw[->,orange!78!black] (middle.south) to[bend left=12] (3.34,1.94);
 \node[fill=white,inner sep=1pt] at (4.6,.55) {$\Lambda$};
 \node[fill=white,inner sep=1pt] (support) at (.45,-1.22)
    {$\operatorname{supp} A$};
 \draw[->,violet!77!black] (support.north) to[bend left=15] (.88,.13);
\end{tikzpicture}%
}
\caption{{The cone configuration used in the proof of
Lemma~\ref{lem:ssc-normality}.}}
\label{fig:normality-fixed-cones}
\label{happa}
\end{figure}
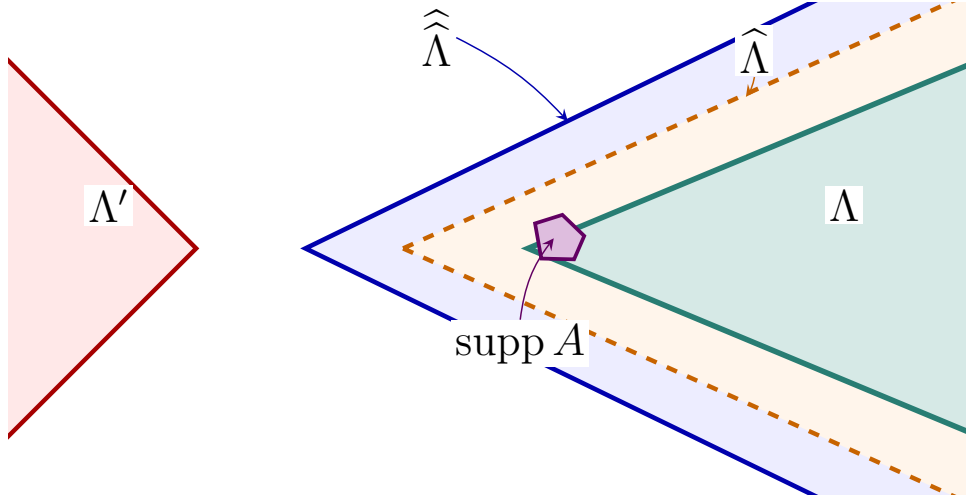
{\it 2.} By this choice, we have
\begin{equation}\label{eq:ssc-exterior-projection}
 P_{\widehat\Lambda^c,\rho}\eta=\eta.
\end{equation}
To see this, let $F\Subset\widehat\Lambda^c$. Since
$\widehat\Lambda^c\subset\Lambda^c$ and
$F\cap\operatorname{supp}A=\varnothing$, we have
\[
 \rho(P_F)\eta=\pi(P_F)\pi(A)\Omega
 =\pi(A)\pi(P_F)\Omega=\pi(A)\Omega=\eta.
\]
Taking $F$ to $\widehat\Lambda^c$ proves
\eqref{eq:ssc-exterior-projection}.

{\it 3.} We next adapt the projection argument in {Proposition}~5.2
{of \cite{BV}} to show that
\begin{equation}\label{eq:ssc-tail-strong}
 P_{\widehat{\widehat\Lambda}\setminus       \widehat{\widehat\Lambda}_N,\rho}\longrightarrow \unit,\quad N\to\infty\quad\text{strongly}.
\end{equation}
By the superselection criterion, there is a unitary $U\in B(\mathcal H)$
such that $\operatorname{Ad}U\circ\rho$ agrees with $\pi$ on
$\A_{(\Lambda')^c}$. Since $\lmk\widehat{\widehat\Lambda}\setminus       \widehat{\widehat\Lambda}_N\rmk\subset(\Lambda')^c$, it follows that
\[
 UP_{\widehat{\widehat\Lambda}\setminus       \widehat{\widehat\Lambda}_N,\rho}U^*=P_{\lmk\widehat{\widehat\Lambda}\setminus       \widehat{\widehat\Lambda}_N\rmk,\pi}.
\]
For any local $B\in \caA$,
\[
P_{\widehat{\widehat\Lambda}\setminus       \widehat{\widehat\Lambda}_N,\pi}\pi(B)\Omega
=\pi(B)\Omega
\]
for $N$ large enough.
Therefore, by the cyclicity of $\Omega$ with respect to $\pi(\caA)$,
$P_{\widehat{\widehat\Lambda}\setminus       \widehat{\widehat\Lambda}_N,\pi}$ converges strongly to $1$.
This implies (\ref{eq:ssc-tail-strong}).

{\it 4.} Set $\xi_N:=P_{\widehat{\widehat\Lambda}\setminus       \widehat{\widehat\Lambda}_N,\rho}\eta$. 
We claim that there is a standard disk $D_N$ of the form $D_N=\{f_N\}^{+s_N}$
such that
\begin{equation}\label{suzume}
 \rho(I-P_X)\xi_N=0
 \quad\text{for all }X\in\mathcal S
 \quad \text{with } X\subset D_N^c
\end{equation}
The ground-space projections commute, so
\eqref{eq:ssc-exterior-projection} implies
\begin{equation}\label{eq:ssc-local-constraints}
 \rho(I-P_X)\xi_N=0
 \quad\text{if }X\in\mathcal S\text{ and either }
 X\subset\widehat\Lambda^c\text{ or }X\subset \lmk\widehat{\widehat\Lambda}\setminus       \widehat{\widehat\Lambda}_N\rmk.
\end{equation}
Therefore, it suffices to show that
the set 
\[
 \mathcal G_N:=\{X\in\mathcal S:
 X\not\subset\widehat\Lambda^c,\ X\not\subset \lmk\widehat{\widehat\Lambda}\setminus       \widehat{\widehat\Lambda}_N\rmk\}
\]
is finite.
To see this, set
\[
 Y_N:=\{z\in\widehat\Lambda:d(z,\lmk\widehat{\widehat\Lambda}\setminus       \widehat{\widehat\Lambda}_N\rmk^c)\le r\}.
\]
Because ${\widehat\Lambda}\ll\widehat{\widehat\Lambda}$,
$Y_N$ is finite. Note that 
if $X\in \mathcal G_N$, then $Y_N\cap X\neq \emptyset$.
In fact, for $X\in\mathcal G_N$, there exist $x\in X\cap\widehat\Lambda$
and $y\in X\cap \lmk\widehat{\widehat\Lambda}\setminus       \widehat{\widehat\Lambda}_N\rmk^c$. Since $d(x,y)\le \operatorname{diam}X\le r$,
we have $x\in Y_N$, and $x\in X\cap Y_N$.
Because {the sets in $\caS$ have} {uniformly bounded} {diameters},
the number of such $X$ is finite. Hence $ \mathcal G_N$ is finite, proving (\ref{suzume}).

{\it 5.}
From {\it 4.} and Lemma \ref{lem:exterior-decomposition},
$(\xi_N, \rho(\cdot)\xi_N)$ is $\sum_{a\in\mathcal C^{\mathrm{EB}} }\varphi_{aD}$-normal.
Because of {\it 3.} (\ref{eq:ssc-tail-strong}), $\lV\xi_N-\eta\rV\to 0$ as $N\to\infty$.
By norm-closedness, therefore $\psi_\eta$ is $\sum_{a\in\mathcal C^{\mathrm{EB}} }\varphi_{aD}$-normal.
\end{proof}

Now we prove Theorem~\ref{anyontheorem}.
\begin{proofof}[Theorem \ref{anyontheorem}]
For each $a\in\ceb$, there is a pure state $\varphi_{aD}$ {satisfying} the properties in
Lemma \ref{lem:pure-charge-states} and Notation \ref{vad}.
By Lemma \ref{lem:superselection-criterion},
a GNS representation $\rho_{a}$ of $\varphi_{aD}$ satisfies the superselection criterion. 
Because $\varphi_{aD}$ is pure, $\rho_a$ is irreducible, hence simple.
{The representations obtained from different choices
of $D$ and the pure state are unitarily equivalent by
Lemma~\ref{lemon}.}
Combining this with the uniqueness of {the} GNS representation up to unitary equivalence, we see {that} the equivalence class of
$\rho_a$ is determined uniquely. This defines a map $\Theta$ from 
 $\ceb$
to the equivalence classes of simple objects in $\caC_\omega^{\ld_0}$. Injectivity follows from Lemma~\ref{lem:annular-separation},
since distinct charges have disjoint representations. 

Conversely, consider an equivalence class {of}
simple objects in $\caC_\omega^{\ld_0}$ with a 
representative $\rho$. Then for any
unit vector $\eta$ in $\caH$, 
$\psi_\eta=\braket{\eta}{\rho(\cdot)\eta}$
is $\sum_{c\in\ceb}\varphi_{cD}$-normal {by} Lemma
\ref{lem:ssc-normality}.
Because $\rho$ is irreducible, there exists a unique $a\in\ceb$ such that the GNS representations of $\psi_\eta$ and $\varphi_D^a$ are unitarily equivalent.
By definition, this $a$ does not depend on the choice of $\eta$ {and} $\rho$. 
Thus $\Theta$ is surjective. 
\end{proofof}

\subsection{GNS representations of \texorpdfstring{$\varphi_{aD}$}{phi(a,D)}}
We show that
when $D$ is deep inside a prescribed cone
$\Lambda$, we can choose a GNS representation
$\rho_a^{D\Lambda}:\caA\to B(\mathcal H)$ of
$\varphi_{aD}$ with cyclic vector $\Omega$ that is
localized in $\Lambda$, meaning that it agrees with
$\pi$ on $\caA_{\Lambda^c}$
(Proposition~\ref{sushi}).

\begin{lemma}\label{lem:cone-split}
Let $ \td \Lambda_0,  \td \Lambda_2$ be lattice cones satisfying
\[
   \td \Lambda_0^{+2}\subset \td \Lambda_2^c.
\]
Then there exists a type I factor $F\subset B(\mathcal H)$ such that
\[
 \pi(\A_{ \td \Lambda_2})''\subset F\subset\pi(\A_{ \td\Lambda_0})'.
\]
\end{lemma}

\begin{proof}
This follows from the proof of \cite[Theorem~{3.31}]{BGNW}
applied to our setting. By \eqref{eq:zeromutualinfo}, our system satisfies the
required condition.
\end{proof}

\begin{prop}\label{sushi}
Let $a\in\mathcal C^{\mathrm{EB}}$.
Let $\td \ld_0$, $\td \ld_1$, $\td\ld_2$ be lattice cones satisfying Condition M.
{Assume also that the complement $\widetilde\Lambda_1^{\,c}$
satisfies Condition D.
}
Let $\varphi_{aD}$ be the pure state
given in Notation \ref{vad} for a standard disk
$D$ of radius $s\ge 6(l_0+2)$.
Suppose that
\begin{align}\label{usagi}
D^{+({l}_0+2)}\subset\td \Lambda_0\subset\td\Lambda_1,
 \qquad \td \Lambda_1^{+2}\subset\leavevmode{\widetilde\Lambda_2}^c, \qquad \td \ld_0^{+5(R+l_0)}\subset \td \ld_1.
\end{align}

Then,
there exists an irreducible representation
$\rho_a^{\td\ld_2^c}$
such that 
\[
 \rho_a^{\td\ld_2^c}
 |_{\mathcal A_{\td\ld_2}}
 =\pi|_{\mathcal A_{\td\ld_2}},
\]
and
\[
 (\Omega,\rho_a^{\td\ld_2^c}(A)\Omega)
 =\varphi_{aD}(A),\qquad A\in\mathcal A.
\]
\end{prop}

\begin{proof}
Note from Lemma \ref{lem:superselection-criterion}
that we have a GNS triple $(\mathcal H,\rho,\xi)$ of $\varphi_{aD}$ with
{an} irreducible representation $\rho$ localized in $\td\ld_0$.

By Lemma~\ref{lem:cone-split}, there exists a type~I factor $F\subset\mathcal B(\mathcal H)$
such that
\[
 \pi(\mathcal A_{\td\ld_2})''\subset F
 \subset\pi(\mathcal A_{\td \Lambda_1})' =\pi(\mathcal A_{\td \Lambda_1^c})''.
\]
In the last equality, we used {Haag} duality for $\td \ld_1$.
Choose Hilbert spaces $\mathcal K_1,\mathcal K_2$ and a unitary
\[
 W:\mathcal H\longrightarrow\mathcal K_1\otimes\mathcal K_2
 \quad\text{such that}\quad
 WFW^*=1\otimes\mathcal B(\mathcal K_2).
\]
For every $A\in\mathcal A_{\td\ld_1^c}$, Lemma \ref{lem:cone-disk-uniqueness} and 
$\td\ld_1^c\subset(D^{+({l}_0+2)})^c$ imply
\[
 \begin{aligned}
 (\Omega,\pi(A)\Omega)
 &=\omega(A)
 =\widetilde\varphi_{aD}(A)
 =\varphi_{aD}(A)\\
 &=(\xi,\rho(A)\xi)
 =(\xi,\pi(A)\xi).
 \end{aligned}
\]
The last equality follows from $\td\ld_1^c\subset\leavevmode{\widetilde\Lambda_0}^c$.
The normal vector states of $\Omega$ and $\xi$ therefore agree on
$\pi(\mathcal A_{\td\ld_1^c})''$, and hence
\[
 (\Omega,x\Omega)=(\xi,x\xi),\qquad x\in F.
\]
Consequently, we may write
\[
 W\Omega=\sum_i\sqrt{\lambda_i}\,e_i\otimes\eta_i,
 \qquad
 W\xi=\sum_i\sqrt{\lambda_i}\,f_i\otimes\eta_i,
\]
where $\lambda_i>0$, $\sum_i\lambda_i=1$,
$\{e_i\}_i$ and $\{f_i\}_i$ are orthonormal systems in
$\mathcal K_1$, and $\{\eta_i\}_i$ is an orthonormal system in
$\mathcal K_2$. Thus there exists a partial isometry
$u\in\mathcal B(\mathcal K_1)$ such that
\[
 (u\otimes 1)W\xi=W\Omega,
\]
with $uu^*$ {and} $u^*u$ {being the}
support projections of $W\Omega$ {and} $W\xi${,} respectively{,}
in $\mathcal B(\mathcal K_1)$.

We claim that $1-u^*u$ and $1-uu^*$ are infinite projections in
$\mathcal B(\mathcal K_1)$. 
By (\ref{usagi}), there exists $X\in\caS$, the
fixed Markov cover, such that
\[
 X\subset\td \Lambda_1\setminus\td \Lambda_0,
 \qquad X\subset\bigl(D^{+({l}_0+2)}\bigr)^c.
\]
Since Lemma~\ref{lem:qxneqI} guarantees $Q_X=P_X\neq I$,
\[
 0\ne\pi(1-P_X)\in\pi(\mathcal A_{\td\ld_1})''\subset F',
\]
there exists a projection $r\in\mathcal B(\mathcal K_1)$ such that
\[
 W\pi(1-P_X)W^*=r\otimes 1.
\]
From $\pi(1-P_X)\Omega=0$, we obtain
\[
 (r\otimes 1)W\Omega=0,
 \qquad r\le 1-uu^*.
\]
Moreover, since $X\subset\td\ld_1\setminus\td \Lambda_0$ and
$X\subset(D^{+({l}_0+2)})^c$,
\[
 \begin{aligned}
 (\xi,\pi(1-P_X)\xi)
 &=(\xi,\rho(1-P_X)\xi)\\
 &=\varphi_{aD}(1-P_X)
 =\widetilde\varphi_{aD}(1-P_X)=0.
 \end{aligned}
\]
Since $\pi(1-P_X)$ is a projection, 
\[
 \pi(1-P_X)\xi=0,
 \qquad (r\otimes 1)W\xi=0,
 \qquad r\le 1-u^*u.
\]
The projection $\pi(1-P_X)$ is infinite in
$\pi(\mathcal A_{\td\ld_1})''\subset F'$, an infinite factor. Therefore $r\otimes 1$
is infinite in
\[
 W\pi(\mathcal A_{\td\ld_1})''W^*
 \subset WF'W^*=\mathcal B(\mathcal K_1)\otimes 1,
\]
and $r$ is infinite in $\mathcal B(\mathcal K_1)$.
It follows that $1-u^*u$ and $1-uu^*$ are infinite projections in
$\mathcal B(\mathcal K_1)$.

 Thus $1-u^*u$ and $1-uu^*$ are equivalent in $\mathcal B(\mathcal K_1)$ and there exists a partial isometry
$v\in\mathcal B(\mathcal K_1)$ such that
\[
 v^*v=1-u^*u,\qquad vv^*=1-uu^*.
\]
Set $w:=u+v\in\mathcal B(\mathcal K_1)$. Then
\[
 \begin{aligned}
 ww^*&=(u+v)(u^*+v^*)=uu^*+vv^*=1,\\
 w^*w&=(u^*+v^*)(u+v)=u^*u+v^*v=1.
 \end{aligned}
\]
Hence $w$ is unitary. Since $(u^*u\otimes 1)W\xi=W\xi$, we have
\[
 (w\otimes 1)W\xi=(u\otimes 1)W\xi=W\Omega.
\]
Define the unitary $U:=W^*(w\otimes 1)W$ on $\mathcal H$.
By Haag duality for $\td\ld_2$,
\[
 U\in W^*(\mathcal B(\mathcal K_1)\otimes 1)W
 =F'\subset\pi(\mathcal A_{\td\ld_2})'
 =\pi(\mathcal A_{\td\ld_2^c})'',
\]
and
\[
 U\xi=W^*(w\otimes 1)W\xi=W^*W\Omega=\Omega.
\]

Set $\rho_a^{\td\ld_2^c}
:=\operatorname{Ad}U\circ\rho$.
Since $\td\ld_2\subset\leavevmode{\widetilde\Lambda_0}^c$ and
$U\in\pi(\mathcal A_{\td\ld_2})'$, we obtain
\[
 \rho_a^{\td\ld_2^c}|_{\mathcal A_{\td\ld_2}}
 =\operatorname{Ad}U\circ
   \rho|_{\mathcal A_{\td\ld_2}}
 =\pi|_{\mathcal A_{\td\ld_2}}.
\]
Thus $\rho_a^{\td\ld_2^c}$ is localized in $\td\ld_2^c$.
Finally, for every $A\in\mathcal A$,
\[
 \begin{aligned}
 (\Omega,\rho_a^{\td\ld_2^c}(A)\Omega)
 &=(U^*\Omega,\rho(A)U^*\Omega)\\
 &=(\xi,\rho(A)\xi)
 =\varphi_{aD}(A).
 \end{aligned}
\]
\end{proof}
\begin{defn}\label{def:deep-inside-cone}
Let $D$ be a standard disk of radius $s\ge 6(l_0+2)$ and {let} $\ld$ {be} a cone.
We say that $D$ is {\it deep inside of $\ld$}
if there are lattice cones  $\td \ld_0$, $\td \ld_1$, $\td\ld_2$ satisfying Condition M
{such that}
\begin{align}\label{kapibara}
D^{+({l}_0+2)}\subset\td \Lambda_0\subset\td\Lambda_1,
 \qquad \td \Lambda_1^{+2}\subset\leavevmode{\widetilde\Lambda_2}^c, \qquad \td \ld_0^{+5(R+l_0)}\subset \td \ld_1,
 \quad \td\ld_2^c\subset \ld.
\end{align}
{We also require that the complement $\widetilde\Lambda_1^{\,c}$
satisfy Condition D.
}
By Proposition \ref{sushi}, 
there exists an irreducible representation
$\rho_{a}^{D\ld}$ localized in $\ld$
such that 
\[
 (\Omega,\rho_{a}^{D\ld}(A)\Omega)
 =\varphi_{aD}(A),\qquad A\in\mathcal A.
\]
For each $(D,\ld)${,} we will fix this representation $\rho_{a}^{D\ld}$.
\end{defn}

The next lemma approximates vectors in $\pi(\caA_\Lambda)''\Omega$  by local vectors in the such choice of representation.

\begin{lemma}\label{lem:cone-local-approximation}
{Let $\Lambda$ be a planar cone and $C$ a closed convex planar cone
with $\Lambda\ll\operatorname{int}C$. 
Set
$\widehat\Lambda:=\mathfrak F(C)$, as in Definition ~{\ref{def:haag-lattice-cone}}.
}
{Let} $D$ {be} a standard disk deep inside of {$\Lambda$}.
Let $a\in\mathcal C^{\mathrm{EB}}$.
Let $\rho$ be an irreducible representation of $\caA$ on $\caH$
such that
\[
\begin{gathered}
 \rho\big|_{\mathcal A_{\Lambda^c}}
 =\pi\big|_{\mathcal A_{\Lambda^c}},\\
 \langle\Omega,\rho(A)\Omega\rangle
 =\varphi_{aD}(A),
 \qquad A\in\mathcal A.
\end{gathered}
\]
Then for every $x\in\pi(\A_\Lambda)''$ and $0<\varepsilon<\frac 12 $,
there exist an $N_0\in\mathbb N$ and
$z\in\A_{(\widehat\Lambda_{N_0})^{+l_0+1}}$ such that
\[
 \|x\Omega-\rho(z)\Omega\|<\varepsilon.
\]
If $x$ is an isometry, then $z$ can be chosen to be unitary.
\end{lemma}
\begin{proof}
We first prove the assertion for an
isometry $u\in\pi(\A_\Lambda)''$.
{By Lemma~\ref{oka}, all sufficiently large
$\widehat\Lambda_N=\widehat\Lambda\cap\Gamma_N$ and all their
graph thickenings are disks.
}
Recall Notation \ref{notation:ground-projections} and set
\[
  \xi_N:=P_{\widehat\Lambda\setminus\widehat\Lambda_N,\rho}u\Omega.
\]
There exists $N_1=N_1(\ld,\widehat\Lambda,D)\in\bbN$ such that
\[
\leavevmode{D^+\subset\widehat\Lambda_N,\qquad}
 d(\Lambda\setminus\Lambda_N,\widehat\Lambda^c)>r,
 \qquad d(\Gamma_N^c,D^+)>r+1,
\]
for all $N\ge N_1$.
Here, $r$ is the range of the fixed Markov cover.

We claim that
\begin{equation}\label{eq:cone-tail-ground}
 \rho(P_X)\xi_N=\xi_N,
\end{equation}
for all $N\ge N_1$, $X\in\mathcal S$ with
$ X\subset(\widehat\Lambda_N)^c$.
To see this, from the choice of $N_1$, note that for any
$X\in\mathcal S$ with
$X\subset(\widehat\Lambda_N)^c$,
we have either $X\cap (\widehat \ld)^c=\emptyset$
or $X\cap \ld=\emptyset$. 
{In} the former case, (\ref{eq:cone-tail-ground})
holds because of the definition of $\xi_N$.
{In} the latter case, we have $\rho(P_X)=\pi(P_X)$.
Commutativity of the ground-space projections gives
\[
 \rho(P_X)P_{\widehat\Lambda\setminus\widehat\Lambda_N,\rho}u\Omega=P_{\widehat\Lambda\setminus\widehat\Lambda_N,\rho}u\pi(P_X)\Omega=P_{\widehat\Lambda\setminus\widehat\Lambda_N,\rho}u\Omega,
\]
proving (\ref{eq:cone-tail-ground}).

Next we prove
that $P_{\widehat\Lambda\setminus\widehat\Lambda_N,\rho}$
converges to $1$ strongly {as} $N\to \infty$.
In fact, for every local observable $A$, its support is disjoint from
$\widehat\Lambda\setminus\widehat\Lambda_N$ for all sufficiently
large $N$. 
Hence $P_{\widehat\Lambda\setminus\widehat\Lambda_N,\rho}\rho(A)\Omega=\rho(A)\Omega$ eventually.
Here we used 
\[
\braket{\Omega}{\rho(P_X)\Omega}=\varphi_{aD}(P_X)=1,
\]
for all $X\in \caS$ with $X\subset \widehat\Lambda\setminus\widehat\Lambda_N$
with $N$ large enough.
Because $\rho$ is irreducible,
this proves the claim. In particular,
there exists $N_0=N_0(\ld,\widehat\Lambda, D,u,\varepsilon)\ge N_1$ such that
\begin{equation}\label{eq:cone-vector-limit}
 \lV \xi_N- u\Omega\rV<\frac{\varepsilon}{16},\quad \text{for all } N\ge N_0.
\end{equation}

For each $N\ge N_0$, set
\[
 \eta_{N}:=\xi_{N}/\|\xi_{N}\|,
\]
and
\begin{align}
\begin{split}
&\varphi_N:=\langle\eta_{N},\rho(\,\cdot\,)\eta_{N}\rangle.
\end{split}
\end{align}
Recall that
\begin{align}
\begin{split}
\varphi_{aD}=\langle\Omega,\rho(\,\cdot\,)\Omega\rangle.
\end{split}
\end{align}

We next show that for any $N\ge N_0$,
\begin{equation}\label{eq:cone-exterior-equality}
 \left.\varphi_N\right|_{\A_{\lmk (\widehat{\Lambda}_{N})^{+l_0+1}\rmk^c}}
 =\left.\varphi_{aD}\right|_{\A_{\lmk (\widehat{\Lambda}_{N})^{+l_0+1}\rmk^c}}.
\end{equation}
In fact, because of (\ref{eq:cone-tail-ground})
and the definition of $\varphi_{aD}$, we have
\begin{align}
\varphi_N(1-P_X)=\varphi_{aD}(1-P_X)=0,
\end{align}
for all $X\in\mathcal S$ with $X\subset(\widehat\Lambda_{N})^c$.
By {Theorem}~\ref{thm:commutingparentH}(iv), this means {that both} $\left.\varphi_N\right|_{\A_{\lmk (\widehat{\Lambda}_{N})^{+l_0+1}\rmk^c}}$
and
 $\left.\varphi_{aD}\right|_{\A_{\lmk (\widehat{\Lambda}_{N})^{+l_0+1}\rmk^c}}$
 belong to $\Sigma\lmk \lmk (\widehat{\Lambda}_{N})^{+l_0+1}\rmk^c\rmk$.
 Because $\rho$ is irreducible,
$\varphi_N$ and $\varphi_{aD}$ are pure states and are unitarily equivalent.
Since $(\widehat\Lambda_N)^{+(l_0+1)}$ and
$\bigl((\widehat\Lambda_N)^{+(l_0+1)}\bigr)^{-1}$ are disks
by Lemma~\ref{oka}(2), choosing standard disks $E$ and then $K$ sufficiently
large makes
\[
 K\setminus E
 \subset K\setminus(\widehat\Lambda_N)^{+(l_0+1)}
\]
an essential inclusion of regular annuli,
hence the restriction
$\Sigma(K\setminus (\widehat{\Lambda}_{N})^{+l_0+1})\to \Sigma(K\setminus E)$
is injective by Lemma~\ref{lem:annular-restriction-injectivity}.
By Lemmas~\ref{lem:exterior-decomposition}
and~\ref{lem:annular-separation}, 
if $E$ is large enough,
\[
\varphi_N\vert_{\caA_{E^c}}
=\varphi_{aD}\vert_{\caA_{E^c}}.
\]
From this we obtain (\ref{eq:cone-exterior-equality}).

Next we show that there exists a unitary $v\in \mathcal A_{\lmk\widehat{\Lambda}_{N_0}\rmk^{+l_0+1}}$
such that
\begin{align}\label{hato}
\rho(v)\Omega=\eta_{N_0}.
\end{align}
To see this, let $e_{11}\in\mathcal A_{\lmk\widehat{\Lambda}_{N_0}\rmk^{+l_0+1}}$ be a minimal
projection of $\mathcal A_{\lmk\widehat{\Lambda}_{N_0}\rmk^{+l_0+1}}$. There exists a unitary
\[
 U:\mathcal H\longrightarrow
   \mathcal H_{\lmk\widehat{\Lambda}_{N_0}\rmk^{+l_0+1}}\otimes
   \rho(e_{11})\mathcal H,
\]
where $\mathcal H_{\lmk\widehat{\Lambda}_{N_0}\rmk^{+l_0+1}}$ is the finite-dimensional local
Hilbert space. If $A\in\mathcal A_{\lmk\widehat{\Lambda}_{N_0}\rmk^{+l_0+1}}$ and
$B\in\mathcal A_{\lmk \lmk\widehat{\Lambda}_{N_0}\rmk^{+l_0+1}\rmk^c}$, this factorization has the form
\[
 U\rho(A B)U^*
 =
 A\otimes\pi_a(B),
 \qquad
 \pi_a(B)
 :=
 \rho(Be_{11})
 \big|_{\rho(e_{11})\mathcal H}.
\]
Since the restrictions {of} $\varphi_{N_0}$ and $\varphi_{aD}$ to 
$\mathcal A_{(\lmk\widehat{\Lambda}_{N_0}\rmk^{+l_0+1})^c}$ agree, {taking}
Schmidt decompositions{,} we have
\[
 U\eta_{N_0}
   =\sum_i\sqrt{\lambda_i}\,e_i\otimes\zeta_i,
 \qquad
 U\Omega
   =\sum_i\sqrt{\lambda_i}\,f_i\otimes\zeta_i,
\]
with $\lambda_i>0$.
Because $\mathcal H_{\lmk\widehat{\Lambda}_{N_0}\rmk^{+l_0+1}}$ is
finite-dimensional, the partial isometry
$\sum_i|e_i\rangle\langle f_i|$ extends to a unitary
$v\in\mathcal A_{\lmk\widehat{\Lambda}_{N_0}\rmk^{+l_0+1}}$
such that $U\eta_{N_0}=(v\otimes \unit)U\Omega$. 
It follows that
\[
 \rho(v)\Omega
 =
 U^*(v\otimes1)U\Omega
 =
 \eta_{N_0},
\]
proving (\ref{hato}).

Thus there {exist} an $N_0=N_0(\ld,\widehat\Lambda, D,u,\varepsilon)\in\bbN$ and a unitary
$v\in \mathcal A_{\lmk\widehat{\Lambda}_{N_0}\rmk^{+l_0+1}}$ such that
\[
 \|u\Omega-\rho(v)\Omega\|
 =
 \|u\Omega-\eta_{N_0}\|
 <\frac{\varepsilon}4.
\]

Because any $x\in\pi(\A_\Lambda)''$
can be written as a linear combination of four unitaries in $\pi(\A_\Lambda)''$,
this completes the proof.
\end{proof}

\section{Merging}\label{sec:merging}

In this section, we introduce merging in infinite systems and connect it
with {the} tensor product in {DHR theory}.

We consider the configuration in
Setting~\ref{setting:merging-inputs}, depicted in
Figures~\ref{fig:merging-regions} and~\ref{fig:three-hole-merging},
with disjoint disk-shaped holes $D_1,D_2,D_3$.

For EB charge labels $a,b$, let $\widetilde\varphi_{aD_1}$
and $\widetilde\varphi_{bD_2}$ be the exterior charge states
constructed in Section~\ref{sec:infinite-volume}, defined on
$\caA_{(D_1^{-l_0})^c}$ and $\caA_{(D_2^{-l_0})^c}$,
respectively.
The merging construction uses their restrictions to
$\caA_{D_1^c}$ and $\caA_{D_2^c}$.
Taking the thermodynamic limit of the two-annulus mergers
from Section~\ref{sougen}, we construct the state
$\widetilde\varphi_{aD_1}\bowtie\widetilde\varphi_{bD_2}$
on $\caA_{(D_1\cup D_2)^c}$
(Lemma~\ref{lem:finite-merging}).
The three-annulus merging of
Lemma~\ref{lem:three-annulus-merging} similarly yields a
three-hole state on $\caA_{(D_1\cup D_2\cup D_3)^c}$
(Proposition~\ref{cor:merging-associativity}).
These states are thermodynamic limits of the unique
finite-volume entropy maximizers with the prescribed
annular marginals.

We can also fill the holes to obtain a state
$\varphi_{aD_1}\bowtie\varphi_{bD_2}$ on $\caA$ whose
marginals on $D_1$ and $D_2$ agree with those of the
chosen pure charge states $\varphi_{aD_1}$ and
$\varphi_{bD_2}$ on $\caA$, respectively
(Lemmas~\ref{lem:fill-two-holes} and~\ref{lem:hole-filling-limit}).
Each chosen pure charge state is recovered from its exterior charge state by a Petz recovery map (Lemma~\ref{ushi}). 
The Petz map fills the hole from its annular collar. It is determined
by the chosen pure state on the hole and collar and acts as the
identity on the rest of the exterior. The state
$\varphi_{aD_1}\bowtie\varphi_{bD_2}$ is obtained from 
$\td \varphi_{aD_1}\bowtie\td \varphi_{bD_2}$,
applying those two Petz recovery maps.

This state $\varphi_{aD_1}\bowtie\varphi_{bD_2}$ can be connected to
the tensor product $\rho_a^{D_1\Lambda_1}\otimes\rho_b^{D_2\Lambda_2}$
of DHR-theory.
Let $\mathcal H$ and $\Omega\in\mathcal H$ be the reference
GNS Hilbert space and its cyclic unit vector, respectively.
Let $\rho_a^{D_1\Lambda_1},\rho_b^{D_2\Lambda_2}:
\caA\to B(\mathcal H)$ be the representations of the chosen
pure charge states supplied by Definition \ref{def:deep-inside-cone},
where  $\Lambda_1,\Lambda_2$ are the separated
localization cones containing $D_1,D_2$ in the setting above.
Using the local Petz recovery representations, together
with localization and Haag duality, we show that
\[
 (\varphi_{aD_1}\bowtie\varphi_{bD_2})(x)
 =\langle\Omega,
   (\rho_a^{D_1\Lambda_1}\otimes\rho_b^{D_2\Lambda_2})(x)
   \Omega\rangle,
 \qquad x\in\caA,
\]
where $\otimes$ denotes the DHR tensor product, acting
on the same Hilbert space $\mathcal H$
(Proposition~\ref{lem:merging-tensor-product}).
This begins to reveal the relation between EB fusion
and DHR fusion.
\subsection{Petz recovery of \texorpdfstring{$\varphi_{aD}$}{phi(a,D)}}
We show
$\varphi_{aD}$
can be recovered from its exterior charge state $\td\varphi_{aD}$, by a Petz recovery map
(Lemma~\ref{ushi}).

\begin{lem}\label{ushi}
Let $\varphi_{aD}$ be the pure state
given in Notation \ref{vad} for a standard disk
$D=\{f_0\}^{+s}$ of radius $s\ge 6(l_0+2)$, and $a\in\ceb$.
Then with the
Petz recovery maps, we have
\begin{align}\label{eq:one-hole-recovery-limits}
\begin{split}
\varphi_{aD}
&=\widetilde\varphi_{aD}\circ
{(R^{(\varphi_{aD})_{DE}}_{E\to DE})}^T
\\
&=\mathop{\mathrm{w}^*\!-\!\lim}_{n\to\infty}\varphi_{aD}\circ
\lmk
\id_{\leavevmode{D}}\otimes
\lmk
R_{\leavevmode{E}\to
  \Gamma_n\setminus\leavevmode{D}}^{
  \td\varphi_{a\leavevmode{D}}}
\rmk^T
\rmk .
\end{split}
\end{align}

where $E:=\{f_0\}^{+t}\setminus \{f_0\}^{+s}$ with $t-s\ge l_0$.

\end{lem}

\begin{proof}
Applying Lemma~\ref{lem:merging-general} to 
\begin{equation}
    D=\{f_0\}^{+s},\;E=\{f_0\}^{+t}\setminus \{f_0\}^{+s},\;
F_N=\{f_0\}^{+N}\setminus \{f_0\}^{+t},\quad N-t\ge l_0
\end{equation}
we see that $I(D:F_N|E)_{\varphi_{aD}}=0$.
Therefore, the Petz recovery maps satisfy
\begin{align}
    \begin{split}
        (\varphi_{aD})_{DEF_N}&=\lmk R_{E\to DE}^{(\varphi_{aD})_{DE}}\otimes \id_{F_N}\rmk((\varphi_{aD})_{EF_N})\,.
        \label{reibou}
    \end{split}
\end{align}
From {\eqref{koke} and
\eqref{hebi}},
\begin{equation}\label{eq:5.4}
    (\varphi_{aD})_{EF_N}=({\tilde \varphi}_{aD})_{EF_N}\,.
\end{equation}
Substituting this identity into \eqref{reibou} and letting
$N\to\infty$ proves the first equality.

Applying Lemma~\ref{lem:merging-general} to 
\begin{equation}
    D=\{f_0\}^{+s},\;E=\{f_0\}^{+t}\setminus \{f_0\}^{+s},\;
G_N=\Gamma_N\setminus \{f_0\}^{+t},
\end{equation}
we see that $I(D:G_N|E)_{\varphi_{aD}}=0$
for $N$ large enough.
Therefore, the Petz recovery maps satisfy
\begin{align}
    \begin{split}
        (\varphi_{aD})_{DEG_N}=\lmk \id_D\otimes R_{E\to EG_N}^{(\varphi_{aD})_{EG_N}}\rmk\lmk (\varphi_{aD})_{DE}\rmk.\label{eq:5.6}
    \end{split}
\end{align}
By the same argument as in \eqref{eq:5.4},
$(\varphi_{aD})_{EG_N}=(\tilde\varphi_{aD})_{EG_N}$.
Substituting this into \eqref{eq:5.6} and passing to the dual maps
proves the second equality, since $DEG_N=\Gamma_N$ exhausts the lattice.

\end{proof}

\begin{figure}[htbp]
\centering
\definecolor{fusionFourA}{RGB}{242,204,217}
\definecolor{fusionFourB}{RGB}{165,228,233}
\definecolor{fusionFourC}{RGB}{214,239,200}
\definecolor{fusionFourD}{RGB}{249,217,183}
\begin{tikzpicture}[x=1cm,y=1cm,line width=0.85pt]
 \fill[fusionFourD] (-2.95,-2.95) rectangle (2.95,2.95);
 \filldraw[fill=fusionFourC,draw=black] (0,0) circle (2.18);
 \filldraw[fill=fusionFourB,draw=black] (0,0) circle (1.38);
 \filldraw[fill=fusionFourA,draw=black] (0,0) circle (0.52);
 \node[font=\Large] at (0,0) {$A$};
 \node[font=\Large] at (0,-0.94) {$B$};
 \node[font=\Large] at (0,-1.79) {$C$};
 \node[font=\Large] at (0,-2.62) {$D$};
\end{tikzpicture}
\caption{A disk $A$, successive annular collars $B$ and $C$,
and the exterior region $D$.
}
\label{fig:fusion-lemma4-regions}
\end{figure}
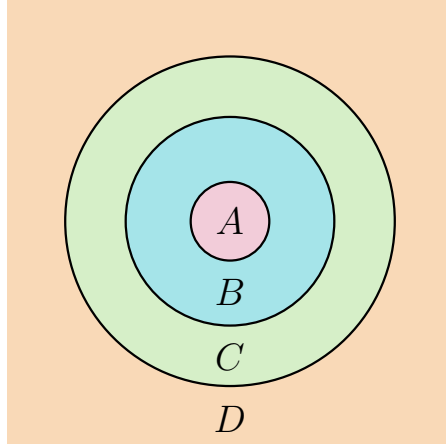
\begin{lem}\label{swan}
Let $A$ be a standard disk and $B$ an annulus around $A$, and $C$ {an} annulus around $AB$
as in Figure \ref{fig:fusion-lemma4-regions}. Set $D:=(ABC)^c$.
{We take $B$ and $C$ to be concentric standard annuli}.

Assume that $A$ has radius at least $6(l_0+2)$,
{and that the minimum widths of $B$ and $C$,
measured in graph-distance layers,
are at least $\max\{25,l_0+1\}$. 
Then the following factorizations hold:
}
\begin{align}
\begin{split}
&\lmk \varphi_{aA}\rmk_{ACD}=\lmk \varphi_{aA}\rmk_{A}\otimes \lmk \varphi_{aA}\rmk_{CD},\\
&\lmk \varphi_{aA}\rmk_{ABD}=\lmk \varphi_{aA}\rmk_{AB}\otimes \lmk \varphi_{aA}\rmk_{D}
\end{split}
\end{align}
\end{lem}
\begin{proof}

This follows from the same observation as in Lemma \ref{tomato}.
\end{proof}
For the rest of this section, we consider the following regions.

\begin{setting}\label{setting:merging-inputs}
{Consider the regions depicted in
Figures~\ref{fig:three-hole-merging}
and~\ref{fig:merging-regions}.
The same disks and collars are used in
Figures~\ref{fig:three-hole-merging}
and~\ref{fig:merging-regions}.
The holes $D_i$ are standard disks satisfying the radius
condition of Notation~\ref{vad}, and $B_i$
are their concentric standard annular collars.

\noindent Figure~\ref{fig:three-hole-merging}:
For $i=1,2,3$,
$H_i=D_i\cup B_i\cup C_i$ is a lattice cone
containing $D_i$ and $B_i$.
Each $H_i$ satisfies Condition M.
The separating strips are $E_1,E_2,E_3$, and we have
\[
 \mathbb L=\bigsqcup_{i=1}^3(H_i\sqcup E_i).
\]
Each $H_i$ contains a convex planar cone $\ld_i$
that contains a lattice cone $\widetilde\ld_i$.
The cones $\widetilde\ld_i$, their thickenings by any
fixed number of graph-distance layers, and the complements
of these regions satisfy Conditions M and D.
For $i=1,2,3$, the disk $D_iB_i=D_i\cup B_i$ is deep inside
the corresponding localization cone $\ld_i$
in the sense of Definition~\ref{def:deep-inside-cone}.

\noindent Figure~\ref{fig:merging-regions}:
The regions $D_1B_1G_{12}$ and $D_2B_2G_{12}'$
(resp.\ $D_2B_2G_{23}$ and $D_3B_3G_{23}'$)
are lattice cones corresponding to half-planes
and satisfying Condition M,
and $I_{12}$, $I'_{12}$
(resp.\ $I_{23}$, $I'_{23}$)
are the intervening strips.

The localization cones satisfy
\[
 \ld_1\subset D_1B_1G_{12},\qquad
 \ld_2\cup\ld_3\subset D_2B_2G_{12}',
\]
\[
 \ld_1\cup\ld_2\subset D_2B_2G_{23},\qquad
 \ld_3\subset D_3B_3G_{23}'.
\]

The disks, collars, and separating strips are chosen
with sufficient size, width, and separation to allow
finite configurations satisfying the geometric hypotheses
of Setting~\ref{setting:two-annulus-merging}
for the two-hole construction, and
Settings~\ref{mimisen} and~\ref{mado}
for the three-hole construction.
These choices also leave sufficient room for the
applications of Lemmas~\ref{lem:merging-general},
\ref{lem:ground-state-information-convex}, and~\ref{swan},
and for the boundary deformations and annular comparisons
satisfying the hypotheses of
Theorem~\ref{thm:information-convex-isomorphism},
Corollary~\ref{nesti}, and
Lemmas~\ref{lem:standard-annulus-comparison}
and~\ref{lem:annular-restriction-injectivity}
needed below.

The configuration is also chosen with sufficient angular
and spatial margins to permit the auxiliary constructions
in Setting~\ref{imo} and~\eqref{somen},
with the geometric hypotheses of
Lemmas~\ref{lem:cone-local-approximation}
and~\ref{lem:annular-separation}
satisfied in their respective applications.
We also assume that all the cones
$\ld_1$, $\ld_2$, $\ld_3$
are in $\caC_{(\theta_0,\varphi_0)}$,
and that the closures of their angular intervals
are pairwise disjoint.

Throughout, Setting~\ref{setting:merging-inputs} refers to the configuration specified in Appendix~\ref{taiyou}. A single such configuration suffices; see Remark~\ref{momiji}.
}

\end{setting}

\begin{figure}[!htbp]
\centering
{\begin{tikzpicture}[
  x=1.25cm,y=1.25cm,
  font=\normalsize,
  every node/.style={text=geomCobalt,inner sep=2pt}
]
\fill[orange!18] (-4.5,-1.6) rectangle (4.5,4.6);
\path[fill=yellow!22,draw=gray!65,line width=.35pt]
  (0.03586,0.13384)--(4.50000,4.59798)--(4.50000,4.60000)
  --(-1.16084,4.60000)--cycle;
\path[fill=violet!18,draw=gray!65,line width=.35pt]
  (4.50000,-1.13405)--(4.50000,4.40202)
  --(0.13384,0.03586)--cycle;
\path[fill=green!32,draw=gray!65,line width=.35pt]
  (-0.07727,0.02071)--(0.03586,0.13384)
  --(-1.16084,4.60000)--(-1.30429,4.60000)--cycle;
\path[fill=red!22,draw=gray!65,line width=.35pt]
  (0.02071,-0.07727)--(4.50000,4.40202)
  --(4.50000,4.59798)--(-0.07727,0.02071)
  --(-0.05657,-0.05657)--cycle;
\path[fill=cyan!22,draw=gray!65,line width=.35pt]
  (4.50000,-1.27750)--(4.50000,-1.13405)
  --(0.13384,0.03586)--(0.02071,-0.07727)--cycle;

\begin{scope}
\clip (-4.5,-1.6) rectangle (4.5,4.6);
\draw[black!80,dashed,line width=.75pt]
  (-12.17818,2.09071)--(-0.36049,0.00694)
  --(-12.17818,-2.07684);
\draw[geomCobalt,densely dotted,line width=.75pt]
  (-12.31782,1.05280)--(-0.36349,0.00694)
  --(-12.31782,-1.03893);
\draw[black!80,dashed,line width=.75pt]
  (2.07684,12.92818)--(-0.00694,1.11049)
  --(-2.09071,12.92818);
\draw[geomCobalt,densely dotted,line width=.75pt]
  (1.03893,13.06782)--(-0.00694,1.11349)
  --(-1.05280,13.06782);
\draw[black!80,dashed,line width=.75pt]
  (12.92818,-2.09071)--(1.11049,-0.00694)
  --(12.92818,2.07684);
\draw[geomCobalt,densely dotted,line width=.75pt]
  (13.06782,-1.05280)--(1.11349,-0.00694)
  --(13.06782,1.03893);
\end{scope}

\filldraw[fill=cyan!28,draw=black!65,line width=.65pt]
  (-3.95000,0.00694) circle (.18);
\filldraw[fill=white,draw=black!65,line width=.65pt]
  (-3.95000,0.00694) circle (.09);
\filldraw[fill=cyan!28,draw=black!65,line width=.65pt]
  (-0.00694,4.00000) circle (.18);
\filldraw[fill=white,draw=black!65,line width=.65pt]
  (-0.00694,4.00000) circle (.09);
\filldraw[fill=cyan!28,draw=black!65,line width=.65pt]
  (4.00000,-0.00694) circle (.18);
\filldraw[fill=white,draw=black!65,line width=.65pt]
  (4.00000,-0.00694) circle (.09);

\node[text=black] at (-3.90,.90) {${D_1}$};
\draw[geomCobalt,line width=.65pt,->]
  (-4.00,.75)--(-3.95000,.00694);
\node[text=black] at (.75,4.15) {${D_2}$};
\draw[geomCobalt,line width=.65pt,->]
  (.87,4.03)--(-.00694,4.00000);
\node[text=black] at (3.90,.85) {${D_3}$};
\draw[geomCobalt,line width=.65pt,->]
  (3.95,.70)--(4.00000,-.00694);

\node[text=black] at (-3.95,-.80) {${B_1}$};
\draw[geomCobalt,line width=.65pt,->]
  (-3.95,-.61)--(-3.95000,-.12806);
\node[text=black] at (.75,3.60) {${B_2}$};
\draw[geomCobalt,line width=.65pt,->]
  (.57,3.60)--(.08852,3.90454);
\node[text=black] at (4.00,-.85) {${B_3}$};
\draw[geomCobalt,line width=.65pt,->]
  (4.00,-.66)--(4.00000,-.14194);

\node[text=black] at (-2.55,1.30) {${C_1}$};
\node[text=black] at (.85,2.60) {${C_2}$};
\node[text=black] at (3.25,1.65) {${C_3}$};
\node at (-2.75,3.50) {$H_1$};
\node at (1.65,4.25) {$H_2$};
\node at (3.60,2.90) {$H_3$};

\node[fill=white,inner sep=3pt,text=black] at (-1.65,2.90) {${E_1}$};
\draw[geomCobalt,line width=.65pt,->]
  (-1.65,2.74)--(-.76,2.85);
\node[fill=white,inner sep=3pt,text=black] at (2.90,2.25) {${E_2}$};
\draw[geomCobalt,line width=.65pt,->]
  (2.90,2.09)--(2.10,2.10);
\node[fill=white,inner sep=3pt,text=black] at (2.20,-1.23) {${E_3}$};
\draw[geomCobalt,line width=.65pt,->]
  (2.20,-1.06)--(2.20,-.59);

\node[fill=white] at (-1.90,.73) {$\Lambda_1$};
\draw[geomCobalt,line width=.65pt,->]
  (-1.90,.59)--(-1.90,.20);
\node[fill=white] at (-.60,1.90) {$\Lambda_2$};
\draw[geomCobalt,line width=.65pt,->]
  (-.42,1.90)--(-.115,1.90);
\node[fill=white] at (1.90,.66) {$\Lambda_3$};
\draw[geomCobalt,line width=.65pt,->]
  (1.90,.50)--(1.90,.10);

\node[fill=white] at (-2.50,-.82) {$\widetilde\Lambda_1$};
\draw[geomCobalt,line width=.65pt,->]
  (-2.50,-.65)--(-2.50,.00694);
\node[fill=white] at (.65,2.05) {$\widetilde\Lambda_2$};
\draw[geomCobalt,line width=.65pt,->]
  (.47,2.05)--(-.00694,2.05);
\node[fill=white] at (1.40,-.90) {$\widetilde\Lambda_3$};
\draw[geomCobalt,line width=.65pt,->]
  (1.58,-.76)--(2.50,-.00694);

\draw[gray!65,line width=.35pt]
  (-4.5,-1.6) rectangle (4.5,4.6);
\end{tikzpicture}
}
\caption{\leavevmode{Three-hole partition from Setting~\ref{setting:merging-inputs}.}}
\label{kawa}
\label{fig:three-hole-merging}
\end{figure}
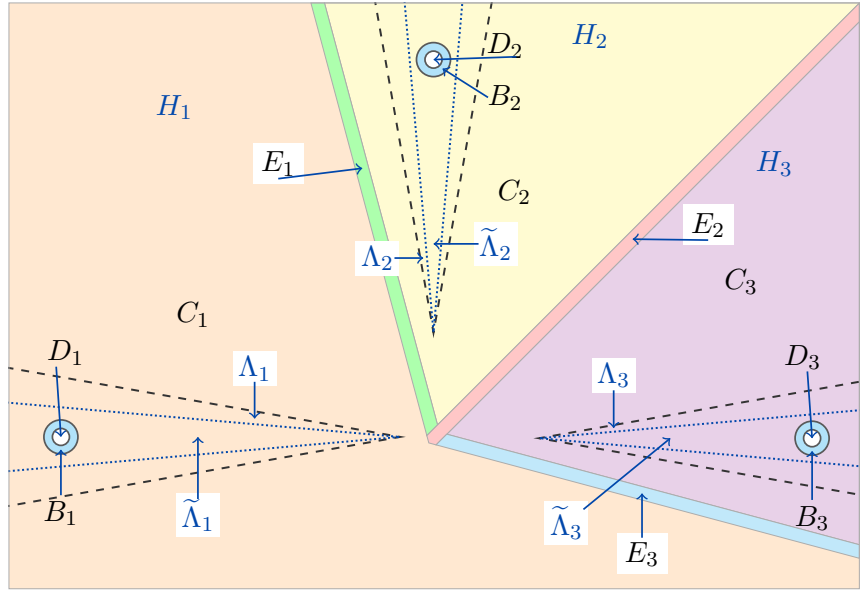

\begin{figure}[p]
\centering
{\begin{tikzpicture}[x=1.22cm,y=1.22cm,font=\normalsize,every node/.style={text=geomCobalt,inner sep=2pt}]
\path[fill=orange!18,draw=gray!65,line width=.35pt] (-4.50000,-1.60000)--(0.23147,-1.60000)--(-1.42981,4.60000)--(-4.50000,4.60000)--cycle;
\path[fill=green!32,draw=gray!65,line width=.35pt] (0.23147,-1.60000)--(0.42872,-1.60000)--(-1.23257,4.60000)--(-1.42981,4.60000)--cycle;
\path[fill=red!22,draw=gray!65,line width=.35pt] (0.42872,-1.60000)--(0.62597,-1.60000)--(-1.03532,4.60000)--(-1.23257,4.60000)--cycle;
\path[fill=violet!18,draw=gray!65,line width=.35pt] (0.62597,-1.60000)--(4.50000,-1.60000)--(4.50000,4.60000)--(-1.03532,4.60000)--cycle;
\filldraw[fill=cyan!28,draw=black!65,line width=.65pt] (-3.60488,0.06935) circle (.50);
\filldraw[fill=white,draw=black!65,line width=.65pt] (-3.60488,0.06935) circle (.32);
\node[font=\normalsize] at (-3.60488,0.06935) {$D_1$};
\filldraw[fill=cyan!28,draw=black!65,line width=.65pt] (-0.06935,3.60488) circle (.50);
\filldraw[fill=white,draw=black!65,line width=.65pt] (-0.06935,3.60488) circle (.32);
\node[font=\normalsize] at (-0.06935,3.60488) {$D_2$};
\node[text=regionPine] at (-2.80000,2.50000) {$G_{12}$};
\node[text=regionPine] at (2.40000,2.50000) {$G'_{12}$};
\node[fill=white,rounded corners=1pt,inner sep=3pt,text=regionPine] at (-1.70000,1.80000) {$I_{12}$};
\draw[geomCobalt,line width=.65pt,->] (-1.70000,1.64000)--(-0.50000,1.55000);
\node[fill=white,rounded corners=1pt,inner sep=3pt,text=regionPine] at (0.85000,1.80000) {$I'_{12}$};
\draw[geomCobalt,line width=.65pt,->] (0.85000,1.64000)--(-0.30000,1.55000);
\node[text=black] at (-3.60000,-0.80000) {${B_1}$};
\draw[geomCobalt,line width=.65pt,->] (-3.60000,-0.62000)--(-3.60000,-0.37000);
\node[text=black] at (0.76000,3.60500) {${B_2}$};
\draw[geomCobalt,line width=.65pt,->] (0.58000,3.60500)--(0.37000,3.60500);
\node[font=\normalsize] at (0.00000,4.90000) {$(r,s)=(1,2)$};
\end{tikzpicture}
\par\vspace{.35cm}
\begin{tikzpicture}[x=1.22cm,y=1.22cm,font=\normalsize,every node/.style={text=geomCobalt,inner sep=2pt}]
\path[fill=orange!18,draw=gray!65,line width=.35pt] (-4.50000,-1.60000)--(-1.86944,-1.60000)--(4.33056,4.60000)--(-4.50000,4.60000)--cycle;
\path[fill=green!32,draw=gray!65,line width=.35pt] (-1.86944,-1.60000)--(-1.60000,-1.60000)--(4.50000,4.50000)--(4.50000,4.60000)--(4.33056,4.60000)--cycle;
\path[fill=red!22,draw=gray!65,line width=.35pt] (-1.60000,-1.60000)--(-1.33056,-1.60000)--(4.50000,4.23056)--(4.50000,4.50000)--cycle;
\path[fill=violet!18,draw=gray!65,line width=.35pt] (-1.33056,-1.60000)--(4.50000,-1.60000)--(4.50000,4.23056)--cycle;
\filldraw[fill=cyan!28,draw=black!65,line width=.65pt] (-0.06935,3.60488) circle (.50);
\filldraw[fill=white,draw=black!65,line width=.65pt] (-0.06935,3.60488) circle (.32);
\node[font=\normalsize] at (-0.06935,3.60488) {$D_2$};
\filldraw[fill=cyan!28,draw=black!65,line width=.65pt] (3.60488,-0.06935) circle (.50);
\filldraw[fill=white,draw=black!65,line width=.65pt] (3.60488,-0.06935) circle (.32);
\node[font=\normalsize] at (3.60488,-0.06935) {$D_3$};
\node[] at (-2.60000,2.50000) {$G_{23}$};
\node[] at (3.15000,2.50000) {$G'_{23}$};
\node[fill=white,rounded corners=1pt,inner sep=3pt] at (-0.40000,1.90000) {$I_{23}$};
\draw[geomCobalt,line width=.65pt,->] (-0.40000,1.74000)--(0.95000,1.10000);
\node[fill=white,rounded corners=1pt,inner sep=3pt] at (3.00000,1.90000) {$I'_{23}$};
\draw[geomCobalt,line width=.65pt,->] (3.00000,1.74000)--(1.15000,1.00000);
\node[text=black] at (-0.89000,3.60500) {${B_2}$};
\draw[geomCobalt,line width=.65pt,->] (-0.69000,3.60500)--(-0.50000,3.60500);
\node[text=black] at (3.60500,-0.85000) {${B_3}$};
\draw[geomCobalt,line width=.65pt,->] (3.60500,-0.67000)--(3.60500,-0.51000);
\node[font=\normalsize] at (0.00000,4.90000) {$(r,s)=(2,3)$};
\end{tikzpicture}
\caption{Two-hole partitions from Setting~\ref{setting:merging-inputs} \label{fig:merging-regions}}}
\end{figure}
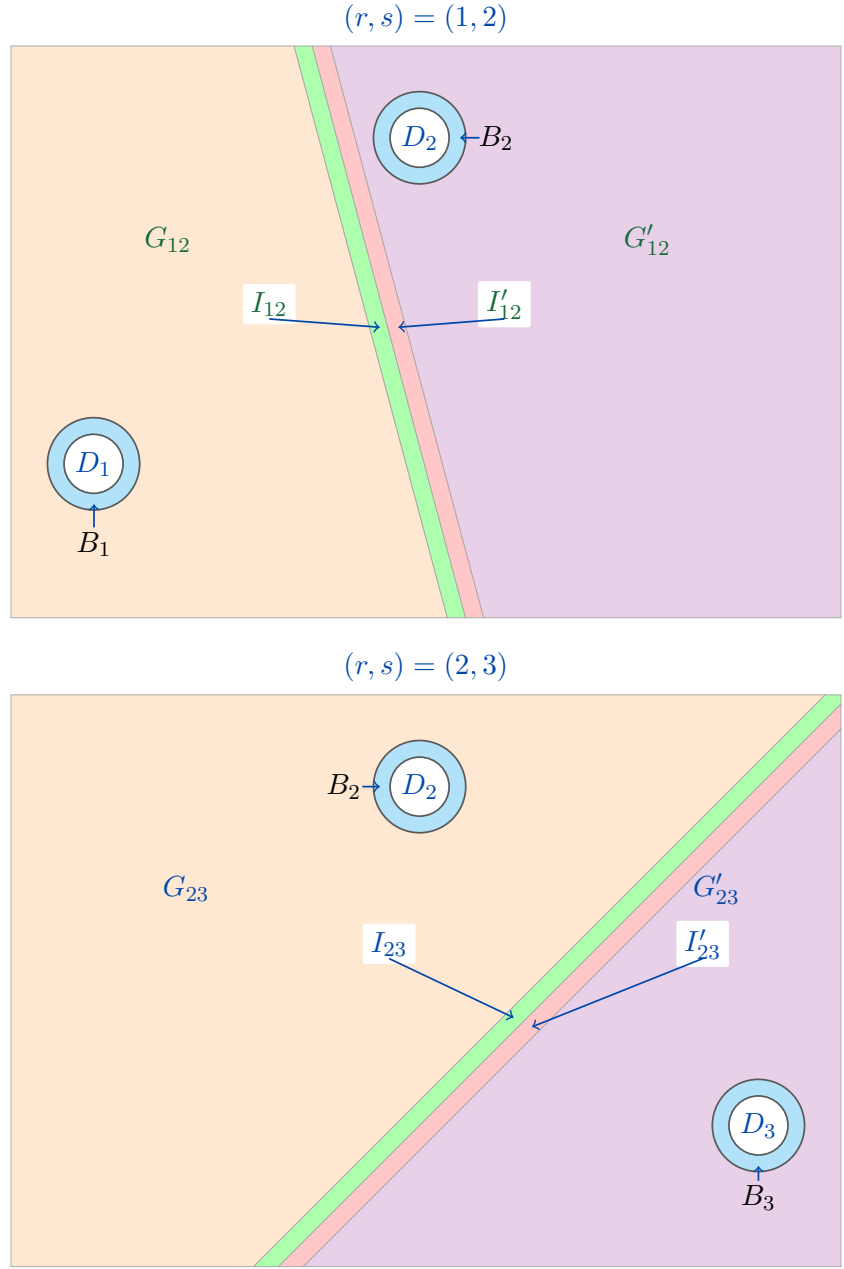

\FloatBarrier

\subsection{Two-hole merging}

{In this subsection, we consider Setting \ref{setting:merging-inputs},  abbreviating
$G_{12}$, $G'_{12}$, $I_{12}$, $I'_{12}$ to
$G$, $G'$, $I$, $I'$, respectively.}

\begin{lem}\label{lem:finite-merging}

Consider Setting \ref{setting:merging-inputs} and
let $a,b\in\mathcal C^{\mathrm{EB}}$.
For all sufficiently large $n$, set
\[
\begin{aligned}
\td\varphi_{aD_1}^{(n)}
&:=
\left.\td\varphi_{aD_1}\right|_{
 \caA_{{{B_1}{G_n}{I_n}{I'_n}}}},\\
\td\varphi_{bD_2}^{(n)}
&:=
\left.\td\varphi_{bD_2}\right|_{
 \caA_{{{I_n}{I'_n}{B_2}{G'_n}}}}.
\end{aligned}
\]

Then the following statements hold.
\begin{enumerate}
\item[(i)]
There is a unique state
$\sigma_n=\td\varphi_{a{D_1}}^{(n)}\bowtie\td\varphi_{b{D_2}}^{(n)}$ on $\caA_{\Gamma_n\setminus({D_1}\cup {D_2})}$
with
\[
 \begin{aligned}
 (\sigma_n)_{{B_1}{G_n}{I_n}{I'_n}}
 &=(\td\varphi_{aD_{\leavevmode{1}}})_{{B_1}{G_n}{I_n}{I'_n}},\\
 (\sigma_n)_{{I_n}{I'_n}{B_2}{G'_n}}
 &=(\td\varphi_{bD_{\leavevmode{2}}})_{{I_n}{I'_n}{B_2}{G'_n}}
 \end{aligned}
\]
and
\[
 \begin{aligned}
 I({B_1}{G_n}:{I'_n}{B_2}{G'_n}| {I_n})_{\sigma_n}&=0,\\
 I({B_1}{G_n}{I_n}:{B_2}{G'_n}| {I'_n})_{\sigma_n}&=0.
 \end{aligned}
\]
Moreover,
\[
 \sigma_n\in\Sigma\bigl(\Gamma_n\setminus({D_1}\cup {D_2})\bigr).
\]
We write $\sigma_n=\td \varphi_{a{D_1}}^{(n)}{\bowtie}\td \varphi_{b{D_2}}^{(n)}$.

\item[(ii)]
Define the convex set
\[
\mathcal S_n
:=
\left\{
  \psi \;\middle|\;
  \begin{aligned}
    &\psi \text{ is a state on }
      \caA_{\Gamma_n\setminus({D_1}\cup {D_2})},\\
    &\psi_{{B_1}{G_n}{I_n}{I'_n}}
      =(\td\varphi_{a{D_1}})_{{B_1}{G_n}{I_n}{I'_n}},\\
    &\psi_{{I_n}{I'_n}{B_2}{G'_n}}
      =(\td\varphi_{b{D_2}})_{{I_n}{I'_n}{B_2}{G'_n}}
  \end{aligned}
\right\}.
\]
Then $\sigma_n$ is {the} unique entropy maximizer in $\mathcal S_n$, and
\[
 S(\sigma_n)
 =S(\td \varphi_{aD_1}^{(n)})+S(\td \varphi_{bD_2}^{(n)})-S(\omega_{{I_n}{I'_n}}).
\]

\item[(iii)]
The states are consistent:
\[
 (\sigma_{n+1})_{\Gamma_n\setminus({D_1}\cup {D_2})}=\sigma_n.
\]
Consequently, they determine a unique state
$\td\varphi_{a{D_1}}{\bowtie}\td\varphi_{b{D_2}}$ on the quasi-local algebra
$\mathcal A_{({D_1}\cup {D_2})^c}$ such that
\[
 (\td\varphi_{a{D_1}}{\bowtie}\td\varphi_{b{D_2}})_{\Gamma_n\setminus({D_1}\cup {D_2})}
 =\sigma_n.
\]

\item[(iv)]
{The merged state has the following two Petz recovery representations:}
\[
 \begin{aligned}
 \sigma_n
 &=
 \td \varphi_{bD_2}\circ\left(
 \lmk R^{(\td \varphi_{aD_1})_{{B_1}{G_n}{I_n}}}_{{I_n}\to {B_1}{G_n}{I_n}}\rmk^T
 \otimes\operatorname{id}_{{I'_n}{B_2}{G'_n}}
 \right)\\
 &=
 \td \varphi_{aD_1}\circ \left(
 \operatorname{id}_{{B_1}{G_n}{I_n}}
 \otimes\lmk
 R^{(\td \varphi_{bD_2})_{{I'_n}{B_2}{G'_n}}}_{{I'_n}\to {I'_n}{B_2}{G'_n}}\rmk^T
 \right).
 \end{aligned}
\]
Here the superscript $T$ denotes the trace adjoint.
\end{enumerate}
\end{lem}

\begin{proof}
Parts (i) and (iv) follow from
Lemmas~\ref{lem:merging-general} and~\ref{lem:merginglemma},
using the face-addition condition for $(I_n,I'_n)$ and
$(I'_n,I_n)$ and local uniqueness on $I_nI'_n$
(Lemma~\ref{lem:localuniqueness}).
For all sufficiently large $n$, the finite configuration satisfies Setting~\ref{setting:two-annulus-merging}, and the two input states are the corresponding extreme points on the annulus. Proposition~\ref{prop:two-annulus-merging} therefore provides $\sigma_n\in\Sigma(\Gamma_n\setminus(D_1\cup D_2))$.

For every $\rho\in\mathcal S_n$, strong subadditivity implies
\[
 \begin{aligned}
 S(\rho)
 &\le
 S(\rho_{{B_1}{G_n}{I_n}{I'_n}})
 +S(\rho_{{I_n}{I'_n}{B_2}{G'_n}})
 -S(\rho_{{I_n}{I'_n}})\\
 &=S(\td \varphi_{aD_1}^{(n)})+S(\td \varphi_{bD_2}^{(n)})-S(\omega_{{I_n}{I'_n}}).
 \end{aligned}
\]
The Markov conditions of $\sigma_n$ in (i) imply
\[
 I({B_1}{G_n}:{B_2}{G'_n}| {I_n}{I'_n})_{\sigma_n}=0
\]
by the chain rule and nonnegativity of conditional mutual
information.
Thus $\sigma_n$ attains the upper bound.
Strict concavity of entropy on the convex set
$\mathcal S_n$ proves (ii).

To prove consistency, set
\[
 \mathcal K_n
 :=\mathcal S_n\cap
 \Sigma\bigl(\Gamma_n\setminus({D_1}\cup {D_2})\bigr).
\]
By Corollary~\ref{nesti}, {there exists a bijection}
\[
 \Phi_n:
 \Sigma\bigl(\Gamma_n\setminus({D_1}\cup {D_2})\bigr)
 \longrightarrow
 \Sigma\bigl(\Gamma_{n+1}\setminus({D_1}\cup {D_2})\bigr)
\]
whose inverse is the restriction.
We claim that
\begin{equation}\label{eq:finite-merging-constraints1}
 \Phi_n(\mathcal K_n)=\mathcal K_{n+1}.
\end{equation}

Indeed, for $\rho\in\mathcal K_n$, the left annular
marginal of $\Phi_n(\rho)$ belongs to
\[
 \Sigma({B_1}{G_{n+1}}{I_{n+1}}{I'_{n+1}})
\]
and restricts to
$\td \varphi_{aD_1}^{(n)}$ on ${B_1}{G_n}{I_n}{I'_n}$.
Injectivity of annular restriction implies
\[
 (\Phi_n(\rho))_{{B_1}{G_{n+1}}{I_{n+1}}{I'_{n+1}}}
 =\td \varphi_{aD_1}^{(n+1)}.
\]
The same argument on the right provides
\[
 (\Phi_n(\rho))_{{I_{n+1}}{I'_{n+1}}{B_2}{G'_{n+1}}}
 =\td \varphi_{bD_2}^{(n+1)}.
\]
Conversely, restriction maps $\mathcal K_{n+1}$ into
$\mathcal K_n$.
This proves~\eqref{eq:finite-merging-constraints1}.

The map $\Phi_n$ preserves entropy differences.
Since $\sigma_n$ is the unique entropy maximizer on
$\mathcal K_n$, and $\sigma_{n+1}$ is the unique entropy
maximizer on $\mathcal K_{n+1}$, it follows that
\[
 \Phi_n(\sigma_n)=\sigma_{n+1}.
\]
Taking the restriction proves
\[
 (\sigma_{n+1})_{\Gamma_n\setminus({D_1}\cup {D_2})}
 =\sigma_n,
\]
 proving (iii).
\end{proof}

We now fill the two holes while preserving the merged exterior state.

\begin{lemma}\label{lem:fill-two-holes}
Consider Setting \ref{setting:merging-inputs}.
For all sufficiently large $n$, define
\[
\begin{aligned}
\zeta_{ab}^n
&:=
(\td\varphi_{a{D_1}}{\bowtie}\td\varphi_{b{D_2}})
_{\Gamma_n\setminus({D_1}\cup {D_2})}\\
&\quad\circ
\left(
 \left(
  R^{(\varphi_{a{D_1}})_{{D_1}{B_1}}}_{{B_1}\to {D_1}{B_1}}
 \right)^T
 \otimes
 \left(
  R^{(\varphi_{b{D_2}})_{{D_2}{B_2}}}_{{B_2}\to {D_2}{B_2}}
 \right)^T
 \otimes
 \operatorname{id}_{{G_n}{I_n}{I'_n}{G'_n}}
\right).
\end{aligned}
\]
Then $\zeta_{ab}^n$ is a state on
$\mathcal A_{\Gamma_n}$ satisfying
\[
 \begin{aligned}
 (\zeta_{ab}^n)_{{D_1}{B_1}{G_n}}
   &=(\varphi_{a{D_1}})_{{D_1}{B_1}{G_n}},\\
 (\zeta_{ab}^n)_{\Gamma_n\setminus({D_1}\cup {D_2})}
   &=\td \varphi_{a{D_1}}^{(n)}{\bowtie}\td \varphi_{b{D_2}}^{(n)},\\
 (\zeta_{ab}^n)_{{D_2}{B_2}{G'_n}}
   &=(\varphi_{b{D_2}})_{{D_2}{B_2}{G'_n}}.
 \end{aligned}
\]
Moreover,
\[
 \begin{aligned}
 I({D_1}{B_1}{G_n}{I_n}{I'_n}:{B_2}{D_2}| {G'_n})_{\zeta_{ab}^n}
   &=0,\\
 I({D_1}{B_1}{G_n}{I_n}{I'_n}{G'_n}:{D_2}| {B_2})_{\zeta_{ab}^n}
   &=0.
 \end{aligned}
\]
\end{lemma}

\begin{proof}
For all sufficiently large $n$, both $(B_1,G_n)$ and $(G_n,B_1)$ satisfy the
face-addition condition. Applying Lemma~\ref{lem:merging-general} to the left hole, its collar, and the exterior region gives

\[
 I({D_1}:{G_n}| {B_1})_{\varphi_{a{D_1}}}=0.
\]
The same argument applied to the exterior merged state provides
\[
 I({B_1}:{I_n}{I'_n}{G'_n}{B_2}| {G_n})_{\td \varphi_{a{D_1}}{\bowtie}\td \varphi_{b{D_2}}}=0.
\]
The marginals on the common region ${B_1}{G_n}$ agree:
\[
 \begin{aligned}
 (\td \varphi_{a{D_1}}{\bowtie}\td \varphi_{b{D_2}})_{{B_1}{G_n}}
 &=(\td \varphi_{a{D_1}})_{{B_1}{G_n}}\\
 &=(\varphi_{a{D_1}})_{{B_1}{G_n}}.
 \end{aligned}
\]
The first equality follows from
Lemma~\ref{lem:finite-merging}, and the second from the
definition of $\varphi_{a{D_1}}$.

Lemma \ref{lem:merginglemma} therefore provides a state
$\gamma_n$ on $\mathcal A_{\Gamma_n\setminus {D_2}}$ such that
\[
 \begin{aligned}
 (\gamma_n)_{{D_1}{B_1}{G_n}}
   &=(\varphi_{a{D_1}})_{{D_1}{B_1}{G_n}},\\
 (\gamma_n)_{\Gamma_n\setminus({D_1}\cup {D_2})}
   &=\td \varphi_{a{D_1}}^{(n)}{\bowtie}\td \varphi_{b{D_2}}^{(n)},
 \end{aligned}
\]
and
\[
 \begin{aligned}
 I({D_1}:{G_n}{I_n}{I'_n}{G'_n}{B_2}| {B_1})_{\gamma_n}
   &=0,\\
 I({D_1}{B_1}:{I_n}{I'_n}{G'_n}{B_2}| {G_n})_{\gamma_n}
   &=0.
 \end{aligned}
\]
Using the recovery map,
\[
 \gamma_n=
 \left(
 R^{(\varphi_{a{D_1}})_{{D_1}{B_1}}}_{{B_1}\to {D_1}{B_1}}
 \otimes
 \operatorname{id}_{{G_n}{I_n}{I'_n}{B_2}{G'_n}}
 \right)(\td \varphi_{a{D_1}}^{(n)}{\bowtie}\td \varphi_{b{D_2}}^{(n)}).
\]

We next fill the right hole ${D_2}$. The corresponding Markov condition for the exterior merged state follows from Lemma~\ref{lem:merging-general}.
Since the left-hole recovery acts only on $B_1$, the data processing inequality provides,
\[
 I({D_1}{B_1}{G_n}{I_n}{I'_n}:{B_2}| {G'_n})_{\gamma_n}=0.
\]
By Lemma~\ref{lem:merging-general},
\[
 I({G'_n}:{D_2}| {B_2})_{\varphi_{b{{D_2}}}}=0.
\]
The marginals on ${B_2}{G'_n}$ agree:
\[
 \begin{aligned}
 (\gamma_n)_{{B_2}{G'_n}}
 &=(
 \td \varphi_{a{D_1}}^{(n)}{\bowtie}\td \varphi_{b{D_2}}^{(n)}
 )_{{B_2}{G'_n}}\\
 &=(\td \varphi_{b{D_2}})_{{B_2}{G'_n}}\\
 &=(\varphi_{b{D_2}})_{{B_2}{G'_n}}.
 \end{aligned}
\]

Applying Lemma \ref{lem:merginglemma} again gives a state $\zeta_{ab}^n$ on $\mathcal A_{\Gamma_n}$
such that
\[
 \begin{aligned}
 (\zeta_{ab}^n)_{\Gamma_n\setminus {D_2}}
   &=\gamma_n,\\
 (\zeta_{ab}^n)_{{D_2}{B_2}{G'_n}}
   &=(\varphi_{b{D_2}})_{{D_2}{B_2}{G'_n}},
 \end{aligned}
\]
and
\[
 \begin{aligned}
 I({D_1}{B_1}{G_n}{I_n}{I'_n}:{B_2}{D_2}| {G'_n})_{\zeta_{ab}^n}
   &=0,\\
 I({D_1}{B_1}{G_n}{I_n}{I'_n}{G'_n}:{D_2}| {B_2})_{\zeta_{ab}^n}
   &=0.
 \end{aligned}
\]
Furthermore, with the recovery map,
\[
 \zeta_{ab}^n=
 \left(
 \operatorname{id}_{{D_1}{B_1}{G_n}{I_n}{I'_n}{G'_n}}
 \otimes
 R^{(\varphi_{b{D_2}})_{{D_2}{B_2}}}_{{B_2}\to {D_2}{B_2}}
 \right)(\gamma_n).
\]
Substituting the recovery formula for $\gamma_n$ implies
\[
 \zeta_{ab}^n=
 \left(
 R^{(\varphi_{a{D_1}})_{{D_1}{B_1}}}_{{B_1}\to {D_1}{B_1}}
 \otimes
 R^{(\varphi_{b{D_2}})_{{D_2}{B_2}}}_{{B_2}\to {D_2}{B_2}}
 \otimes
 \operatorname{id}_{{G_n}{I_n}{I'_n}{G'_n}}
 \right)(\td \varphi_{a{D_1}}^{(n)}{\bowtie}\td \varphi_{b{D_2}}^{(n)}).
\]
Finally, since the second merging preserves the entire
marginal on $\Gamma_n\setminus {D_2}$, we obtain
\[
 (\zeta_{ab}^n)_{{D_1}{B_1}{G_n}}
 =(\gamma_n)_{{D_1}{B_1}{G_n}}
 =(\varphi_{a{{D_1}}})_{{D_1}{B_1}{G_n}}
\]
and
\[
 (\zeta_{ab}^n)_{\Gamma_n\setminus({D_1}\cup {D_2})}
 =(\gamma_n)_{\Gamma_n\setminus({D_1}\cup {D_2})}
 =\td \varphi_{a{D_1}}^{(n)}{\bowtie}\td \varphi_{b{D_2}}^{(n)}.
\]
This proves all the assertions.
\end{proof}
As a corollary, we obtain
\begin{lemma}\label{lem:hole-filling-limit}
Consider Setting~\ref{setting:merging-inputs} and
 $\td\varphi_{a{D_1}}{\bowtie}\td\varphi_{b{D_2}}$ from Lemma~\ref{lem:finite-merging} and
$\zeta_{ab}^n$ from Lemma~\ref{lem:fill-two-holes}. 
Then
\[
\varphi_{a{D_1}}{\bowtie}\varphi_{b{D_2}}:=(\td\varphi_{a{D_1}}{\bowtie}\td\varphi_{b{D_2}})\circ
 \lmk
  \lmk R^{(\varphi_{a{D_1}})_{{D_1}{B_1}}}_{{B_1}\to {D_1}{B_1}}\rmk^T
 \otimes
 \lmk R^{(\varphi_{b{D_2}})_{{D_2}{B_2}}}_{{B_2}\to {D_2}{B_2}}\rmk^T
 \otimes\operatorname{id}_{\A_{({D_1}{B_1}\cup {D_2}{B_2})^c}}
 \rmk
\]
is the unique state on $\A$ whose restriction to $\A_{\Gamma_n}$ is
$\zeta_{ab}^n$ for every $n$ large enough.
\end{lemma}

For later use, we record the following consequence of
the local Markov property and local uniqueness.
\begin{lemma}\label{washi}
Consider  Setting~\ref{setting:merging-inputs}.
Then
\[
 \widetilde\varphi_{a{D_1}}
 =
 \mathop{\mathrm{weak}^*\!-\!\lim}_{n\to\infty}
 \omega\circ
 \left[
 \left(
 R^{(\varphi_{a{D_1}})_{({B_1}{G}{I})_n}}_{{I_n}\to({B_1}{G}{I})_n}
 \right)^T
 \otimes
 \operatorname{id}_{{I'_n}{D_2}{B_2} {G'_n}}
 \right].
\]
{This equality is on $\mathcal A_{{D_1}^c}$.}
\end{lemma}
\begin{proof}
Apply Lemma~\ref{lem:merging-general} across the separating strip $I_n$.
The marginal on $I_nI_n'D_2B_2G_n'$ equals that of $\omega$
by disk uniqueness. Petz recovery proves the asserted
identity on local observables.
\end{proof}

\subsection{Tensor product and merging}

{In this subsection, we consider Setting~\ref{setting:merging-inputs}, abbreviating
$G_{12}$, $G'_{12}$, $I_{12}$, $I'_{12}$ to
$G$, $G'$, $I$, $I'$, respectively.
}

In this subsection we relate the tensor product in the DHR theory {to} merging.
\begin{prop}\label{lem:merging-tensor-product}
Consider Setting~\ref{setting:merging-inputs}.
For any $a,b\in\mathcal C^{\mathrm{EB}}$,
\[
 (\varphi_{a{D_1}}{\bowtie}\varphi_{b{D_2}})(x)
 =\langle\Omega,(\rho_a^{{D_1}\ld_1}\otimes\rho_b^{{D_2}\ld_2})(x)\Omega\rangle,
 \qquad x\in\A,
\]
where $\otimes$ denotes the categorical tensor product of DHR-theory.
\end{prop}

\begin{proof}
Set
\[
 \widehat C:={D_1}{B_1}{G},\qquad
 \widehat C':={G'}{B_2}{D_2}.
\]
We consider local observables
\[
 X_{\widehat C}\in\mathcal A_{\widehat C},\qquad
 X_{{I}}\in\mathcal A_{{I}},\qquad
 X_{{I'}}\in\mathcal A_{{I'}},\qquad
 X_{\widehat C'}\in\mathcal A_{\widehat C'},
\]
and put
\[
 X:=X_{\widehat C}\otimes X_{{I}}\otimes
 X_{{I'}}\otimes X_{\widehat C'}.
\]

{{
By the definition of the tensor product and the same
transportability and normality argument as in
\cite[Section~2]{Ogata}, using lattice Haag duality
and Lemma~\ref{rem:lattice-cone-approximation},
}}
\[
 \begin{aligned}
 &(\rho_{a}^{{D_1}\Lambda_1}\otimes\rho_{b}^{{D_2}\Lambda_2})(X)\\
 &\qquad=
 \rho_{a}^{{D_1}\Lambda_1}(X_{\widehat C})
 \pi(X_{{I}}\otimes X_{{I'}})
 \rho_{b}^{{D_2}\Lambda_2}(X_{\widehat C'}).
 \end{aligned}
\]

Define the recovered local observables
\[
 \begin{aligned}
 \widehat X_{{B_1}{G}}
 &:=
 \left(
 (R^{(\varphi_{a{D_1}})_{{D_1}{B_1}}}_{{B_1}\to {D_1}{B_1}})^{T}
 \otimes\operatorname{id}_{{G}}
 \right)(X_{\widehat C}),\\
 \widetilde X_{{B_2}{G'}}
 &:=
 \left(
 (R^{(\varphi_{b{D_2}})_{{D_2}{B_2}}}_{{B_2}\to {D_2}{B_2}})^{T}
 \otimes\operatorname{id}_{{G'}}
 \right)(X_{\widehat C'}).
 \end{aligned}
\]
They belong to $\mathcal A_{{B_1}{G}}$ and 
$\mathcal A_{{B_2}{G'}}$, respectively, and because the {adjoints} of the recovery maps are UCP,
\[
 \|\widehat X_{{B_1}{G}}\|\le\|X_{\widehat C}\|,
 \qquad
 \|\widetilde X_{{B_2}{G'}}\|\le\|X_{\widehat C'}\|.
\]

Let $\td\varphi_{a{D_1}}^{(n)}$ and $\td\varphi_{b{D_2}}^{(n)}$ be the annular charge
states of {Lemma~\ref{lem:finite-merging}} on
\[
 ({B_1}{G}{I}{I'})_n
 \quad\text{and}\quad
 ({I}{I'}{B_2}{G'})_n.
\]
By Lemma \ref{lem:hole-filling-limit}
\[
 \begin{aligned}
 &(\varphi_{a{D_1}}{\bowtie}\varphi_{b{D_2}})(X)\\
 &\quad=
 \lim_{n\to\infty}
 (\td\varphi_{a{D_1}}^{(n)}{\bowtie}\td\varphi_{b{D_2}}^{(n)})
 \bigl(
 \widehat X_{{B_1}{G}}\otimes X_{{I}}\otimes
 X_{{I'}}\otimes\widetilde X_{{B_2}{G'}}
 \bigr).
 \end{aligned}
\]
Using the recovery representation of the finite-volume
merged state, we obtain 
\[
 \begin{aligned}
 &(\varphi_{a{D_1}}{\bowtie}\varphi_{b{D_2}})(X)\\
 &\quad=
 \lim_{n\to\infty}
 \td\varphi_{a{D_1}}^{(n)}\left(
 \widehat X_{{B_1}{G}}\otimes X_{{I}}\otimes
 (R^{
   (\td\varphi_{b{D_2}}^{(n)})_{({I'}{B_2}{G'})_n}
 }_{{I'_n}\to({I'}{B_2}{G'})_n})^{T}
 \bigl(X_{{I'}}\otimes\widetilde X_{{B_2}{G'}}\bigr)
 \right).
 \end{aligned}
\]
Substituting the definition of $\widehat X_{{B_1}{G}}$ and
using the finite-volume recovery identity of Lemma~\ref{ushi},
we obtain
\[
 \begin{aligned}
 &(\varphi_{a{D_1}}{\bowtie}\varphi_{b{D_2}})(X)\\
 &\quad=
 \lim_{n\to\infty}
 \varphi_{a{D_1}}\left(
 X_{\widehat C}\otimes X_{{I}}\otimes
 (R^{
   (\td\varphi_{b{D_2}}^{(n)})_{({I'}{B_2}{G'})_n}
 }_{{I'_n}\to({I'}{B_2}{G'})_n})^{T}
 \bigl(X_{{I'}}\otimes\widetilde X_{{B_2}{G'}}\bigr)
 \right)\\
 &\quad=
 \lim_{n\to\infty}
 \Bigl\langle
 \Omega,\,
 \rho_{a}^{{D_1}\Lambda_1}(X_{\widehat C})\\
 &\qquad\quad \pi\left(
 X_{{I}}\otimes
 (R^{
   (\td\varphi_{b{D_2}}^{(n)})_{({I'}{B_2}{G'})_n}
 }_{{I'_n}\to({I'}{B_2}{G'})_n})^{T}
 \bigl(X_{{I'}}\otimes\widetilde X_{{B_2}{G'}}\bigr)
 \right)\Omega
 \Bigr\rangle\\
 &\quad=
 \lim_{n\to\infty}
 \Bigl\langle
 \rho_{a}^{{D_1}\Lambda_1}(X_{\widehat C})^*\Omega,\,\\
 &\qquad\quad \pi\left(
 X_{{I}}\otimes
 (R^{
   (\td\varphi_{b{D_2}}^{(n)})_{({I'}{B_2}{G'})_n}
 }_{{I'_n}\to({I'}{B_2}{G'})_n})^{T}
 \bigl(X_{{I'}}\otimes\widetilde X_{{B_2}{G'}}\bigr)
 \right)\Omega
 \Bigr\rangle.
 \end{aligned}
\]
In the second equality, we used that
$\rho_{a}^{{D_1}\Lambda_1}$ agrees with $\pi$ on the central
strips, which lie outside its localization cone.

Let
\[
 Y_{\widehat C}\in\mathcal A_{\widehat C},
 \qquad
 Y_{\widehat C'}\in\mathcal A_{\widehat C'}
\]
be local observables.
Adding and subtracting the term with $\pi(Y_{\widehat C})\Omega$ in the first argument of the preceding inner product provides
\[
 \begin{aligned}
 &(\varphi_{a{D_1}}{\bowtie}\varphi_{b{D_2}})(X)\\
 &=
 \lim_{n\to\infty}
 \omega\left(
 Y_{\widehat C}^*\otimes X_{{I}}\otimes
 (R^{
   (\td\varphi_{b{D_2}}^{(n)})_{({I'}{B_2}{G'})_n}
 }_{{I'_n}\to({I'}{B_2}{G'})_n})^{T}
 \bigl(X_{{I'}}\otimes\widetilde X_{{B_2}{G'}}\bigr)
 \right)\\
 &\quad+
 \lim_{n\to\infty}
 \Bigl\langle
 \bigl(
 \rho_{a}^{{D_1}\Lambda_1}(X_{\widehat C})^*
 -\pi(Y_{\widehat C})
 \bigr)\Omega,\,\\
 &\qquad\quad \pi\left(
 X_{{I}}\otimes
 (R^{
   (\td\varphi_{b{D_2}}^{(n)})_{({I'}{B_2}{G'})_n}
 }_{{I'_n}\to({I'}{B_2}{G'})_n})^{T}
 \bigl(X_{{I'}}\otimes\widetilde X_{{B_2}{G'}}\bigr)
 \right)\Omega
 \Bigr\rangle.
 \end{aligned}
\]

The right-hand version of Lemma~\ref{washi} identifies the
first limit with the exterior charge state
$\widetilde\varphi_{b{D_2}}$ on
$\mathcal A_{({D_2})^c}$:
\[
 \begin{aligned}
 &\lim_{n\to\infty}
 \omega\left(
 Y_{\widehat C}^*\otimes X_{{I}}\otimes
 (R^{
   (\td\varphi_{b{D_2}}^{(n)})_{({I'}{B_2}{G'})_n}
 }_{{I'_n}\to({I'}{B_2}{G'})_n})^{T}
 \bigl(X_{{I'}}\otimes\widetilde X_{{B_2}{G'}}\bigr)
 \right)\\
 &\quad=
 \widetilde\varphi_{b{D_2}}
 \bigl(
 Y_{\widehat C}^*\otimes X_{{I}}\otimes
 X_{{I'}}\otimes\widetilde X_{{B_2}{G'}}
 \bigr)\\
 &\quad=
 \widetilde\varphi_{b{D_2}}
 \left(
 Y_{\widehat C}^*\otimes X_{{I}}\otimes X_{{I'}}
 \otimes
 \left(
 (R^{
   (\varphi_{b{D_2}})_{{D_2}{B_2}}
 }_{{B_2}\to {D_2}{B_2}})^{T}
 \otimes\operatorname{id}_{{G'}}
 \right)(X_{\widehat C'})
 \right)\\
 &\quad=
 \varphi_{b{D_2}}
 \bigl(
 Y_{\widehat C}^*\otimes X_{{I}}\otimes
 X_{{I'}}\otimes X_{\widehat C'}
 \bigr)\\
 &\quad=
 \left\langle
 \Omega,\,
 \pi(Y_{\widehat C}^*\otimes X_{{I}}\otimes X_{{I'}})
 \rho_{b}^{{D_2}\Lambda_2}(X_{\widehat C'})\Omega
 \right\rangle.
 \end{aligned}
\]
The third equality is Lemma~\ref{ushi}.
The last equality follows from localization of
$\rho_{b}^{{D_2}\Lambda_2}$ in $\ld_2$.

We then add and subtract $\pi(Y_{\widehat C'})$:
\[
 \begin{aligned}
 &\left\langle
 \Omega,\,
 \pi(Y_{\widehat C}^*\otimes X_{{I}}\otimes X_{{I'}})
 \rho_{b}^{{D_2}\Lambda_2}(X_{\widehat C'})\Omega
 \right\rangle\\
 &=
 \left\langle
 \Omega,\,
 \pi(Y_{\widehat C}^*\otimes X_{{I}}\otimes X_{{I'}})
 \pi(Y_{\widehat C'})\Omega
 \right\rangle\\
 &\quad+
 \left\langle
 \Omega,\,
 \pi(Y_{\widehat C}^*\otimes X_{{I}}\otimes X_{{I'}})
 \bigl(
 \rho_{b}^{{D_2}\Lambda_2}(X_{\widehat C'})
 -\pi(Y_{\widehat C'})
 \bigr)\Omega
 \right\rangle.
 \end{aligned}
\]
The difference between the first term and the desired
tensor-product expectation is
\[
 \begin{aligned}
 &\left\langle
 \Omega,\,
 \pi(Y_{\widehat C}^*\otimes X_{{I}}\otimes X_{{I'}})
 \pi(Y_{\widehat C'})\Omega
 \right\rangle
 -
 \left\langle
 \Omega,\,
 (\rho_{a}^{{D_1}\Lambda_1}\otimes\rho_{b}^{{D_2}\Lambda_2})(X)
 \Omega
 \right\rangle\\
 &=
 \left\langle
 \bigl(
 \pi(Y_{\widehat C})
 -\rho_{a}^{{D_1}\Lambda_1}(X_{\widehat C})^*
 \bigr)\Omega,\,
 \pi(X_{{I}}\otimes X_{{I'}}\otimes Y_{\widehat C'})
 \Omega
 \right\rangle\\
 &\quad+
 \left\langle
 \rho_{a}^{{D_1}\Lambda_1}(X_{\widehat C})^*\Omega,\,
 \pi(X_{{I}}\otimes X_{{I'}})
 \bigl(
 \pi(Y_{\widehat C'})
 -\rho_{b}^{{D_2}\Lambda_2}(X_{\widehat C'})
 \bigr)\Omega
 \right\rangle.
 \end{aligned}
\]

The adjoint recovery maps are unital and completely
positive, and hence contractive. Therefore,
\[
 \begin{aligned}
 &\left\|
 (R^{
   (\td\varphi_{b{D_2}}^{(n)})_{({I'}{B_2}{G'})_n}
 }_{{I'_n}\to({I'}{B_2}{G'})_n})^{T}
 \bigl(X_{{I'}}\otimes\widetilde X_{{B_2}{G'}}\bigr)
 \right\|\\
 &\qquad\le
 \|X_{{I'}}\|\,\|\widetilde X_{{B_2}{G'}}\|
 \le
 \|X_{{I'}}\|\,\|X_{\widehat C'}\|.
 \end{aligned}
\]
Combining the preceding identities and applying
the Cauchy--Schwarz inequality yields
\[
 \begin{aligned}
 &\left|
 (\varphi_{a{D_1}}{\bowtie}\varphi_{b{D_2}})(X)
 -
 \left\langle
 \Omega,\,
 (\rho_{a}^{{D_1}\Lambda_1}\otimes\rho_{b}^{{D_2}\Lambda_2})(X)
 \Omega
 \right\rangle
 \right|\\
 &\le
 \left\|
 \bigl(
 \pi(Y_{\widehat C})
 -\rho_{a}^{{D_1}\Lambda_1}(X_{\widehat C})^*
 \bigr)\Omega
 \right\|
 \|X_{{I}}\|\,\|X_{{I'}}\|\,
 \|\pi(Y_{\widehat C'})\Omega\|\\
 &\quad+
 \|X_{\widehat C}\|\,\|X_{{I}}\|\,\|X_{{I'}}\|
 \left\|
 \bigl(
 \pi(Y_{\widehat C'})
 -\rho_{b}^{{D_2}\Lambda_2}(X_{\widehat C'})
 \bigr)\Omega
 \right\|\\
 &\quad+
 \|\pi(Y_{\widehat C})\Omega\|\,
 \|X_{{I}}\|\,\|X_{{I'}}\|
 \left\|
 \bigl(
 \rho_{b}^{{D_2}\Lambda_2}(X_{\widehat C'})
 -\pi(Y_{\widehat C'})
 \bigr)\Omega
 \right\|\\
 &\quad+
 \left\|
 \bigl(
 \rho_{a}^{{D_1}\Lambda_1}(X_{\widehat C})^*
 -\pi(Y_{\widehat C})
 \bigr)\Omega
 \right\|
 \|X_{{I}}\|\,\|X_{{I'}}\|\,\|X_{\widehat C'}\|.
 \end{aligned}
\]

Finally, we use the inclusions
\[
 \rho_{a}^{{D_1}\Lambda_1}(X_{\widehat C})^*
 \in\pi(\mathcal A_{\widehat C})'',
 \qquad
 \rho_{b}^{{D_2}\Lambda_2}(X_{\widehat C'})
 \in\pi(\mathcal A_{\widehat C'})'',
\]
{which follow} from Haag duality for $\widehat C$ and $\widehat C'$. Recall that the lattice cones $\widehat C$ and $\widehat C'$ satisfy Condition M.
By the Kaplansky density theorem and the norm density
of local observables, bounded nets of local observables
$Y_{\widehat C}^\alpha\in \caA_{\widehat C}$ and $Y_{\widehat C'}^\alpha\in\caA_{\widehat C'}$ can be chosen,
so that
\[
 \begin{aligned}
 \pi(Y_{\widehat C}^\alpha)\Omega
 &\longrightarrow
 \rho_{a}^{{D_1}\Lambda_1}(X_{\widehat C})^*\Omega,\\
 \pi(Y_{\widehat C'}^\alpha)\Omega
 &\longrightarrow
 \rho_{b}^{{D_2}\Lambda_2}(X_{\widehat C'})\Omega.
 \end{aligned}
\]
Substituting these approximating nets into the preceding bound,
we find that every term tends to zero. Consequently,
\[
 \begin{aligned}
 &(\varphi_{a{D_1}}{\bowtie}\varphi_{b{D_2}})
 \bigl(
 X_{\widehat C}\otimes X_{{I}}\otimes
 X_{{I'}}\otimes X_{\widehat C'}
 \bigr)\\
 &\qquad=
 \left\langle
 \Omega,\,
 (\rho_{a}^{{D_1}\Lambda_1}\otimes\rho_{b}^{{D_2}\Lambda_2})
 \bigl(
 X_{\widehat C}\otimes X_{{I}}\otimes
 X_{{I'}}\otimes X_{\widehat C'}
 \bigr)\Omega
 \right\rangle.
 \end{aligned}
\]
The linear span of these local tensor products is
norm dense in $\mathcal A$.
Both sides are bounded linear functionals on
$\mathcal A$, so
\[
 \varphi_{a{D_1}}{\bowtie}\varphi_{b{D_2}}
 =
 \left\langle
 \Omega,\,
 (\rho_{a}^{{D_1}\Lambda_1}\otimes\rho_{b}^{{D_2}\Lambda_2})(\,\cdot\,)
 \Omega
 \right\rangle.
\]
\end{proof}

\subsection{Three-hole merging}

For the three-hole configuration of
Setting~\ref{setting:merging-inputs} (Figure~\ref{fig:three-hole-merging}), the next proposition extends the maximum-entropy construction of Lemma~\ref{lem:finite-merging}(ii)--(iii) to iterated merging and associativity. 
\begin{prop}\label{cor:merging-associativity}
Consider Setting~{\ref{setting:merging-inputs}}. {Then}, for any $a,b,c\in\mathcal C^{\mathrm{EB}}$,
there exists a unique state
$\td\varphi_{a{D_1}}{\bowtie}\td\varphi_{b{D_2}}{\bowtie}\td\varphi_{c{D_3}}$
on $\caA_{{D_1}^c\cap{{D_2}}^c\cap{{D_3}}^c}$
such that
\begin{enumerate}
\item
\begin{align}
\begin{split}
&\left.
\td\varphi_{a{D_1}}{\bowtie}\td\varphi_{b{D_2}}{\bowtie}\td\varphi_{c{D_3}}
\right\vert_{\caA_{H_1\setminus {D_1}}}
=
\left.\td\varphi_{a{D_1}}
\right\vert_{\caA_{H_1\setminus {D_1}}},\\
&\left.
\td\varphi_{a{D_1}}{\bowtie}\td\varphi_{b{D_2}}{\bowtie}\td\varphi_{c{D_3}}
\right\vert_{\caA_{H_2\setminus {D_2}}}
=
\left.\td\varphi_{b{D_2}}
\right\vert_{\caA_{H_2\setminus {D_2}}},\\
&\left.
\td\varphi_{a{D_1}}{\bowtie}\td\varphi_{b{D_2}}{\bowtie}\td\varphi_{c{D_3}}
\right\vert_{\caA_{H_3\setminus {D_3}}}
=
\left.\td\varphi_{c{D_3}}
\right\vert_{\caA_{H_3\setminus {D_3}}}.
\end{split}
\end{align}

\item
For each $n$ large enough,
$(\td\varphi_{a{D_1}}{\bowtie}\td\varphi_{b{D_2}}{\bowtie}\td\varphi_{c{D_3}})_
{\Gamma_n\cap {D_1}^c\cap{{D_2}}^c\cap{{D_3}}^c}$
is the unique von Neumann entropy maximizer among states
\[
\eta\in\Sigma\lmk
\Gamma_n\cap {D_1}^c\cap{{D_2}}^c\cap{{D_3}}^c
\rmk
\]
satisfying
\begin{align}\label{eq:5.10}
\begin{split}
&\left.\eta
\right\vert_{\caA_{H_1\cap\Gamma_n\setminus {D_1}}}
=
\left.\td\varphi_{a{D_1}}
\right\vert_{\caA_{H_1\cap\Gamma_n\setminus {D_1}}},\\
&\left.\eta
\right\vert_{\caA_{H_2\cap\Gamma_n\setminus{{D_2}}}}
=
\left.\td\varphi_{b{D_2}}
\right\vert_{\caA_{H_2\cap\Gamma_n\setminus{{D_2}}}},\\
&\left.\eta
\right\vert_{\caA_{H_3\cap\Gamma_n\setminus{{D_3}}}}
=
\left.\td\varphi_{c{D_3}}
\right\vert_{\caA_{H_3\cap\Gamma_n\setminus{{D_3}}}}.
\end{split}
\end{align}
\end{enumerate}

 \end{prop}
\begin{proof}
For all sufficiently large $n$, choose an annular cover $X_{1,n},X_{2,n},X_{3,n}$ of
$Z_n:=\Gamma_n\setminus(D_1\cup D_2\cup D_3)$ satisfying Setting~\ref{mimisen}, with each $X_{i,n}$ essentially containing $B_i$, as permitted by Setting~\ref{setting:merging-inputs}. By Lemma~\ref{lem:annular-restriction-injectivity}, within $\Sigma(Z_n)$, fixing the marginals
in \eqref{eq:5.10} is equivalent to fixing the corresponding marginals
on $B_i$, and hence on $X_{i,n}$.
Lemma~\ref{lem:three-annulus-merging} therefore provides the unique entropy maximizer.
As in Lemma~\ref{lem:finite-merging}(iii), restriction preserves these constraints
and entropy differences, so the maximizers are consistent
and determine the asserted state.
\end{proof}

{By an argument analogous to the proof of
 Proposition \ref{lem:merging-tensor-product}, we obtain the following.}

\begin{prop}\label{mizuumi}
Consider Setting {\ref{setting:merging-inputs}}.
Let $a,b,c\in\mathcal C^{\mathrm{EB}}$.
The state given by
\[
 \begin{aligned}
 &(\varphi_{a{D_1}}{\bowtie}\varphi_{b{D_2}})
     {\bowtie}\varphi_{c{D_3}}\\
 &\quad:=\bigl((\td \varphi_{a{D_1}}{\bowtie}\td \varphi_{b{D_2}}){\bowtie}\td \varphi_{c{D_3}}\bigr)
    \circ\Bigl(
       (R^{(\varphi_{a{D_1}})_{{D_1}{B_1}}}_{{B_1}\to {D_1}{B_1}})^{T}
       \otimes
       (R^{(\varphi_{b{D_2}})_{{D_2}{B_2}}}_{{B_2}\to {D_2}{B_2}})^{T}\\
 &\hspace{47mm}\otimes
       (R^{(\varphi_{c{D_3}})_{{D_3}{B_3}}}_{{B_3}\to {D_3}{B_3}})^{T}
       \otimes\operatorname{id}_{({D_1}{B_1}{D_2}{B_2}{D_3}{B_3})^c}
       \Bigr)
 \end{aligned}
\]
is represented as
\[
 \begin{aligned}
 &(\varphi_{a{D_1}}{\bowtie}\varphi_{b{D_2}})
       {\bowtie}\varphi_{c{D_3}}=
 \bigl\langle\Omega,
   \bigl((\rho_a^{{D_1}\Lambda_1}\otimes\rho_b^{{D_2}\Lambda_2})
      \otimes\rho_c^{{D_3}\Lambda_3}\bigr)(\,\cdot\,)\Omega
 \bigr\rangle.
 \end{aligned}
\]

\end{prop}

\section{Existence of conjugates}\label{sec:cone-approximation}

For simple objects $\rho,\bar\rho:\mathcal{A}\to B(\mathcal{H})$ in the DHR-category,
a solution of the conjugate equations consists of intertwiners
\[
 r\in(\pi,\bar\rho\otimes\rho),
 \qquad
 \bar r\in(\pi,\rho\otimes\bar\rho)
\]
satisfying
\[
 \begin{aligned}
  (\bar r^*\otimes1_\rho)(1_\rho\otimes r)
    &=1_\rho,\\
  (r^*\otimes1_{\bar\rho})(1_{\bar\rho}\otimes\bar r)
    &=1_{\bar\rho}.
 \end{aligned}
\]
The existence of such a pair means that $\bar\rho$ is a
conjugate of $\rho$. (See \cite{NT})

In this section, we show that the irreducible representations
corresponding to $a\in\ceb$ and its anti-particle $\bar a$ form such a pair.
\begin{proposition}\label{prop:conjugate-isometries}
Let $a\in\mathcal C^{\mathrm{EB}}$, and let $\bar a\in\mathcal C^{\mathrm{EB}}$ denote its
antiparticle label defined in Sec.~\ref{tuchi}.
Let $\rho_a:=\rho_a^{D_0\ld_0}$,
$\rho_{\bar a}:=\rho_{\bar a}^{D_0\ld_0}\in {\caO_{\ld_0}}$ 
be representations given in Definition~\ref{def:deep-inside-cone} with respect to $a$, $\bar a$ and
a pair of a disk and a cone
$(D_0, \ld_0)$ with
$\ld_0\in \mathcal C_{(\theta_0,\varphi_0)}$.
Then there exist isometries
\[
R_a\in
(\pi,\rho_{\bar a}\otimes\rho_a),
\qquad
\bar R_a\in
(\pi,\rho_a\otimes\rho_{\bar a}),
\]
and a positive number $\kappa_a$ such that
\[
\begin{aligned}
\bar R_a^*T_{\rho_a}(R_a)
 =\kappa_a\,\mathrm{id}_{\mathcal H},\quad R_a^*T_{\rho_{\bar a}}(\bar R_a)
=\kappa_a\,\mathrm{id}_{\mathcal H}.
\end{aligned}
\]

\end{proposition}
Consequently,
\[
 r=\kappa_a^{-1/2}R_a,
 \qquad
 \bar r=\kappa_a^{-1/2}\bar R_a
\]
solve the conjugate equations for
$\rho=\rho_a$ and
$\bar\rho=\rho_{\bar a}$.

The starting point of the proof is the decomposition of
$\rho_a\otimes\rho_{\bar a}$ into copies of the irreducible
charged representations
$\rho_c:\mathcal{A}\to B(\mathcal{H})$, with
$c\in\mathcal{C}^{\mathrm{EB}}$
(Corollary~\ref{lem:sector-product-decomposition}).
Choose the representative of the vacuum charge $\mathbf{1}$
to be $\rho_{\mathbf{1}}=\pi$.
The projection onto the sum of the vacuum summands is
\[
 Q_{a\bar a}^{\mathbf{1}}
 :=\sum_m
 u_{a\bar a}^{\mathbf{1}}(m)
 u_{a\bar a}^{\mathbf{1}}(m)^*,
\]
where the isometric intertwiners
$u_{a\bar a}^{\mathbf{1}}(m)\in
(\pi,\rho_a\otimes\rho_{\bar a})$
have mutually orthogonal ranges and enumerate all vacuum
copies in the decomposition; the sum converges strongly.
Define $Q_{\bar a a}^{\mathbf{1}}$ analogously for the reverse
tensor product; see~\eqref{qdef}.

We then use the three-hole configuration of
Settings~\ref{setting:merging-inputs} and~\ref{imo},
with charges $a,\bar a,a$.
For the two adjacent pairs,
Lemma~\ref{lem:annular-sector-projection} identifies the
action of their vacuum-sector projection
$Q_{a\bar a}^{\mathbf{1}}$ on $\Omega$
as limits of represented annular charge detectors
$S_{C_N^{12}}^{\bf 1}$ (\ref{ame}).
Then, from the finite-volume calculation in Proposition~\ref{concon}
of annular charge detectors, 
we obtain
\[
 Q_{a\bar a}^{\mathbf{1}}
 T_{\rho_a}\lmk Q_{\bar a a}^{\mathbf{1}}\rmk\ne0.
\]
(Lemma~\ref{lem:vacuum-projection-nonzero}).
Because those
projections are given in terms of
$u_{a\bar a}^{\mathbf{1}}(m)\in
(\pi,\rho_a\otimes\rho_{\bar a})$
and $u_{\bar a a}^{\mathbf{1}}(m)\in
(\pi,\rho_{\bar a}\otimes\rho_{a})$,
we obtain a solution of conjugate equations.

\subsection{DHR direct sums}
As a corollary of Lemma \ref{lem:ssc-normality}, we obtain 
\begin{cor}\label{lem:sector-product-decomposition}
Let $\Lambda_1,\Lambda_2$ be cones,
\leavevmode{containing the standard disks}
$D_1$, $D_2$ deep inside, respectively.
Let $a,b\in\mathcal C^{\mathrm{EB}}$.
Then there are multiplicities
\[
 N_{ab}^{cDHR}\in\{0,1,2,\ldots,\infty\},\qquad
 c\in\mathcal C^{\mathrm{EB}},
\]
and isometries
\[
 u_{ab}^{c}(m)\in(\rho_c^{D_1\Lambda_1},\rho_a^{D_1\Lambda_1}\otimes\rho_b^{D_2\Lambda_2}),\qquad
 1\le m\le N_{ab}^{cDHR},
\]
with mutually orthogonal ranges, such that
\[
 \sum_{c\in\mathcal C^{\mathrm{EB}}}
 \sum_{m=1}^{N_{ab}^{cDHR}}
 u_{ab}^{c}(m)u_{ab}^{c}(m)^*=I
\]
in the strong operator topology. 
\end{cor}

\begin{proof}
Since the superselection category is closed under its categorical tensor
product, $\rho_a^{D_1\Lambda_1}\otimes\rho_b^{D_2\Lambda_2}$ satisfies the superselection criterion.
Applying Lemma~\ref{lem:ssc-normality}, we obtain the result.

\end{proof}

Next we identify the sector projections using finite annular collars
(see Figure~\ref{fig:paired-detector-regions}).
\begin{setting}\label{imo}
Fix $w_t\in\bbN$ large enough.
Under Setting~\ref{setting:merging-inputs},
one can check that
there exist a lattice
cone $\widetilde\Lambda$ satisfying Condition M, an open planar
cone $\Lambda$, a closed convex planar cone $C_{\mathrm{aux}}$,
and a planar cone $\mathcal T_{12}$
such that
\begin{align}\label{ramune}
 \begin{split}
 & \Lambda_1\cup\Lambda_2\subset|\widetilde\Lambda|\subset\Lambda,
 \qquad \Lambda\ll\operatorname{int}C_{\mathrm{aux}}
 \ll\mathcal T_{12},\\
  &\mathcal T_{12}\perp_{3w_t} \Lambda_3
  \end{split}
\end{align}
Set $\widetilde \ld':= \mathfrak F\lmk C_{\mathrm{aux}}\rmk$ and 
$T_{12}:=
 \mathfrak F(\overline{\mathcal T_{12}})$.
 We assume that $T_{12}$ satisfies Condition M.

 For each $N\in\bbN$ large enough, set
 $ J_{12}(N):=\lmk (T_{{12}})_N\rmk^{-2w_t}$.
 We consider the annular collars
\[
C_N^{12}:=(T_{12})_N\setminus J_{{12}}(N),
\]
surrounding $D_1$,$D_2$.
{We choose $T_{23}$ and
$C_N^{23}$ analogously.}
See Figure~\ref{fig:paired-detector-regions}.
Again, a concrete realization is given in Appendix~\ref{taiyou}.
\end{setting}

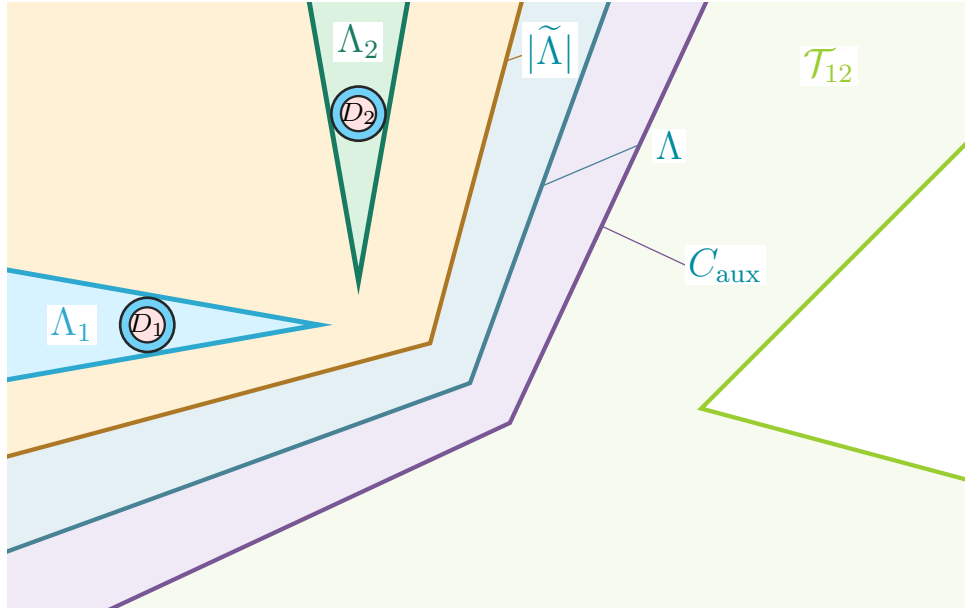
\begin{figure}[htbp]
\centering
\resizebox{\linewidth}{!}{%
\begin{tikzpicture}[x=.85cm,y=.85cm,font=\large]
 \definecolor{lemFourOuterFill}{HTML}{F0EAF7}
 \definecolor{lemFourOuterLine}{HTML}{785794}
 \definecolor{lemFourMiddleFill}{HTML}{E4F0F4}
 \definecolor{lemFourMiddleLine}{HTML}{488295}
 \definecolor{lemFourInnerFill}{HTML}{FFF1D5}
 \definecolor{lemFourInnerLine}{HTML}{AD7826}
 \definecolor{lemFourLeftFill}{HTML}{D6F3FF}
 \definecolor{lemFourLeftLine}{HTML}{2DA8CE}
 \definecolor{lemFourUpperFill}{HTML}{DAF2DF}
 \definecolor{lemFourUpperLine}{HTML}{197A62}
 \clip (0,0) rectangle (12,7.6);
 \begin{scope}[shift={(8.70,2.50)}]
  \fill[regionYellowGreen!8]
    (0,0)--(45:30) arc[start angle=45,end angle=345,radius=30]--cycle;
  \draw[draw=regionYellowGreen,line width=1.3pt]
    (45:30)--(0,0)--(345:30);
 \end{scope}
 \begin{scope}[shift={(6.30,2.32)}]
  \fill[lemFourOuterFill]
    (0,0)--(65:30) arc[start angle=65,end angle=205,radius=30]--cycle;
  \draw[draw=lemFourOuterLine,line width=1.3pt]
    (65:30)--(0,0)--(205:30);
 \end{scope}
 \begin{scope}[shift={(5.80,2.82)}]
  \fill[lemFourMiddleFill]
    (0,0)--(70:30) arc[start angle=70,end angle=200,radius=30]--cycle;
  \draw[draw=lemFourMiddleLine,line width=1.3pt]
    (70:30)--(0,0)--(200:30);
 \end{scope}
 \begin{scope}[shift={(5.30,3.32)}]
  \fill[lemFourInnerFill]
    (0,0)--(75:30) arc[start angle=75,end angle=195,radius=30]--cycle;
  \draw[draw=lemFourInnerLine,line width=1.3pt]
    (75:30)--(0,0)--(195:30);
 \end{scope}
 \begin{scope}[shift={(3.9,3.55)}]
  \fill[lemFourLeftFill] (0,0)--(170:30)--(190:30)--cycle;
  \draw[draw=lemFourLeftLine,line width=1.5pt]
    (170:30)--(0,0)--(190:30);
 \end{scope}
 \begin{scope}[shift={(4.4,4.10)}]
  \fill[lemFourUpperFill] (0,0)--(80:30)--(100:30)--cycle;
  \draw[draw=lemFourUpperLine,line width=1.5pt]
    (80:30)--(0,0)--(100:30);
 \end{scope}
 \filldraw[fill=cyan!48,draw=black!85,line width=.8pt]
   (1.75,3.55) circle (.34);
 \filldraw[fill=red!11,draw=black!85,line width=.7pt]
   (1.75,3.55) circle (.22);
 \node[font=\scriptsize] at (1.75,3.55) {$D_1$};
 \filldraw[fill=cyan!48,draw=black!85,line width=.8pt]
   (4.40,6.20) circle (.34);
 \filldraw[fill=red!11,draw=black!85,line width=.7pt]
   (4.40,6.20) circle (.22);
 \node[font=\scriptsize] at (4.40,6.20) {$D_2$};
 \node[text=lemFourLeftLine,fill=white,inner sep=1pt] at (0.8,3.55)
   {$\Lambda_1$};
 \node[text=lemFourUpperLine,fill=white,inner sep=1pt] at (4.4,7.1)
   {$\Lambda_2$};
 \draw[draw=lemFourInnerLine] (6.65,7.00)--(6.25,6.86);
 \node[text=coneCyan,fill=white,inner sep=1pt] at (6.80,7.00)
   {$|\widetilde\Lambda|$};
 \draw[draw=lemFourMiddleLine] (7.90,5.80)--(6.70,5.29);
 \node[text=coneCyan,fill=white,inner sep=1pt] at (8.30,5.80)
   {$\Lambda$};
 \draw[draw=lemFourOuterLine] (8.50,4.30)--(7.45,4.79);
 \node[text=coneCyan,fill=white,inner sep=1pt] at (9.00,4.30)
   {$C_{\mathrm{aux}}$};
 \node[text=regionYellowGreen,fill=white,inner sep=1pt] at (10.3,6.8)
   {$\mathcal T_{12}$};
\end{tikzpicture}%
}
\caption{{The auxiliary cones for the pair $(1,2)$ in
Lemma~\ref{lem:annular-sector-projection}.
}
}
\label{fig:lemma4-configuration}
\end{figure}

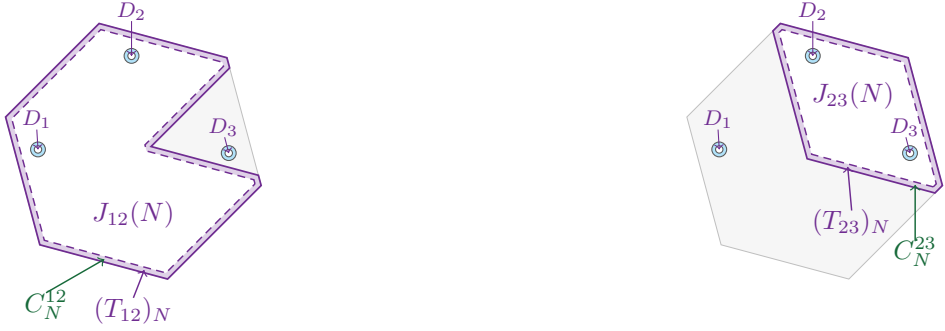
\begin{figure}[!htbp]
\centering
\begin{tikzpicture}[x=.35cm,y=.35cm,font=\scriptsize,
  every node/.style={text=detectorPlum,inner sep=1pt}]
\path[use as bounding box] (-5.2,-6.7) rectangle (5.2,5.7);
\path[fill=gray!7,draw=gray!50,line width=.35pt] (4.82963,-1.29410)--(3.53553,3.53553)--(-1.29410,4.82963)--(-4.82963,1.29410)--(-3.53553,-3.53553)--(1.29410,-4.82963)--cycle;
\path[fill=detectorPlum!22,draw=detectorPlum,line width=.65pt] (4.82963,-1.29410)--(4.72610,-0.90772)--(0.66921,0.17932)--(3.63906,3.14916)--(3.53553,3.53553)--(-1.29410,4.82963)--(-4.82963,1.29410)--(-3.53553,-3.53553)--(1.29410,-4.82963)--cycle;
\path[fill=white,draw=detectorPlum,densely dashed,line width=.55pt] (4.59781,-1.23198)--(4.55640,-1.07743)--(0.49951,0.00961)--(0.43739,0.24143)--(3.40724,3.21128)--(3.36583,3.36583)--(-1.23198,4.59781)--(-4.59781,1.23198)--(-3.36583,-3.36583)--(1.23198,-4.59781)--cycle;
\filldraw[fill=cyan!28,draw=black!65,line width=.4pt] (-3.60488,0.06935) circle (.28);
\filldraw[fill=white,draw=black!65,line width=.4pt] (-3.60488,0.06935) circle (.14);
\node at (-3.65000,1.20000) {$D_1$};
\draw[detectorPlum,line width=.4pt,->] (-3.65000,0.82000)--(-3.60488,0.06935);
\filldraw[fill=cyan!28,draw=black!65,line width=.4pt] (-0.06935,3.60488) circle (.28);
\filldraw[fill=white,draw=black!65,line width=.4pt] (-0.06935,3.60488) circle (.14);
\node at (-0.06935,5.30000) {$D_2$};
\draw[detectorPlum,line width=.4pt,->] (-0.06935,4.92000)--(-0.06935,3.60488);
\filldraw[fill=cyan!28,draw=black!65,line width=.4pt] (3.60488,-0.06935) circle (.28);
\filldraw[fill=white,draw=black!65,line width=.4pt] (3.60488,-0.06935) circle (.14);
\node at (3.40000,0.90000) {$D_3$};
\draw[detectorPlum,line width=.4pt,->] (3.45000,0.53000)--(3.60488,-0.06935);
\node[font=\small] at (0.00000,-2.40000) {$J_{12}(N)$};
\node[font=\small,text=regionPine] at (-3.3,-5.8) {$C_N^{12}$};
\draw[regionPine,line width=.55pt,->] (-3.3,-5.35)--(-1.0938,-4.0817);
\node[font=\small] at (0.00000,-6.00000) {$(T_{12})_N$};
\draw[detectorPlum,line width=.55pt,->] (0.00000,-5.54000)--(0.41302,-4.48596);
\end{tikzpicture}
\hfill
\begin{tikzpicture}[x=.35cm,y=.35cm,font=\scriptsize,
  every node/.style={text=detectorPlum,inner sep=1pt}]
\path[use as bounding box] (-5.2,-6.7) rectangle (5.2,5.7);
\path[fill=gray!7,draw=gray!50,line width=.35pt] (4.82963,-1.29410)--(3.53553,3.53553)--(-1.29410,4.82963)--(-4.82963,1.29410)--(-3.53553,-3.53553)--(1.29410,-4.82963)--cycle;
\path[fill=detectorPlum!22,draw=detectorPlum,line width=.65pt] (4.54679,-1.57694)--(4.82963,-1.29410)--(3.53553,3.53553)--(-1.29410,4.82963)--(-1.57694,4.54679)--(-0.28284,-0.28284)--cycle;
\path[fill=white,draw=detectorPlum,densely dashed,line width=.55pt] (4.48467,-1.34512)--(4.59781,-1.23198)--(3.36583,3.36583)--(-1.23198,4.59781)--(-1.34512,4.48467)--(-0.11314,-0.11314)--cycle;
\filldraw[fill=cyan!28,draw=black!65,line width=.4pt] (-3.60488,0.06935) circle (.28);
\filldraw[fill=white,draw=black!65,line width=.4pt] (-3.60488,0.06935) circle (.14);
\node at (-3.65000,1.20000) {$D_1$};
\draw[detectorPlum,line width=.4pt,->] (-3.65000,0.82000)--(-3.60488,0.06935);
\filldraw[fill=cyan!28,draw=black!65,line width=.4pt] (-0.06935,3.60488) circle (.28);
\filldraw[fill=white,draw=black!65,line width=.4pt] (-0.06935,3.60488) circle (.14);
\node at (-0.06935,5.30000) {$D_2$};
\draw[detectorPlum,line width=.4pt,->] (-0.06935,4.92000)--(-0.06935,3.60488);
\filldraw[fill=cyan!28,draw=black!65,line width=.4pt] (3.60488,-0.06935) circle (.28);
\filldraw[fill=white,draw=black!65,line width=.4pt] (3.60488,-0.06935) circle (.14);
\node at (3.40000,0.90000) {$D_3$};
\draw[detectorPlum,line width=.4pt,->] (3.45000,0.53000)--(3.60488,-0.06935);
\node[font=\small] at (1.45000,2.10000) {$J_{23}(N)$};
\node[font=\small,text=regionPine] at (3.8,-3.8) {$C_N^{23}$};
\draw[regionPine,line width=.55pt,->] (3.8,-3.4)--(3.8,-1.2692);
\node[font=\small] at (1.40000,-2.70000) {$(T_{23})_N$};
\draw[detectorPlum,line width=.55pt,->] (1.40000,-2.24000)--(1.25090,-0.58622);
\end{tikzpicture}
\caption{{Paired detector regions from Setting~\ref{imo}.}
{{The two panels show}} $J_{rs}(N)\subset (T_{rs})_N\subset\Gamma_N$
{for $(r,s)=(1,2),(2,3)$, from left to right.}
Solid and dashed purple contours bound $(T_{rs})_N$ and
$J_{rs}(N)$, respectively; the shaded band is
$C_N^{rs}=(T_{rs})_N\setminus J_{rs}(N)$.
}
\label{fig:paired-detector-regions}
\end{figure}

\FloatBarrier

\begin{lemma}\label{lem:annular-sector-projection}
{Consider Setting~\ref{setting:merging-inputs} and Setting \ref{imo}.
For distinct $i,j\in\{1,2,3\}$ and
$a,b\in\mathcal C^{\mathrm{EB}}$, let
$\{u_{ab}^{d,\Lambda_i\Lambda_j}(m)\}_{d,m}$
be the isometries of
Corollary~\ref{lem:sector-product-decomposition}
in the decomposition of
$\rho_a^{D_i\Lambda_i}\otimes\rho_b^{D_j\Lambda_j}$,
where $d\in\mathcal C^{\mathrm{EB}}$ and $m$ ranges over
its multiplicity.
Define the charge-$c$ sector projection by
\begin{equation}\label{qdef}
 Q_{ab}^{c,\Lambda_i\Lambda_j}
 :=\sum_m
 u_{ab}^{c,\Lambda_i\Lambda_j}(m)
 u_{ab}^{c,\Lambda_i\Lambda_j}(m)^*,
 \qquad c\in\mathcal C^{\mathrm{EB}},
\end{equation}
with strong operator convergence.

Then for $(i,j)\in\{(1,2),(2,3)\}$,
\[
 \lim_{N\to\infty}
 \left\|
 \left(
  (\rho_a^{D_i\Lambda_i}\otimes
   \rho_b^{D_j\Lambda_j})(S_{C_N^{ij}}^c)
  -Q_{ab}^{c,\Lambda_i\Lambda_j}
 \right)\Omega
 \right\|=0
 \qquad(c\in\mathcal C^{\mathrm{EB}}).
\]
Here $S_{C_N^{ij}}^c$ is the support projection of the
charge-$c$ annular state, and $\Omega$ is the reference GNS vector.
}
\end{lemma}

\begin{proof}
{\it 1.} {We prove the assertion for $(i,j)=(1,2)$; the same argument
{applies to $(2,3)$.}
Write $C_N:=C_N^{12}$ and omit the cone labels from
$u_{ab}^{d,\Lambda_1\Lambda_2}(m)$ and
$Q_{ab}^{c,\Lambda_1\Lambda_2}$.
{\ignorespaces 

See Figure \ref{fig:lemma4-configuration}.
Note that every fixed finite subset of
${\widetilde{\Lambda'}}^{+(l_0+2)}$ 
(with  $\widetilde \ld':= \mathfrak F\lmk C_{\mathrm{aux}}\rmk$)is disjoint from $C_N$
for all sufficiently large $N$.
}
}

{\it 2. }
For each $u_{ab}^{d}(m)$, we claim
\begin{align}\label{zou}
\lim_{N\to\infty}\bigl\|\bigl(\rho_d^{D_1\Lambda_1}(S_{C_N}^{c})-\delta_{dc}I\bigr)
 \lmk u_{ab}^{d}(m)\rmk^*\Omega\bigr\|=0.
\end{align}
For the moment, we abbreviate $u:=u_{ab}^{d}(m)$.
Since $\rho_a^{D_1\Lambda_1}\otimes\rho_b^{D_2\Lambda_2}$ and $\rho_d^{D_1\Lambda_1}$ 
agree with $\pi$ on
$\A_{\leavevmode{(\Lambda_1\cup\Lambda_2)^c}}$, the intertwining relation and Haag duality imply
\[
 \lmk u_{ab}^{d}(m)\rmk^*\in\pi\lmk  \caA_{\lmk \ld_1\cup\ld_2\rmk^c}\rmk'\subset
 \pi(\A_{\td\ld^c})'
 =\pi(\A_{\td\ld})''\subset
 \pi(\A_\Lambda)''.
\]
(Recall that $\td \ld$ satisfies Condition M).
{Lemma~\ref{lem:cone-local-approximation}
applies to $\rho_d^{D_1\Lambda_1}$, which is localized in
$\Lambda$ and has vector state $\varphi_{dD_1}$ at $\Omega$.
Thus,} for every $\delta>0$ there is a local observable
$z_{dm}^\delta$, supported in
${\widetilde{\Lambda'}}^{+(l_0+2)}$, such that
\begin{align}\label{nuts}
\|\lmk u_{ab}^{d}(m)\rmk^*\Omega-\rho_d^{D_1\Lambda_1}(z_{dm}^\delta)\Omega\|<\delta.
\end{align}

Since $z_{dm}^\delta$ is local, its support
is contained in $\lmk {\widetilde{\Lambda'}}^{+(l_0+2)}\rmk_{l_{dm}^\delta}$ 
for a sufficiently large $l_{dm}^\delta$. 
Hence from the observation in {\it 1. }
\begin{align}\label{ginger}
 [z_{dm}^{\delta},S_{C_N}^{c}]=0,
\end{align}
{for all sufficiently large $N$}. 
Note that
\[
 \langle\Omega,\rho_d^{D_1\Lambda_1}(S_{C_N}^{c})\Omega\rangle
 =\varphi_{dD_1}\lmk S_{C_N}^{c}\rmk
 =\delta_{dc}
\]
for all $N$ large enough, by Lemma \ref{lem:annular-separation} and (\ref{koke}).
Since $S_{C_N}^{c}$ is a projection, it follows that
\[
 \rho_d^{D_1\Lambda_1}(S_{C_N}^{c})\Omega=\delta_{dc}\Omega.
\]
Consequently, using (\ref{ginger}), for all $N$ large enough
\[
 \bigl(\rho_d^{D_1\Lambda_1}(S_{C_N}^{c})-\delta_{dc}I\bigr)
 \rho_d^{D_1\Lambda_1}(z_{dm}^{\delta})\Omega=0.
\]
The operator in parentheses is a contraction. Thus, for any $\delta>0$, from (\ref{nuts}), 
\[
 \bigl\|\bigl(\rho_d^{D_1\Lambda_1}(S_{C_N}^{c})-\delta_{dc}I\bigr)
 u^*\Omega\bigr\|<\delta
\]
for all sufficiently large $N$. Since $\delta$ is arbitrary, 
we have (\ref{zou}).

{\it 3.} We choose increasing finite index sets $I_{l}$ that exhaust all pairs
$(d,m)$, and set
\[
 F_{l}:=\sum_{(d,m)\in I_{l}}
 u_{ab}^{d}(m)u_{ab}^{d}(m)^*.
\]
Then $F_{l}\to I$ strongly. Orthogonality of the ranges and the
intertwining relations yield
\[
\begin{aligned}
 &\bigl\|\left((\rho_a^{D_1\Lambda_1}\otimes\rho_b^{D_2\Lambda_2})(S_{C_N}^{c})-Q_{ab}^{c}\right)F_{l}\Omega\bigr\|^2\\
 &\quad=\sum_{(d,m)\in I_{l}}
 \bigl\|\bigl(\rho_d^{D_1\Lambda_1}(S_{C_N}^{c})-\delta_{dc}I\bigr)
 u_{ab}^{d}(m)^*\Omega\bigr\|^2
 \longrightarrow0
\end{aligned}
\]
for each fixed ${l}$, by {\it 2}{.}

Finally, $(\rho_a^{D_1\Lambda_1}\otimes\rho_b^{D_2\Lambda_2})(S_{C_N}^{c})$ and $Q_{ab}^{c}$ are projections, so their
difference has norm at most one. Therefore,
\[
\begin{aligned}
 \bigl\|\left((\rho_a^{D_1\Lambda_1}\otimes\rho_b^{D_2\Lambda_2})(S_{C_N}^{c})-Q_{ab}^{c}\right)\Omega\bigr\|
 &\le\bigl\|\left((\rho_a^{D_1\Lambda_1}\otimes\rho_b^{D_2\Lambda_2})(S_{C_N}^{c})-Q_{ab}^{c}\right)F_{l}\Omega\bigr\|\\
 &\quad+\|(I-F_{l})\Omega\|.
\end{aligned}
\]
First taking $N\to\infty$, and then ${l}\to\infty$, proves the assertion.
\end{proof}

\subsection{Two-hole merging}

\begin{lemma}\label{lem:paired-charge-expectation}
Consider {Setting~\ref{setting:merging-inputs}}
and Setting \ref{imo}.
Let $a,b,c\in\mathcal C^{\mathrm{EB}}$, and
let $Q_{ab}^{d\ld_1\ld_2}$ and $Q_{bc}^{f\ld_2\ld_3}$, $d,f\in\mathcal C^{\mathrm{EB}}${,}
be the sector projections  for
$\rho_a^{{D_1}\ld_1}\otimes\rho_b^{{D_2}\ld_2}$ and $\rho_b^{{D_2}\ld_2}\otimes\rho_c^{{D_3}\ld_3}$, respectively, as in
 (\ref{qdef}) {in} Lemma~\ref{lem:annular-sector-projection}. Then
\[
\begin{aligned}
 \langle\Omega,Q_{ab}^{d\ld_1\ld_2}Q_{bc}^{f\ld_2\ld_3}\Omega\rangle
 &=\lim_{N\to\infty}\lmk \td\varphi_{a{D_1}}{\bowtie}\td\varphi_{b{D_2}}{\bowtie}\td\varphi_{c{D_3}}\rmk
       (S_{C_N^{12}}^{d}S_{C_N^{23}}^{f})
\end{aligned}
\]
for all $d,f\in\mathcal C^{\mathrm{EB}}$.
\end{lemma}
\begin{proof}
We omit {the cone labels} $\ld_i$ {from} $Q_{ab}^{d\ld_1\ld_2}$ and $Q_{bc}^{f\ld_2\ld_3}$.
Note from the setting that
\[
 x_N:=S_{C_N^{12}}^{d}\in \A_{\Lambda_3^c},\qquad y_N:=S_{C_N^{23}}^{f}\in\A_{\Lambda_1^c}.
\]

{{
By localization and lattice Haag duality (recall that $T_{23}$ satisfies Condition M),
\[
 (\rho_b^{{D_2}\Lambda_2}\otimes
  \rho_c^{{D_3}\Lambda_3})(y_N)
 \in\pi(\A_{{T_{23}}^c})'
 =\pi(\A_{{T_{23}}})''.
\]
The normal extension of $\rho_a^{{D_1}\Lambda_1}$
acts identically on this algebra,
by localization and normality.
}}
This and the definition of the
DHR tensor product imply
\begin{align}
\begin{split}
&(\rho_{a}^{{D_1}\ld_1}\otimes\rho_b^{{D_2}\ld_2}\otimes\rho_c^{{D_3}\ld_3})(x_N)=(\rho_{a}^{{D_1}\ld_1}\otimes\rho_b^{{D_2}\ld_2})(x_N),\\ &(\rho_{a}^{{D_1}\ld_1}\otimes\rho_b^{{D_2}\ld_2}\otimes\rho_c^{{D_3}\ld_3})(y_N)=(\rho_{b}^{{D_2}\ld_2}\otimes\rho_c^{{D_3}\ld_3})(y_N).
\end{split}
\end{align}
Multiplicativity therefore yields
\[
 (\rho_{a}^{{D_1}\ld_1}\otimes\rho_b^{{D_2}\ld_2}\otimes\rho_c^{{D_3}\ld_3})(x_Ny_N)=(\rho_{a}^{{D_1}\ld_1}\otimes\rho_b^{{D_2}\ld_2})(x_N)(\rho_{b}^{{D_2}\ld_2}\otimes\rho_c^{{D_3}\ld_3})(y_N).
\]

Applying  Lemma~\ref{lem:annular-sector-projection} to each pair yields
\[
 (\rho_{a}^{{D_1}\ld_1}\otimes\rho_b^{{D_2}\ld_2})(x_N)\Omega\longrightarrow Q_{ab}^{d}\Omega,\qquad
 (\rho_{b}^{{D_2}\ld_2}\otimes\rho_c^{{D_3}\ld_3})(y_N)\Omega\longrightarrow Q_{bc}^{f}\Omega.
\]
These operators are projections, and hence
\[
\begin{aligned}
 &\left|\langle\Omega,(\rho_{a}^{{D_1}\ld_1}\otimes\rho_b^{{D_2}\ld_2}\otimes\rho_c^{{D_3}\ld_3})(x_Ny_N)\Omega\rangle
       -\langle\Omega,Q_{ab}^{d}Q_{bc}^{f}\Omega\rangle\right|\\
 &\quad\le\|((\rho_{a}^{{D_1}\ld_1}\otimes\rho_b^{{D_2}\ld_2})(x_N)-Q_{ab}^{d})\Omega\|
          +\|((\rho_{b}^{{D_2}\ld_2}\otimes\rho_c^{{D_3}\ld_3})(y_N)-Q_{bc}^{f})\Omega\|\longrightarrow0.
\end{aligned}
\]
By Proposition~\ref{mizuumi},
we have
\begin{align*}
\langle\Omega,(\rho_{a}^{{D_1}\ld_1}\otimes\rho_b^{{D_2}\ld_2}\otimes\rho_c^{{D_3}\ld_3})(x_Ny_N)\Omega\rangle
=\lmk \td\varphi_{a{D_1}}{\bowtie}\td\varphi_{b{D_2}}{\bowtie}\td\varphi_{c{D_3}}\rmk
       (S_{C_N^{12}}^{d}S_{C_N^{23}}^{f}).
\end{align*}
Substituting this completes the proof.
\end{proof}

Note from Proposition \ref{concon} {that} the weight considered here
for $b=\bar a$, $c=a$ is positive.
\begin{lemma}\label{lem:vacuum-pair-weight}
{Consider Setting~\ref{setting:merging-inputs} and Setting \ref{imo}.
Let $a\in\mathcal C^{\mathrm{EB}}$, and let $\bar a$
be its antiparticle as in Subsection~\ref{tuchi}.}
Then, for every sufficiently large $N$,
\[
 (\td\varphi_{a{D_1}}{\bowtie}\td\varphi_{\bar a {D_2}}{\bowtie}\td\varphi_{a{D_3}})\bigl(S_{C_N^{12}}^{\mathbf1}S_{C_N^{23}}^{\mathbf1}\bigr)
=e^{-4f(a)}
\]
In particular, the left-hand side is independent of $N$ once $N$ is
sufficiently large.
\end{lemma}

{\begin{proof}
This follows from {Proposition} \ref{concon} and {Proposition} \ref{cor:merging-associativity}.
\end{proof}
}

\begin{lemma}\label{lem:vacuum-projection-nonzero}
Consider {Setting~\ref{setting:merging-inputs}} and Setting \ref{imo}.
Let $a\in\mathcal C^{\mathrm{EB}}${,} and {let} $\bar a\in \mathcal C^{\mathrm{EB}}$ be {its} antiparticle
given in Subsection~\ref{tuchi}.
Let $Q_{a\bar a}^{\mathbf1\ld_1\ld_2}$ and $Q_{\bar a a}^{\mathbf1\ld_2\ld_1}$
be the sector projections  for
$\rho_a^{D_1\ld_1}\otimes\rho_{\bar a}^{D_2\ld_2}$ and $\rho_{\bar a}^{D_2\ld_2}\otimes\rho_{a}^{D_1\ld_1}$, respectively, as in
 (\ref{qdef}) {in} Lemma~\ref{lem:annular-sector-projection}. 

Then
\[
 Q_{a\bar a}^{\mathbf1,\Lambda_1\Lambda_2}\,
 T_{\rho_a^{D_1\Lambda_1}}
 \!\left(Q_{\bar a a}^{\mathbf1,\Lambda_2\Lambda_1}\right)
 \ne0.
\]
\end{lemma}
\begin{proof}
By Lemmas~\ref{lem:paired-charge-expectation}
and~\ref{lem:vacuum-pair-weight},
\[
 \langle\Omega,Q_{a\bar a}^{\mathbf1,\Lambda_1\Lambda_2}Q_{\bar a a}^{\mathbf1,\Lambda_2\Lambda_3}\Omega\rangle
 =e^{-4f(a)}>0.
\]
In particular, $Q_{a\bar a}^{\mathbf1,\Lambda_1\Lambda_2}Q_{\bar a a}^{\mathbf1,\Lambda_2\Lambda_3}\ne0$.

Choose a unitary intertwiner $V$ from $\rho_{a}^{D_1\ld_1}$ to $\rho_{a}^{D_3\ld_3}$, so that
\[
 \rho_{a}^{D_3\ld_3}=\operatorname{Ad}V\circ\rho_{a}^{D_1\ld_1}.
\]
By the unitary covariance of the vacuum-sector projections,
\[
 Q_{\bar a a}^{\mathbf1,\Lambda_2\Lambda_1}=T_{\rho_{\bar a}^{D_2\Lambda_2}}(V^*)\,Q_{\bar a a}^{\mathbf1,\Lambda_2\Lambda_3}\,T_{\rho_{\bar a}^{D_2\Lambda_2}}(V).
\]
{{
Since $Q_{\bar a a}^{\mathbf1,\Lambda_2\Lambda_3}$
commutes with
$(\rho_{\bar a}^{D_2\Lambda_2}\otimes
  \rho_a^{D_3\Lambda_3})(\A)$,
localization and lattice Haag duality imply
\[
 Q_{\bar a a}^{\mathbf1,\Lambda_2\Lambda_3}
 \in\pi(\A_{{T_{23}}^c})'
 =\pi(\A_{{T_{23}}})''.
\]
Recall that $T_{23}$ satisfies Condition M.
By localization and normality, 
\[
 T_{\rho_a^{D_1\Lambda_1}}
 \bigl(Q_{\bar a a}^{\mathbf1,\Lambda_2\Lambda_3}\bigr)
 =Q_{\bar a a}^{\mathbf1,\Lambda_2\Lambda_3}.
\]
Moreover, the commutation relation of
$Q_{a\bar a}^{\mathbf1,\Lambda_1\Lambda_2}$
with
$(\rho_a^{D_1\Lambda_1}\otimes
  \rho_{\bar a}^{D_2\Lambda_2})(\A)$
extends to the common auxiliary algebra  $\caB$; see
\cite[Lemmas~2.4 and~3.5]{Ogata}.
In particular,
\[
 \left[
 Q_{a\bar a}^{\mathbf1,\Lambda_1\Lambda_2},
 T_{\rho_a^{D_1\Lambda_1}}
 T_{\rho_{\bar a}^{D_2\Lambda_2}}(V^*)
 \right]=0.
\]
}}
Hence
\[
\begin{aligned}
&Q_{a\bar a}^{1,\Lambda_1\Lambda_2}
 T_{\rho_a^{D_1\ld_1}}
 \bigl(Q_{\bar a a}^{1,\Lambda_2\Lambda_1}\bigr)\\
&=
 Q_{a\bar a}^{1,\Lambda_1\Lambda_2}
 T_{\rho_a^{D_1\ld_1}}
 \bigl(
 T_{\rho_{\bar a}^{D_2\Lambda_2}}(V^{*})
 Q_{\bar a a}^{1,\Lambda_2\Lambda_3}
 T_{\rho_{\bar a}^{D_2\Lambda_2}}(V)
 \bigr)\\
&=
 T_{\rho_a^{D_1\ld_1}}
 T_{\rho_{\bar a}^{D_2\Lambda_2}}(V^{*})
 Q_{a\bar a}^{1,\Lambda_1\Lambda_2}
 T_{\rho_a^{D_1\ld_1}}
 \bigl(Q_{\bar a a}^{1,\Lambda_2\Lambda_3}\bigr)
 T_{\rho_a^{D_1\ld_1}}
 T_{\rho_{\bar a}^{D_2\Lambda_2}}(V)\\
 &=
 T_{\rho_a^{D_1\ld_1}}
 T_{\rho_{\bar a}^{D_2\Lambda_2}}(V^{*})
 Q_{a\bar a}^{1,\Lambda_1\Lambda_2}
 Q_{\bar a a}^{1,\Lambda_2\Lambda_3}
 T_{\rho_a^{D_1\ld_1}}
 T_{\rho_{\bar a}^{D_2\Lambda_2}}(V)
 \ne 0,
\end{aligned}
\]
{by the unitarity of} $V$ and {Lemma~\ref{lem:paired-charge-expectation}}.

\end{proof}
\subsection{Proof of Proposition~\ref{prop:conjugate-isometries}}
{By unitary transport, it suffices to consider the case where $\ld_0$ is equal to $\ld_1$ in 
Setting~\ref{setting:merging-inputs}.}
We omit {the disk labels} $D_i$ from $\rho_a^{D_1\ld_1}$ {and similar notation}.

\noindent{\it 1.}
We claim that 
there are isometries
\[
\bar R_a'\in
(\pi,\rho_a^{\Lambda_1}\otimes\rho_{\bar a}^{\Lambda_2}),
\qquad
R_a'\in
(\pi,\rho_{\bar a}^{\Lambda_2}\otimes\rho_a^{\Lambda_1})
\]
such that
\[
\bar R_a'^*T_{\rho_a^{\Lambda_1}}(R_a')\ne0.
\]
Indeed, by Lemma \ref{lem:vacuum-projection-nonzero},

\[
Q_{a\bar a}^{1,\Lambda_1\Lambda_2}
T_{\rho_a^{\Lambda_1}}
\bigl(Q_{\bar a a}^{1,\Lambda_2\Lambda_1}\bigr)\ne0,
\]
{so} there are \(\mu,\nu\) such that
\[
u_{a\bar a}^{1,\Lambda_1\Lambda_2}(\mu)^*
T_{\rho_a^{\Lambda_1}}
\bigl(u_{\bar a a}^{1,\Lambda_2\Lambda_1}(\nu)\bigr)\ne0.
\]
(Recall (\ref{qdef}).)
Setting
\[
\bar R_a'
 :=u_{a\bar a}^{1,\Lambda_1\Lambda_2}(\mu)
 \in(\pi,\rho_a^{\Lambda_1}\otimes\rho_{\bar a}^{\Lambda_2}),
\qquad
R_a'
 :=u_{\bar a a}^{1,\Lambda_2\Lambda_1}(\nu)
 \in(\pi,\rho_{\bar a}^{\Lambda_2}\otimes\rho_a^{\Lambda_1})
\]
proves the claim.

\noindent{\it 2.}
We next claim that the isometries  \(R_a',\bar R_a'\) in {\it 1.}  satisfy
\[
\bar R_a'^*T_{\rho_a^{\Lambda_1}}(R_a'),
\quad
R_a'^*T_{\rho_{\bar a}^{\Lambda_2}}(\bar R_a')
\in\mathbb C\,\mathrm{id}_{\mathcal H}.
\]
To see this, note for $A\in\caA$ that
\[
\begin{aligned}
\bar R_a'^*T_{\rho_a^{\Lambda_1}}(R_a')
 \rho_a^{\Lambda_1}(A)
&=\bar R_a'^*T_{\rho_a^{\Lambda_1}}
   \bigl(R_a'\pi(A)\bigr)\\
&=\bar R_a'^*T_{\rho_a^{\Lambda_1}}
   \bigl((\rho_{\bar a}^{\Lambda_2}
         \otimes\rho_a^{\Lambda_1})(A)R_a'\bigr)\\
&=\bar R_a'^*
   \bigl((\rho_a^{\Lambda_1}
         \otimes\rho_{\bar a}^{\Lambda_2}
         \otimes\rho_a^{\Lambda_1})(A)\bigr)
   T_{\rho_a^{\Lambda_1}}(R_a')\\
&=\rho_a^{\Lambda_1}(A)
   \bar R_a'^*T_{\rho_a^{\Lambda_1}}(R_a').
\end{aligned}
\]
Hence
\[
\bar R_a'^*T_{\rho_a^{\Lambda_1}}(R_a')
\in\operatorname{End}\rho_a^{\Lambda_1}
 =\mathbb C\,\mathrm{id}_{\mathcal H}.
\]
The second property can be proven analogously.

\noindent {\it 3.}
We claim that there is \(\lambda\ne0\) such that
\[
\bar R_a'^*T_{\rho_a^{\Lambda_1}}(R_a')
 =\bar\lambda\,\mathrm{id}_{\mathcal H},\qquad
R_a'^*T_{\rho_{\bar a}^{\Lambda_2}}(\bar R_a')
 =\lambda\,\mathrm{id}_{\mathcal H},
\]
The first equality follows from {\it 2.}, and \(\lambda\ne0\)
follows from {\it 1}{.} Moreover,
\[
\begin{aligned}
\bigl(
 R_a'^*T_{\rho_{\bar a}^{\Lambda_2}}(\bar R_a')
\bigr)^*R_a'
&=
T_{\rho_{\bar a}^{\Lambda_2}}(\bar R_a'^*)R_a'R_a'\\
&=
T_{\rho_{\bar a}^{\Lambda_2}}(\bar R_a'^*)
T_{\rho_{\bar a}^{\Lambda_2}}
T_{\rho_a^{\Lambda_1}}(R_a')\,R_a'\\
&=
T_{\rho_{\bar a}^{\Lambda_2}}
\bigl(\bar R_a'^*T_{\rho_a^{\Lambda_1}}(R_a')\bigr)
R_a'\\
&=\bar\lambda  R_a'.
\end{aligned}
\]

Combining {this} with {\it 2.} {provides}
\[
{{R_a'^*}}T_{{{\rho_{\bar a}^{\Lambda_2}}}}
({{\bar R_a'}})
 ={{\lambda}}\,\mathrm{id}_{\mathcal H}.
\]

\noindent{\it 4.} 
Write \(\lambda=e^{i\theta}|\lambda|\), and set
\[
\td {R}_a:=e^{i\theta}R_a',
\qquad \td {\bar R}_a:=\bar R_a',
\qquad
\kappa_a:=|\lambda|.
\]
Then
\[
\td {\bar R}_a^*T_{\rho_a^{\Lambda_1}}(\td R_a)
 =\kappa_a\,\mathrm{id}_{\mathcal H},
\qquad
\td {R}_a^*T_{\rho_{\bar a}^{\Lambda_2}}(\td {\bar R}_a)
 =\kappa_a\,\mathrm{id}_{\mathcal H}.
\]

\noindent {\it 5.}
Take \(V_{\bar a}\in\mathcal U(\mathcal B)\) such that
\[
\rho_{\bar a}^{\Lambda_2}
 =\operatorname{Ad}V_{\bar a}\circ\rho_{\bar a}^{\Lambda_1}.
\]
Set
\[
R_a:=(V_{\bar a}^*\otimes1)\td R_a,
\qquad
\bar R_a:=
 (1_{\rho_a^{\Lambda_1}}\otimes V_{\bar a}^*)\td{\bar R}_a.
\]
Then
\[
R_a\in
(\pi,\rho_{\bar a}^{\Lambda_1}\otimes\rho_a^{\Lambda_1}),
\qquad
\bar R_a\in
(\pi,\rho_a^{\Lambda_1}\otimes\rho_{\bar a}^{\Lambda_1}),
\]
and
\[
\begin{aligned}
\bar R_a^*T_{\rho_a^{\Lambda_1}}(R_a)
&=
(\td{\bar R}_a)^*
(1_{\rho_a^{\Lambda_1}}\otimes V_{\bar a})
T_{\rho_a^{\Lambda_1}}(V_{\bar a}^*\td R_a)\\
&=
(\td{\bar R}_a)^*T_{\rho_a^{\Lambda_1}}(\td R_a)
 =\kappa_a\,\mathrm{id}_{\mathcal H},
\end{aligned}
\]
while
\[
\begin{aligned}
R_a^*T_{\rho_{\bar a}^{\Lambda_1}}(\bar R_a)
&=
\td R_a^*V_{\bar a}
T_{\rho_{\bar a}^{\Lambda_1}}
\bigl(
 T_{\rho_a^{\Lambda_1}}(V_{\bar a}^*)\td{\bar R}_a
\bigr)\\
&=
\td R_a^*T_{\rho_{\bar a}^{\Lambda_2}}
\bigl(
 T_{\rho_a^{\Lambda_1}}(V_{\bar a}^*)\td{\bar R}_a
\bigr)V_{\bar a}\\
&=
\td{R}_a^*
T_{\rho_{\bar a}^{\Lambda_2}}
T_{\rho_a^{\Lambda_1}}(V_{\bar a}^*)
T_{\rho_{\bar a}^{\Lambda_2}}(\td{\bar R}_a)V_{\bar a}\\
&=
V_{\bar a}^*\td R_a^*
{{T_{\rho_{\bar a}^{\Lambda_2}}}}(\td{\bar R}_a)
V_{\bar a}\\
&=\kappa_a\,\mathrm{id}_{\mathcal H}.
\end{aligned}
\]

\clearpage
\section{Cyclicity}\label{sec:cyclicity}

{{In this section we prove the following cyclicity result.}}

\begin{prop}\label{cyclicity}
{{Consider {Setting~\ref{setting:merging-inputs}}.}}
{{Let $a,b\in\mathcal C^{\mathrm{EB}}$.}}
 Then
\[
\overline{
 (\rho_a^{D_1\Lambda_1}\otimes\rho_b^{D_2\Lambda_2})
 (\mathcal A)\Omega
}
=\mathcal H.
\]
\end{prop}

This cyclicity is an ingredient in the proof of the main theorem,
Theorem~\ref{thm:fusion-endomorphisms}.

For this, it suffices to show
\[
  \pi(\mathcal A_{\Lambda_1^c\cap\Lambda_2^c})\pi(\mathcal A_{\Lambda_1})
 \pi(\mathcal A_{\Lambda_2})
\Omega\subset \overline{
 (\rho_a^{D_1\Lambda_1}\otimes\rho_b^{D_2\Lambda_2})
 (\mathcal A)\Omega.
}
\]
By localization of $\rho_a^{D_1\Lambda_1}\otimes\rho_b^{D_2\Lambda_2}$, we have
\[
 (\rho_a^{D_1\Lambda_1}\otimes\rho_b^{D_2\Lambda_2})
 \lmk \mathcal A_{\Lambda_1^c\cap\Lambda_2^c}\rmk
 =
 \pi\lmk \mathcal A_{\Lambda_1^c\cap\Lambda_2^c}\rmk.
\]
It therefore suffices to show
\[
\begin{aligned}
 \pi(\mathcal A_{\Lambda_1})\Omega
 \subset
 \overline{\rho_a^{D_1\Lambda_1}(\mathcal A_{K_1})\Omega},\qquad
 \pi(\mathcal A_{\Lambda_2})\Omega
 \subset
 \overline{\rho_b^{D_2\Lambda_2}(\mathcal A_{K_2})\Omega},
\end{aligned}
\]
for the suitably enlarged lattice cones $K_1,K_2$ chosen below,
with $\Lambda_i\subset K_i$ and $K_1\cap K_2=\varnothing$.

The key observation, inspired by subfactor theory
(Step~3 below), is
\[
\begin{aligned}
 \pi(\mathcal A_{\Lambda_1})''
 \subset
 \rho_a^{D_1\Lambda_1}
 (\mathcal A_{\widehat\Lambda_1})''\bar R_a,\qquad
 \pi(\mathcal A_{\Lambda_2})''
 \subset
 \rho_b^{D_2\Lambda_2}
 (\mathcal A_{\widehat\Lambda_2})''\bar R_b,
\end{aligned}
\]
for the isometries from
Proposition~\ref{prop:conjugate-isometries}. Here,
$\widehat\Lambda_i$, $i=1,2$, are the intermediate
planar cones chosen below, with
$\Lambda_i\subset \widehat\Lambda_i\subset K_i$.
From this, it suffices to show
\[
\bar R_a\Omega\in \overline{\rho_a^{D_1\Lambda_1}(\mathcal A_{K_1})\Omega},\qquad
\bar R_b\Omega\in  \overline{\rho_b^{D_2\Lambda_2}(\mathcal A_{K_2})\Omega}.
\]
For our chosen representations,
Lemma~\ref{lem:cone-local-approximation} provides these approximations,
which locality and Haag duality allow us to combine.

\begin{proofof}[Proposition \ref{cyclicity}]
In Setting~\ref{setting:merging-inputs}, for each $i=1,2$, we may choose convex
planar cones $\ld_i$, $\widehat\ld_i$, a closed convex cone $C_i'$
and a lattice cone $\widetilde\Lambda_i'$ satisfying Condition M
such that
\begin{align}\label{somen}
\begin{split}
 &\Lambda_i\ll\widehat\Lambda_i\ll\operatorname{int}C_i',
 \qquad
 \Lambda_i^{+2(l_0+2)}\subset\widehat\Lambda_i,
 \qquad
 \operatorname{int}C_i'\in\mathcal C_{(\theta_0,\varphi_0)},\\
 &\Lambda_i
 \subset \bigl|\widetilde\Lambda_i'\bigr|
 \subset \widehat\Lambda_i,\\
 &d\bigl(\widetilde\Lambda_1'',\widetilde\Lambda_2''\bigr)
 \ge 4(l_0+2).
\end{split}
\end{align}
The set
${\widetilde\Lambda_i''}=\mathfrak F({C_i'})$
is a lattice cone satisfying Condition M.

\noindent {\it 1.}
Let
\begin{align}
\begin{split}
&R_a\in
(\pi,\rho_{\bar a}^{D_1\Lambda_1}
 \otimes\rho_a^{D_1\Lambda_1}),\qquad
\bar R_a\in
(\pi,\rho_a^{D_1\Lambda_1}
 \otimes\rho_{\bar a}^{D_1\Lambda_1}),
\\
&R_b\in
(\pi,\rho_{\bar b}^{{{D_2\Lambda_2}}}
 \otimes\rho_b^{D_2\Lambda_2}),\qquad
\bar R_b\in
(\pi,\rho_b^{D_2\Lambda_2}
 \otimes\rho_{\bar b}^{D_2\Lambda_2})
\end{split}
\end{align}
be the {solutions} to the conjugation {equations} for the {pairs}
$(\rho_a^{D_1\Lambda_1},\rho_{\bar a}^{D_1\Lambda_1})$
and
$(\rho_b^{D_2\Lambda_2},\rho_{\bar b}^{D_2\Lambda_2})${,}
respectively{,} given in
Proposition
~\ref{prop:conjugate-isometries}.

{{
Localization and Haag duality for
$\widetilde\Lambda_i'$, $i=1,2$, imply 

\begin{align}\label{sakana}
\begin{split}
R_a,\bar R_a
&\in\pi(\mathcal A_{\Lambda_1^c})'
 \subset
 \pi\bigl(
   \mathcal A_{(\widetilde\Lambda_1')^c}
 \bigr)'
\\
&=
 \pi(\mathcal A_{\widetilde\Lambda_1'})''
 \subset
 \pi(\mathcal A_{\widehat\Lambda_1})'',
\\
R_b,\bar R_b
&\in\pi(\mathcal A_{\Lambda_2^c})'
 \subset
 \pi\bigl(
   \mathcal A_{(\widetilde\Lambda_2')^c}
 \bigr)'
\\
&=
 \pi(\mathcal A_{\widetilde\Lambda_2'})''
 \subset
 \pi(\mathcal A_{\widehat\Lambda_2})''.
 \end{split}
\end{align}
}}

\noindent{\it 2.}
We claim that 
\begin{align}
\bar R_a\bar R_b\Omega
\in
\overline{
 (\rho_a^{D_1\Lambda_1}\otimes\rho_b^{D_2\Lambda_2})
 (\mathcal A)\Omega
}.
\end{align}
To see this, fix any $\epsilon>0$.
{We apply Lemma~\ref{lem:cone-local-approximation} with inner planar
cone $\widehat\Lambda_i$ and outer lattice cone
${\widetilde\Lambda_i''}=\mathfrak F({C_i'})$.
Writing $\mathcal U(\cdot)$ for the unitary group,
}
there exist
$N_\epsilon\in\mathbb N$ and
$u_{i\epsilon}\in
{{
\mathcal U\!\left(
 \mathcal A_{
  \left(
   ({\widetilde\Lambda_i''})_{N_\epsilon}
  \right)^{+(l_0+1)}
 }
\right)
}}$,
$i=1,2$, such that

\begin{align}
\begin{split}
\bigl\|
 \bar R_a\Omega
 -\rho_a^{D_1\Lambda_1}(u_{1\epsilon})\Omega
\bigr\|<\epsilon,\qquad
\bigl\|
 \bar R_b\Omega
 -\rho_b^{D_2\Lambda_2}(u_{2\epsilon})\Omega
\bigr\|<\epsilon.
\end{split}
\end{align}
{Because $K_i:=({\widetilde\Lambda_i''})^{+(l_0+1)}$ satisfies Condition M,
we have
\begin{align}\label{heya}
\begin{split}
&\rho_a^{D_1\Lambda_1}(\mathcal A_{K_1})\subset \rho_a^{D_1\Lambda_1}(\mathcal A_{K_1^c})'=
 \pi(\mathcal A_{K_1^c})'
=\pi(\mathcal A_{K_1})''\\
&\rho_b^{D_2\Lambda_2}(\mathcal A_{K_2})\subset \rho_b^{D_2\Lambda_2}(\mathcal A_{K_2^c})'=
 \pi(\mathcal A_{K_2^c})'
=\pi(\mathcal A_{K_2})''.
\end{split}
\end{align}
Note by (\ref{somen}) that $K_1\cap K_2=\emptyset$.
Since $\bar R_a\in\pi(\mathcal A_{K_1})''$ by (\ref{sakana}), 
 we obtain
\[
 [\bar R_a,\rho_b^{D_2\Lambda_2}(u_{2\epsilon})]=0,\qquad
 (\rho_a^{D_1\Lambda_1}\otimes\rho_b^{D_2\Lambda_2})(u_{1\epsilon}u_{2\epsilon})
 =\rho_a^{D_1\Lambda_1}(u_{1\epsilon})\rho_b^{D_2\Lambda_2}(u_{2\epsilon}).
\]
The two factors on the right are commuting unitaries.
}
Then
\[
\begin{aligned}
&
\bigl\|
 \bar R_a\bar R_b\Omega
 -(\rho_a^{D_1\Lambda_1}\otimes\rho_b^{D_2\Lambda_2})
  (u_{1\epsilon}u_{2\epsilon})\Omega
\bigr\|\\
&\leq
\bigl\|
 \bar R_a\bigl(
  \bar R_b\Omega
  -\rho_b^{D_2\Lambda_2}(u_{2\epsilon})\Omega
 \bigr)
\bigr\|\\
&\quad+
\bigl\|
 \bar R_a\rho_b^{D_2\Lambda_2}
  (u_{2\epsilon})\Omega
 -(\rho_a^{D_1\Lambda_1}\otimes\rho_b^{D_2\Lambda_2})
  (u_{1\epsilon}u_{2\epsilon})\Omega
\bigr\|\\
&\leq
\bigl\|
 \bar R_b\Omega
 -\rho_b^{D_2\Lambda_2}(u_{2\epsilon})\Omega
\bigr\|\\
&\quad+
\bigl\|
 \rho_b^{D_2\Lambda_2}(u_{2\epsilon})
 \bar R_a\Omega
 -(\rho_a^{D_1\Lambda_1}\otimes\rho_b^{D_2\Lambda_2})
  (u_{2\epsilon}u_{1\epsilon})\Omega
\bigr\|\\
&=
\bigl\|
 \bar R_b\Omega
 -\rho_b^{D_2\Lambda_2}(u_{2\epsilon})\Omega
\bigr\|\\
&\quad+
\bigl\|
 \bar R_a\Omega
 -\rho_a^{D_1\Lambda_1}(u_{1\epsilon})\Omega
\bigr\|\\
&<2\epsilon.
\end{aligned}
\]
Because $u_{1\epsilon}u_{2\epsilon}\in\caA$,
this proves the claim.

\noindent{\it 3.}
Next we claim 
\[
\pi(\mathcal A_{\Lambda_1})''\subset \rho_a^{D_1\Lambda_1}(\mathcal A_{\widehat\ld_1})''
 \bar R_a,
\qquad
\pi(\mathcal A_{\Lambda_2})''\subset \rho_b^{D_2\Lambda_2}(\mathcal A_{\widehat\ld_2})''
 \bar R_b
\]
To see this, we use the argument in \cite{LongoRehren1995}, subsection 2.7.
Note that for every
$A\in\mathcal A_{\ld_1}$, we have
\begin{align}
\bar R_a\pi(A)
 =T_{\rho_a^{D_1\Lambda_1}}T_{\rho_{\bar a}^{D_1\Lambda_1}}\pi (A)
  \bar R_a, \quad\text{and}\quad
  T_{\rho_a^{D_1\Lambda_1}}T_{\rho_{\bar a}^{D_1\Lambda_1}}\pi (A)\in
  \rho_a^{D_1\ld_1}(\mathcal A_{\widehat\Lambda_1})''.
\end{align}
We also have 
\begin{align*}
T_{\rho_a^{D_1\Lambda_1}}(R_a^*)\bar R_a=\kappa_a.
\end{align*}
Thus for every
$A\in\mathcal A_{\Lambda_1}$
\[
\kappa_a\pi(A)=T_{\rho_a^{D_1\ld_1}}(R_a^*)\bar R_a\pi(A)
 =
T_{\rho_a^{D_1\ld_1}}(R_a^*)
T_{\rho_a^{D_1\ld_1}}T_{\rho_{\bar a}^{D_1\Lambda_1}}\pi(A)
\bar R_a\in \rho_a^{D_1\ld_1}(\mathcal A_{\widehat\ld_1})''\bar R_a
\]
{{
}}
Hence
\[
\pi(\mathcal A_{\Lambda_1})''
 \subset
\rho_a^{D_1\ld_1}(\mathcal A_{\widehat\ld_1})''
 \bar R_a.
\]
The proof for the second one is the same.

\noindent{\it 4.}
\begin{align}
\begin{split}
&\pi(\mathcal A_{\Lambda_1})''
\pi(\mathcal A_{\Lambda_2})''\Omega
\overset{\text{\it 3.}}{\subset}
\rho_a^{D_1\ld_1}(\mathcal A_{\widehat\Lambda_1})''
 \bar R_a\,
\rho_b^{D_2\Lambda_2}(\mathcal A_{\widehat\Lambda_2})''
 \bar R_b\Omega\\
&=
\rho_a^{D_1\ld_1}(\mathcal A_{\widehat\Lambda_1})''
\rho_b^{D_2\Lambda_2}(\mathcal A_{\widehat\Lambda_2})''
\bar R_a\bar R_b\Omega\\
&\overset{\text{\it 2.}}\subset
\rho_a^{D_1\ld_1}(\mathcal A_{\widehat\Lambda_1})''
\rho_b^{D_2\Lambda_2}(\mathcal A_{\widehat \Lambda_2})''
\overline{
 (\rho_a^{D_1\ld_1}\otimes\rho_b^{D_2\Lambda_2})
 (\mathcal A)\Omega
}.
\end{split}
\end{align}
Combining {this} with
$\pi\vert_{\mathcal A_{(\Lambda_1\cup\Lambda_2)^c}}
=(\rho_a^{D_1\ld_1}\otimes\rho_b^{D_2\Lambda_2})\vert_{\mathcal A_{(\Lambda_1\cup\Lambda_2)^c}}$,
we have
\[
\pi(\mathcal A_{(\Lambda_1\cup\Lambda_2)^c})
\pi(\mathcal A_{\Lambda_1})''
\pi(\mathcal A_{\Lambda_2})''\Omega
\subset
\overline{
 (\rho_a^{D_1\ld_1}\otimes\rho_b^{D_2\Lambda_2})
 (\mathcal A)\Omega
}.
\]
Therefore
\[
\mathcal H
 =
\overline{
 \pi(\mathcal A)\Omega
}
\subset
\overline{
 (\rho_a^{D_1\ld_1}\otimes\rho_b^{D_2\Lambda_2})
 (\mathcal A)\Omega
}.
\]
Consequently,
\[
\mathcal H
 =
\overline{
 (\rho_a^{D_1\ld_1}\otimes\rho_b^{D_2\Lambda_2})
 (\mathcal A)\Omega
}.
\]
\end{proofof}

\section{Fusion in Entanglement Bootstrap and DHR Theory}\label{sec:fusion}

In this section, we relate fusion in entanglement bootstrap
to fusion in DHR theory.

We begin with an outline of the argument.
The finite-volume fusion-space isomorphisms and consistency under
restriction provide, in the thermodynamic limit, an affine bijection
\[
 \iota_Y^{abc}:\mathcal D(V_{ab}^c)
 \longrightarrow\Sigma_{ab}^c(Y),
 \qquad Y=D_1^c\cap D_2^c,
\]
from density matrices on the EB fusion space $V_{ab}^c$ of
Theorem~\ref{thm:fusionspace} to states on the common exterior of
the two holes with inner charges $a,b$ and outer charge $c$
(Notation~\ref{Hello} and Lemma~\ref{turn}).

Each state $\iota_Y^{abc}(\sigma)$ is dominated by a positive
multiple of the $c$-component of
$\widetilde\varphi_{aD_1}\bowtie\widetilde\varphi_{bD_2}$.
The fixed Petz maps fill both holes and send the full merged state
$\widetilde\varphi_{aD_1}\bowtie\widetilde\varphi_{bD_2}$
to the tensor-product vector state
\[
 \varphi_{aD_1}\bowtie\varphi_{bD_2}=\langle\Omega,
 (\rho_a^{D_1\Lambda_1}\otimes\rho_b^{D_2\Lambda_2})
 (\,\cdot\,)\Omega\rangle
\]
on the full observable algebra $\mathcal A$
(Proposition~\ref{lem:merging-tensor-product}), where $\Omega$
is the reference GNS vector. Filling both holes in $\iota_Y^{abc}(\sigma)$ with the same
Petz recovery maps provides the state
$\eta_{ab}^{cD_1D_2}(\sigma)$ on $\mathcal A$ (Lemma \ref{lem:fusion-recovered-states}).
The recovered state $\eta_{ab}^{cD_1D_2}(\sigma)$ is therefore
dominated by a positive multiple of
$\varphi_{aD_1}\bowtie\varphi_{bD_2}$.
Together with the cyclicity established in
Proposition~\ref{cyclicity}, this provides a unique positive operator
$T'(\sigma)\in\operatorname{End}
(\rho_a^{D_1\Lambda_1}\otimes\rho_b^{D_2\Lambda_2})$ satisfying
\[
 \eta_{ab}^{cD_1D_2}(\sigma)(x)
 =\langle\Omega,T'(\sigma)
 (\rho_a^{D_1\Lambda_1}\otimes\rho_b^{D_2\Lambda_2})(x)
 \Omega\rangle,
 \qquad x\in\mathcal A.
\]
This map $T'$ connects fusion in entanglement bootstrap
to fusion in DHR theory.

We now turn to the details.
\begin{nota}\label{Hello}
{Consider Setting \ref{setting:merging-inputs}.}
Let $a,b,c\in\mathcal C^{\mathrm{EB}}$.
Set $Y:=D_1^c\cap D_2^c$ and 
\[
{\partial D_i:=D_i^{+25}\setminus D_i,}
\qquad i=1,2.
\]

{Define $\Sigma_{ab}^c(Y)$ to be the set of all
$\psi\in\Sigma(Y)$ such that
\[
 \psi({S_{\partial D_1}^a})
 =\psi({S_{\partial D_2}^b})=1,
\]
and
\[
 \psi\left({S_{\Gamma_N^{+25}\setminus\Gamma_N}^c}\right)=1
 \quad\text{for all sufficiently large }N.
\]
}

\end{nota}
{{
By \eqref{eq:fusion-space-isomorphism},
\eqref{kumo}, and \eqref{sora},
}}
we obtain the following.
\begin{lem}\label{turn}
There exists an affine bijection
\[
 \iota_{Y}^{abc}:\mathcal D(V_{ab}^c)
 \longrightarrow\Sigma_{ab}^c(Y)
\]
implemented by a completely positive trace-preserving map.
Furthermore,
\begin{equation}\label{eq:fusion-exterior-maximal}
 \td\varphi_{aD_1}{\bowtie}\td\varphi_{bD_2}
 =
 \sum_{c:N_{ab}^c>0}
 p_c\,
 \iota_Y^{abc}
 \left(\frac{I_{V_{ab}^c}}{N_{ab}^c}\right).
\end{equation}
 Here $p_c$ is the weight given by
\eqref{eq:abmergedweight}.
\end{lem}
\begin{proof}
{{
The first assertion follows from
\eqref{eq:fusion-space-isomorphism},
\eqref{kumo}, and \eqref{sora},
by the consistency argument in Section~ {\ref{sec:state-consistency}}:
the compatible finite-volume CPTP maps and their
inverses determine the asserted affine bijection
in the infinite-volume limit.
}}

The finite-volume decomposition in Subsection~\ref{sougen} and Lemma \ref{lem:finite-merging} provide the
maximum-entropy decomposition of the finite-volume
merged state:
\[
 \left(
 \widetilde\varphi_{aD_1}{\bowtie}
 \widetilde\varphi_{bD_2}
 \right)|_{\mathcal A_{Y_N}}
 =
 \sum_{c:N_{ab}^c>0}
 p_c\,
 \iota_{Y_N}^{abc}
 \left(\frac{I_{V_{ab}^c}}{N_{ab}^c}\right).
\]
{{
Here $p_c$ is again the weight given by
\eqref{eq:abmergedweight}, independent of $N$.
Evaluating this identity on local observables
and using \eqref{sora} proves
\eqref{eq:fusion-exterior-maximal}.
}}
\end{proof}

The main theorem is the following.
\begin{theorem}\label{thm:fusion-endomorphisms}
Consider Setting \ref{setting:merging-inputs} and {Notation}~\ref{Hello}.
Let $a,b,c\in\ceb$ with $N_{ab}^c>0$.
Let $\operatorname{End}(\rho_a^{D_1\Lambda_1}\otimes\rho_b^{D_2\Lambda_2})$
be the set of all endomorphisms of $\rho_a^{D_1\Lambda_1}\otimes\rho_b^{D_2\Lambda_2}$.
Let $Q_{ab}^c$ be the projection onto the $c$-component of 
$\rho_a^{D_1\Lambda_1}\otimes\rho_b^{D_2\Lambda_2}$, {{as defined in \eqref{qdef}}}.
Define a convex set
\[
 \mathcal S^c:=\left\{
 T\in\operatorname{End}(\rho_a^{D_1\Lambda_1}\otimes\rho_b^{D_2\Lambda_2})\mid T\ge 0,\quad
 \langle\Omega,T\Omega\rangle=1,\quad
 \operatorname{supp}T\le Q_{ab}^c\right\}.
\]
Then there is a unique affine bijection
\[
 T':\mathcal D(V_{ab}^c)\longrightarrow\mathcal S^c
\]
such that
\begin{equation}\label{eq:fusion-endomorphism-correspondence}
\begin{split}
 &\lmk\iota_Y^{abc}(\sigma)\circ 
 \lmk {\lmk R_{B_1\to B_1D_1}^{\varphi_{aD_1}}\rmk}^T\otimes
{\lmk R_{B_2\to B_2D_2}^{\varphi_{bD_2}}\rmk}^T\rmk\rmk(x)\\
& =\langle\Omega,T'(\sigma)(\rho_a^{D_1\Lambda_1}\otimes\rho_b^{D_2\Lambda_2})(x)\Omega\rangle
\end{split}
\end{equation}
for every $\sigma\in\mathcal D(V_{ab}^c)$ and $x\in\A$.
\end{theorem}
\begin{rem}\label{momiji}
Note, due to the unitary equivalence, the linear space
$\operatorname{End}(\rho_a^{D_1\Lambda_1}\otimes\rho_b^{D_2\Lambda_2})$
is isomorphic to
$\operatorname{End}(\rho_a^{\ld_0}\otimes\rho_b^{\ld_0})$
for any representations
$\rho_a^{\ld_0}$, $\rho_b^{\ld_0}$
that are equivalent to
$\rho_a^{D_1\Lambda_1}$ and
$\rho_b^{D_2\Lambda_2}$ respectively.

\end{rem}

\begin{lem}\label{lem:fusion-recovered-states}
Consider Setting \ref{setting:merging-inputs} and {Notation}~\ref{Hello}.
For $\sigma\in\mathcal D(V_{ab}^c)$, define the state
$\eta_{ab}^{cD_1D_2}(\sigma)$ on $\mathcal A$ by
\begin{equation}\label{eq:fusion-recovered-state}
 \eta_{ab}^{cD_1D_2}(\sigma)
 :=
 \iota_Y^{abc}(\sigma)\circ
 \left(
 \left(R_{\bone\to D_1\bone}^{\varphi_{aD_1}}\right)^T
 \otimes
 \left(R_{\btwo\to D_2\btwo}^{\varphi_{bD_2}}\right)^T
 \right).
\end{equation}
Then
\begin{equation}\label{eq:fusion-product-maximal}
\begin{aligned}
 &\langle\Omega,
   (\rho_a^{D_1\Lambda_1}\otimes\rho_b^{D_2\Lambda_2})
   (\,\cdot\,)\Omega\rangle
   =
 \sum_{c:N_{ab}^c>0}
 p_c\,
 \eta_{ab}^{cD_1D_2}
 \left(\frac{I_{V_{ab}^c}}{N_{ab}^c}\right).
\end{aligned}
\end{equation}
Furthermore,
$
 \eta_{ab}^{cD_1D_2}
 \left(\sigma\right)
$
is $\rho_{c}^{D_1\ld_1}$-normal for any $\sigma\in\mathcal D(V_{ab}^c)$.
\end{lem}

\begin{proof}
By Lemma~\ref{lem:hole-filling-limit},
Proposition~\ref{lem:merging-tensor-product} and Lemma \ref{turn},
we have
\[
\begin{aligned}
 &\langle\Omega,
   (\rho_a^{D_1\Lambda_1}\otimes\rho_b^{D_2\Lambda_2})
   (\,\cdot\,)\Omega\rangle\\
 &=
 (\td\varphi_{aD_1}{\bowtie}\td\varphi_{bD_2})\circ
 \left(
 \left(R_{\bone\to D_1\bone}^{\varphi_{aD_1}}\right)^T
 \otimes
 \left(R_{\btwo\to D_2\btwo}^{\varphi_{bD_2}}\right)^T
 \right)\\
&=\sum_{c:N_{ab}^c>0}
 p_c\,
 \iota_Y^{abc}
 \left(\frac{I_{V_{ab}^c}}{N_{ab}^c}\right)\circ
 \left(
 \left(R_{\bone\to D_1\bone}^{\varphi_{aD_1}}\right)^T
 \otimes
 \left(R_{\btwo\to D_2\btwo}^{\varphi_{bD_2}}\right)^T
 \right)\\
 &= \sum_{c:N_{ab}^c>0}
 p_c\,
 \eta_{ab}^{cD_1D_2}
 \left(\frac{I_{V_{ab}^c}}{N_{ab}^c}\right).
\end{aligned}
\]

By the outer-charge condition and compatibility of annular restrictions, $\eta_{ab}^{cD_1D_2}(\sigma)$ and $\varphi_{cD_1}$ agree outside a sufficiently large disk. The finite-region extension argument of Lemma~\ref{lem:state-extensions}(iii) therefore gives the asserted normality.

\end{proof}

\begin{lem}\label{lem:fusion-one-hole-support}
Let $A$ be a standard disk
with radius $s\ge 6(l_0+2)$
with $B$ a standard annulus around $A$,
with width at least $\max\{25, l_0+1\}$.
Set $C:=(AB)^c$.
Let $\ld$ be a cone with $A$ deep inside of $\ld$.

Let $a\in\ceb$. Let 
$\caH_C$ {be} a Hilbert space, $\pi_a$ an irreducible representation 
of $\caA_{C}$ on $\caH_C$
and 
$U$ be a unitary
 such that
\[
 U:\mathcal H
 \longrightarrow
 \mathcal H_A\otimes\mathcal H_B
 \otimes\mathcal H_C
\]
and
\begin{align}
\begin{split}
 &U\rho_{a}^{A\ld}(x_Ax_Bx_C)U^*=
 x_A\otimes x_B\otimes \pi_a(x_C),\\
 &\quad x_A\in \caA_{A},\;
 x_B\in \caA_{B},\; x_C\in \caA_{C}.
\end{split}
\end{align}

Then there are finite-dimensional Hilbert spaces
$\mathcal H_{B_L},\mathcal H_{B_R}$, an isometry
\[
 U_B:\mathcal H_{B_L}\otimes\mathcal H_{B_R}
 \longrightarrow\mathcal H_B,
\]
and unit vectors
\[
 \xi_{AB_L}\in\mathcal H_A\otimes\mathcal H_{B_L},
 \qquad
 \xi_{B_RC}\in
 \mathcal H_{B_R}\otimes\mathcal H_C
\]
such that
\begin{equation}\label{eq:fusion-first-schmidt-splitting}
 (1_A\otimes U_B^*\otimes1_{C})U\Omega
 =
 \xi_{AB_L}\otimes\xi_{B_RC}.
\end{equation}
Moreover, $U_BU_B^*$ is the support of
$(\varphi_{aA})_B$, and
\[
 \tr_A
 |\xi_{AB_L}\rangle\langle\xi_{AB_L}|,
 \qquad
 \tr_{C}
 |\xi_{B_RC}\rangle\langle\xi_{B_RC}|
\]
are invertible on $\mathcal H_{B_L}$ and
$\mathcal H_{B_R}$, respectively.
The density matrix of
$(\varphi_{aA})_{AB}$ is 
\[
\begin{aligned}
(1_A\otimes U_B)
 \left(
 |\xi_{AB_L}\rangle\langle\xi_{AB_L}|
 \otimes
 \tr_{C}
 |\xi_{B_RC}\rangle\langle\xi_{B_RC}|
 \right)
 (1_A\otimes U_B)^*.
\end{aligned}
\]
\end{lem}

\begin{proof}
This follows from the structure of pure quantum Markov states.
Set
\[
 \sigma_X
 :=\tr_{X^c}|U\Omega\rangle\langle U\Omega|,
\]
for $ X=A,C, AB,  AC$ etc.
Recall from Lemma \ref{swan} that
\begin{align}
\begin{split}
&\lmk \varphi_{aA}\rmk_{AC}=\lmk \varphi_{aA}\rmk_{A}\otimes \lmk \varphi_{aA}\rmk_{C}
\end{split}
\end{align}

Choose spectral decompositions on the supports:
\[
 \sigma_A
 =\sum_{i=1}^{n}\lambda_i
   |e_i^A\rangle\langle e_i^A|,
 \qquad
 \sigma_{C}
 =\sum_{j=1}^{m}\mu_j
   |f_j^{C}\rangle\langle f_j^{C}|,
\]
where $\lambda_i,\mu_j>0$.
Both ranks are finite, since $A$ and $B$ are finite.

We use the same observation as in
\cite[proof of Theorem~1.3]{CarlenLieb}.
As $U\Omega$ purifies $\sigma_A\otimes\sigma_{C}$,
there is an orthonormal family
$\{\beta_{ij}^B\}_{i,j}$ in $\mathcal H_B$ such that
\[
 U\Omega
 =
 \sum_{i=1}^{n}\sum_{j=1}^{m}
 \sqrt{\lambda_i\mu_j}\,
 e_i^A\otimes\beta_{ij}^B\otimes f_j^{C}.
\]
Take
\[
 \mathcal H_{B_L}:=\mathbb C^n,
 \qquad
 \mathcal H_{B_R}:=\mathbb C^m,
\]
with their standard orthonormal bases $\{e_i^{(n)}\}$, $\{e_j^{(m)}\}$, 
and define an isometry
$
 U_B:\mathcal H_{B_L}\otimes\mathcal H_{B_R}
 \longrightarrow\mathcal H_B,
$ by
\[
 U_B(e_i^{(n)}\otimes e_j^{(m)})
 :=\beta_{ij}^B.
\]
Then
\[
\begin{aligned}
 &(1_A\otimes U_B^*\otimes1_{C})U\Omega\\
 &\quad=
 \left(\sum_i\sqrt{\lambda_i}\,
       e_i^A\otimes e_i^{(n)}\right)
 \otimes
 \left(\sum_j\sqrt{\mu_j}\,
       e_j^{(m)}\otimes f_j^{C}\right).
\end{aligned}
\]
Define the two factors on the right-hand side as
$\xi_{AB_L}$ and $\xi_{B_RC}$, respectively.
This proves
\eqref{eq:fusion-first-schmidt-splitting}.
Note that \[
 \tr_A
 |\xi_{AB_L}\rangle\langle\xi_{AB_L}|,
 \qquad
 \tr_{C}
 |\xi_{B_RC}\rangle\langle\xi_{B_RC}|
\]
are invertible on $\mathcal H_{B_L}$ and
$\mathcal H_{B_R}$, respectively, because $\lambda_i,\mu_j>0$.

Substituting this, the density matrix of
$(\varphi_{aA})_{AB}$ is 
\[
\begin{aligned}
(1_A\otimes U_B)
 \left(
 |\xi_{AB_L}\rangle\langle\xi_{AB_L}|
 \otimes
 \tr_{C}
 |\xi_{B_RC}\rangle\langle\xi_{B_RC}|
 \right)
 (1_A\otimes U_B)^*.
\end{aligned}
\]

Clearly,  $U_BU_B^*$ is the support of
$(\varphi_{aA})_B$.
\end{proof}

\begin{lem}\label{lem:fusion-dominated-support}
Consider Setting \ref{setting:merging-inputs}
and {Notation}~\ref{Hello}.
Apply Lemma~\ref{lem:fusion-one-hole-support} to
$\varphi_{a{D_1}}$ and $\varphi_{b{D_2}}$, and
obtain $\xi_{{D_i}B_{iL}}$, $U_{B_i}$, $\caH_{B_iL}$,
{and} $\caH_{B_iR}$ for $i=1,2$
as given there.
Define projections
\[
r_{{D_i}B_i}
:=
(1_{{D_i}}\otimes U_{B_i})
\left(
 |\xi_{{D_i}B_{iL}}\rangle
 \langle\xi_{{D_i}B_{iL}}|
 \otimes1_{B_{iR}}
\right)
(1_{{D_i}}\otimes U_{B_i})^*
\in\caB(\caH_{{D_i}B_i}),
\quad i=1,2.
\]
Let $\varphi$ be a state on $\mathcal A$ satisfying
\begin{align}\label{tori}
\varphi
\leq C\,
\langle\Omega,
(\rho_a^{D_1\Lambda_1}\otimes\rho_b^{D_2\Lambda_2})
(\,\cdot\,)\Omega\rangle
\end{align}
for some finite constant $C>0$.
Let $\sigma_N^\varphi$ be the density matrix of
$\varphi\vert_{\caA_{\Gamma_N}}$.

Then, we have
\[
\varphi(r_{{D_1}{B_1}})
=
\varphi(r_{{D_2}{B_2}})
=1.
\]
Furthermore,
for every sufficiently large $N$, there is a density
matrix $\tau'_N$ on
\[
\mathcal H_{B_{1R}}\otimes
\mathcal H_{B_{2R}}\otimes
\mathcal H_{Z_N}
\]
with $Z_N:=\Gamma_N\setminus{{D_1}B_1{D_2}B_2}$
such that, after the canonical rearrangement of tensor
factors,
\begin{equation}\label{eq:fusion-dominated-density}
\begin{aligned}
&\left(
 (1_{{D_1}}\otimes U_{{B_1}}^*)
 \otimes
 (1_{{D_2}}\otimes U_{{B_2}}^*)
 \otimes1_{Z_N}
\right)
\sigma_N^\varphi\\
&\qquad\cdot
\left(
 (1_{{D_1}}\otimes U_{{B_1}})
 \otimes
 (1_{{D_2}}\otimes U_{{B_2}})
 \otimes1_{Z_N}
\right)\\
&\quad=
|\xi_{{D_1}B_{1L}}\rangle
\langle\xi_{{D_1}B_{1L}}|
\otimes
|\xi_{{D_2}B_{2L}}\rangle
\langle\xi_{{D_2}B_{2L}}|
\otimes\tau'_N.
\end{aligned}
\end{equation}
\end{lem}

\begin{proof}
Note that
\begin{align}
\begin{split}
&\left.
\langle\Omega,
(\rho_a^{D_1\Lambda_1}\otimes\rho_b^{D_2\Lambda_2})
(\,\cdot\,)\Omega\rangle
\right\vert_{\caA_{{D_1}{B_1}}}
=
\varphi_{a{D_1}}
\vert_{\caA_{{D_1}{B_1}}}\\
&\left.
\langle\Omega,
(\rho_a^{D_1\Lambda_1}\otimes\rho_b^{D_2\Lambda_2})
(\,\cdot\,)\Omega\rangle
\right\vert_{\caA_{{D_2}{B_2}}}
=
\varphi_{b{D_2}}
\vert_{\caA_{{D_2}{B_2}}}.
\end{split}
\end{align}
Their support projections are precisely
$r_{{D_1}{B_1}}$
and $r_{{D_2}{B_2}}$.
Therefore
\[
\langle\Omega,
(\rho_a^{D_1\Lambda_1}\otimes\rho_b^{D_2\Lambda_2})
(1-r_{{D_i}B_i})\Omega\rangle=0.
\]
Substituting this {into} (\ref{tori})
provides $\varphi(r_{{D_i}B_i})=1$.

The two projections $r_{D_1B_1}$ and $r_{D_2B_2}$
 have disjoint {spatial} supports and commute.
Thus, for sufficiently large $N$,
\[
\sigma_N^\varphi
=
r_{{D_1}{B_1}}
r_{{D_2}{B_2}}\,
\sigma_N^\varphi\,
r_{{D_1}{B_1}}
r_{{D_2}{B_2}}.
\]
In the coordinates given by
$U_{{B_1}}$ and
$U_{{B_2}}$,
the first two support factors are one-dimensional.
Consequently, the density matrix has the form
\eqref{eq:fusion-dominated-density}.
\end{proof}

\begin{lem}\label{lem:fusion-central-components}
Consider Setting \ref{setting:merging-inputs} and {Notation}~\ref{Hello}.
Let $Q_{ab}^c$ be the projection onto the $c$-component of 
$\rho_a^{D_1\Lambda_1}\otimes\rho_b^{D_2\Lambda_2}$, {{as defined in \eqref{qdef}}}.

Then, for every $c\in\ceb$ with $N_{ab}^c>0$ and
$x\in\mathcal A$,
\begin{equation}\label{eq:fusion-central-component}
\begin{aligned}
 &p_c\,
 \eta_{ab}^{cD_1D_2}
 \left(\frac{I_{V_{ab}^c}}{N_{ab}^c}\right)(x)\\
 &\qquad=
 \langle\Omega,
 Q_{ab}^c
 (\rho_a^{D_1\Lambda_1}\otimes\rho_b^{D_2\Lambda_2})(x)
 \Omega\rangle.
\end{aligned}
\end{equation}
If $N_{ab}^c=0$, then $Q_{ab}^c=0$.
\end{lem}

\begin{proof}
This follows by taking $c$-component of Lemma \ref{lem:fusion-recovered-states}.

\end{proof}

\begin{proofof}[Theorem \ref{thm:fusion-endomorphisms}]

\noindent{\it 1. Existence of $T'(\sigma)$.}
Let $\sigma\in\mathcal D(V_{ab}^c)$.
Since $0\leq\sigma\leq I_{V_{ab}^c}$,
Lemma~\ref{turn} implies
\[
 \iota_Y^{abc}(\sigma)
 \leq N_{ab}^c\,
 \iota_Y^{abc}\left(\frac{I_{V_{ab}^c}}{N_{ab}^c}\right)
 \leq\frac{ N_{ab}^c}{p_c}\,
 (\widetilde\varphi_{aD_1}{\bowtie}\widetilde\varphi_{bD_2}).
\]
Applying 
Lemma~\ref{lem:fusion-recovered-states}, Lemma{~\ref{lem:hole-filling-limit}},
{and} Proposition~\ref{lem:merging-tensor-product},
we obtain
\begin{equation}\label{eq:fusion-sector-domination}
 \eta_{ab}^{cD_1D_2}(\sigma)
 \leq \frac{ N_{ab}^c}{p_c}\,
 \langle\Omega,
 (\rho_a^{D_1\Lambda_1}\otimes\rho_b^{D_2\Lambda_2})
 (\,\cdot\,)\Omega\rangle.
\end{equation}
Therefore,  
we get
\[
 T'(\sigma)\in
 \operatorname{End}
 (\rho_a^{D_1\Lambda_1}\otimes\rho_b^{D_2\Lambda_2}),
 \qquad 0\leq T'(\sigma)\leq  \frac{ N_{ab}^c}{p_c}\,1,
\]
such that
\[
 \eta_{ab}^{cD_1D_2}(\sigma)(x)
 =\langle\Omega,T'(\sigma)
 (\rho_a^{D_1\Lambda_1}\otimes\rho_b^{D_2\Lambda_2})(x)
 \Omega\rangle,
 \qquad x\in\mathcal A.
\]
In particular, $\langle\Omega,T'(\sigma)\Omega\rangle=1$.

{{By Lemma~\ref{lem:fusion-recovered-states},}}
this state is $\rho_c^{D_1\Lambda_1}$-normal.
Moreover, $\Omega$ is cyclic for
$\rho_a^{D_1\Lambda_1}\otimes\rho_b^{D_2\Lambda_2}$
by Proposition~\ref{cyclicity}.
Therefore, 
\[
 (\operatorname{supp}T'(\sigma))\mathcal H=
 \overline{
 (\rho_a^{D_1\Lambda_1}\otimes\rho_b^{D_2\Lambda_2})
 (\mathcal A)T'(\sigma)^{1/2}\Omega}
 \subset Q_{ab}^c\mathcal H.
\]
Thus $\operatorname{supp}T'(\sigma)\leq Q_{ab}^c$.

\noindent{\it 2. Uniqueness of $T'(\sigma)$}

To prove uniqueness, let $T'_1,T'_2\in \mathcal S^c$ 
 {implement} the same state.
Then for every $x\in\mathcal A$,
\[
 \langle(T'_1-T'_2)\Omega,
 (\rho_a^{D_1\Lambda_1}\otimes\rho_b^{D_2\Lambda_2})(x)
 \Omega\rangle=0.
\]
Cyclicity in Proposition~\ref{cyclicity} implies $(T'_1-T'_2)=0$.
This uniqueness also shows that
$\sigma\mapsto T'(\sigma)$ is affine.

\noindent{\it 3.  Injectivity of $T'$}

Suppose that $T'(\sigma_1)=T'(\sigma_2)$.
Then
\[
 \eta_{ab}^{cD_1D_2}(\sigma_1)
 =\eta_{ab}^{cD_1D_2}(\sigma_2).
\]
Set
\[
 \widehat B_i:={D_i}B_i,\qquad Z:= \widehat B_1^c\cap\widehat B_2^c,
 \qquad
 Y_N:=Y\cap\Gamma_N,\qquad
 Z_N:=Z\cap \Gamma_N.
\]
Both recovery maps in (\ref{eq:fusion-recovered-state}) act identically on
$\mathcal A_{\widehat B_1^c\cap\widehat B_2^c}$.
Consequently,
\[
 \iota_Y^{abc}(\sigma_1)|_{\mathcal A_{Z_N}}
 =\iota_Y^{abc}(\sigma_2)|_{\mathcal A_{Z_N}}.
\]
{For all sufficiently large $N$, restriction
$\Sigma(Y_N)\to\Sigma(Z_N)$ is injective by
Lemma~\ref{lem:annular-restriction-injectivity}, applied successively
at the two holes. Thus the two states agree on $\mathcal A_{Y_N}$,
and hence on $\mathcal A_Y$.
}
By the injectivity of $\iota_Y^{abc}$ in
Lemma~\ref{turn}, we obtain $\sigma_1=\sigma_2$.

\noindent{\it 4. Surjectivity of $T'$.}

Let
$S'\in\mathcal S^c$
and define the state on $\mathcal A$
\[
 \varphi:=\langle\Omega,
 S'(\rho_a^{D_1\Lambda_1}\otimes\rho_b^{D_2\Lambda_2})
 (\,\cdot\,)\Omega\rangle.
\]
Then
\begin{equation}\label{eq:fusion-reverse-domination}
 \varphi\leq\|S'\|\,
 \langle\Omega,
 (\rho_a^{D_1\Lambda_1}\otimes\rho_b^{D_2\Lambda_2})
 (\,\cdot\,)\Omega\rangle.
\end{equation}

\smallskip
\noindent{(i)} 
We show that $\varphi|_{\mathcal A_Y}\in\Sigma_{ab}^c(Y)$.

{{Once this is shown, Lemma~\ref{turn} provides}}
a $\sigma_\varphi\in\mathcal D(V_{ab}^c)$
such that
$\iota_Y^{abc}(\sigma_\varphi)
=\varphi|_{\mathcal A_Y}$.

For each
\[
 X\in\mathcal S \quad\text{with}\quad
 X\subset \lmk {D_1}^{-2l_0}\rmk^c
 \cap \lmk {D_2}^{-2l_0}\rmk^c,
\]
the inequality (\ref{eq:fusion-reverse-domination}) implies
$\varphi(1-P_X)=0$.

{By Lemma~\ref{lem:ground-state-information-convex},
the preceding zero-energy condition implies
\[
 \varphi|_{\mathcal A_{D_f}}
 =\omega|_{\mathcal A_{D_f}}
 \qquad(f\in Y).
\]
Hence, by Notation~\ref{notation:regions},
\[
 \varphi|_{\mathcal A_Y}\in\Sigma(Y).
\]
}

Let $L_i:=\widehat B_i^{+25}\setminus \widehat B_i$ {for} $i=1,2$.
{{
The regions in Setting~\ref{setting:merging-inputs} are chosen so that 
$L_1\subset {D_1}{B_1}{G_{12}}$ and $L_2\subset {D_2}{B_2}{G'_{12}}$.
Since Proposition~\ref{lem:merging-tensor-product}
implies
}}

\begin{align}
\begin{split}
\langle\Omega,
 (\rho_a^{D_1\Lambda_1}\otimes\rho_b^{D_2\Lambda_2})
 \lmk {S_{L_1}^a}\rmk
 \Omega\rangle
 =\varphi_{aD_1}\lmk {S_{L_1}^a}\rmk=1,
\end{split}
\end{align}
the inequality (\ref{eq:fusion-reverse-domination})
implies
$\varphi(1-{S_{L_1}^a})=0$.
Similarly, we have $\varphi(1-{S_{L_2}^b})=0$.

Because
$0\leq S'\leq\|S'\|Q_{ab}^c$, by the definition of $Q_{ab}^c$,
$\varphi$ is $\rho_c^{D_1\ld_1}$-normal.
Hence $\varphi|_{\mathcal A_Y}\in\Sigma_{ab}^c(Y)$.

\noindent{(ii)} We verify the Markov property at each hole
${D_i}-B_i-F$ with $F$ a finite subset {of} $\widehat B_i^c$.
Apply Lemma~\ref{lem:fusion-dominated-support} using
\eqref{eq:fusion-reverse-domination}.
For every sufficiently large $N$, the density matrix
of $\varphi_{\Gamma_N}$ has the form
\eqref{eq:fusion-dominated-density}.
In particular,
\[
 \varphi(r_{{D_1}B_1})=\varphi(r_{{D_2}B_2})=1.
\]

Let $i\in\{1,2\}$ and let $F\subset({D_i}B_i)^c$
be any finite region.
Choose $N$ so large that ${D_1}B_1{D_2}B_2F\subset\Gamma_N$.
In the support coordinates given by
\[
 U_{B_i}:\mathcal H_{B_{iL}}\otimes\mathcal H_{B_{iR}}
 \longrightarrow\mathcal H_{B_i}
\]
of Lemma~\ref{lem:fusion-one-hole-support},
taking a partial trace of
\eqref{eq:fusion-dominated-density} implies
\begin{align}\label{enpitsu}
     \varphi_{{D_i}B_iF}
 =(1_{D_i}\otimes U_{B_i}\otimes 1_F)
 \lmk
 |\xi_{{D_i}B_{iL}}\rangle\langle\xi_{{D_i}B_{iL}}|
   \otimes\tau_{B_{iR}F}\rmk(1_{D_i}\otimes U_{B_i}^*\otimes 1_F).
\end{align}
Here $\tau_{B_{iR}F}$ is a density matrix on
$\mathcal H_{B_{iR}}\otimes\mathcal H_F$.
Set
\[
 \rho_{B_{iL}}:=\tr_{{D_i}}
 |\xi_{{D_i}B_{iL}}\rangle\langle\xi_{{D_i}B_{iL}}|,
 \qquad
 \tau_{B_{iR}}:=\tr_F\tau_{B_{iR}F}.
\]
Since $\xi_{{D_i}B_{iL}}$ is a unit vector and entropy
is invariant under isometries,
\begin{align*}
 I({D_i}:F| B_i)_\varphi
 &:=S(\varphi_{{D_i}B_i})+S(\varphi_{B_iF})
     -S(\varphi_{B_i})-S(\varphi_{{D_i}B_iF})\\
 &=S(\tau_{B_{iR}})
   +S(\rho_{B_{iL}})+S(\tau_{B_{iR}F})\\
 &\quad-S(\rho_{B_{iL}})-S(\tau_{B_{iR}})
       -S(\tau_{B_{iR}F})=0.
\end{align*}
Thus $\varphi$ is Markov across either collar,
with any finite part of the exterior on the other side.

\noindent{(iii)}
Next we show
\[
\varphi\vert_{\caA_{D_1B_1}}=\varphi_{aD_1}\vert_{\caA_{D_1B_1}},\quad
\varphi\vert_{\caA_{D_2B_2}}=\varphi_{bD_2}\vert_{\caA_{D_2B_2}}.
\]
To see this, recall from Lemma \ref{lem:fusion-one-hole-support}
that the density matrix of
$(\varphi_{aD_1})_{D_1B_1}$ is 
\begin{align}\label{pen}
(1_{D_1}\otimes U_{B_1})
 \left(
 |\xi_{D_1B_{1L}}\rangle\langle\xi_{D_1B_{1L}}|
 \otimes
 \zeta_{B_{1R}}
 \right)
 (1_{D_1}\otimes U_{B_1})^*,
\end{align}
with a density matrix 
$\zeta_{B_{1R}}$ on
$\caH_{B_{1R}}$,
and 
the density matrix of
$(\varphi_{aD_1})_{B_1}$ is 
\[
\begin{aligned}
 U_{B_1}
 \left(
\Tr_{D_1} \lmk |\xi_{D_1B_{1L}}\rangle\langle\xi_{D_1B_{1L}}|\rmk\otimes
 \zeta_{B_{1R}}
 \right)
 U_{B_1}^*.
\end{aligned}\]
On the other hand, by (\ref{enpitsu}),
the density matrix of $\varphi_{{D_1}B_1}$ is
\begin{align}\label{kamemushi}
    (1_{D_1}\otimes U_{B_1})
 \lmk
 |\xi_{{D_1}B_{1L}}\rangle\langle\xi_{{D_1}B_{1L}}|
   \otimes\tau_{B_{1R}}\rmk(1_{D_1}\otimes U_{B_1}^*),
\end{align}
and
the density matrix of $\varphi_{B_1}$ is
\begin{align*}
    U_{B_1}
 \lmk
\Tr_{D_1}\lmk|\xi_{{D_1}B_{1L}}\rangle
\langle\xi_{{D_1}B_{1L}}|\rmk
   \otimes\tau_{B_{1R}}\rmk U_{B_1}^*.
\end{align*}
Since $\varphi|_{\mathcal A_Y}\in\Sigma_{ab}^c(Y)$ by (i), injectivity of annular restriction (Lemma~\ref{lem:standard-annulus-comparison}) gives $\varphi_{B_1}=(\varphi_{aD_1})_{B_1}$. 
This implies $\zeta_{B_{1R}}=\tau_{B_{1R}}$.
Substituting this into  (\ref{pen}) and (\ref{kamemushi}),
we obtain the claim.

\noindent{(iv)}
By the Markov property in (ii) and the marginal identities in (iii),
\[
 \varphi_{\Gamma_N}
 =
 \varphi_{Y_N}\circ \left(
 {(R_{B_1\to D_1B_1}^{\varphi_{aD_1}})}^T
 \otimes {(R_{B_2\to D_2B_2}^{\varphi_{bD_2}})}^T
 \otimes\operatorname{id}_{Z_N}
 \right).
\]
Since the regions $\Gamma_N$ exhaust the lattice,
using (i),
we obtain the following equalities:
\[
\begin{aligned}
 \varphi
 &=\varphi|_{\mathcal A_Y}\circ
 \left(
 (R_{B_1\to D_1B_1}^{\varphi_{aD_1}})^T
 \otimes(R_{B_2\to D_2B_2}^{\varphi_{bD_2}})^T
 \right)\\
 &=\iota_Y^{abc}(\sigma_\varphi)\circ
 \left(
 (R_{B_1\to D_1B_1}^{\varphi_{aD_1}})^T
 \otimes(R_{B_2\to D_2B_2}^{\varphi_{bD_2}})^T
 \right)\\
 &=\eta_{ab}^{cD_1D_2}(\sigma_\varphi)\\
 &=\langle\Omega,T'(\sigma_\varphi)
 (\rho_a^{D_1\Lambda_1}\otimes\rho_b^{D_2\Lambda_2})
 (\,\cdot\,)\Omega\rangle.
\end{aligned}
\]
Comparing {this} with the definition of $\varphi$ and using
the uniqueness proved in  {\textup{(2)}}, we obtain
$S'=T'(\sigma_\varphi)$.
This proves surjectivity.

\end{proofof}

Consider the setting of Theorem~\ref{thm:fusion-endomorphisms}.
Let $\ld_0\in\caC_{(\theta_0,\varphi_0)}$ be a cone including $\ld_1$ and $\ld_2$.
Then $\rho_a:=\rho_a^{D_1\ld_1}$, $\rho_b:=\rho_b^{D_2\ld_2}$
are localized in $\ld_0$.
We regard them as simple objects of $\caC_{\omega}^{\ld_0}$.

Let $a,b,c\in\ceb$ with $N_{ab}^c>0$. The homomorphism space
\[
W:=\lmk \rho_c,\rho_a\otimes\rho_b\rmk
\]
is a Hilbert space
with inner product $S^*T=\langle S,T\rangle_W I_{\mathcal H}$.
Introduce the algebra
\[
\caM:=Q_{ab}^c \operatorname{End}(\rho_a\otimes\rho_b)Q_{ab}^c.
\]
The left action of $\caM$ on $W$ gives the canonical $*$-isomorphism
\[
 j:\mathcal M\longrightarrow\mathcal B(W),\qquad j(T)S:=TS.
\]

The convex set $\caS_c$ of
Theorem~\ref{thm:fusion-endomorphisms}
can be represented as
\[
\caS_c=\{x\in\caM_+\mid  \langle\Omega,x\Omega\rangle=1\}.
\]
Because $\Omega$ is separating for $\caM$, any element of
$\caM$ can be written as a complex linear combination of
elements in $\caS_c$.
Hence the affine bijection $T'$ given in
Theorem~\ref{thm:fusion-endomorphisms} extends to
a linear bijection
\[
\hat T': \caB(V_{ab}^c)\to \caM.
\]
In particular
$
\dim \caB(V_{ab}^c)
=\dim \caM=\dim\caB(W).
$
and we obtain
\[
N_{ab}^c=\dim\lmk \rho_c,\rho_a\otimes\rho_b\rmk.
\]

Using $\hat T'$, we construct a positive trace preserving
map $\Phi:\caB(V_{ab}^c)\to \caB(W)$:
Because $\Omega$ is separating for $\caM$,
\[
\caB(W)\ni x\mapsto  \langle\Omega,j^{-1}(x)\Omega\rangle\in\bbC
\]
is positive and faithful.
Hence there exists a positive invertible operator
$D'\in\caB(W)$ such that
\[
\tr_W x D'= \langle\Omega,j^{-1}(x)\Omega\rangle,\quad x\in \caB(W).
\]
Define $\Phi: \caB(V_{ab}^c)\to \caB(W)$ by
\[
\Phi(y):={D'}^{\frac 12} j\lmk \hat T'(y)\rmk{D'}^{\frac 12},\quad
y\in\caB(V_{ab}^c) .
\]
Then $\Phi$ is a positive trace preserving linear bijection.

\begin{cor}\label{cor:fusion-space-antiunitary}
There is 
an antiunitary operator $U:V_{ab}^c\to W$:
such that
\begin{align}\label{rep}
   \Phi(\sigma)=U\sigma U^*,\qquad \sigma\in\mathcal D(V_{ab}^c). 
\end{align}
\end{cor}

\begin{proof}
From Lemma \ref{lem:affine-density-matrices},
there exists $U:V_{ab}^c\to W$, either unitary or anti-unitary
satisfying (\ref{rep}).
We exclude the possibility of unitary.

Let $r\ge1$, $\sigma=[\sigma_{ij}]\in M_r(\mathcal B(V_{ab}^c))_+$.
For any local $A_1,\ldots,A_r\in\mathcal A$,
consider the positive element
 $[A_i^*A_j]\ge0$ in $M_{r}(\caA)$.
Pairing the finite-volume CPTP realization of Lemma~\ref{turn},
followed by the fixed Petz recoveries, with $[A_i^*A_j]\ge0$
and using Theorem~\ref{thm:fusion-endomorphisms}, we obtain
\begin{align*}
 0&\le\sum_{i,j=1}^r
 \langle\Omega,\hat T'(\sigma_{ij})\rho(A_j^*A_i)\Omega\rangle\\
 &=\sum_{i,j=1}^r
 \langle\rho(A_j)\Omega,\hat T'(\sigma_{ij})\rho(A_i)\Omega\rangle,
\end{align*}
with $\rho:=\rho_a\otimes\rho_b$.
By the cyclicity of $\Omega$ for $\rho(\mathcal A)$ and norm
density of local observables, this implies 
$(\hat T'\otimes \id_r)(\sigma^T)\ge0$, with $T$ transpose.
Applying $j$ and sandwiching with $D'^{1/2}$ gives
$(\Phi\otimes \id_r)(\sigma^T)\ge0$.
If $U$ were unitary, this would imply $\sigma^T\ge0$
for every positive block matrix $\sigma$.
For $\dim V_{ab}^c\ge2$, take $r=2$ and
$\sigma_{ij}=|v_i\rangle\langle v_j|$ with $v_1,v_2\in V_{ab}^c$
orthonormal: $[\sigma_{ij}]\ge0$, whereas $[\sigma_{ji}]$
has eigenvector $(v_2,-v_1)$ with eigenvalue $-1$.
Thus $U$ is antiunitary in this case; in dimension one an antiunitary
implementer may also be chosen.
\end{proof}
Combining this and Lemma \ref{lem:fusion-central-components},
we see that 
$ D'=\frac{p_c}{N_{ab}^c}$.
In fact, from Lemma \ref{lem:fusion-central-components}
and Theorem~\ref{thm:fusion-endomorphisms}, we have
\[
Q_{ab}^c=p_cT'\lmk \frac{\unit_{V_{ab}^c}}{N_{ab}^c}\rmk.
\]
Applying $j$, this implies
\[
\unit_W=p_c{D'}^{-\frac 12} \Phi\lmk \frac{\unit_{V_{ab}^c}}{N_{ab}^c}\rmk {D'}^{-\frac 12}
=\frac{p_c}{N_{ab}^c}{D'}^{-1}.
\]
In the last equality, we used Corollary \ref{cor:fusion-space-antiunitary}.
Hence we obtain $ D'=\frac{p_c}{N_{ab}^c}$.

Substituting this, we conclude
\[
U\sigma U^*=\frac{p_c}{N_{ab}^c} j\lmk \hat T'(\sigma)\rmk,\quad
 \sigma\in \caD(V_{ab}^c) .
\]

\section*{Acknowledgements}

Kohtaro Kato acknowledges support from
MEXT/JSPS Grant-in-Aid for Transformative Research Areas~(B)
under Grant No.~JP24H00829, JSPS KAKENHI Grant No. JP23K17668,
and the MEXT Quantum Leap Flagship Program (MEXT Q-LEAP)
under Grant No.~JPMXS0120319794.

Yoshiko Ogata acknowledges support from
JSPS KAKENHI under Grant Nos.~JP19K03534 and JP22H01127,
and JST CREST under Grant No.~JPMJCR19T2.
\section*{Data availability}
No datasets were generated or analysed during the current study.
\section*{AI use}
All new mathematical content in the main text was developed by the authors and provided in the form of handwritten notes and handwritten LaTeX source files. AI tools were used to transcribe the handwritten notes into LaTeX, combine the LaTeX files, and assist with editorial tasks, including English-language editing and the preparation of figures in LaTeX.
The detailed lattice-geometric setting in the appendix was generated
by AI and checked by the authors. It was not treated in detail
in~\cite{SKK}.

} %

\appendix
\section{Basic facts from entanglement bootstrap}\label{basiceb}
We give lattice versions of the entanglement-bootstrap lemmas from \cite{SKK} used in the main text. The basic arguments already appear in \cite{SKK}, but the proofs do not explicitly take the lattice structure into account.
For completeness, we include the details in this appendix.

\subsection{Properties of disks}\label{discsec}
We first prove the face-ordering lemma used below. 
\begin{lemma}[Ordering the faces of a disk]\label{lem:disk-face-ordering}
Let $K\Subset\mathbb L$ be a disk containing $m$ faces.
There is an enumeration
\[
 K=\{f_1,\ldots,f_m\}
\]
such that every region
\[
 R_j:=\{f_1,\ldots,f_j\},\qquad 1\le j\le m,
\]
is a disk. Moreover, for every $1\le j<m$, the sets
\[
 B_j:=\mathcal N(f_{j+1})\cap R_j,
 \qquad C_j:=\mathcal N(f_{j+1})\setminus R_j
\]
form two nonempty consecutive arcs in the cyclic ordering of the six
neighbors of $f_{j+1}$. More generally, for any disk $B\subset K$, there is an ordering
$K\setminus B=\{g_1,\ldots,g_n\}$ such that
$K_j:=B\cup\{g_1,\ldots,g_j\}$ is a disk for $0\le j\le n$ and
\[
  \varnothing\ne \cN(g_j)\cap K_{j-1}\subsetneq \cN(g_j),
  \qquad
  \cN(g_j)\cap K_{j-1}\text{ is connected}
  \quad (1\le j\le n).
\]
\end{lemma}
\begin{proof}
We first show that, for any disks $B\subsetneq K$, there is a boundary face $f\in K\setminus B$ whose removal leaves a disk. Choose a boundary face $f$ of $K$ outside $B$, which exists since $B$ has no holes.

If $S\setminus\{f\}$ is connected, its face complement is also connected: that complement is obtained from the connected complement of $K$ by adjoining the adjacent face $f$. Thus $S\setminus\{f\}$ is a disk.

Suppose instead that removing $f$ disconnects $K$, and choose a connected component $T$ of $K\setminus\{f\}$ disjoint from $B$. Every adjacency between $T$ and $K\setminus T$ passes through $f$. Choose a curve from an interior point of $T$ to infinity which avoids the closed hexagon $f$, avoids lattice vertices, and crosses lattice edges transversely. Such a curve must leave $T$ directly into the complement of $K$. Hence $T$ contains a boundary face $g$ of $K$.

We now replace $f$ by $g$. If removing $g$ preserves connectedness, we are done. Otherwise, a component of $K\setminus\{g\}$ disjoint from $B$ is contained in $T\setminus\{g\}$. Repeating the preceding argument strictly decreases the number of faces in the selected component. The process must terminate, yielding a boundary face outside $B$ whose removal leaves a disk.

Apply this removal procedure while keeping $B$ fixed, and reverse the removal order. This provides the required ordering of $K\setminus B$. Taking $B$ to be a single face proves the first assertion.

For each added face $f$, let $S$ and $S'$ denote the regions before and after its addition. Since $S$, $\{f\}$, and $S'$ are disks, additivity of the Euler characteristic yields
\[
\chi(|S|\cap|f|)=\chi(|S|)+\chi(|f|)-\chi(|S'|)=1.
\]
The intersection is a union of edges of the added hexagon, with no isolated vertices. Such a union has Euler characteristic one precisely when it is a single nonempty proper arc. Its complementary edges form another nonempty arc. For $(S,f,S')=(R_j,f_{j+1},R_{j+1})$, these arcs correspond to $B_j$ and $C_j$, respectively. Taking $(S,f,S')=(K_{j-1},g_j,K_j)$ shows that $\mathcal N(g_j)\cap K_{j-1}$ is a nonempty proper arc of $\mathcal N(g_j)$.
\end{proof}

\begin{proofof}[Lemma \ref{lem:finite-disk-local-uniqueness}]
Choose the enumeration and regions $R_j,B_j,C_j$ from
Lemma~\ref{lem:disk-face-ordering}. 
We prove by induction that every state $\tau$ on $\mathcal A_{K^+}$
satisfying the stated local constraints has
\[
 \tau_{R_j}=\omega_{R_j}.
\]
The assertion for $j=1$ follows by restricting the local constraint on
$D_{f_1}$.

Suppose that it holds for $j$, and set $f:=f_{j+1}$.
Since $B_j\subset R_j$, strong subadditivity leads
\[
 S(f| R_j)_\tau\le S(f| B_j)_\tau.
\]
The regions $\{f\},R_j,C_j$ are pairwise disjoint, so weak
monotonicity implies
\[
 S(f| R_j)_\tau\ge -S(f| C_j)_\tau.
\]
All regions appearing here are contained in $K^+$.

The sets $B_j,C_j$ form the partition into two nonempty consecutive arcs required by A1.
Since $\tau_{D_f}=\omega_{D_f}$, we have
\[
 S(f| B_j)_\tau+S(f| C_j)_\tau
 =S(f| B_j)_\omega+S(f| C_j)_\omega=0.
\]
Consequently,
\[
 S(f| R_j)_\tau=S(f| B_j)_\omega.
\]
Using the induction hypothesis, we obtain
\[
 S(\tau_{R_{j+1}})
 =S(\omega_{R_j})+S(f| B_j)_\omega
 =S(\omega_{R_{j+1}}).
\]

Now fix an admissible state $\rho$, and define the state
\[
 \mu:=\tfrac12\bigl(\rho+\omega_{K^+}\bigr)
\]
on $\mathcal A_{K^+}$. Both $\rho$ and $\mu$ satisfy the same local
constraints. The entropy equality just proved therefore yields
\[
 S\!\left(\frac{\rho_{R_{j+1}}+\omega_{R_{j+1}}}{2}\right)
 =\frac{S(\rho_{R_{j+1}})+S(\omega_{R_{j+1}})}{2}.
\]
Strict concavity of the von Neumann entropy implies
\[
 \rho_{R_{j+1}}=\omega_{R_{j+1}}.
\]
This completes the induction and proves the assertion.
\end{proofof}

\subsection{Isomorphism theorems}\label{isomsec}

For a finite lattice region $F$, let $J(F)$ consist of the states
$\rho\in\mathcal D_F$ satisfying
\begin{align}
\rho_{D_g\cap F}&=\omega_{D_g\cap F} \quad(g\in F),\label{eq:J-local}\\
 I(g:F\setminus D_g| V)_\rho&=0\quad
 \left(g\in F,\ \varnothing\ne V\subsetneq \cN(g),\
       V\subset F\text{ a connected arc}\right).
 \label{eq:J-Markov}
\end{align}
This auxiliary set is not the intrinsic set of
\cite[Definition C.1]{SKK}.
In the recovery formulas below, $R^\eta_{U\to UV}:\caA_U\to\caA_{UV}$
is the full-matrix CPTP extension of the Petz map determined by
$\eta_{UV}$ in Eq.~\eqref{eq:Petzrecoverymap}.

\begin{lem}[Intrinsic constraints and exterior witnesses]
\label{lem:information-convex-bridge}
For every finite regular $F$, $\Sigma(F)\subset J(F)$.
Restriction preserves $J$, and every state of $J(F)$ has marginal
$\omega_K$ on every regular disk $K\subset F$.
For every shell domain $F$,
\[
 J(F)=\Sigma(F).
\]
\end{lem}
\begin{proof}
A1, strong subadditivity, and weak monotonicity provide $\Sigma(F)\subset J(F)$ and $\omega_F\in J(F)$. Restriction preserves $J$ by the consistency of the reduced states and the data processing inequality. Lemma~\ref{lem:disk-face-ordering} and Petz recovery imply $J(K)=\{\omega_K\}$ for every disk $K$. Applying the local Markov conditions $I(g:ABC\setminus D_g| V)_\rho=0$ along the face ordering from $B$ to $BC$ and using the chain rule imply $I(A:C| B)_\rho=0$ whenever $\rho\in J(ABC)$, $B$ and $BC$ are disks, and $d(A,C)>1$. 

Consider a shell addition $X'=X\cup\{d\}$, and set
\begin{equation}
    C:=X\cap K_d(3),\qquad
    B:=X\cap\bigl(K_d(4)\setminus K_d(3)\bigr),\qquad
    A:=X\setminus K_d(4).
\end{equation}
By definition, $B$ and $BC$ are disks, and $V:={\mathcal N}(d)\cap X$ is a nonempty proper consecutive arc. Moreover, $d(A,C)>1$ and $d(AB,d)>3$.

For $\rho\in J(X)$, disk uniqueness, $I(d:BC\setminus V| V)_\omega=0$, and the preceding Markov condition imply, as in the proof of \cite[Proposition~C.2]{SKK},
\begin{equation}
    \tau
    :=
    \mathcal R^{\omega_{dV}}_{V\to dV}(\rho_X)
    =
    \mathcal R^{\rho_{AB}}_{B\to AB}(\omega_{BCd}).
\end{equation}
Taking the partial trace over $d$ provides $\tau_X=\rho_X$, and the same locality and data-processing argument shows that $\tau\in J(X')$.

Iterating this construction to complete the current layer and then add one layer along every boundary provides, for every $\rho\in J(F)$, an extension $\tau\in J(G)$ of $\rho$, where $G\supseteq F^+$. For every $g\in F$, condition (A.1) provides $\tau_{D_g}=\omega_{D_g}$. Thus $\tau_{F^+}$ is a defining extension for $\rho$, proving $J(F)\subseteq\Sigma(F)$.
\end{proof}

\begin{proofof}[Theorem \ref{thm:information-convex-isomorphism}]
For each shell addition $X'=X\cup\{d\}$, the construction in the proof of Lemma~\ref{lem:information-convex-bridge} provides a fixed reference Petz map from $\Sigma(X)$ to $\Sigma(X')$. The local Markov condition at $d$ makes this map inverse to restriction and provides the state-independent entropy increment $S(\omega_{dV})-S(\omega_V)$, where $V:=\cN(d)\cap X$. This map is CPTP and its entropy increment is independent of the input state. Compose these maps along the chosen path and their inverses in the reverse order, as in \cite[Theorem~3.10]{SKK}.
The maps and their inverses are contractive, so they preserve trace
distance. Their entropy increments are state-independent, so they
preserve entropy differences.
\end{proofof}

\begin{lemma}[Injectivity of restriction]
\label{lem:annular-restriction-injectivity}
Let $X\subset Y$ be an essential inclusion of regular annuli.
Then restriction $J(Y)\to J(X)$ has a fixed CPTP left
inverse {on its image and satisfies}
\[
 S(\rho_Y)-S(\rho_X)=S(\omega_Y)-S(\omega_X),
\]
for every $\rho_Y\in J(Y)$, where
$\rho_X=\tr_{Y\setminus X}\rho_Y$.
{In particular these assertions hold for $\Sigma(Y)\to\Sigma(X)$;
restriction} is injective and preserves trace distance.

The same conclusions hold for the following pairs of finite
regular regions $X\subset Y$.
Let $K$ be a finite disk and let $H_1,\ldots,H_m\subset K$,
$m\ge1$, be pairwise nonadjacent disk holes, each nonadjacent
to $K^c$.

\begin{enumerate}
\item Enlarge each hole $H_i$ to a disk $H_i'\subset K$,
with the enlarged holes pairwise nonadjacent and nonadjacent
to $K^c$, and set
\[
 Y=K\setminus\bigcup_{i=1}^{m}H_i,
 \qquad
 X=K\setminus\bigcup_{i=1}^{m}H_i'.
\]

\item Choose a disk $K'\subset K$ containing all the holes,
with each $H_i$ nonadjacent to $(K')^c$, and set
\[
 Y=K\setminus\bigcup_{i=1}^{m}H_i,
 \qquad
 X=K'\setminus\bigcup_{i=1}^{m}H_i.
\]
\end{enumerate}
No surjectivity is asserted here.

\end{lemma}
\begin{proof}
Apply A1 and the Petz recovery criterion successively along the
{relative} boundary face orderings from Lemma~\ref{lem:disk-face-ordering}.
The constraints defining J are preserved under restriction, including
at intermediate regions. Reversing
these removals constructs the common CPTP left inverse; the entropy
increments are fixed by the reference state. Contractivity of the restriction map and its CPTP left inverse implies that restriction preserves trace distance. The same argument applies to the separated holes and outer hull.
\end{proof}
\begin{lem}
\label{lem:intrinsic-sandwich}
 {Let $P\subset E\subset B$ be finite regular regions. Suppose
$J(P)=\Sigma(P)$ and $J(B)=\Sigma(B)$, and that restriction
$J(B)\to J(P)$ is a bijection with a CPTP inverse $R$.
Suppose also that the restriction $J(E)\to J(P)$ is injective.
Then $J(E)=\Sigma(E)$, and both intermediate restrictions are
bijections with CPTP inverses.
If both inclusions $P\subset E$ and $E\subset B$
satisfy the hypotheses of
Lemma~\ref{lem:annular-restriction-injectivity},
the corresponding restriction maps preserve entropy differences.
}

\end{lem}
\begin{proof}
Recover a state of $J(E)$ from its
$P$ marginal using the stated inverse $R:J(P)\to J(B)$.
Injectivity of restriction $J(E)\to J(P)$ identifies the recovered
$E$ marginal with the original state. The recovered state on $B$ has an extension to $B^+$ satisfying the
defining marginal constraints. Restricting this extension to $E^+$ proves $J(E)=\Sigma(E)$. Factoring the bijection $J(B) \rightarrow J(P)$ through $J(E)$ gives the two inverse maps and shows that both restrictions preserve entropy differences.
\end{proof}

\begin{proofof}[Lemma \ref{lem:standard-annulus-comparison}]

For an essential inclusion $X\subset Y$, write
\[
r_{Y\to X}:\Sigma(Y)\longrightarrow\Sigma(X)
\]
for the partial trace.

Fix a graph annulus $A$.
Choose a graph annulus $Q$ concentric with $A_0$ whose
inner disk contains the outer disks of both $A_0$ and $A$.
Choose its width sufficiently large that $Q$ essentially
contains a graph annulus concentric with $A$.
By enlarging only the outer radii of $A_0$ and $A$,
obtain graph annuli $B_0$ and $B$, respectively, containing $Q$.
Thus
\begin{align}\label{denwa}
    A_0\subset B_0\supset Q\subset B\supset A,
\end{align}
with all four inclusions essential.

The restrictions
\[
r_{B_0\to A_0},\qquad r_{B_0\to Q},\qquad r_{B\to A}
\]
are concentric restriction bijections by
Corollary~\ref{nesti}.
Moreover, there is a graph annulus $P$ concentric with $B$ such that $P \subset Q \subset B$ and both inclusions are essential. By Lemma~\ref{lem:annular-restriction-injectivity}, restriction $J(Q)\to J(P)$ is injective.
Lemma~\ref{lem:intrinsic-sandwich} therefore makes
$r_{B\to Q}$ a bijection as well.
All four bijections have CPTP inverses.

Define
\[
\Phi_A
:=
r_{B\to A}\circ
(r_{B\to Q})^{-1}\circ
r_{B_0\to Q}\circ
(r_{B_0\to A_0})^{-1}
:\Sigma(A_0)\longrightarrow\Sigma(A).
\]
The shell isomorphisms and
Lemma~\ref{lem:annular-restriction-injectivity}
show that this affine bijection and its inverse are
restrictions of CPTP maps, preserve the reference states,
and preserve entropy differences.

Enlarging the auxiliary outer radii leaves $\Phi_A$
unchanged, by factorization of partial traces and
cancellation of the added restriction bijections.
Any two choices of $Q$ are essentially contained in a
common graph annulus concentric with $A_0$.
After enlarging the outer radii, factoring through this
common annulus and cancelling its restriction bijections
proves independence of the choices.
Taking $A=A_0$ and $B=B_0$ gives
$\Phi_{A_0}=\operatorname{id}$.

For an essential inclusion $A\subset A'$, choose the same
$Q$ and $B_0$, and outer enlargements $B\subset B'$
of $A$ and $A'$, respectively.
The same factorization implies
\[
r_{A'\to A}\circ\Phi_{A'}=\Phi_A.
\]

To prove uniqueness, let $(\Psi_A)_A$ be another family
satisfying these conditions, and fix a graph annulus $A$. Consider the diagram  (\ref{denwa}).
By
Lemma~\ref{lem:annular-restriction-injectivity},
each restriction in this diagram is injective.
By normalization and compatibility, 
\[
r_{B_0\to A_0}\circ\Phi_{B_0}
=\operatorname{id}
=r_{B_0\to A_0}\circ\Psi_{B_0}.
\]
Injectivity therefore implies
$\Phi_{B_0}=\Psi_{B_0}$, and restriction to $Q$ provides
$\Phi_Q=\Psi_Q$.
Consequently,
\[
r_{B\to Q}\circ\Phi_B
=\Phi_Q
=\Psi_Q
=r_{B\to Q}\circ\Psi_B.
\]
Injectivity again implies $\Phi_B=\Psi_B$.
Restricting to $A$ yields $\Phi_A=\Psi_A$.
Since $A$ was arbitrary, the two families coincide.

\end{proofof}
\subsection{Charges for general regular annuli}

\begin{lem}[Charge states from a reference annulus]
\label{lem:detector-annuli}
Fix a graph annulus
$A_0=A_{f_0}(s_0,t_0)$.
By Theorem~\ref{thm:annlusconvex}, its information convex
$\Sigma(A_0)$
has a finite set
$\mathcal C^{\mathrm{EB}}$,
labelling the extremal points of $\Sigma(A_0)$,
with $1\in\ceb$ corresponding to $\omega_{A_0}$.
For each graph annulus $B=A_f(s,T)$, let
\[
\Phi_B:\Sigma(A_0)\longrightarrow\Sigma(B)
\]
be the canonical isomorphism of
Lemma~\ref{lem:standard-annulus-comparison}.

Let $A=K\setminus H$ be a finite regular annulus,
where $K,H$ are disks and
\[
K_f(s)\subset H
\]
for some face $f$ and integer $s\ge19$.
Choose $T$ sufficiently large that
\[
K\subset K_f(T)
\]
 and $T-s\ge25$.
Then the inclusion  $A\subset B:=A_f(s,T)$ is essential.
Define
\[
\sigma_A^a
:=
\tr_{B\setminus A}
\bigl(\Phi_B(\sigma_{A_0}^a)\bigr),
\qquad a\in\mathcal C^{\mathrm{EB}}.
\]
These states are independent of the choices of $f,s,T$.
They satisfy
\[
\sigma_A^a\in\Sigma(A),\qquad
\sigma_A^a\sigma_A^b=0\quad(a\ne b),\qquad
\sigma_A^1=\omega_A,
\]
and
\[
S(\sigma_A^a)-S(\sigma_A^b)
=
S(\sigma_{A_0}^a)-S(\sigma_{A_0}^b).
\]
For any essential inclusion $A\subset A'$ between
annuli of this class,
\begin{align}\label{eq:charge-restriction}
\tr_{A'\setminus A}\sigma_{A'}^a
=\sigma_A^a.
\end{align}
\end{lem}
\begin{proof}
Since $\Phi_B(\sigma_{A_0}^a)\in\Sigma(B)$, its marginal $\sigma_A^a$ belongs to  $\Sigma(A)$.
By Lemmas~\ref{lem:standard-annulus-comparison}
and~\ref{lem:annular-restriction-injectivity},
the comparison followed by restriction preserves
trace distance and entropy differences.
For distinct labels $a,b\in\mathcal C^{\mathrm{EB}}$,
the states $\sigma_{A_0}^a$ and $\sigma_{A_0}^b$ on the
reference annulus $A_0$ have orthogonal supports.
Since comparison and restriction preserve trace distance,
\[
\|\sigma_A^a-\sigma_A^b\|_1
=
\|\sigma_{A_0}^a-\sigma_{A_0}^b\|_1
=2,
\]
where $\|\cdot\|_1$ denotes the trace norm.
Two density matrices have trace-norm distance $2$
if and only if their supports are orthogonal.
Hence
\[
\sigma_A^a\sigma_A^b=0
\qquad(a\ne b).
\]
Thus the states $\sigma_A^a$ have mutually orthogonal
supports and the stated entropy differences.
Both maps preserve the reference states, so
$\sigma_A^1=\omega_A$.

Now we prove the independence of $\tr_{B\setminus A}
\bigl(\Phi_B(\sigma_{A_0}^a)\bigr)$ on $B$.
Fix two choices $B_1,B_2$.
Choose a graph disk $O$ containing their outer disks,
large enough that $W:=O\setminus H$ is regular and
essentially contains a graph annulus $Q$.
For $i=1,2$, let $\widetilde B_i$ be obtained from $B_i$
by enlarging only its outer radius so that its outer
disk contains $O$.
Thus
\[
A\subset B_i\subset\widetilde B_i,
\qquad
Q\subset W\subset\widetilde B_i,
\]
with all annular inclusions essential.

By Lemma~\ref{lem:standard-annulus-comparison}, 
\[
\tr_{\widetilde B_i\setminus A}
\circ\Phi_{\widetilde B_i}
=
\tr_{B_i\setminus A}\circ\Phi_{B_i},
\qquad
\tr_{\widetilde B_i\setminus Q}
\circ\Phi_{\widetilde B_i}
=\Phi_Q.
\]
From the second equality, for any fixed state $\rho\in \Sigma(A_0)$,
the two induced states $(\tr_{\widetilde B_i\setminus W}
\circ\Phi_{\widetilde B_i})(\rho)$
 on $W$ have the same restriction
to $Q$.
By Lemma~\ref{lem:annular-restriction-injectivity},
restriction $\Sigma(W)\to\Sigma(Q)$ is injective,
so these states $(\tr_{\widetilde B_i\setminus W}
\circ\Phi_{\widetilde B_i})(\rho)$ coincide.
Restricting to $A$ and using the first identity proves
independence of the original choices $B_1,B_2$.

Compatibility under essential annular restriction
follows immediately from the definition.

\end{proof}

\subsection{Fusion}
\label{app:two-hole-extension}

We extend Theorem~\ref{thm:fusionspace} to the regular two-hole disks specified in
Definition~\ref{def:two-hole-sandwich}. 
\begin{defn}
\label{def:two-hole-sandwich}
Let
\[
Y=K\setminus(H_1\cup H_2)
\]
be a finite regular two-hole disk, with outer disk $K$ and ordered disk holes $H_1, H_2$.

Suppose there exist two-hole shell domains
\[
P=K^P\setminus(H_1^P\cup H_2^P),
\qquad
B=K^B\setminus(H_1^B\cup H_2^B),
\]
each satisfying
Definition~\ref{def:two-hole-shell-family}, such that
\[
K^P\subseteq K\subseteq K^B,
\qquad
H_i^B\subseteq H_i\subseteq H_i^P
\quad(i=1,2).
\]
Thus
\[
P\subseteq Y\subseteq B,
\]
with corresponding boundary components.
Assume that $P \subseteq Y \subseteq B$ satisfies the hypotheses of Lemma~\ref{lem:intrinsic-sandwich},
including those for preservation of entropy differences. 

Choose the three detector annuli $A_1,A_2,A_3$ of $P$
as in Definition~\ref{def:two-hole-shell-family}.
They are contained in $Y$.
For $a,b,c\in\mathcal C^{\mathrm{EB}}$, define
\[
\begin{aligned}
\Sigma_{ab}^c(Y;A_1,A_2,A_3)
:=\{\rho_Y\in\Sigma(Y):\;&
(\rho_Y)_{A_1}=\sigma_{A_1}^a,\quad
(\rho_Y)_{A_2}=\sigma_{A_2}^b,\\
&(\rho_Y)_{A_3}=\sigma_{A_3}^c\}.
\end{aligned}
\]
\end{defn}

\begin{lem}[Consistency of charge conditions]
\label{lem:two-hole-sector-independence}
Let $Y$ satisfy Definition~\ref{def:two-hole-sandwich},
and let $(A_1,A_2,A_3)$ and $(A_1',A_2',A_3')$
be two choices of detector annuli allowed in that
definition, possibly arising from different inner
shell domains.
The indices $1,2$ correspond to the ordered holes,
and $3$ to the outer boundary.

For every state $\rho_Y\in\Sigma(Y)$, every charge
label $x\in\mathcal C^{\mathrm{EB}}$, and $i=1,2,3$,
\[
(\rho_Y)_{A_i}=\sigma_{A_i}^{x}
\quad\Longleftrightarrow\quad
(\rho_Y)_{A_i'}=\sigma_{A_i'}^{x}.
\]
Consequently, the fixed-charge sector $\Sigma_{ab}^c(Y;A_1,A_2,A_3)$ is independent
of the allowed detector choices and is denoted by
$\Sigma_{ab}^c(Y)$, where $a,b$ are the charges at
the ordered holes and $c$ is the outer charge.
\end{lem}

\begin{proof}
We fix an admissible outer shell domain $B$ and let
$C_1,C_2,C_3$ be its width-$25$ boundary collars.
By Lemma~\ref{lem:intrinsic-sandwich}, each
$\rho_Y\in\Sigma(Y)$ extends to a state
$\rho_B\in\Sigma(B)$.

For any allowed detector $A_i$, form an annulus $D_i$
by replacing its hole with the corresponding hole
of $B$ when $i=1,2$, and by replacing its outer disk
with the outer disk of $B$ when $i=3$.
The disk inclusions and width bounds make $D_i$
a one-hole shell domain satisfying
\[
A_i,C_i\subseteq D_i\subseteq B,
\]
with $A_i\subseteq D_i$ and $C_i\subseteq D_i$
essential.

Compatibility of the canonical states and injectivity
of annular restriction, by
Lemmas~\ref{lem:detector-annuli}
and~\ref{lem:annular-restriction-injectivity}, imply
\[
(\rho_Y)_{A_i}=\sigma_{A_i}^{x}
\quad\Longleftrightarrow\quad
(\rho_B)_{D_i}=\sigma_{D_i}^{x}
\quad\Longleftrightarrow\quad
(\rho_B)_{C_i}=\sigma_{C_i}^{x}.
\]
Applying the same argument to $A_i'$ proves the
equivalence.
Applying it at all three boundaries proves the
independence of the fixed-charge sector.
\end{proof}

\begin{rem}\label{lem:two-hole-sector-isomorphism}
In the setting of Definition~\ref{def:two-hole-sandwich},
the conclusions of Theorem~\ref{thm:fusionspace}
for the inner shell domain $P$ extend to the
intermediate region $Y$, with the same fusion
multiplicities, by Lemma~\ref{lem:intrinsic-sandwich}.
\end{rem}

\subsection{Proof and extension of the merging proposition}
\label{app:merging-entropy}

{\ignorespaces

\begin{proofof}[Proposition~\ref{prop:two-annulus-merging}]
Disk uniqueness gives agreement on the overlaps. Applying A1
along the relative disk orderings shows that the four conditional
mutual informations vanish. 
The separation hypothesis then allows the two-recovery
argument of \cite[Propositions~C.2 and C.5]{SKK}, showing
that the merger satisfies the intrinsic local constraints.
Since $Y$ is a shell domain, the merged state belongs to $\Sigma(Y)$ by
Lemma~\ref{lem:information-convex-bridge}.  
The shell comparisons and the collar identities established
in the proof of Theorem~\ref{thm:fusionspace} allow us to apply
the sector-entropy argument of \cite[Lemma~4.8]{SKK}.
The entropy-maximization argument of
\cite[Section~4.3, Eqs.~(44)--(47)]{SKK}
then provides the decomposition, weights, and sum identity.
The maximizer is unique by strict concavity of entropy. 
The expression for each sector maximizer follows from
entropy-difference preservation in
Theorem~\ref{thm:fusionspace}.
\end{proofof}

\begin{cor}[Extension to the regions of Definition~\ref{def:two-hole-sandwich} ]
\label{cor:two-annulus-merging-sandwich}
Let $Y$ be a finite regular two-hole region satisfying
Definition~\ref{def:two-hole-sandwich}.
Using the detector annuli chosen in that definition,
assume the annular-partition, detector-containment,
and separation conditions of
Setting~\ref{setting:two-annulus-merging}.

Then all conclusions of
Proposition~\ref{prop:two-annulus-merging}
hold for the canonical input states and their merger
defined as in Setting~\ref{setting:two-annulus-merging}.
\end{cor}

\begin{proof}
By Lemma~\ref{lem:intrinsic-sandwich},
the intrinsic local constraints characterize the
information convex $\Sigma(Y)$.
The argument using the two recovery maps in the proof of Proposition~\ref{prop:two-annulus-merging} therefore shows that the merged state belongs to $\Sigma(Y)$. 
The sector comparison in Remark~\ref{lem:two-hole-sector-isomorphism}
transfers the fusion-space structure and sector-entropy
formula from the inner shell domain to $Y$.
The remaining entropy-maximization argument applies unchanged.
\end{proof}

 }

\section{Lattice-cone geometry}\label{app:good-lattice-cones}
We prove Lemma~\ref{oka}.

\begin{proofof}[Lemma~\ref{oka}]
We prove (1). The lattice-cone assertion follows from Definition \ref{def:haag-lattice-cone}. Fix $j\ge0$, and write $X_n=X\cap\Gamma_n$ and $\partial\Gamma_n=\Gamma_n\setminus\Gamma_{n-1}$.

Every point of $|L|$ can be joined within its tile to a point of $C$. Since $C\subset|L|$ is convex, $L$ is connected.
Hence $L^{+j}$ is also connected and unbounded.

Consider parallel straight lines of tiles, with consecutive tiles sharing an edge. By convexity of $C$, the tiles of $L$
on each line form an interval, possibly empty. When one layer is added, the enlarged original interval,
if nonempty, overlaps the intervals of neighbours from adjacent lines, by connectedness. At the step from $L^{+k}$ to $L^{+(k+1)}$, if a line contains no tiles of $L^{+k}$, at most one adjacent line contains such tiles, by connectedness.
Thus the same property holds for $L^{+j}$. Every tile outside $L^{+j}$ can therefore follow its line
indefinitely without entering $L^{+j}$, so every component of $(L^{+j})^c$ is unbounded.

For fixed $j$, sufficiently far from the apex of $C$, thickening by $j$ layers has the same effect as shifting
each boundary line of $C$ outward. The exceptional region near the apex is bounded for fixed $j$.
Thus, on every sufficiently large layer $\partial\Gamma_n$, the tiles change between $L^{+j}$ and its complement
exactly twice, once near each shifted boundary half-line. Both portions are therefore nonempty and connected. Choose $\ell_C(j)\ge1$ so that this holds for every $n\ge\ell_C(j)$. Since every component of $(L^{+j})^c$ intersects every
sufficiently large layer, $(L^{+j})^c$ is connected. 

Fix $X\in\{L^{+j},(L^{+j})^c\}$ and $N>l\ge\ell_C(j)$. Connectedness and unboundedness of $X$ imply that every
component of $X_l$ intersects the connected set $X\cap\partial\Gamma_l$, so $X_l$ is connected.
They also ensure an edge between $X\cap\partial\Gamma_n$ and $X\cap\partial\Gamma_{n+1}$. Consequently,
\begin{equation*}
X_N\setminus X_l
=\bigcup_{n=l+1}^{N}(X\cap\partial\Gamma_n)
\end{equation*}
is connected.

Their complements are
\begin{equation*}
(X_l)^c=X^c\cup\Gamma_l^c,
\qquad
(X_N\setminus X_l)^c
=X^c\cup\Gamma_l\cup\Gamma_N^c.
\end{equation*}
These are connected because $X^c$ is connected, unbounded, and intersects $\Gamma_l$.
Since exactly three hexagons meet at each vertex, no vertex pinches occur. Thus both finite regions are disks.

We next prove (2): for all sufficiently large $n$, $L_n^{+j}$ is a disk for every $j\ge0$, and
$(L_n^{+j})^{-1}$ is a disk for every $j\ge1$. Set $n_C=\ell_C(0)$ and fix $n\ge n_C$.  In each case, we prove that both the region and its
complement are connected. Since $L_n$ is connected, so is every thickening $L_n^{+j}$: each of its tiles can be joined to $L_n$ by a shortest path contained in $L_n^{+j}$. Consider parallel straight lines of tiles, with consecutive
tiles sharing an edge. By convexity of $C$ and the shape of $\Gamma_n$, the tiles of $L_n$ on each line form
an interval, possibly empty. Thickening preserves this property: contributions from adjacent lines join the existing interval, while an empty line receives tiles from at most one side, by connectedness. Thus every tile outside $L_n^{+j}$ can escape along its line. Joining these paths outside a sufficiently large graph ball shows that the complement is connected. Hence $L_n^{+j}$ is a disk.

For $j\ge1$, put $E=(L_n^{+j})^{-1}$. Since
\begin{equation*}
L_n^{+(j-1)}\subset E,
\qquad
E^c=\bigl((L_n^{+j})^c\bigr)^{+1},
\end{equation*}
and every tile of $E\setminus L_n^{+(j-1)}$ is adjacent
to $L_n^{+(j-1)}$, both $E$ and $E^c$ are connected.
Thus $E$ is also a disk.

We next prove (3). Condition D follows from the disk property in (1).
To verify Condition M, set
\begin{equation*}
A=X_l,\qquad B=X_N\setminus X_l,\qquad
D=X^c\cap\Gamma_N.
\end{equation*}
The regions $A,B,D$ and $A\cup B\cup D=\Gamma_N$
are nonempty disks. Moreover,
$B\cup D=\Gamma_N\setminus A$ is an annulus because
$A$ lies strictly inside $\Gamma_N$.
The extended Axiom A1 provides
\begin{equation*}
S(A| B)_\omega+S(A| D)_\omega=0.
\end{equation*}
These identities, for both choices of $X$, imply Condition M.

Finally, we prove (4). Since $\Lambda\subset\operatorname{int}C$, we have $\Lambda\subset L$ and $\Lambda\setminus L_n=\Lambda\setminus\Gamma_n$. Let $a_\Lambda$ be the apex of $\Lambda$. Strict angular containment from $\Lambda\ll\operatorname{int}C$ provides a constant $c>0$ such that 
\begin{equation*}
d_{\mathrm E}(x,C^c)\ge c\,|x-a_\Lambda| \qquad (x\in\Lambda).
\end{equation*}
Comparing Euclidean distance with the tile graph distance, and using that $\Gamma_n$ is a graph ball of radius $n$,
we obtain constants $a>0$ and $b\ge0$ such that
\begin{equation*}
d(\Lambda\setminus L_n,L^c)\ge an-b \longrightarrow\infty \qquad (n\to\infty).
\end{equation*}

\end{proofof}

\section{Setting \ref{setting:merging-inputs} and Setting \ref{imo}}\label{taiyou}
Set $\ell_t=\sqrt3\,d_{\mathrm{hex}}/2$.
Define the face-center map
$\bm x:\mathbb Z^2\to\mathbb R^2$ by
\[
 \bm x(i,j)
 ={\ell_t}\bigl(i\bm e_{-\pi/12}+j\bm e_{\pi/4}\bigr),
\]
and let $f_{i,j}\in\mathbb L$ denote the face with center
$\bm x(i,j)$.
We choose constants $R_{\pi/3,\pi/36}$ and
$R_{14\pi/45,\pi/180}$ as in
Lemma~\ref{rem:lattice-cone-approximation}, and set
\[
 R_*:=\max\{R_{\pi/3,\pi/36},R_{14\pi/45,\pi/180}\}.
\]
We fix
\[
 w_t=\left\lceil100(l_0+R)+\frac{R_*}{\ell_t}\right\rceil+25,
 \qquad r_t=10000w_t.
\]
\noindent{(i)}
We take cones
\begin{align*}
 \Lambda_1
 &=\Lambda_{\bm x(-30r_t,-10r_t),\,\pi,\,\pi/18},\\
 \Lambda_2
 &=\Lambda_{\bm x(-30r_t,40r_t),\,\pi/2,\,\pi/18},\\
 \Lambda_3
 &=\Lambda_{\bm x(30r_t,10r_t),\,0,\,\pi/18}.
\end{align*}
We consider lattice cones
\begin{align*}
 \widetilde\Lambda_1
 &=\mathfrak F\!\left(
   \overline{\Lambda_{
     \bm x(-30r_t,-10r_t)+3000w_t\ell_t\bm e_\pi,\,
     \pi,\,\pi/36}}
   \right),\\
 \widetilde\Lambda_2
 &=\mathfrak F\!\left(
   \overline{\Lambda_{
     \bm x(-30r_t,40r_t)+3000w_t\ell_t\bm e_{\pi/2},\,
     \pi/2,\,\pi/36}}
   \right),\\
 \widetilde\Lambda_3
 &=\mathfrak F\!\left(
   \overline{\Lambda_{
     \bm x(30r_t,10r_t)+3000w_t\ell_t\bm e_0,\,
     0,\,\pi/36}}
   \right).
\end{align*}

\noindent{(ii)}
Choose the faces
\[
 f_1=f_{-300r_t,-100r_t},\qquad
 f_2=f_{-300r_t,400r_t},\qquad
 f_3=f_{300r_t,100r_t}.
\]
{For $i=1,2,3$, define the disks $D_i$, their annular collars $B_i$,
by
\[
 D_i=\{f_i\}^{+w_t},\qquad
 B_i=D_i^{+w_t}\setminus D_i.
\]
}
\noindent{(iii)}
For three-hole merging, 
define the following lattice cones  $H_i$, $i=1,2,3$,
by
\begin{align*}
 H_1&=\{f_{i,j}:j\le-r_t-1\ \text{or}\ i+j\le-r_t-1\},\\
 H_2&=\{f_{i,j}:i\le-r_t,\ i+j\ge r_t\},\\
 H_3&=\{f_{i,j}:i\ge r_t,\ j\ge r_t\}.
\end{align*}
Each $H_i$  contains  $\Lambda_i$ with a sufficiently large buffer. Set $C_i=H_i\setminus D_i^{+w_t}$.

The three intervening strips with a width of $2h$ (or $2h-1$) are
\begin{align*}
 {E_1}&=\{f_{i,j}:i\le-r_t,\ -r_t\le i+j\le r_t-1\},\\
 {E_2}&=\{f_{i,j}:-r_t<i<r_t,\ j\ge-r_t,\ i+j\ge-r_t\},\\
 {E_3}&=\{f_{i,j}:i\ge r_t,\ -r_t\le j\le r_t-1\}.
\end{align*}
It holds that ${\mathcal L}=H_1\sqcup H_2\sqcup H_3 \sqcup E_1 \sqcup E_2 \sqcup E_3$.
\noindent{(iv)}
For two-hole merging, define the face-coordinate functions
$u_{12},u_{23}:\mathbb Z^2\to\mathbb Z$ by
\[
 u_{12}(i,j):=i+j,\qquad u_{23}(i,j):=i.
\]
For each ordered pair $(r,s)\in\{(1,2),(2,3)\}$,
using the fixed holes $D_r,D_s\subset\mathbb L$
and collar width $w_t$, define the remaining regions
\begin{align*}
 G_{rs}
 &:=\{f_{i,j}:u_{rs}(i,j)\le-100w_t-1\}
       \setminus D_r^{+w_t},\\
 G'_{rs}
 &:=\{f_{i,j}:u_{rs}(i,j)\ge100w_t+1\}
       \setminus D_s^{+w_t},
\end{align*}
and the intervening strips
\begin{align*}
 I_{rs}
 &:=\{f_{i,j}:-100w_t\le u_{rs}(i,j)\le0\},\\
 I'_{rs}
 &:=\{f_{i,j}:1\le u_{rs}(i,j)\le100w_t\}.
\end{align*}

\noindent{(v)}
For Section~\ref{sec:cone-approximation},
{we also use the open planar cones
\begin{align*}
 \mathcal T_{12}
 &:=\Lambda_{\bm x(100w_t-1,100w_t-1),\,13\pi/12,\,5\pi/6},\\
 \mathcal T_{23}
 &:=\Lambda_{\bm x(0,-100w_t),\,\pi/4,\,\pi/3}.
\end{align*}
Set
\[
T_{rs}:=
 \mathfrak F(\overline{\mathcal T_{rs}})
 \qquad\bigl((r,s)\in\{(1,2),(2,3)\}\bigr).
\]
}

{For $(r,s)\in\{(1,2),(2,3)\}$, and $N$ large enough, put}
\[
 J_{rs}(N):=\lmk (T_{rs})_N\rmk^{-2w_t},\qquad
 C_N^{rs}:=(T_{rs})_N\setminus J_{rs}(N).
\]

\noindent{(vi)}
Section~\ref{sec:cone-approximation} :
For each $(r,s)\in\{(1,2),(2,3)\}$, choose the open planar
cones $\Lambda_{rs},\Lambda\subset\mathbb R^2$ and the closed
convex planar cone $C_{\mathrm{aux}}\subset\mathbb R^2$, and define
the lattice cones $\widetilde\Lambda,\widetilde\Lambda'\subset\mathbb L$ 
using the corresponding column of the table below:
\[
\begin{array}{c|c|c}
 (r,s)&(1,2)&(2,3)\\ \hline
 \Lambda_{rs}
 &\Lambda_{\bm 0,\,3\pi/4,\,\pi/3}
 &\Lambda_{\bm 0,\,\pi/4,\,14\pi/45}\\[3pt]
 \Lambda
 &\Lambda_{\bm x(10w_t,-5w_t),\,3\pi/4,\,3\pi/8}
 &\Lambda_{\bm x(0,-10w_t),\,\pi/4,\,19\pi/60}\\[3pt]
 C_{\mathrm{aux}}
 &\overline{\Lambda_{\bm x(20w_t,-10w_t),\,3\pi/4,\,5\pi/12}}
 &\overline{\Lambda_{\bm x(0,-20w_t),\,\pi/4,\,29\pi/90}}\\[3pt]
 \widetilde\Lambda
 &\mathfrak F(\overline{\Lambda_{12}})
 &\mathfrak F(\overline{\Lambda_{23}})\\[3pt]
 \widetilde\Lambda'
 &\mathfrak F(C_{\mathrm{aux}})
 &\mathfrak F(C_{\mathrm{aux}})
\end{array}
\]

\noindent{(vii)} For Section \ref{sec:cyclicity} (\ref{somen}), set
\begin{align*}
 \widetilde\Lambda_1'
 &:=
 \mathfrak F\!\left(
 \overline{\Lambda_{
   \bm x(-30r_t,-10r_t)-50w_t\ell_t\bm e_\pi,\,
   \pi,\,7\pi/120}}
 \right),\\
 \widetilde\Lambda_2'
 &:=
 \mathfrak F\!\left(
 \overline{\Lambda_{
   \bm x(-30r_t,40r_t)-50w_t\ell_t\bm e_{\pi/2},\,
   \pi/2,\,7\pi/120}}
 \right).
\end{align*}
\begin{align*}
 \widehat\Lambda_1
 &:=
 \Lambda_{
   \bm x(-30r_t,-10r_t)-100w_t\ell_t\bm e_\pi,\,
   \pi,\,11\pi/180},\\
 \widehat\Lambda_2
 &:=
 \Lambda_{
   \bm x(-30r_t,40r_t)-100w_t\ell_t\bm e_{\pi/2},\,
   \pi/2,\,11\pi/180}.
\end{align*}
\begin{align*}
 C_1'
 &:=
 \overline{\Lambda_{
   \bm x(-30r_t,-10r_t)-200w_t\ell_t\bm e_\pi,\,
   \pi,\,\pi/15}},\\
 C_2'
 &:=
 \overline{\Lambda_{
   \bm x(-30r_t,40r_t)-200w_t\ell_t\bm e_{\pi/2},\,
   \pi/2,\,\pi/15}}.
\end{align*}
\[
 \widetilde\Lambda_i''
 :=\mathfrak F(C_i')\subset\mathbb L,
 \qquad i=1,2.
\]

\section{Basic Lemma}

\begin{lemma}\label{lem:affine-density-matrices}
Let $V$ and $W$ be nonzero finite-dimensional complex Hilbert spaces.
Every affine bijection $\Phi:D(V)\to D(W)$ extends uniquely to a
complex-linear bijection $\widehat{\Phi}:B(V)\to B(W)$.
Moreover, there exists a unitary or antiunitary operator $U:V\to W$
such that
\[
\widehat{\Phi}(x)=
UxU^*,  
\qquad x\in D(V).
\]
\end{lemma}
\begin{proof}
This is a finite-dimensional consequence of Kadison's theorem
\cite{Kadison1951}, by duality.
\end{proof}


\begin{thebibliography}{9}
\bibitem{BR1}
O.~Bratteli and D.~W.~Robinson,
\emph{Operator Algebras and Quantum Statistical Mechanics~1},
second ed., Springer-Verlag, Berlin, 1987.

\bibitem{BV}
A.~Bols and S.~Vadnerkar,
\emph{Classification of the Anyon Sectors of Kitaev's Quantum Double Model},
Commun. Math. Phys. \textbf{406} (2025), Article~188.
\href{https://doi.org/10.1007/s00220-025-05363-w}{doi:10.1007/s00220-025-05363-w}.

\bibitem{BK1}
A.~Bols and B.~Kj{\ae}r,
\emph{Sector Theory of Levin--Wen Models I: Classification of Anyon Sectors},
\href{https://arxiv.org/abs/2511.21521}{arXiv:2511.21521} (2025).

\bibitem{BK2}
A.~Bols and B.~Kj{\ae}r,
\emph{Sector Theory of Levin--Wen Models II: Fusion and Braiding},
\href{https://arxiv.org/abs/2603.01936}{arXiv:2603.01936} (2026).
\bibitem{SKK}
B.~Shi, K.~Kato, and I.~H.~Kim,
\emph{Fusion rules from entanglement},
Ann. Physics \textbf{418} (2020), 168164,
\href{https://arxiv.org/abs/1906.09376}{arXiv:1906.09376}.
\bibitem{BGNW}
S.~Bachmann, A.~Getz, P.~Naaijkens, and N.~Wray,
\emph{Stacking and the triviality of invertible phases},
preprint (2025),
\href{https://arxiv.org/abs/2511.08382}{arXiv:2511.08382}.
\bibitem{KLRS}
I.~H.~Kim, T.-C.~Lin, D.~Ranard, and B.~Shi,
\emph{Strict area law implies commuting parent Hamiltonian},
preprint (2024),
\href{https://arxiv.org/abs/2404.05867}{arXiv:2404.05867}.

\bibitem{Ogata}
Y.~Ogata,
\emph{A derivation of braided $C^*$-tensor categories from gapped
 ground states satisfying the approximate Haag duality}, J. Math. Phys. \textbf{63}, 011902 (2022),  
\href{https://arxiv.org/abs/2106.15741}{arXiv:2106.15741}.

\bibitem{OPR}
Y.~Ogata, D.~P\'erez-Garc\'ia, and A.~Ruiz-de-Alarc\'on,
\emph{Haag Duality for 2D Quantum Spin Systems},
\href{https://arxiv.org/abs/2509.23734}
{arXiv:2509.23734} (2025).





\bibitem{CarlenLieb}
E. A. Carlen and E. H. Lieb,
\emph{Bounds for Entanglement via an Extension of Strong Subadditivity of Entropy},
Lett. Math. Phys. \textbf{101}, 1–11 (2012).


\bibitem{HJPW}
P.~Hayden, R.~Jozsa, D.~Petz, and A.~Winter,
\emph{Structure of States Which Satisfy Strong Subadditivity of Quantum
Entropy with Equality},
Commun. Math. Phys. \textbf{246} (2004), 359--374,
\href{https://arxiv.org/abs/quant-ph/0304007}{arXiv:quant-ph/0304007}.

\bibitem{DPModular}
J.~Derezi\'nski and C.-A.~Pillet,
\emph{Tomita--Takesaki theory}, Section~4, ``The finite dimensional case,''
\href{https://pillet.univ-tln.fr/data/pdf/Tomita-Takesaki_theory.pdf}
{author-posted version}.

\bibitem{Kitaev2006}
A.~Kitaev,
\emph{Anyons in an exactly solved model and beyond},
Ann. Physics \textbf{321} (2006), 2--111,
\href{https://arxiv.org/abs/cond-mat/0506438}
{arXiv:cond-mat/0506438}.

\bibitem{Kitaev2003}
A.~Yu.~Kitaev,
\emph{Fault-tolerant quantum computation by anyons},
Ann. Physics \textbf{303} (2003), 2--30,
\href{https://arxiv.org/abs/quant-ph/9707021}
{arXiv:quant-ph/9707021}.

\bibitem{LevinWen2005}
M.~A.~Levin and X.-G.~Wen,
\emph{String-net condensation:
A physical mechanism for topological phases},
Phys. Rev. B \textbf{71} (2005), 045110,
\href{https://arxiv.org/abs/cond-mat/0404617}
{arXiv:cond-mat/0404617}.

\bibitem{MolnarEtAl2026}
A.~Moln\'ar, A.~Ruiz de Alarc\'on, J.~Garre-Rubio,
N.~Schuch, J.~I.~Cirac, and D.~P\'erez-Garc\'ia,
\emph{Matrix product operator algebras I:
representations of weak Hopf algebras and projected
entangled pair states},
Commun. Math. Phys. \textbf{407} (2026), Article~203,
\href{https://arxiv.org/abs/2204.05940}
{arXiv:2204.05940}.

\bibitem{ShiLu2019}
B.~Shi and Y.-M.~Lu,
\emph{Characterizing topological order by the information convex},
Phys. Rev. B \textbf{99} (2019), 035112,
\href{https://arxiv.org/abs/1801.01519}
{arXiv:1801.01519}.

\bibitem{Shi2019}
B.~Shi,
\emph{Seeing topological entanglement through the information convex},
Phys. Rev. Research \textbf{1} (2019), 033048,
\href{https://arxiv.org/abs/1810.01986}
{arXiv:1810.01986}.

\bibitem{Haah2016}
J.~Haah,
\emph{An invariant of topologically ordered states
under local unitary transformations},
Commun. Math. Phys. \textbf{342} (2016), 771--801,
\href{https://arxiv.org/abs/1407.2926}
{arXiv:1407.2926}.

\bibitem{DHR1971}
S.~Doplicher, R.~Haag, and J.~E.~Roberts,
\emph{Local observables and particle statistics I},
Commun. Math. Phys. \textbf{23} (1971), 199--230,
\href{https://doi.org/10.1007/BF01877742}
{doi:10.1007/BF01877742}.

\bibitem{BF1982}
D.~Buchholz and K.~Fredenhagen,
\emph{Locality and the structure of particle states},
Commun. Math. Phys. \textbf{84} (1982), 1--54,
\href{https://doi.org/10.1007/BF01208370}
{doi:10.1007/BF01208370}.

\bibitem{Naaijkens2011}
P.~Naaijkens,
\emph{Localized endomorphisms in Kitaev's toric code on the plane},
Rev. Math. Phys. \textbf{23} (2011), 347--373,
\href{https://arxiv.org/abs/1012.3857}
{arXiv:1012.3857}.

\bibitem{KimRanard2024}
I.~H.~Kim and D.~Ranard,
\emph{Classifying 2D topological phases:
mapping ground states to string-nets},
preprint (2024),
\href{https://arxiv.org/abs/2405.17379}
{arXiv:2405.17379}.

\bibitem{Petz2008}
D.~Petz,
\textit{Quantum Information Theory and Quantum Statistics},
Theoretical and Mathematical Physics,
Springer, Berlin Heidelberg, 2008.

\bibitem{Kato2016PRA}
K.~Kato, F.~Furrer, and M.~Murao,
Information-theoretical analysis of topological entanglement entropy and multipartite correlations,
Phys.\ Rev.\ A \textbf{93}, 022317 (2016),
doi:10.1103/PhysRevA.93.022317.

\bibitem{LiuVanLuijk2026}
R.~Liu and L.~van Luijk,
\emph{An entropic characterization of Haag duality},
preprint (2026),
\href{https://arxiv.org/abs/2609.13550}
{arXiv:2609.13550}.
\bibitem{NPW}
{{
P.~Naaijkens, D.~Penneys, and D.~Wallick,
\emph{Local topological order, Haag duality, and
reflection positivity},
preprint (2026),
\href{https://arxiv.org/abs/2605.10693v1}
{arXiv:2605.10693v1}.
}}
\bibitem{NSYBulkGap}
{{
B.~Nachtergaele, R.~Sims, and A.~Young,
\emph{Stability of the bulk gap for frustration-free
topologically ordered quantum lattice systems},
Lett. Math. Phys. \textbf{114} (2024), Article~24,
\href{https://doi.org/10.1007/s11005-023-01767-8}
{doi:10.1007/s11005-023-01767-8}.
}}
\bibitem{NT}
{S.~Neshveyev and L.~Tuset,
\emph{Compact Quantum Groups and Their Representation Categories},
Cours Sp\'ecialis\'es, vol.~20,
Soci\'et\'e Math\'ematique de France, 2013.
}
\bibitem{LongoRehren1995}
R.~Longo and K.-H.~Rehren,
\emph{Nets of subfactors},
Rev. Math. Phys. \textbf{7} (1995), 567--597,
\href{https://arxiv.org/abs/hep-th/9411077}{arXiv:hep-th/9411077}.

\bibitem{BKprep}
A.~Bols and B.~Kj{\ae}r,
In progress.
\bibitem{Kadison1951}
R.~V.~Kadison,
\emph{Isometries of operator algebras},
Ann. of Math. (2) \textbf{54} (1951), no.~2, 325--338.

\end{thebibliography}
\end{document}